\documentclass[letterpaper, 12pt]{article}
\usepackage{subcaption}
\usepackage[margin=1in]{geometry}
\usepackage{float}
\usepackage{graphicx}%
\usepackage{multirow}%
\usepackage{amsmath,amssymb,amsfonts}%
\usepackage{xcolor}%
\usepackage{authblk}
\usepackage{changes}
\usepackage{array}
\usepackage[style=chem-acs, articletitle=true]{biblatex} 

\title{Learning noisy phase transition dynamics from stochastic partial differential equations}

\usepackage{hyperref}
\hypersetup{
  colorlinks=true,
  linkcolor=blue,
  filecolor=magenta,
  urlcolor=red,
  pdfpagemode=FullScreen,
}

\newcommand{\mc}[1]{\mathcal{#1}} %mathcal
\newcommand{\LRp}[1]{\left( #1 \right)} % adaptive left and right parentheses
\newcommand{\eval}[2][\right]{\relax \ifx#1\right\relax \left.\fi#2#1\rvert}
\renewcommand{\epsilon}{\varepsilon}

\newcommand{\F}{\mc{F}} %mathcal
\newcommand{\cozero}{$c_0 = \mc{U} \LRp{-0.01,0.01}$}
\newcommand{\cofour}{$c_0 = 0.4 + \mc{U} \LRp{-0.01,0.01}$}

\newcommand{\nflux}{\texttt{non-flux}}
\newcommand{\femv}{\texttt{FE-MV}}
\newcommand{\nfemv}{\texttt{nonFE-MV}}
\newcommand{\fem}{\texttt{FE-M}}
\newcommand{\nfem}{\texttt{nonFE-M}} 

\author[1]{Luning Sun\textsuperscript{$\dagger$}}
\author[2]{Van Hai Nguyen\textsuperscript{$\dagger$}}
\author[1]{Shusen Liu}
\author[1]{John Klepeis}
\author[1]{Fei Zhou\thanks{zhou6@llnl.gov}}
\affil[1]{
  Lawrence Livermore National Laboratory, Livermore, CA, 94550,USA}

\affil[2]{
  The University of Texas at Austin, Austin, TX, 78712, USA}

\date{}

\begin{document}

\maketitle
\begingroup
\renewcommand\thefootnote{} % suppress numbering
\footnotetext{\textsuperscript{$\dagger$}Equal contribution.}
\endgroup
% \begin{abstract}
% The non-equilibrium dynamics of mesoscale phase transitions is closely and often decisively affected by thermal fluctuations and molecular-scale perturbations, or ``noise''.
% A significant challenge in modeling these systems lies in effectively handling noise.
% We study the stochastic Cahn-Hilliard Equation for spinodal decomposition and nucleation
% %partial differential equations (SPDEs) 
% as an exemplar noisy dynamical system, and develop both deterministic and stochastic (generative) surrogate machine-learning models. %to bridge the gap in computational modeling of noisy dynamical systems. Our results 
% Our results demonstrate excellent agreement with the training data
% %. SPDE for both deterministic and stochastic dynamics. Remarkably, the surrogate models 
% and exhibit strong generalization capabilities, successfully scaling to large systems despite being trained exclusively on much smaller sizes. Lessons learned on learning from stochastic dynamics will be discussed.
% %Additionally, the surrogate models can achieve speedup over SDE solver.
% \end{abstract}

\begin{abstract}
% The non-equilibrium dynamics of mesoscale phase transitions are decisively influenced by thermal fluctuations and molecular-scale perturbations. In systems such as spinodal decomposition and nucleation, stochasticity is not merely a modeling artifact but a fundamental driver of pattern formation and kinetic pathways. Accurately learning such noisy dynamics remains challenging, particularly for machine-learning surrogate models that must reproduce both ensemble variability and underlying conservation laws.

% In this work, we study the stochastic Cahn–Hilliard equation as a prototypical example of noise-driven phase separation and develop physics-aware surrogate models, including deterministic and stochastic variants. Unlike existing data-driven or generative approaches that introduce randomness at the state level, we model stochasticity explicitly through inter-cell fluxes, enabling exact mass conservation during rollout by construction. The proposed framework combines learnable free energy representations with neural flux models that parameterize both mobility and stochastic flux amplitudes.

% Numerical experiments in one- and three-dimensional systems demonstrate that the proposed surrogates accurately reproduce ensemble statistics, capture noise-accelerated coarsening behavior, and generalize to larger spatial domains despite being trained on significantly smaller systems. These results highlight the importance of structure-preserving stochastic modeling for learning physically consistent surrogate dynamics of noisy phase transitions.
The non-equilibrium dynamics of mesoscale phase transitions are fundamentally shaped by thermal fluctuations, which not only seed instabilities but actively control kinetic pathways, including rare, barrier-crossing events such as nucleation that are entirely inaccessible to deterministic models. Machine-learning surrogates for such systems must therefore represent stochasticity explicitly, enforce conservation laws by construction, and expose physically interpretable structure. We develop physics-aware surrogate models for the stochastic Cahn–Hilliard equation in 3D that satisfy all three requirements simultaneously. The key innovation is to parameterize the surrogate at the level of inter-cell fluxes, decomposing each flux into a deterministic mobility-weighted chemical-potential gradient and a learnable noise amplitude. This design guarantees exact mass conservation at every  step and adds physical fluctuations to inter-cell mass transport. A learnable free energy functional provides thermodynamic interpretability, validated by independent recovery of the bulk double-well landscape, interfacial excess energy, and curvature-independent interfacial tension. Tests demonstrate accurate reproduction of ensemble statistics and noise-accelerated coarsening, with generalization to 64 larger spatial domains in volume and 160x longer temporal domain than those seen during training. Critically, the stochastic surrogate captures thermally activated nucleation in the metastable regime, a qualitative capability that no deterministic surrogate can provide regardless of training, thus establishing flux-level stochasticity as an architectural necessity rather than an optional enhancement.
\end{abstract}

\section{Introduction}
\label{sec:intro}

Non-equilibrium phase transitions are central to materials science and chemistry,
governing the formation, transformation, and stability of matter across a wide range
of conditions and compositions.
A decisive, and often underappreciated, factor in these processes is stochasticity:
thermal fluctuations and atomistic perturbations that are not mere background noise
but are fundamental drivers of kinetic pathways and emergent phenomena.
In spinodal decomposition, an initially uniform, thermodynamically unstable mixture
spontaneously separates into coexisting phases through continuous amplification of
composition fluctuations at all wavelengths \cite{cahn1958free}; stochastic noise not only seeds
this instability but actively accelerates coarsening and sets the characteristic
length scale of the resulting microstructure.
In nucleation and precipitation from a metastable state, a thermodynamic barrier must
be surmounted by a rare thermal fluctuation---a process entirely absent in a
deterministic description---and the crossing probability is exponentially sensitive
to the noise amplitude, as captured by classical nucleation theory \cite{Becker1935AP-CNT}.
Understanding and accurately reproducing such noise-driven dynamics is therefore not
optional but required for quantitative mesoscale modeling, with implications ranging
from the design of advanced alloys and polymer blends to self-organization in soft
matter and biological assemblies.

Computational methods for mesoscale phase transitions can be classified by how they
represent stochasticity.
Particle-based methods, such as molecular dynamics (MD)~\cite{binder1985monte} and
kinetic Monte Carlo (KMC)~\cite{voter2007introduction}, resolve thermal fluctuations
explicitly through individual particle trajectories.
They are physically complete but prohibitively expensive at the spatiotemporal scales
relevant to microstructural evolution.
Deterministic phase-field models~\cite{cahn1958free,chen2002phase} overcome the
mesoscale barrier through a continuum order-parameter description, but entirely
suppress stochasticity and consequently miss fluctuation-driven phenomena such as
noise-accelerated coarsening, the stochastic spread of nucleation times, and the
sensitivity of pattern selection to early-time fluctuations.
Stochastic partial differential equations (SPDEs), of which the stochastic
Cahn--Hilliard equation (CHE)~\cite{cook1970brownian,Bronchart2008PRL-Finel} is the
canonical mesoscale example, offer a more faithful description by coupling a
deterministic driving force with a noise field whose statistics are grounded in the
fluctuation-dissipation theorem.
However, SPDEs remain computationally expensive at scales relevant to microstructural
evolution and require \textit{a priori} specification of noise statistics that may not
be readily available outside of well-characterized model systems.

The rapid development of machine-learning (ML) surrogate models for physical
simulations offers an attractive path toward cost-effective mesoscale dynamics, but
existing approaches are largely restricted to the deterministic
regime~\cite{Yang2021P, MontesdeOcaZapiain2021nCM, Fan2024MLST, Bertin2024nCM-Learning, Ji2025-Scalable, Tian2025-Scaling}.
Physics-informed neural networks~\cite{raissi2019physics,sun2020surrogate,wang2021understanding}
and residual-based methods can incorporate stochastic forcing terms in principle,
but the optimization process effectively averages over noise realizations, yielding
a smooth mean-field solution rather than a genuine stochastic ensemble.
Operator-learning architectures~\cite{lu2019deeponet,li2020fourier} and sequence
models~\cite{vaswani2017attention,han2022predicting,sun2023unifying} trained on
stochastic trajectories learn state-to-state transition maps that approximate the
conditional mean of the next state; they implicitly absorb stochasticity as part
of the prediction target rather than modeling it explicitly, and provide no mechanism
for enforcing conservation laws during roll-out~\cite{sun2023unifying}.
Generative approaches---diffusion models~\cite{gao2024bayesian,du2024conditional} and
normalizing flows~\cite{sun2023unifying}---introduce randomness directly at the level
of the predicted state field, enabling diverse trajectory sampling, but they decouple
noise from its physical origin in inter-cell transport.
As a consequence, even when trained on strictly conserved data, state-level generative
models do not guarantee that individual samples conserve mass, because the injected
noise is not constrained to arise from a divergence-free flux.

Related works in chemistry and biophysics addresses the inverse problem of learning effective
stochastic dynamics, including potential of mean force (PMF) and the diffusion coefficient,
%---the two kinetic coefficients of the overdamped Langevin equation---
from single-molecule
or particle-tracking trajectories \cite{Haas2013JPCB,Frishman2020PRX-SFI}.
% Frishman and Ronceray
% showed that there is a fundamental information-theoretic bound on how rapidly these coefficients can be extracted from Brownian trajectories, and 
% proposed the Stochastic Force Inference (SFI) framework to saturate that bound via quasi-maximum-likelihood projection~\cite{Frishman2020PRX-SFI}.
The present work deals with an analogous identification problem of learning free
energy, mobility, and noise amplitude from stochastic phase-field trajectories,
though in a qualitatively different context.
The concentration field $c(\mathbf{x},t)$ plays the role of the collective
variable, yet it is either a continuous field (function) or discretized at a very high-dimensional mesh rather
than atomic coordinates.
Consequently, the learned objects are field-level functionals (free energy
$A[c]$, mobility field $M_{ij}$, noise amplitude $B_{ij}$) rather than functions of coordinates.
%Moreover, our goal is a surrogate that generates faithful trajectories at coarsened time steps, rather than post-hoc parameter inference from already-collected data.
% Bridging this gap---structured physical parameterization at the level of
% individual inter-cell fluxes, combined with neural functional approximation
% of the free energy and kinetic coefficients---is the central architectural
% contribution of this work.

The above literature survey points to three requirements for a physically faithful surrogate for
stochastic conserved microstructure dynamics.
First, \textit{stochasticity} must be represented explicitly and at the correct scale:
randomness should be introduced at the level of inter-cell fluxes, where it has a
well-defined physical meaning tied to microscopic fluctuations of current density,
rather than at the level of the predicted concentration field.
Second, \textit{mass conservation} must be guaranteed by construction during every
step of an autoregressive roll-out, not merely encouraged through a soft loss term.
Third, the surrogate should encode \textit{physically interpretable structure}: the
free energy landscape, mobility, and noise amplitude that govern the dynamics should
be learnable as distinct, explainable modules rather than absorbed into an opaque
black-box map.
No existing ML surrogate framework fully satisfies all three requirements simultaneously.

In this work we address this gap by developing a conservative and stochastic surrogate
for the stochastic Cahn--Hilliard equation, chosen as a prototypical and
well-characterized model of noise-driven conserved dynamics.
The key architectural innovation is to parameterize the surrogate at the level of
inter-cell fluxes rather than states.
At each autoregressive step, the predicted concentration update is computed as the
divergence of a learned flux field comprising a deterministic mobility-weighted
chemical-potential gradient and a stochastic noise term, directly mirroring the
Bronchart--Finel discretization of the CHE~\cite{Bronchart2008PRL-Finel}.
Mass conservation is guaranteed by the antisymmetry of the learned flux,
$F_{ij} = -F_{ji}$, enforced by construction rather than by a penalty term.
The chemical potential is obtained by automatic differentiation of a learnable
scalar free energy functional, encoding the correct thermodynamic structure while
learning the free energy landscape entirely from trajectory data.
Mobility and noise amplitude are parameterized as symmetric functions of
learned latent cell representations, sharing the chemical-potential backbone.
We evaluate the framework across model variants obtained by ablating the
free energy constraint and the stochastic flux component, enabling a controlled
analysis of each physical ingredient.

The scope of the present work is deliberately a study of model design and
architectural feasibility, or the \textit{inner loop} of a broader surrogate modeling program.
Our primary training and test regime is spinodal decomposition, in which the
mean concentration lies within the thermodynamically unstable region
($|\langle c \rangle| < 0.5$ spinodal; $|\langle c \rangle| < 0.85$ solubility limit).
This choice is intentional: in the spinodal regime, the SPDE generates abundant,
information-rich training trajectories, and the physical effects of stochasticity are immediate and direct: 
noise-accelerated coarsening dynamics
and inter-trajectory divergence.
Spinodal decomposition therefore provides stringent tests of the architecture.
Nucleation and precipitation from metastable initial conditions are qualitatively
harder: the rare-event nature of nucleation means that training data near the
barrier is intrinsically sparse, and reproducing the full distribution of nucleation
waiting times requires sufficient exposure to barrier-crossing configurations.
We include a preliminary extrapolation test near the metastable boundary
($\langle c \rangle = -0.51$, just outside the spinodal), which demonstrates that the
learned free-energy based surrogate can capture mean nucleation behavior in
near-boundary cases, while also revealing that, as a direct consequence of
training data concentrated in the spinodal region, the predicted distribution of
nucleation times is narrower than the ground truth .
Extending surrogate performance into the deeply metastable precipitation regime
requires a qualitatively different \textit{outer loop}: systematic data curation
and active learning strategies that adaptively sample rare-event trajectories and
expand the training distribution toward and beyond the solubility limit.
Such a data-curation strategy, while critical for a comprehensive and quantitative study across the whole concentration spectrum, is beyond the architectural questions addressed here and is deferred to future work.

Numerical experiments in one and three spatial dimensions demonstrate that the
stochastic, free energy constrained variant (\femv) accurately reproduces ensemble
statistics, noise-accelerated coarsening dynamics, and spatial microstructure on
the training grid, and generalizes to spatial domains 4x larger in linear dimension (64x larger volume) than those
seen during training.
The learned free energy function correctly recovers the double-well bulk energy
landscape, interfacial excess energy, and the curvature-independence of the
ground-truth interfacial tension---providing stringent, interpretable validation
of the thermodynamic consistency of the surrogate.
The remainder of this paper is organized as follows.
Section~\ref{sec:method} introduces the stochastic CHE, the flux-preserving
surrogate architecture, and the training procedure.
Section~\ref{sec:results_sPDE} analyzes the noise-driven dynamics of the reference
SPDE system in one and three dimensions.
Section~\ref{sec:results_ml} evaluates surrogate model performance, covering
in-distribution inference, spatial extrapolation, free energy verification, and
concentration-space extrapolation.
Section~\ref{sec:conclusion} summarizes findings and outlines directions for future work.

\begin{figure}[htp]
%Plot[ c^4/4 -c^2/2 + 0.2(((1+c)/2)*log((1+c)/2) + ((1-c)/2)*log((1-c)/2)), {c, -1,1}]
    \centering
    \includegraphics[width=0.7\textwidth]{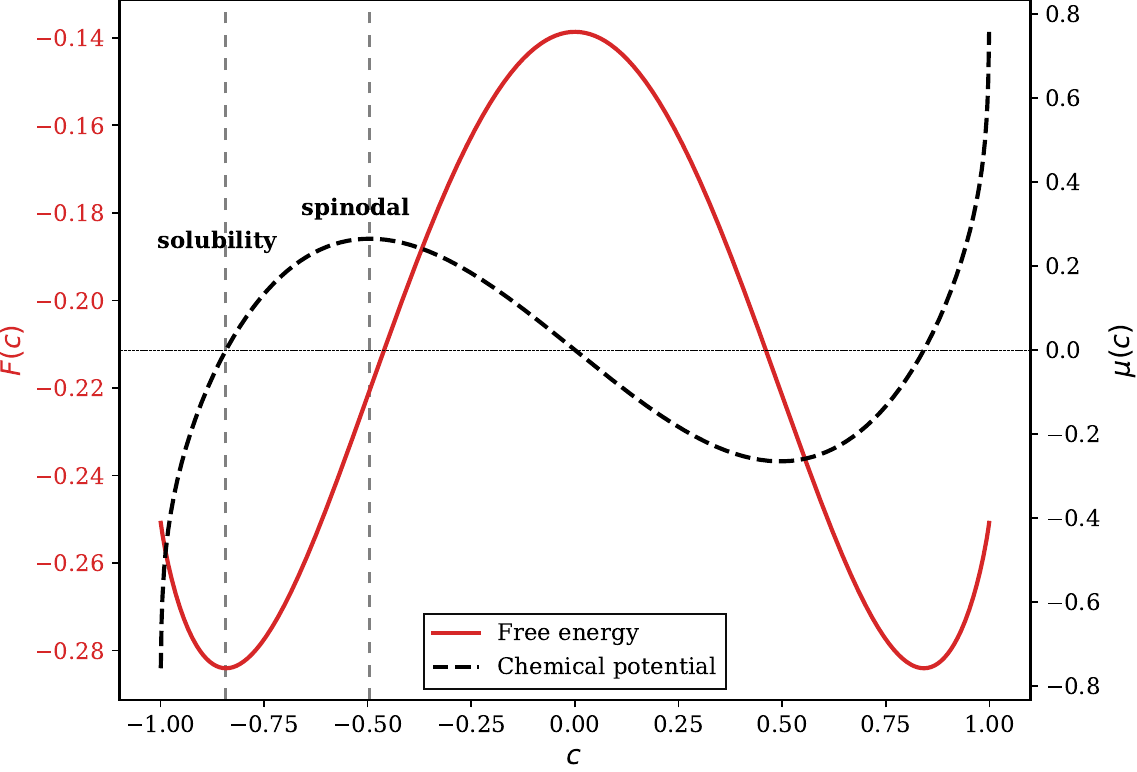}
% %%%%%%%%%%%%%%%%%%%%%%%%%%% move below to SI
%     % \includegraphics[width=0.7\textwidth]{fig/toy2d-free-energy.png}
%     \begin{tabular}{c| c c c c c}
%         \multicolumn{3}{c}{Free energy curves at different temperature T} & 
%         \multicolumn{3}{c}{Free energy landscape and solubility limit}
%         \\
%         \multicolumn{3}{c}{\includegraphics[width=.47\textwidth]{fig/Final/free_energy_versus_temperature.png}} 
%         & 
%         \multicolumn{3}{c}{\includegraphics[width=.47\textwidth]{fig/Final/energy_contour_with_solubility.png}} 
%     \end{tabular}
% %%%%%%%%%%%%%%%%%%%%%%%%%%% END
    \caption{Bulk free energy and chemical potential at $T=0.2$. The concentrations of solubility ($\pm 0.85$) and spinodal ($\pm 0.5$) are highlighted.
 % \added{[Suggestion: move original figures (bottom rows) to appendix ]}
    %, and replace with: subfig (a), FE curve with T=0.2 only (these whole T range is not really considered and only distraction.) (b) a schematic highlighting that we are using GT of SPDE, and surrogate are spatiotemporally downsampled generative NN predicting mean and variance.]}
    }
    \label{fig:bulk_free_energy}
\end{figure}

\begin{figure}[htp!]
\centering
\includegraphics[width=0.8\textwidth]{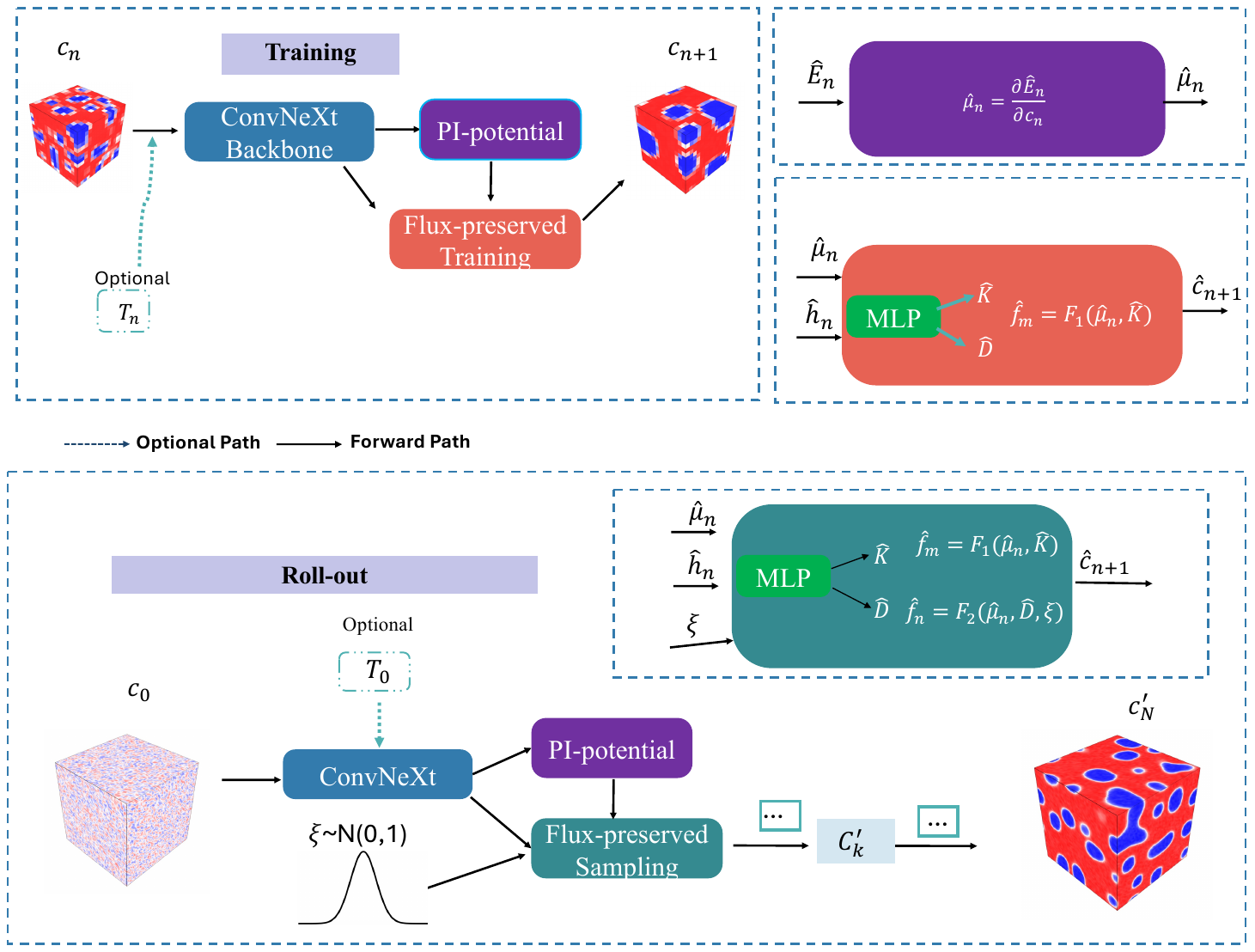}
\caption{Flux-preserved training and sampling framework. }%\LS{don't use the red blocks, seperate the constraints from the MLP block}}
\label{fig:scheme}
\end{figure}

\section{Methods}
\label{sec:method}

\subsection{Stochastic Cahn--Hilliard equation}
\label{sec:method-SPDE}

The stochastic CHE for the concentration field $c(\mathbf{x},t)$ is
\begin{equation}
  \frac{\partial c}{\partial t}
  = -\nabla \cdot \mathbf{f}
  = -\nabla \cdot \!\left( M\nabla\mu + \sigma\,\frac{d\mathbf{W}}{dt} \right),
  \label{eq:SPDE}
\end{equation}
where $\mathbf{f}$ is the concentration flux, $M$ is the mobility,
$\sigma$ is the noise amplitude, and $d\mathbf{W}/dt$ is a vector-valued
white noise field acting on the fluxes.
The chemical potential $\mu$ is the functional derivative of the total free energy $A[c]$:
\begin{align}
  \mu &= \frac{\delta A[c]}{\delta c},
  \label{eq:mu} \\
  A   &= A_\text{b}[c] + A_\text{i}[c]
       = \int \!\left(a_\text{b} + a_\text{i}\right) dV,
  \label{eq:def_energy}
\end{align}
with bulk and interfacial free energy densities
\begin{align}
  a_\text{b} &= \frac{c^4}{4} - \frac{c^2}{2}
               + T\!\left(\frac{1+c}{2}\log\frac{1+c}{2}
                          + \frac{1-c}{2}\log\frac{1-c}{2}\right),
  \label{eq:ab} \\
  a_\text{i} &= \frac{\kappa}{2}|\nabla c|^2,
  \label{eq:ai}
\end{align}
parameterized by temperature $T$ and gradient-penalty stiffness $\kappa$.
% Fig.~\ref{fig:bulk_free_energy_app} shows the free energy curves at different temperature $T$ and free energy landscape with the corresponding solubility limit. It can be seen that as temperature increases, the solubility limit decreases. We chose $T=0.2$ for futher analysis. 
As shown in Fig.~\ref{fig:bulk_free_energy}, the double-well structure of $a_\text{b}(c)$ at $T < 1$ drives phase separation, with spinodal boundaries at $|c| = 0.5$ and equilibrium solubility limits at $|c| \approx 0.85$ for $T = 0.2$.

We solve Eq.~(\ref{eq:SPDE}) numerically following the coarse-grained lattice
discretization of Bronchart et al.~\cite{Bronchart2008PRL-Finel}, using
Euler--Maruyama time integration on a regular mesh.
The flux between neighboring cells $i$ and $j$ is
\begin{equation}
  f_{ij} = \frac{D_{ij}}{T}\left(\mu_j - \mu_i\right)\delta t
           + \left(2D_{ij}\right)^{1/2}\!\epsilon\,\sqrt{\delta t}\;\xi,
  \label{eq:flux_spde}
\end{equation}
where
$D_{ij} = D_{ji}
  = \left(\frac{1}{2}\left|1 - c_i^2\right|\left|1 - c_j^2\right| + 0.5\right)
    \left(1 + 0.1T\right)$
is a symmetric, concentration-dependent diffusivity derived from the
fluctuation-dissipation theorem~\cite{Bronchart2008PRL-Finel},
$\xi \sim \mathcal{N}(0,1)$ is drawn independently for each ordered pair $(i,j)$,
and $\epsilon$ is an adjustable noise amplitude hyperparameter.
The factor $\sqrt{2D_{ij}}$ in the stochastic term ensures that the
variance of the discrete flux is proportional to the local diffusivity,
consistent with the fluctuation-dissipation relation at the lattice level.
The interfacial contribution to the chemical potential is discretized as
\begin{equation}
  \mu_i^\text{i} = -S \sum_{\Delta\mathbf{r}}
    \left(c_{i+\Delta\mathbf{r}} + c_{i-\Delta\mathbf{r}} - 2c_i\right),
  \label{eq:mu_interface}
\end{equation}
where the sum runs over the neighbor lattice directions and the constant
$S = 0.002$ is fixed throughout.
Mass conservation follows from the antisymmetry constraint $f_{ji} = -f_{ij}$:
each undirected bond is visited exactly once, so $\sum_i c_i$ is
conserved at every step.
The concentration update is
\begin{equation}
  c_i(t + \delta t) = c_i(t) - \sum_{j \in \mathcal{N}(i)} f_{ij},
  \label{eq:update_spde}
\end{equation}
where $\mathcal{N}(i)$ denotes all neighbors $j$ for which $f_{ij}$ is computed
with $i$ as the source cell.

\subsection{Flux-preserving surrogate model}
\label{sec:surrogate_model}

\paragraph{Overall structure.}
We develop autoregressive neural network surrogates that predict the concentration
field at a coarsened time step $\Delta t \gg \delta t$, skipping
$N = \Delta t / \delta t$ solver steps per surrogate evaluation and thereby
achieving computational speedup proportional to $N$.
Mirroring the structure of Eq.~(\ref{eq:update_spde}), the concentration is updated as a flux divergence:
\begin{equation}
  \hat{c}_i(t + \Delta t) = c_i(t) - \sum_{j \in \mathcal{N}(i)} F_{\theta,ij},
  \label{eq:surrogate_update}
\end{equation}
where $F_{\theta,ij} = -F_{\theta,ji}$ is the learned inter-cell flux.
This antisymmetric constraint is enforced by construction (see below),
guaranteeing exact mass conservation during roll-out.

\paragraph{Flux decomposition.}
The learned flux mirrors the structure of the SPDE flux in Eq.~(\ref{eq:flux_spde}),
decomposing into a deterministic mean and a stochastic noise component:
\begin{equation}
  F_{\theta,ij}
  = {\hat{M}_{\theta_1,ij}
               \left(\hat{\mu}_{\theta_2,i} - \hat{\mu}_{\theta_2,j}\right)\Delta t}
              % _{\text{flux mean}}
  + {\hat{B}_{\theta_3,ij}\sqrt{2\Delta t} \xi}
              % _{\text{flux noise}},
  \label{eq:flux_surrogate}
\end{equation}
where $\hat{M}_{\theta_1,ij}$ is the learned mobility,
$\hat{\mu}_{\theta_2}$ is the learned chemical potential,
$\hat{B}_{\theta_3,ij}$ is the learned noise amplitude,
and $\xi \sim \mathcal{N}(0,1)$.
The $\sqrt{2\Delta t}$ scaling of the stochastic term reflects the
incoherent accumulation of $N$ independent solver-step noise contributions:
the standard deviation of the summed noise scales as
$\sqrt{N} \cdot \sqrt{2 D_{ij}}\,\epsilon\,\sqrt{\delta t}
 = \sqrt{2 D_{ij}}\,\epsilon\,\sqrt{\Delta t}$,
so $\hat{B}_{\theta_3,ij}$ absorbs the mobility-dependent prefactor and is
learned from data without being pre-specified.
The network parameters $\theta = \{\theta_1, \theta_2, \theta_3\}$ are
trained end-to-end.

\paragraph{Chemical potential and free energy module.}
We propose two variants for the chemical potential surrogate $\hat{\mu}_{\theta_2}$.
In the first, a neural network directly predicts the chemical potential field
from the concentration and temperature,
$\hat{\mu}_{\theta_2} = \hat{F}_{\theta_2}(c, T)$,
without imposing any thermodynamic structure.
In the second, which we term the \emph{free energy (FE) variant}, the network
instead learns a scalar free-energy functional:
\begin{align}
  \hat{E}_{\theta_2},\,\hat{h}_{\theta_2} &= \hat{F}_{\theta_2}(c, T),
  \label{eq:FE_network} \\
  \hat{\mu}_{\theta_2} &= \frac{\partial \hat{E}_{\theta_2}}{\partial c},
  \label{eq:AD_mu}
\end{align}
where $\hat{h}_{\theta_2} \in \mathbb{R}^d$ is a latent encoding of the local
thermodynamic state and $\hat{\mu}_{\theta_2}$ is obtained by automatic
differentiation through $\hat{E}_{\theta_2}$.
This formulation embeds the correct thermodynamic structure---the chemical potential
is by construction the gradient of a scalar potential---and exposes the learned
free energy landscape as an independently verifiable quantity.
A ConvNeXt backbone~\cite{convnext} is used in all cases to map the
concentration field and temperature to $\hat{E}_{\theta_2}$ and
$\hat{h}_{\theta_2}$.

\paragraph{Mobility and noise amplitude modules.}
The mobility and noise amplitude are modeled as functions of the latent
representations of the two cells sharing a diffusion edge $(i,j)$, together with
the inter-cell distance $d_{ij}$:
\begin{align}
  M_{\theta_1,ij}       &= \Psi_{\theta_1}
    \!\left(\hat{h}_{\theta_2,i},\,\hat{h}_{\theta_2,j},\,d_{ij}\right),
  \label{eq:mobility_raw} \\
  \hat{M}_{\theta_1,ij} &= \tfrac{1}{2}
    \!\left(M_{\theta_1,ij} + M_{\theta_1,ji}\right),
  \label{eq:mobility_sym}
\end{align}
and analogously for the noise amplitude:
\begin{align}
  B_{\theta_3,ij}       &= \Psi_{\theta_3}
    \!\left(\hat{h}_{\theta_2,i},\,\hat{h}_{\theta_2,j},\,d_{ij}\right),
  \label{eq:noise_raw} \\
  \hat{B}_{\theta_3,ij} &= \tfrac{1}{2}
    \!\left(B_{\theta_3,ij} + B_{\theta_3,ji}\right).
  \label{eq:noise_sym}
\end{align}
The explicit symmetrization enforces $\hat{M}_{ij} = \hat{M}_{ji}$ and
$\hat{B}_{ij} = \hat{B}_{ji}$, which combined with the antisymmetric flux
in Eq.~(\ref{eq:flux_surrogate}) ensures that Eq.~(\ref{eq:surrogate_update})
conserves mass exactly at every step.
Both $\Psi_{\theta_1}$ and $\Psi_{\theta_3}$ are multi-layer perceptrons (MLPs).
The distance $d_{ij}$ is provided as an input because each surrogate step
spans many solver steps, so the effective transport range can extend beyond
the nearest-neighbor shell; including $d_{ij}$ allows the MLP to modulate
contributions from bonds at different ranges.
We include bonds up to the third-neighbor shell in the present study;
the dependence of surrogate accuracy on this cutoff as a function of
$\Delta t / \delta t$ is an open question left for future work.

\paragraph{Model variants and baseline.}
We study four structured variants obtained by switching the free energy
formulation and the stochastic flux independently.
\femv{} (free energy, mean+variance) uses the FE variant for the chemical
potential and the full flux in Eq.~(\ref{eq:flux_surrogate}); this is the primary
proposed method.
\fem{} (free energy, mean only) uses the FE variant but retains only the
deterministic flux mean, setting $\hat{B}_{\theta_3,ij} = 0$ during both
training and roll-out.
\nfemv{} (non-FE, mean+variance) uses the full stochastic flux but replaces
the AD-derived chemical potential with a direct network prediction.
\nfem{} (non-FE, mean only) combines direct chemical-potential prediction with
the deterministic flux mean only.
As an additional baseline, we include a black-box surrogate \nflux{}, in which
a single network $\Phi_\theta$ directly predicts the concentration update:
\begin{equation}
  \hat{c}_i(t + \Delta t) = c_i(t) + \Phi_{\theta,i}(c(t)).
  \label{eq:nonflux}
\end{equation}
Unlike the four structured variants, \nflux{} does not decompose the
update into a flux divergence, and therefore neither enforces mass conservation
nor exposes physically interpretable modules. A summary can be found in Table \ref{tab:models}.
The general framework is illustrated in Fig.~\ref{fig:scheme}.
\begin{table}[htbp]
\centering
\caption{Comparison of surrogate model variants.}
\label{tab:models}
\renewcommand{\arraystretch}{1.4}
\begin{tabular}{p{2cm}p{2.5cm}p{2.5cm}p{2.5cm}p{2.5cm}p{2.5cm}}
\hline
 & \textbf{FE-MV} & \textbf{FE-M} & \textbf{nonFE-MV} & \textbf{nonFE-M} & \textbf{non-flux} \\
\hline
\textbf{model}
  & Free energy, mean+variance
  & Free energy, mean only
  & Non-FE, mean+variance
  & Non-FE, mean only
  & Non-flux black-box \\
\textbf{Conserv.}
  & Yes & Yes & Yes & Yes & No \\
\textbf{Value update}
  & $\hat{c}_i(t{+}\Delta t) = c_i(t) - \sum_{j} F_{ij}$
  & same & same & same
  & $\hat{c}_i(t{+}\Delta t) = c_i(t) + \Phi_{\theta,i}(c(t))$ \\
\textbf{$F_{ij}$ eval.}
  & $\hat{M}_{ij}(\hat{\mu}_i {-} \hat{\mu}_j)\Delta t + \hat{B}_{ij}\sqrt{2\Delta t}\,\xi$
  % & $\hat{M}_{ij}(\hat{\mu}_i {-} \hat{\mu}_j)\Delta t$, $\hat{B}{=}0$
  & same but $\hat{B}{=}0$
  & same as FE-MV
  % & $\hat{M}_{ij}(\hat{\mu}_i {-} \hat{\mu}_j)\Delta t$, $\hat{B}{=}0$
  & same but $\hat{B}{=}0$
  & N/A \\
\textbf{Chemical potential}
  & autograd $\hat{\mu} = \partial \hat{E}_{\theta_2}/ \partial c$
  & same
  & direct $\hat{F}_{\theta_2}(c,T)$
  & same % as nonFE-MV
  & N/A \\
\textbf{Stochastic}
  & Yes & No & Yes & No & No \\
\textbf{Free en.}
  & Yes & Yes & No & No & No \\
\textbf{Primary}
  & Yes & No & No & No & No \\
\hline
\end{tabular}
\end{table}

\subsection{Data generation and model training}
\label{sec:data_generation}

\paragraph{SPDE simulations.}
All SPDE simulations are run at fixed temperature $T = 0.2$. The primary results use noise level $\epsilon = 0.2$ unless otherwise noted.
Training trajectories are generated from 100 independent SPDE trajectories on a
$16^3$ mesh. Each is initialized with $\mathbf{c}_0 = \langle c \rangle  + \mathbf{u}_0$, including tiny initial perturbations $\mathbf{u}_0 \sim \mathcal{U}(-0.01,0.01)$ to start the demixing and a mean concentration
$\langle c \rangle  \sim \mathcal{U}(-0.5, 0.5)$ chosen randomly within the spinodal region
($|c_0| < 0.5$). Therefore all training trajectories correspond to
spinodal decomposition.
%, each initialized from a different randomly drawn mean
%concentration with a small per-cell perturbation
%$c_0 = c + \mathcal{U}(-0.01, 0.01)$.
They are integrated for $N_t = 24{,}000$ solver steps at
$\delta t = 10^{-3}$, and frames are saved every $n_t = 100$ steps,
yielding 240 frames per trajectory, of which frames 25 through 240 are used in training. 
% from which frames 25 through 240 are retained; the first 24 frames are
% excluded because the transient spinodal instability at very short times
% produces anomalously large validation loss that is unrepresentative of
% the long-time behavior we seek to learn.
A validation set, 10 trajectories randomly split from the training set, was used for choosing the best surrogate model but otherwise not presented.
Finally, separate test sets were obtained in much longer rollouts to assess the long-term stability and accuracy of the chosen model. These tests will be discussed in detail.

\paragraph{Training procedure.}
% Surrogate models are trained independently at each noise level $\epsilon$;
The model is trained on an NVIDIA H100 GPU using the AdamW
optimizer~\cite{adamw} with a learning rate of $10^{-4}$ and a batch size of 32.
To encourage temporal consistency and suppress roll-out error accumulation,
the training loss is evaluated over five consecutive autoregressive steps,
from $\hat{c}_{t+1}$ through $\hat{c}_{t+5}$ conditioned on $c_t$.
To ensure numerical stability of the stochastic models, the output of the
noise amplitude module is passed through a sigmoid activation function,
which bounds $\hat{B}_{\theta_3,ij}$ from above at the level
corresponding to $\epsilon = 1$ in the reference SPDE.
This prevents the learned noise amplitude from drifting to unphysically large
values during early training, where the NLL loss is sensitive to the variance prediction.

\paragraph{Loss functions.}
The deterministic variants \nflux{}, \nfem{}, and \fem{} are trained with the
mean squared error (MSE) loss:
\begin{equation}
  \mathcal{L}_{\text{MSE}}
  = \frac{1}{N_s} \sum_{i=1}^{N_s} \left\| c_i - \hat{c}_i \right\|_2^2,
  \label{eq:loss_mse}
\end{equation}
where $i$ indexes grid cells and $c_i$, $\hat{c}_i$ are the ground-truth and
predicted concentrations, respectively.
The stochastic variants \nfemv{} and \femv{} are trained with the Gaussian
negative log-likelihood (NLL):
\begin{equation}
  \mathcal{L}_{\text{NLL}}
  = \frac{1}{N_s} \sum_{i=1}^{N_s}
    \left[ \log \hat{v}_i
           + \frac{\left(c_i - \hat{c}_i\right)^2}{2\hat{v}_i} \right],
  \label{eq:loss_nll}
\end{equation}
where $\hat{v}_i$ is the predicted variance of the concentration update at
cell $i$, obtained by summing the independent edge contributions:
\begin{equation}
  \hat{v}_i = \sum_{\langle ij \rangle} 2\Delta t\,\hat{B}_{\theta_3,ij}^2.
  \label{eq:predicted_var}
\end{equation}
This expression assumes that the noise contributions on different edges are
statistically independent, which matches the ground-truth SPDE discretization
in Eq.~(\ref{eq:flux_spde}) where a separate $\xi$ is drawn for each bond.
The resulting diagonal covariance structure is the natural leading-order
approximation; off-diagonal correlations between edges could in principle
arise from multi-step dynamics within a single surrogate step, but are
neglected here as a simplifying assumption.
The NLL loss simultaneously optimizes the predicted mean $\hat{c}_i$ and the
predicted variance $\hat{v}_i$, penalizing both systematic bias and
miscalibrated uncertainty.

\section{Results on simulating stochastic CHE}
\label{sec:results_sPDE}
In this section, we present a brief demonstration of the SPDE results before delving into surrogate models in the next one.
\subsection{One-dimensional case}
\label{sec:1D_SPDE_main}
\begin{figure}[htp]
    \centering
    \begin{tabular}{c c}
        no noise $\epsilon=0$ & noise $\epsilon=0.2$ \\
        \raisebox{-0.5\height}{\includegraphics[width = 0.5\textwidth]{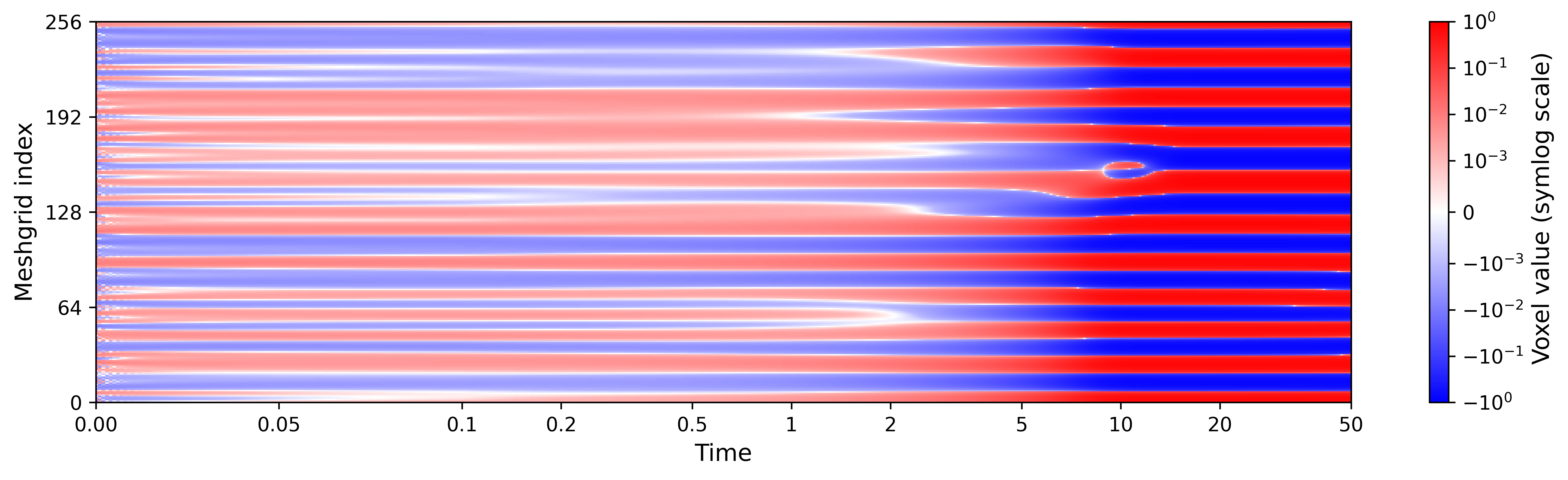}} &
        \raisebox{-0.5\height}{\includegraphics[width = 0.5\textwidth]{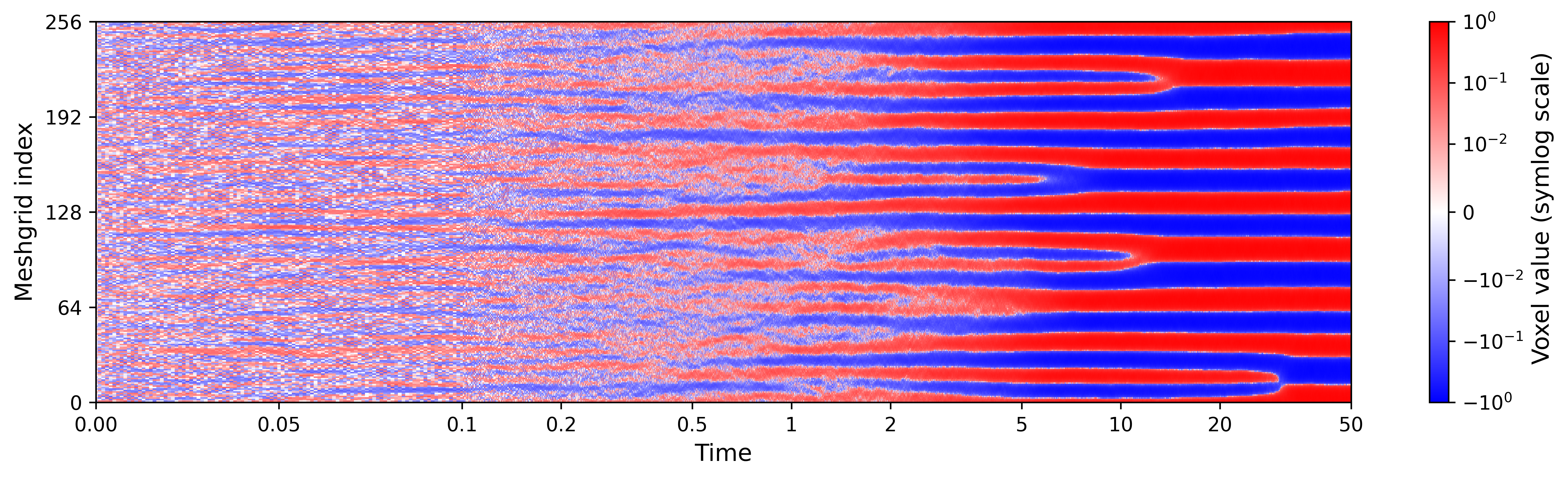}} \\
        \multicolumn{2}{c}{(a)} \\[0.3em]
        \multicolumn{2}{c}{\includegraphics[width=1\textwidth]{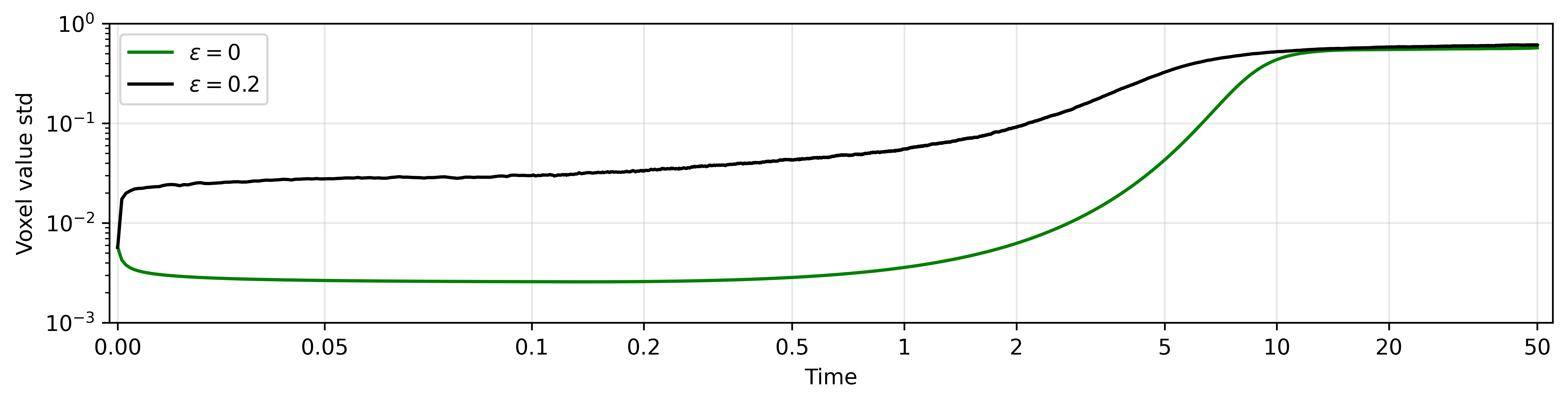}} \\
        \multicolumn{2}{c}{(b)} \\[0.3em]
        no noise $\epsilon=0$ & noise $\epsilon=0.2$ \\
        \raisebox{-0.5\height}{\includegraphics[width = 0.5\textwidth]{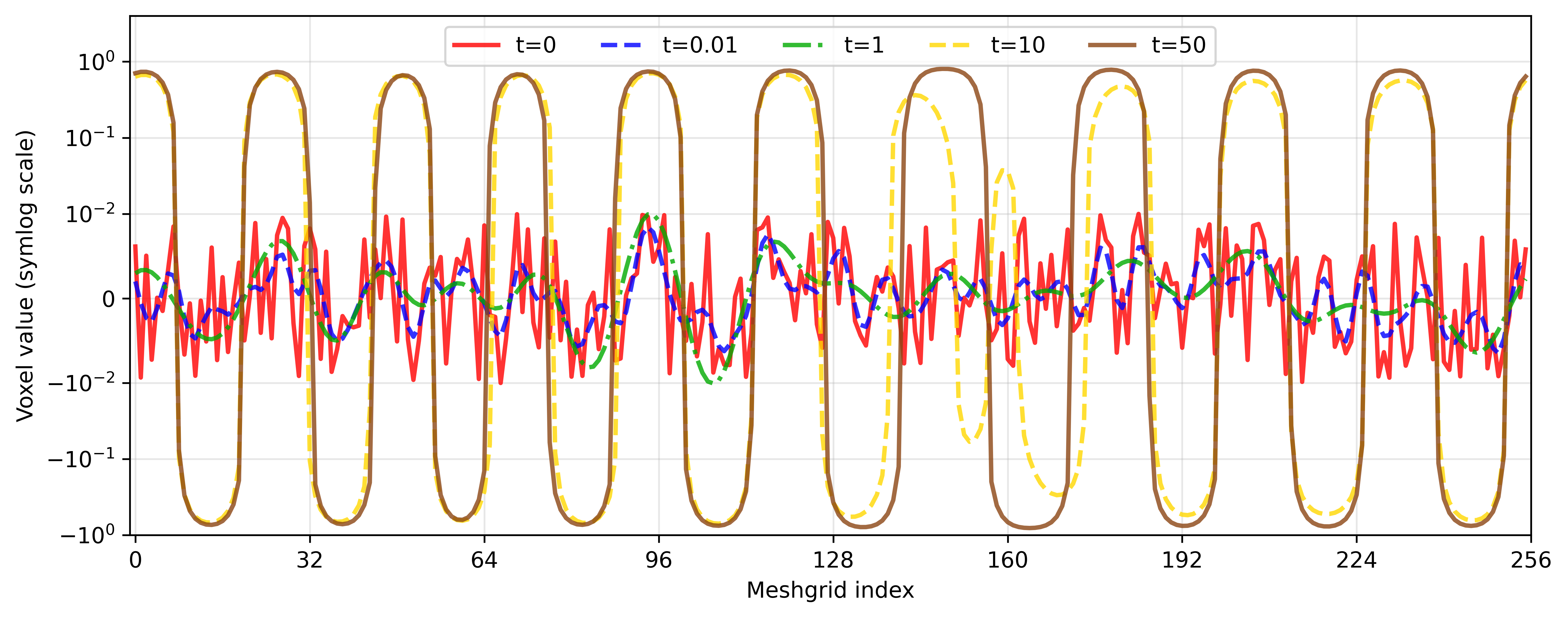}} &
        \raisebox{-0.5\height}{\includegraphics[width = 0.5\textwidth]{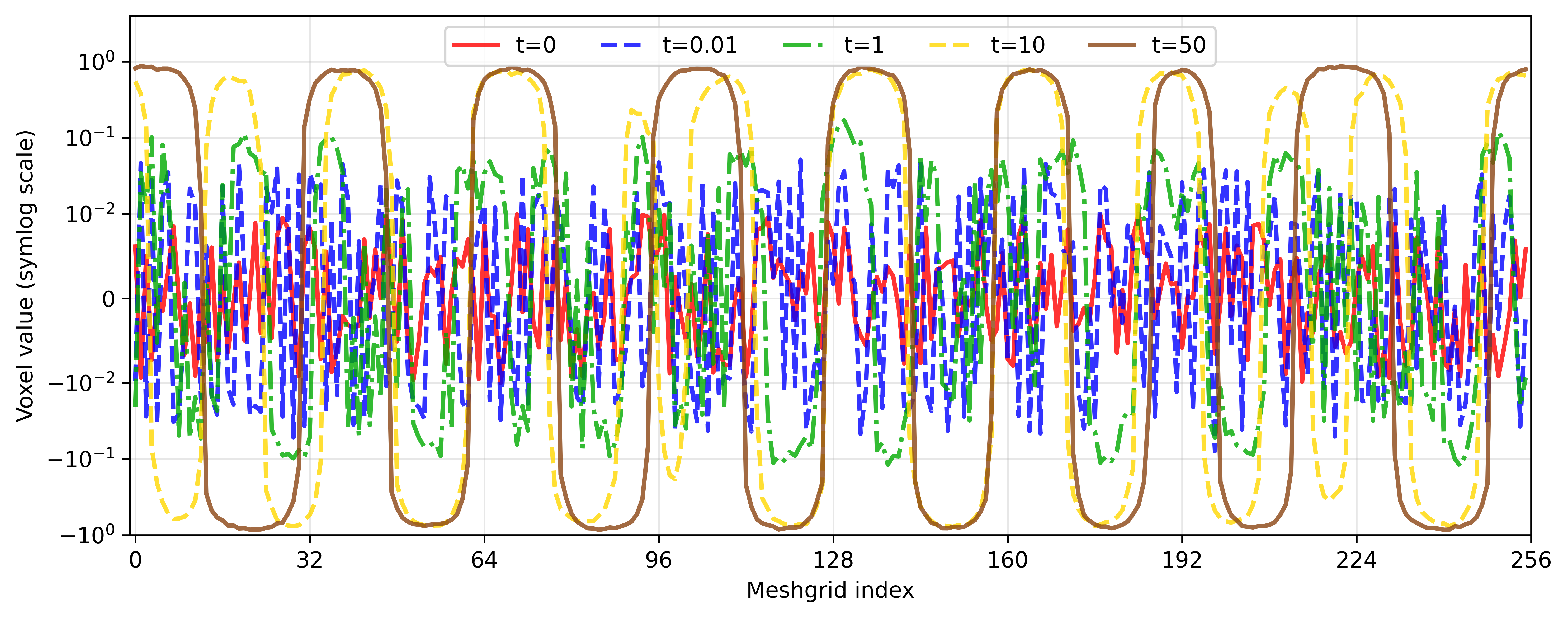}} \\
        \multicolumn{2}{c}{(c)}
    \end{tabular}
    \caption{
    Results of 1D CHE initialized with $c \sim 0 + \mathcal{U}(-0.01,0.01)$. (a) Time evolution of $c$ without and with noise. (b) Standard deviation of $c$ over time. (c) Representative frames without and with noise.
    }
    \label{fig:1d_frames_fourier_256}
\end{figure}

% \added{[We should say SPDE instead of SDE throughout the paper]}\hai{will fixed at the end}
We qualitatively analyzed the stochastic dynamics by a one-dimensional simulation result of CHE systems. We run simulations in a periodic one-dimensional system with $N_x=256$ grid points and $\delta t=10^{-3}$. The concentration field $c$ was initialized with average $\langle c \rangle=0$ and random uniform noise from $\mathcal{U}(-0.01,0.01)$. We compare the results without noise $\epsilon=0$ and with noise $\epsilon=0.2$.  In Fig.~\ref{fig:1d_frames_fourier_256}, the subplot (a) shows the spatial value evolution versus time. It has a clear trend that adding noise will make the trajectory much noisy. Subplot (b) shows the standard deviation (std) of all voxel values in a frame versus time. The noisy trajectory has a higher std overall, suggesting that the presence of thermal fluctuations accelerates the demixing dynamics. 
%The initial dip with $\epsilon=0$ was due to reduction of interfacial energy from the artificial fluctuations. 
This is seen more quantitatively from the representative frames at different time stamps (subplot (c) of Fig.~\ref{fig:1d_frames_fourier_256}).
% The third row shows the snapshot at different time steps from $t=0$ to $t=50$. It also shows the trend that adding noise makes the simulation result more noisier in the beginning. Therefore, the conclusion is adding the stochastic noise would accelerate the dynamics evolution of the system.
% % We have move the 1d 256 for \cozero{} here for showing that the golomeration gets stuck at local minamal energy model. 
More results corresponding to one-dimensional analysis for different grid size and grid size \cofour{} is presented in the appendix 
 \ref{sec:1D_SPDE_32} and \ref{sec:1D_SPDE_256}.

The rest of the paper will focus on 3D results.

% First, we start by qualitatively analyze the stochastic dynamics by a one-dimensional simulation result. The conclusion is adding stochastic forces will acclerate the evolution of the system. 

\subsection{Three-dimensional case}

%Furthermore, we carry out a more comprehensive analysis of the stochastic partial differential equation (SPDE) result on three dimensional spatial grids. Similarly to the previous section, we solve the SPDE with $\epsilon=0.2$ and without $\epsilon=0$ noise, while the initial concentration conditions are initialized by \cozero. 
% To ensure fairness, we use the same random seed for initial and stochastic random noise realizations. \added{[FZ: why is this fair? Do we even need to use same random seed? They are supposed to be random anyway.]} \hai{we originally want to show same noise seed leads to the same equilibrium, but it seems not true, so we can ignore this}
%The time step-size for all SPDE simulations is set to $\delta t = 0.001$ and 
3D simulations were performed in similar settings except a 3D periodic grid. Step-size convergence tests are shown in the appendix, \ref{sec:convergence_dt}. Here we consider two representative examples, 50:50 mixture ($\langle c \rangle = 0$) with large, interconnected domains of the two immiscible phases, and 70:30 mixture ($\langle c \rangle = 0.4$) which demixes many disconnected droplets of the minority phase merging into larger droplets, i.e.\ Ostwald ripening.

\begin{figure}[htp]
    \centering
    \begin{tabular}{c| c c c c c}
        % \multicolumn{3}{c}{Free energy curves at different temperature T} & 
        % \multicolumn{3}{c}{Free energy landscape and solubility limit}
        % \\
        % \multicolumn{3}{c}{\includegraphics[width=.47\textwidth]{fig/Final/free_energy_versus_temperature.png}} 
        % & 
        % \multicolumn{3}{c}{\includegraphics[width=.47\textwidth]{fig/Final/energy_contour_with_solubility.png}} 
        % % \\
        % % Initial condition & & & Sample 1 & Sample 2 & Sample 3
        % % \\
        % % \raisebox{-0.5\height}{\includegraphics[width=0.155\textwidth]{fig/Final/SDE_different_samples_from_same_int_initial_conditions.png}}&
        % % \raisebox{-0.5\height}{\includegraphics[width=0.155\textwidth]{fig/Final/SDE_different_samples_from_same_int_initial_conditions.png}}&
        % % \raisebox{-0.5\height}{\includegraphics[width=0.155\textwidth]{fig/Final/SDE_different_samples_from_same_int_initial_conditions.png}}&
        % % \raisebox{-0.5\height}{\includegraphics[width=0.155\textwidth]{fig/Final/SDE_different_samples_from_same_init_0.png}}&
        % % \raisebox{-0.5\height}{\includegraphics[width=0.155\textwidth]{fig/Final/SDE_different_samples_from_same_init_1.png}}&
        % % \raisebox{-0.5\height}{\includegraphics[width=0.155\textwidth]{fig/Final/SDE_different_samples_from_same_init_2.png}}
        % \\
        % t=0 & t=0.2 & t=1 & t=2 & t=5 & t=10
        $t$=0 & $t$=0.2 & $t$=1 & $t$=2 & $t$=5 & $t$=10
        \\
        \raisebox{-0.5\height}{}&
        \raisebox{-0.5\height}{\includegraphics[width=0.155\textwidth]{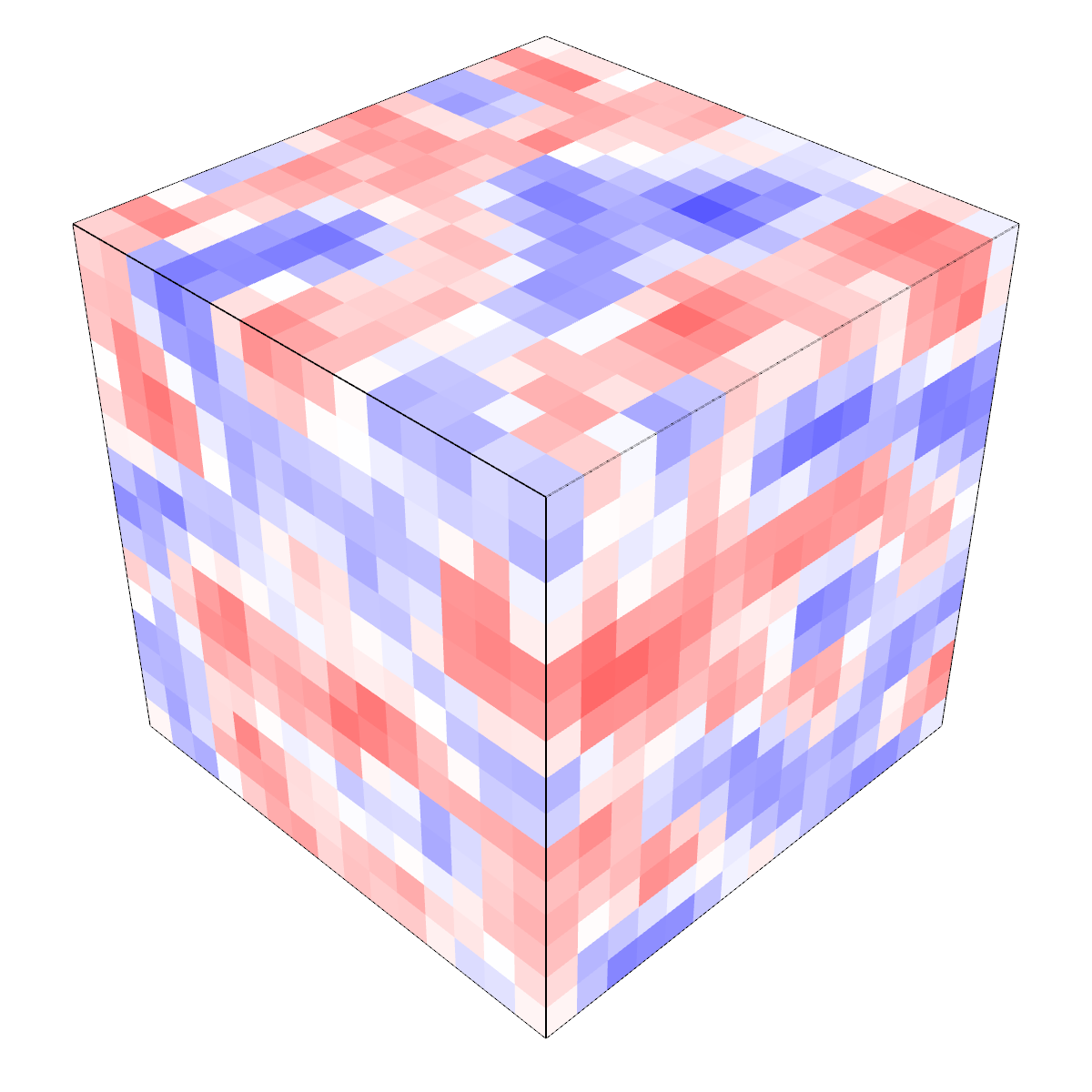}}&
        \raisebox{-0.5\height}{\includegraphics[width=0.155\textwidth]{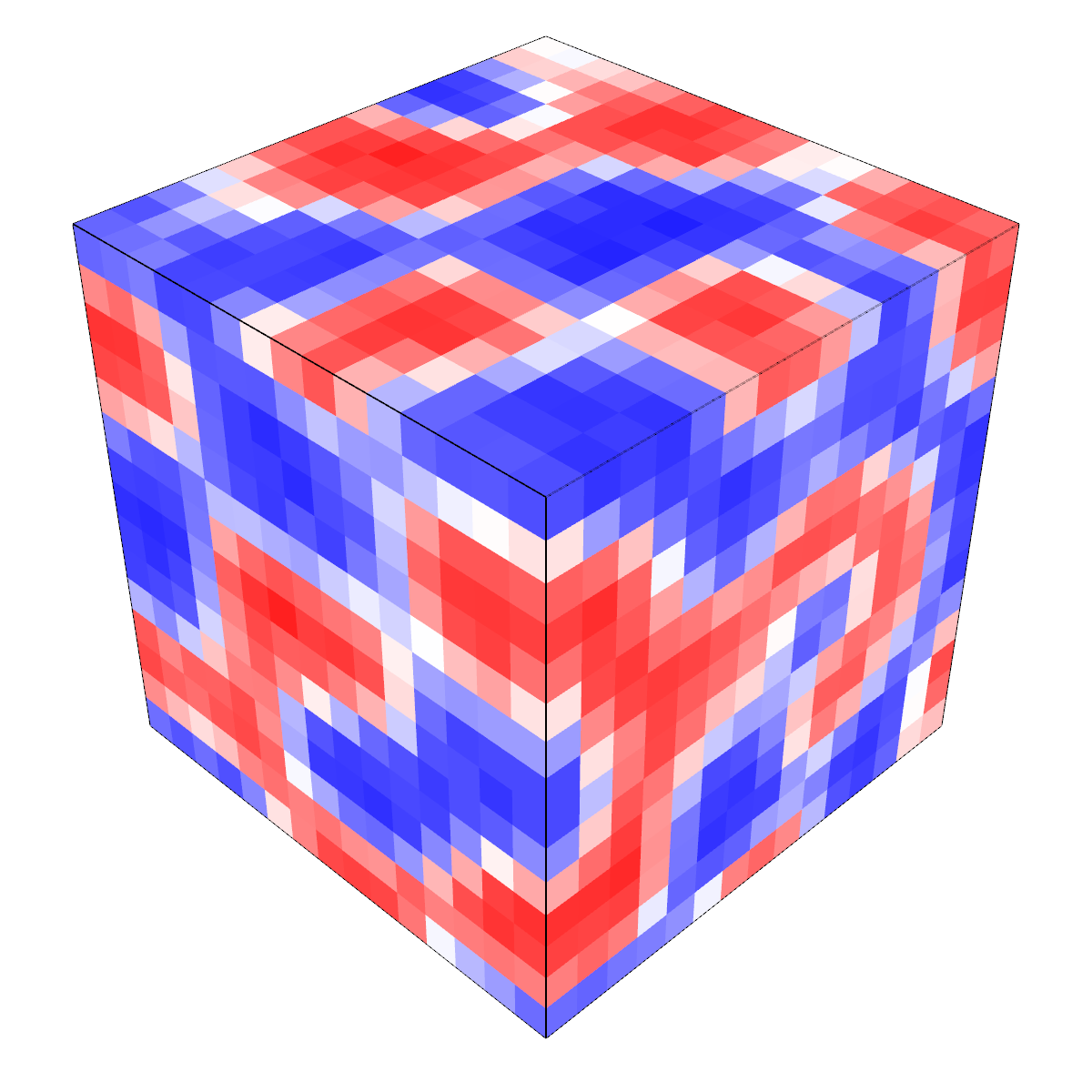}}&
        \raisebox{-0.5\height}{\includegraphics[width=0.155\textwidth]{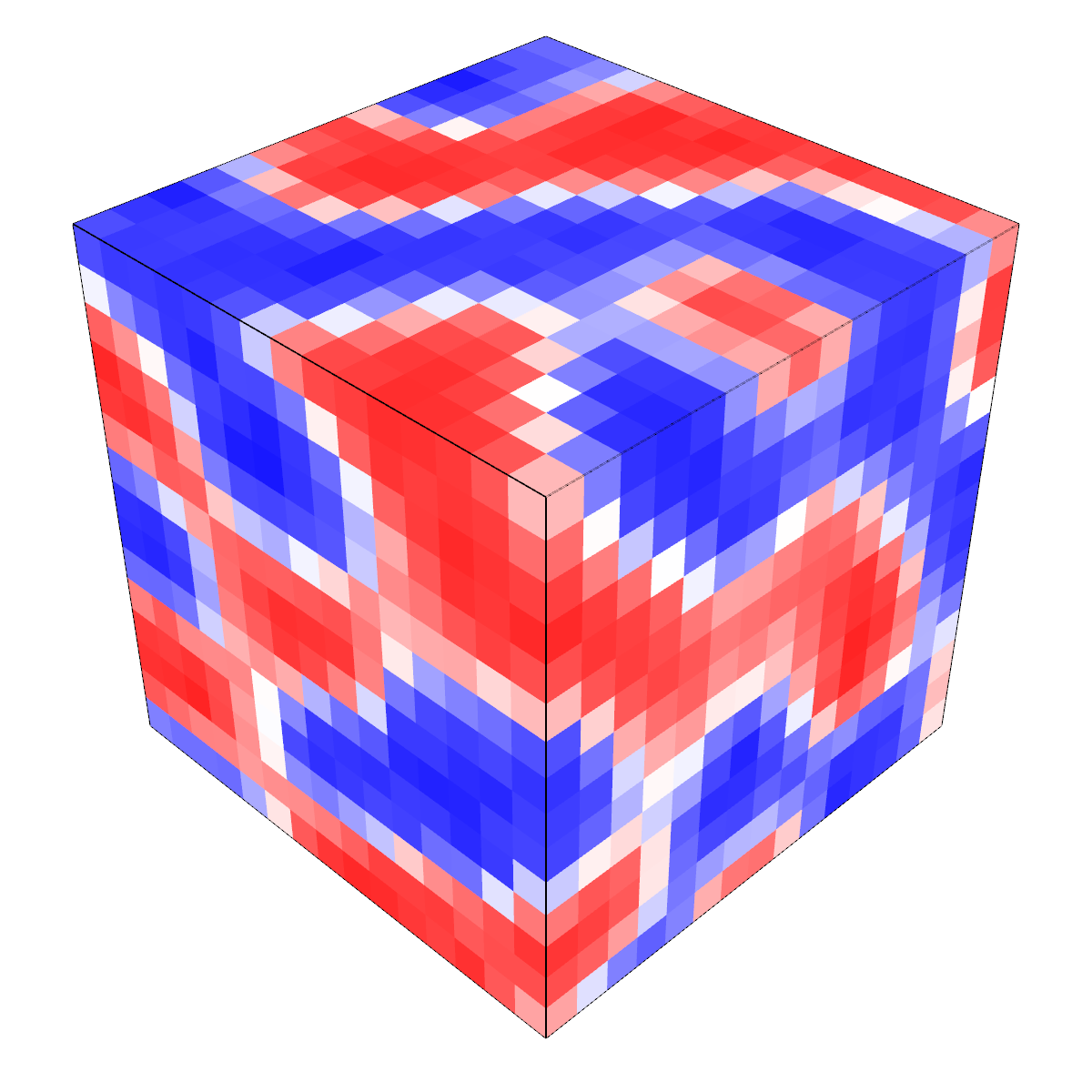}}&
        \raisebox{-0.5\height}{\includegraphics[width=0.155\textwidth]{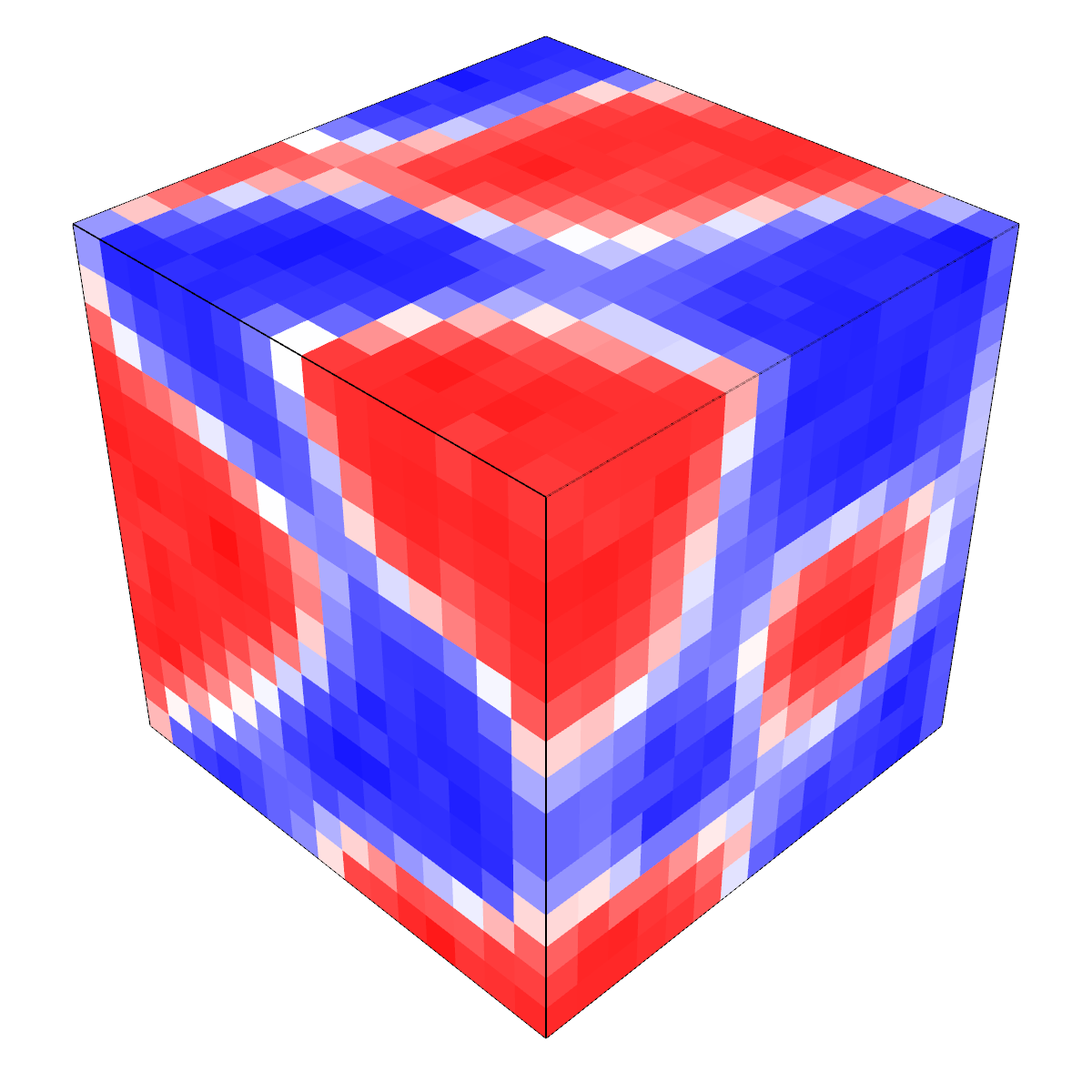}}&
        \raisebox{-0.5\height}{\includegraphics[width=0.155\textwidth]{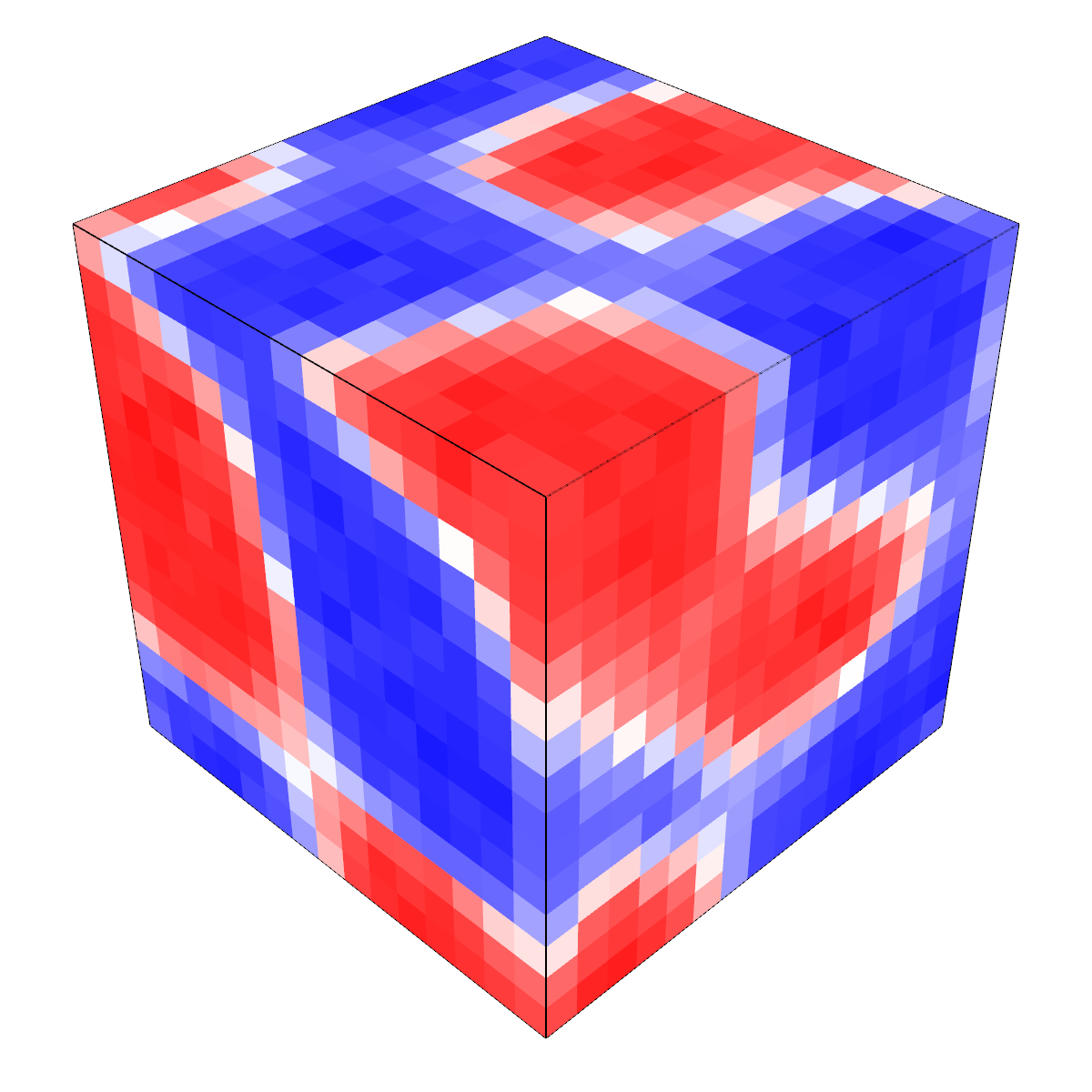}}
        \vspace{-2ex}
        \\
        \raisebox{-0.5\height}{\includegraphics[width=0.155\textwidth]{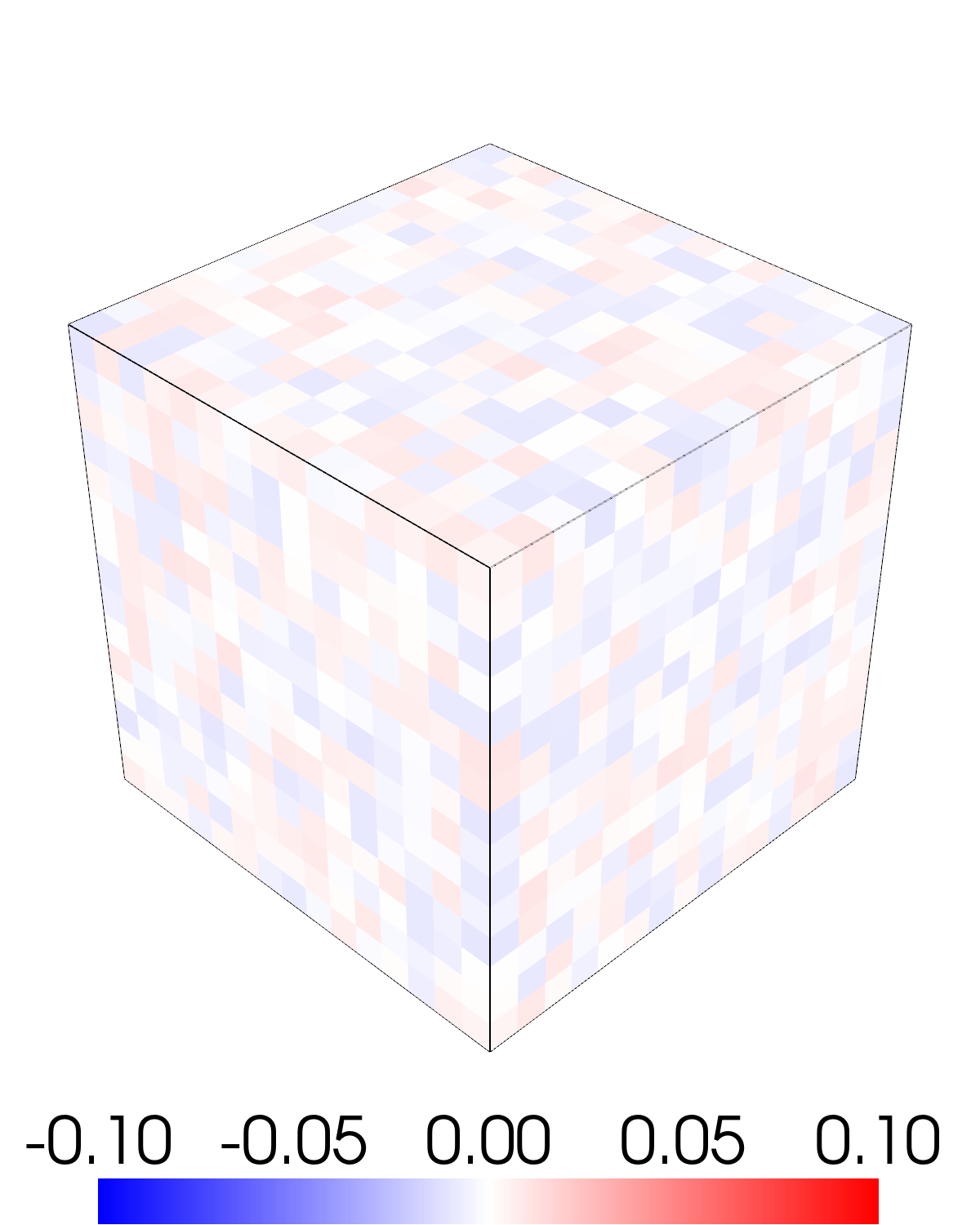}}&
        \raisebox{-0.5\height}{\includegraphics[width=0.155\textwidth]{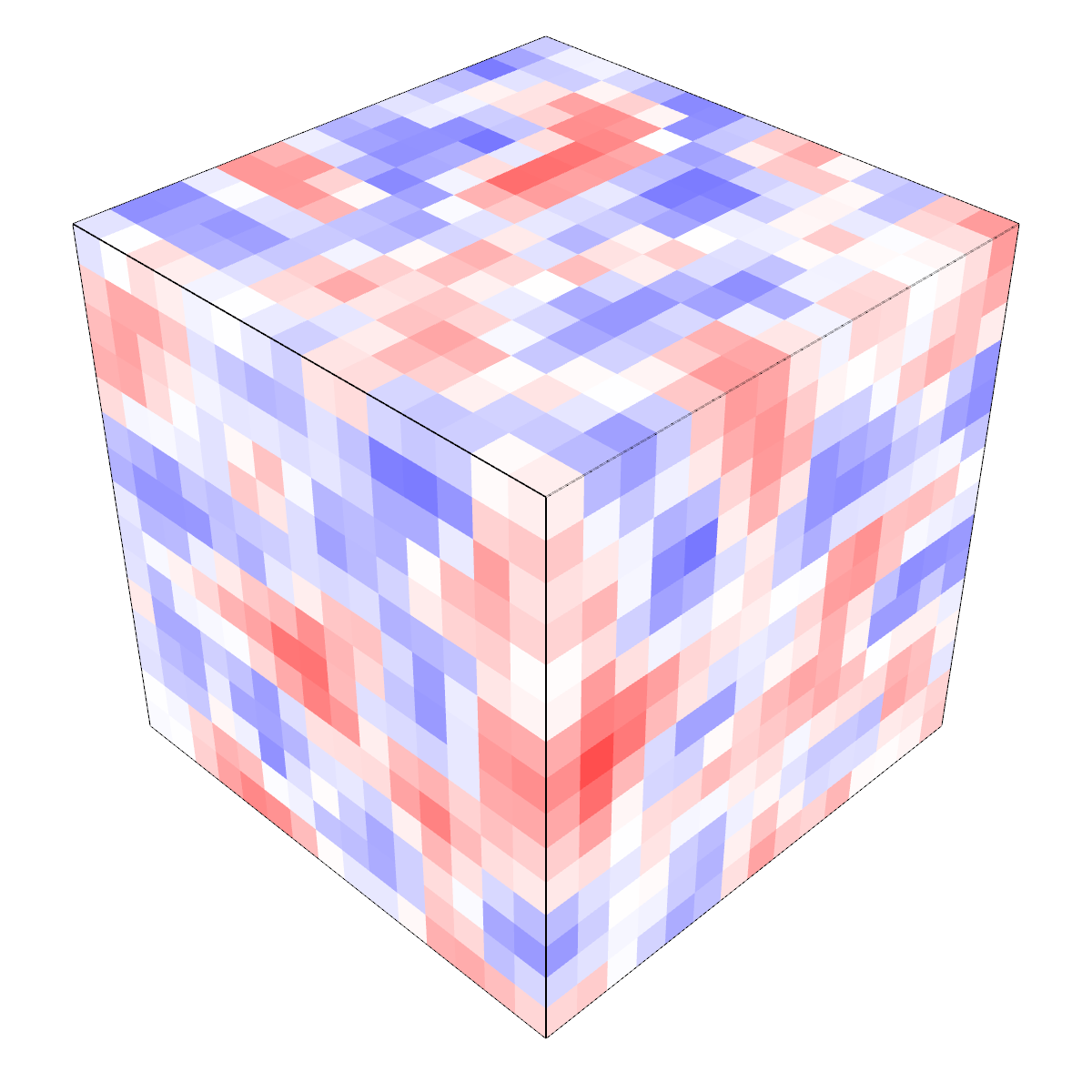}}&
        \raisebox{-0.5\height}{\includegraphics[width=0.155\textwidth]{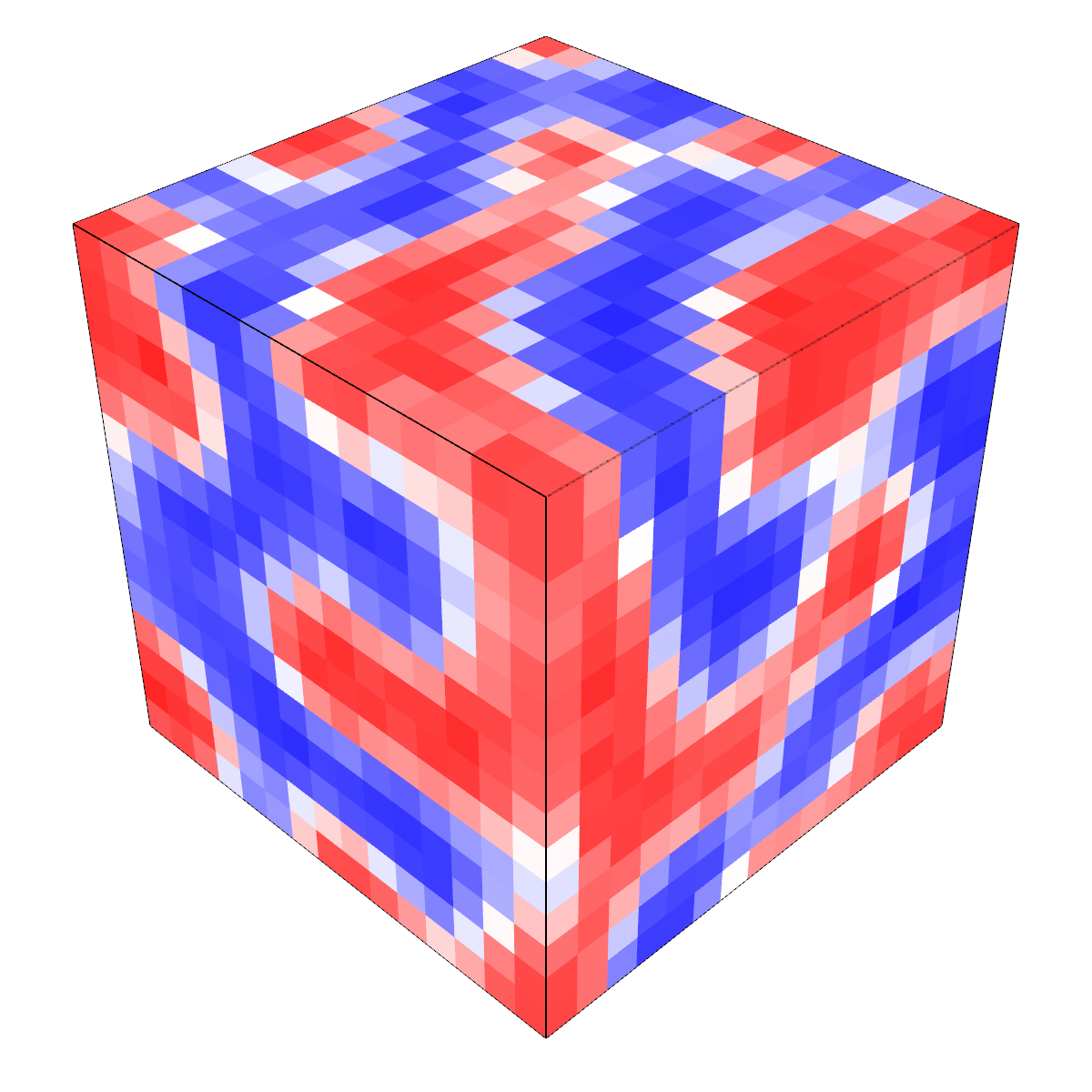}}&
        \raisebox{-0.5\height}{\includegraphics[width=0.155\textwidth]{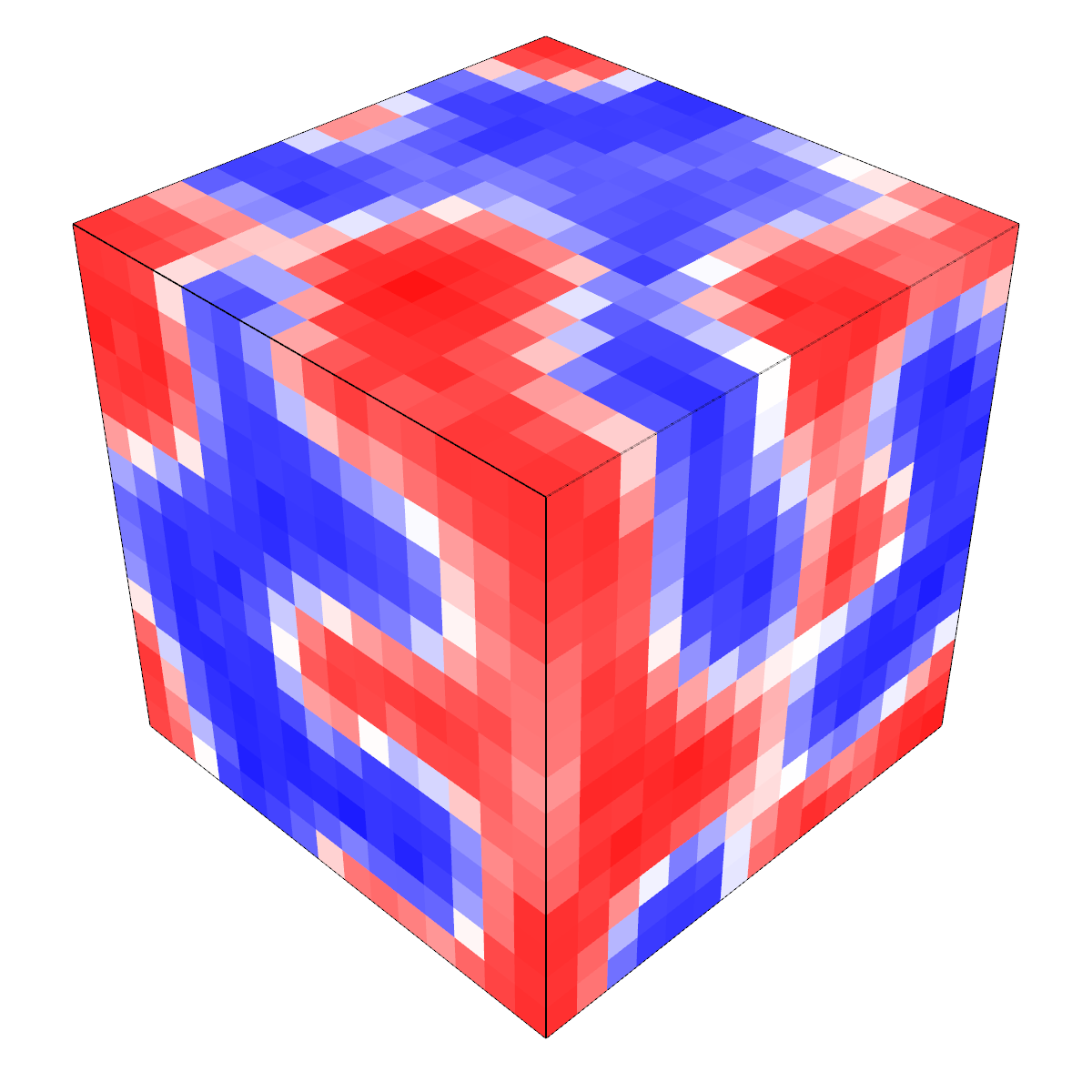}}&
        \raisebox{-0.5\height}{\includegraphics[width=0.155\textwidth]{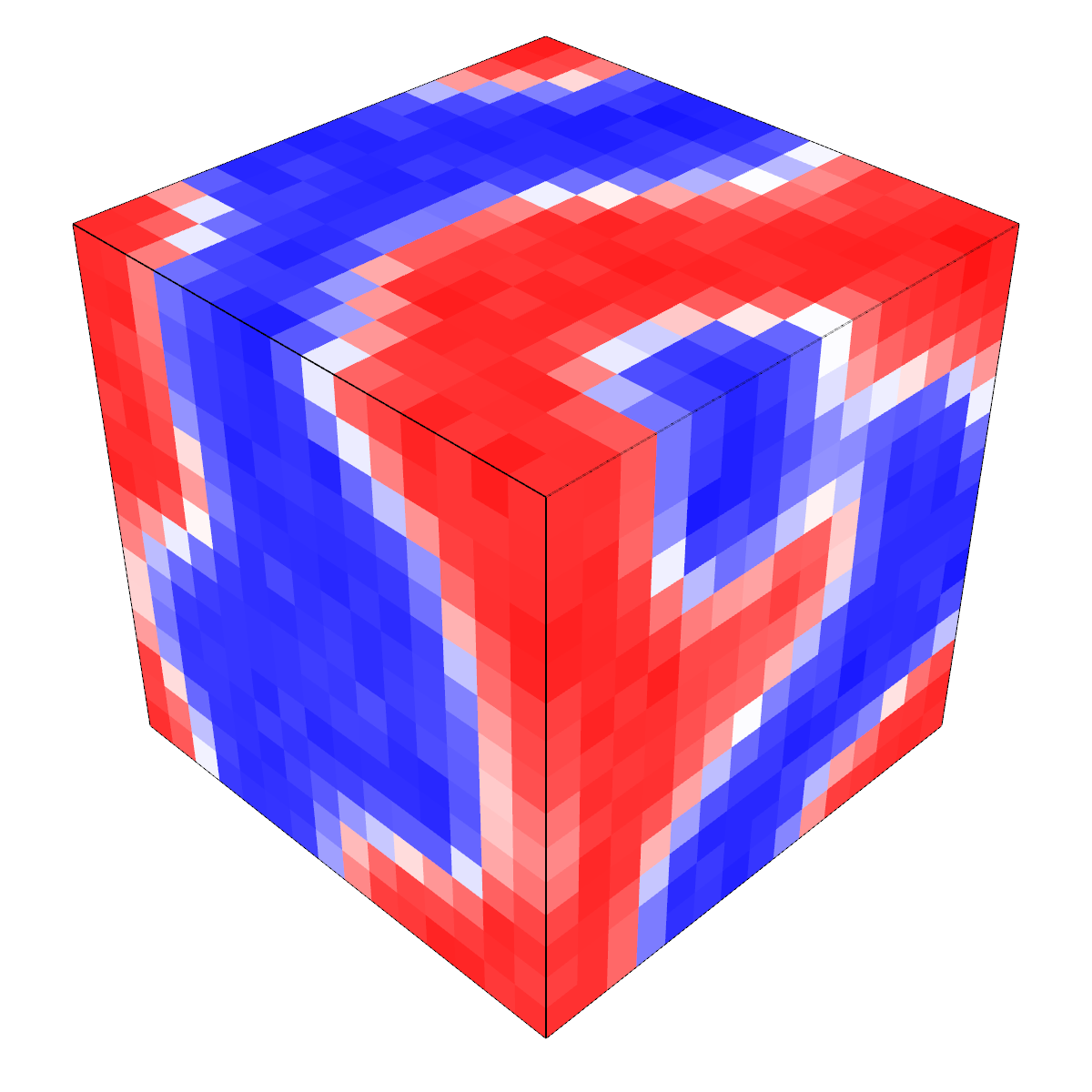}}&
        \raisebox{-0.5\height}{\includegraphics[width=0.155\textwidth]{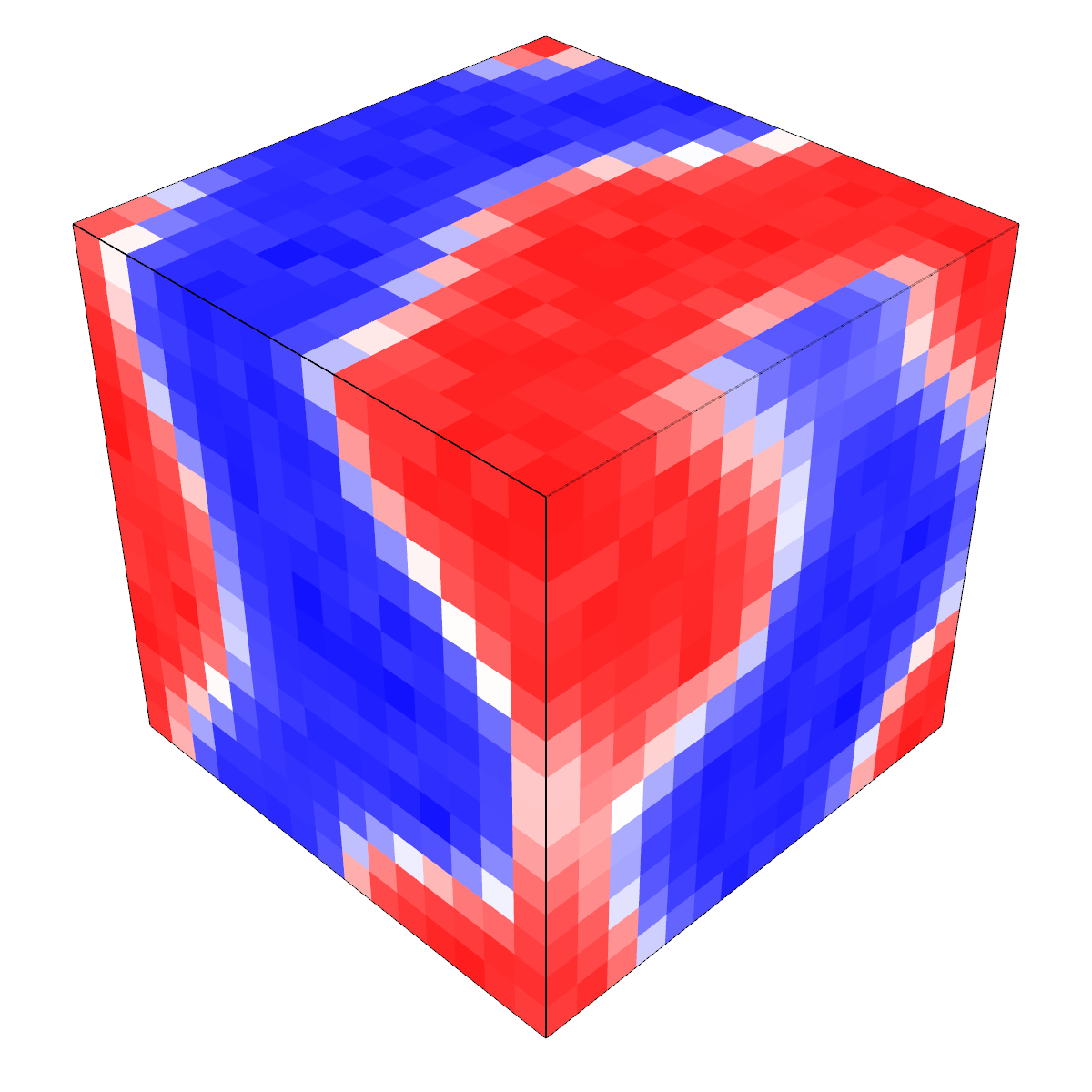}}
        \vspace{-2ex}
        \\
        \raisebox{-0.5\height}{}&
        \raisebox{-0.5\height}{\includegraphics[width=0.155\textwidth]{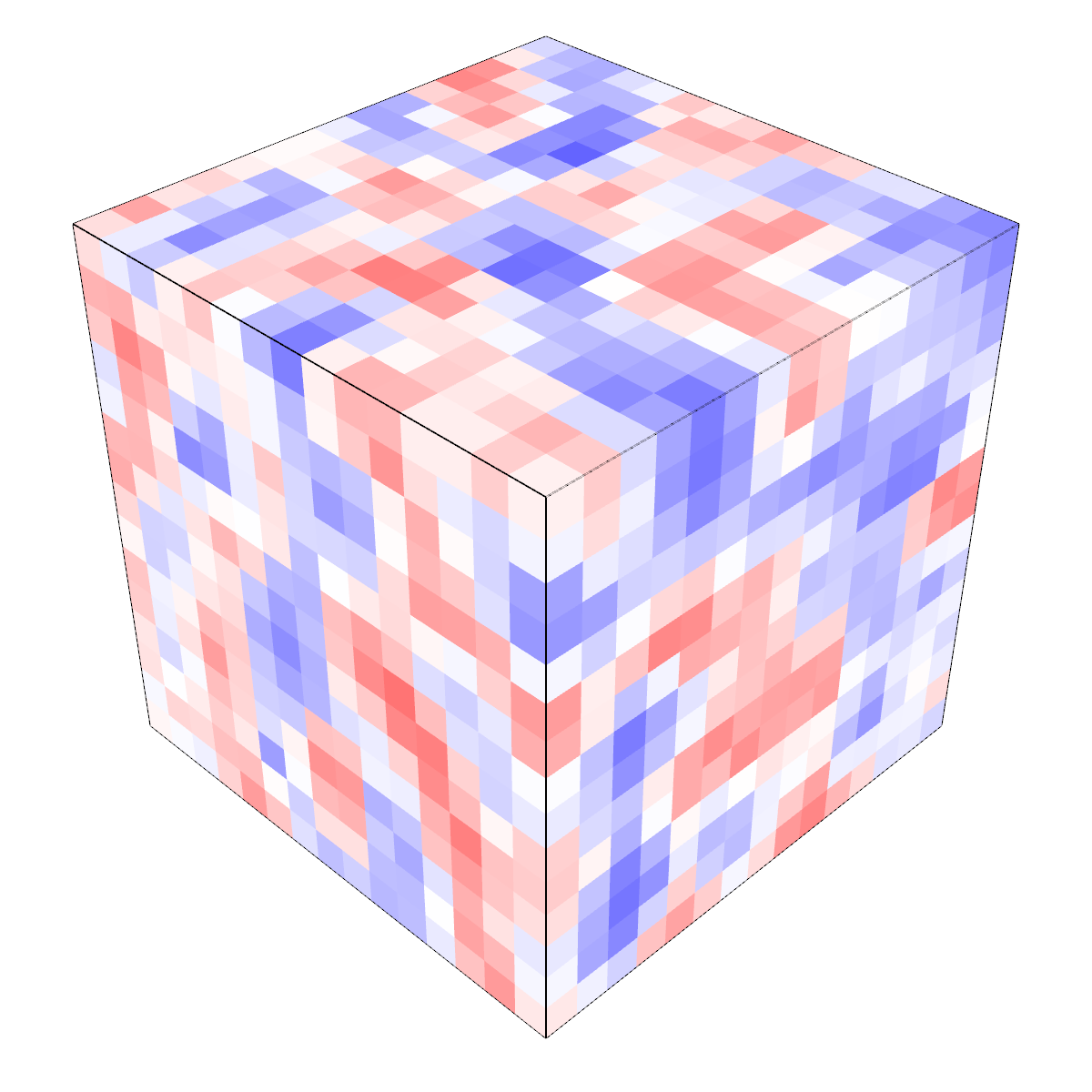}}&
        \raisebox{-0.5\height}{\includegraphics[width=0.155\textwidth]{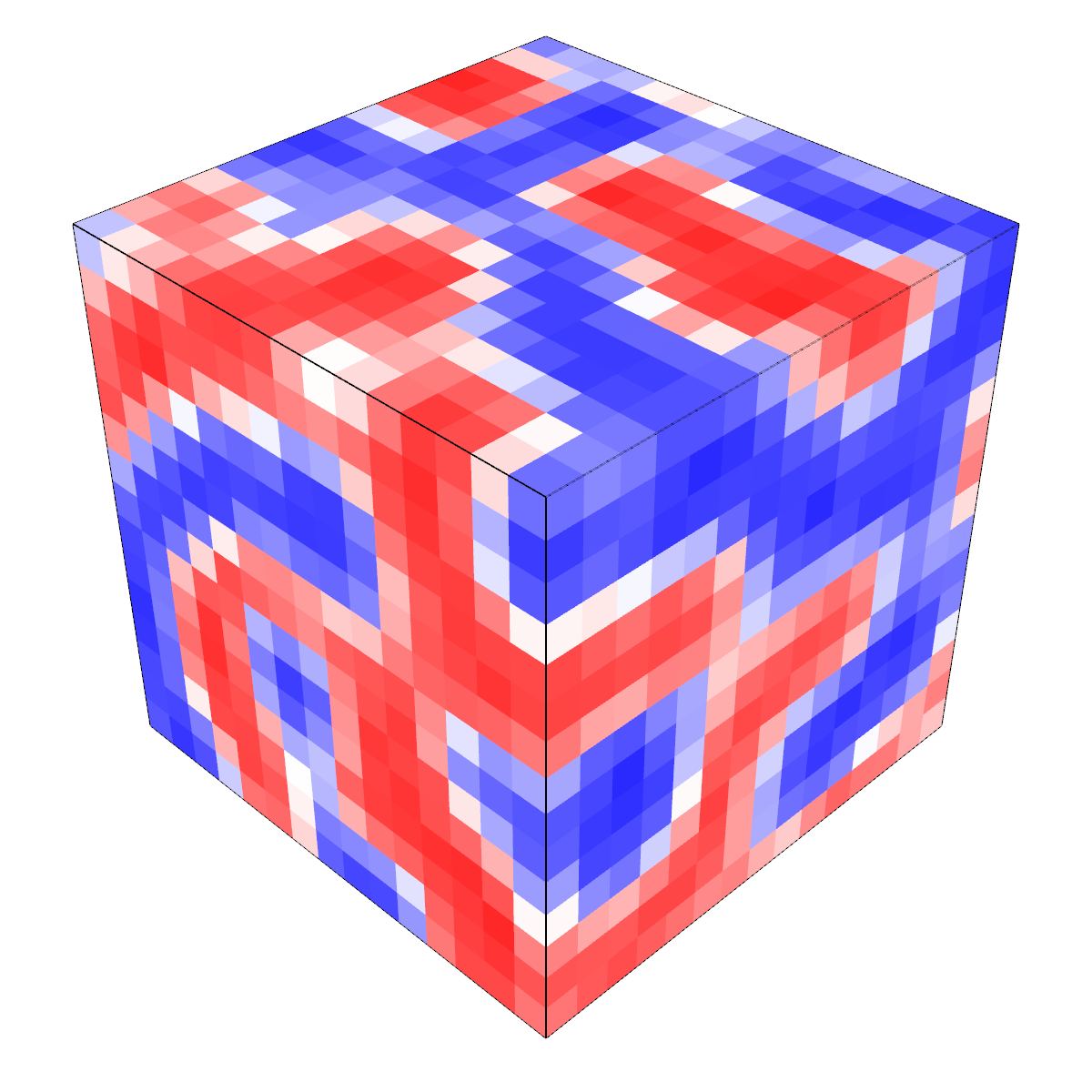}}&
        \raisebox{-0.5\height}{\includegraphics[width=0.155\textwidth]{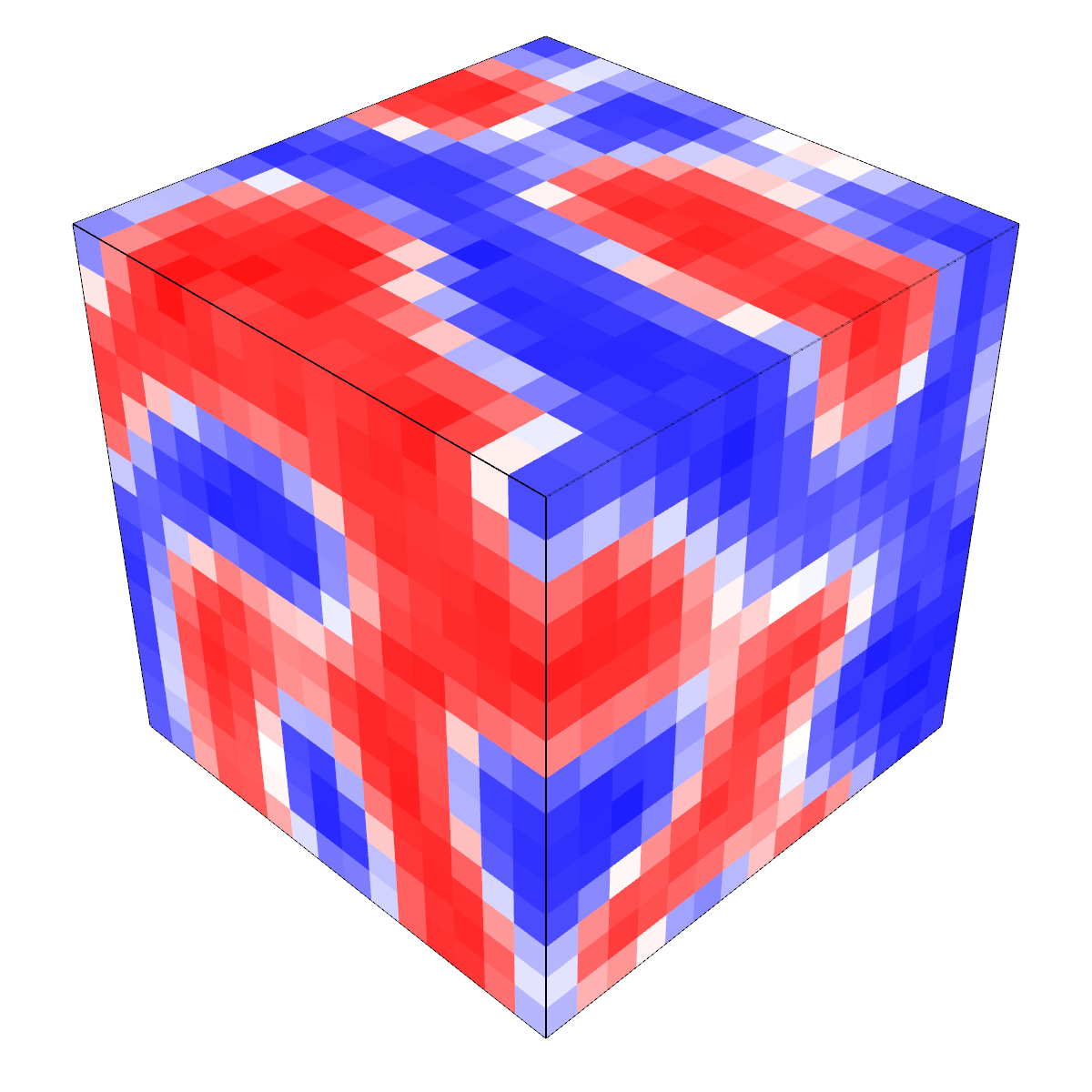}}&
        \raisebox{-0.5\height}{\includegraphics[width=0.155\textwidth]{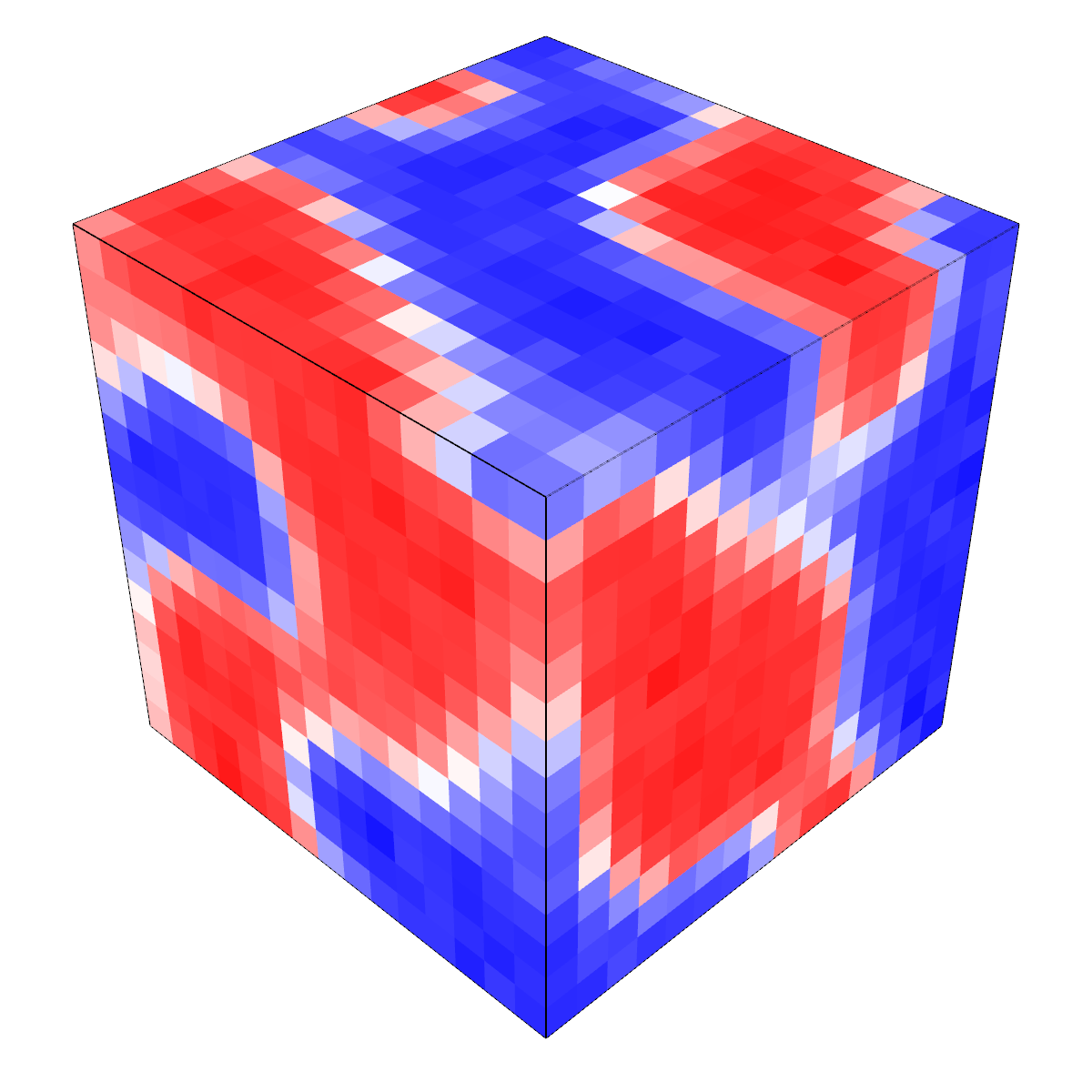}}&
        \raisebox{-0.5\height}{\includegraphics[width=0.155\textwidth]{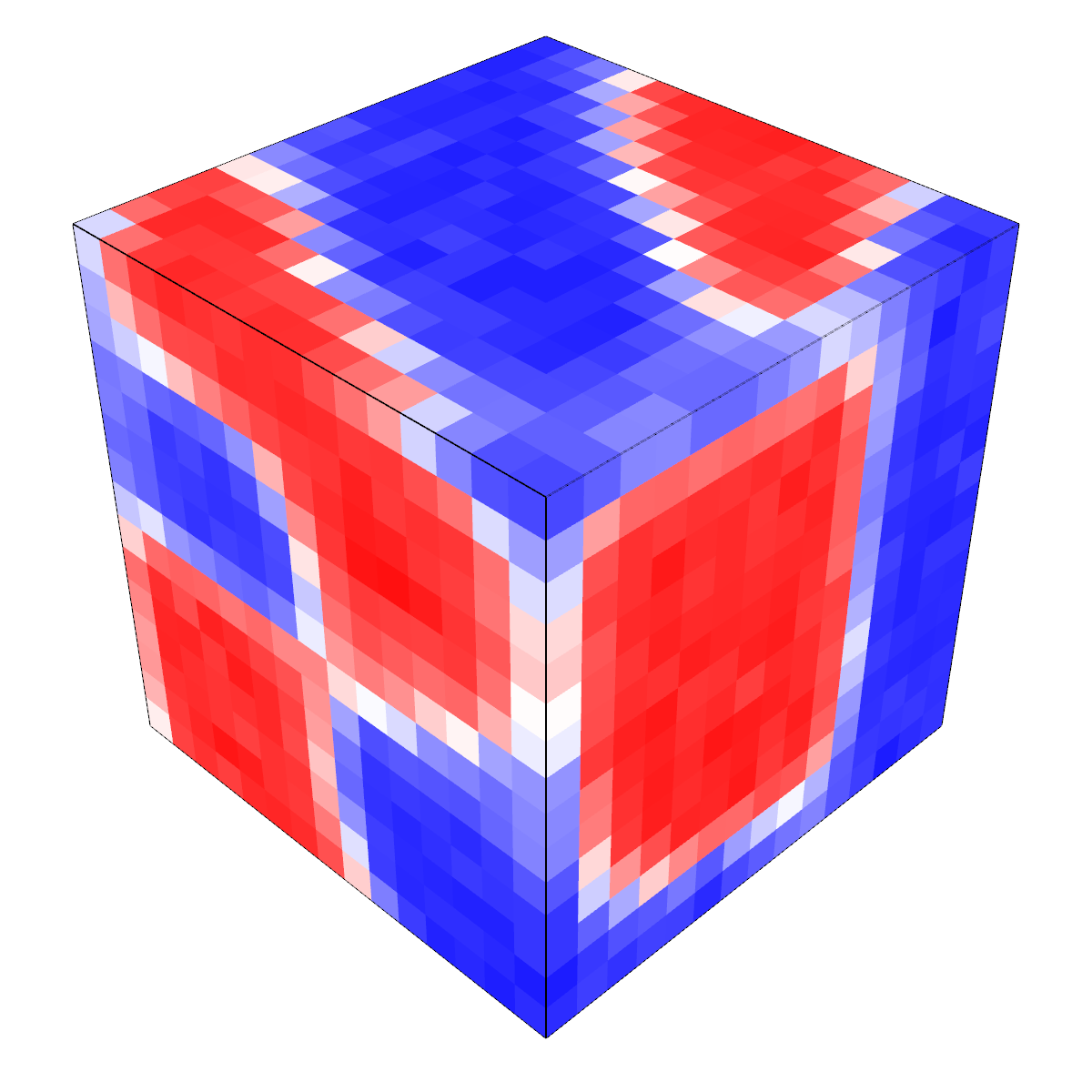}}
        \\
        &
        \multicolumn{5}{c}{\includegraphics[width=.75\textwidth]{fig/Final/SDE_different_samples_from_same_init_colormap.png}} 
    \end{tabular}
     \caption{
    %\added{[For submitted final version, we should save a complex tabular figure like this into a single PNG or PDF. The color bar is very small. Can you enlarge it and maybe put it as a separate figure in the empty space of the table below the left most figure? Can you adjust the color bar, too? Like a non-linear scale, such that it has very light colors from 0 to 0.01 linearly, then 0.01 to 0.8 linearly, like two piecewise linear scale as the color bar? I'd like to see the initial state being paler than this. And the current color is saturated at 0.02 making it hard to determine the true value of late evolution.]}
    % \hai{sorry, the right figures's colormap are already in the range of -1,1, the left figure need to be adjusted, otherwise it is almost blank. Cannot find a way to have piecewise linear colormap.}
    %{\bf Top:} Bulk free energy and solubility limit of the studied system. 
    % [FZ: show the big picture of the free energy landscape]
    %{\bf Bottom:
    % [FZ: can you instead show, from left to right, a single initial structure, then 2 or 3 rows of snapshots, each row representing a different simulation trajectory? It should start from uniform on the left, then many fine structures, then coarser structures. Each snapshot can be small]
    %  }
    Example 3D  SPDE trajectories with $16^3$ grid of 50:50 mixture $(\langle c \rangle=0)$ with $\epsilon=0.2$ starting from the same initial condition.
    %, a perturbed uniform concentration field {\cozero}.% The noise level for SPDE is $\epsilon=0.2$. 
    % \added{[FZ:check periodic boundary condition? Check the $\sigma^2=10^{-2}$ or $\sigma=10^{-1}$ above? I thought it was $10^{-2}$ or -3? ]}
    }
    \label{fig:SDE_different_equilibrium}
\end{figure}

In the first case, given the same initial configuration $\langle c \rangle = 0$ including slight perturbations drawn from $ \mathcal{U}(-0.01,0.01)$, the SPDE with noise level $\epsilon=0.2$ in three independent runs leads to different trajectories due to stochasticity, as shown Fig.~\ref{fig:SDE_different_equilibrium}. The clear differences justify our need of building probabilistic machine learning surrogates. The trajectories demonstrate typical demixing behaviors of spinodal decomposition. 
Additionally, supplementary Fig.~\ref{fig:SDE_different_noise_levels_variance_voxel_new_app} in Sec.~\ref{sec:3D_SPED_app} shows the effect of noise on the evolution of the SPDE system. Finite noise ($\epsilon=0.2$) notably accelerates the evolution of the system compared to deterministic $\epsilon=0$ in 3D, similar to 1D.

\setlength{\tabcolsep}{2pt}

\begin{figure}[htp]
    \centering
    \begin{tabular}{c c c c c c c}
        & \multicolumn{3}{c}{\includegraphics[width=.47\textwidth]{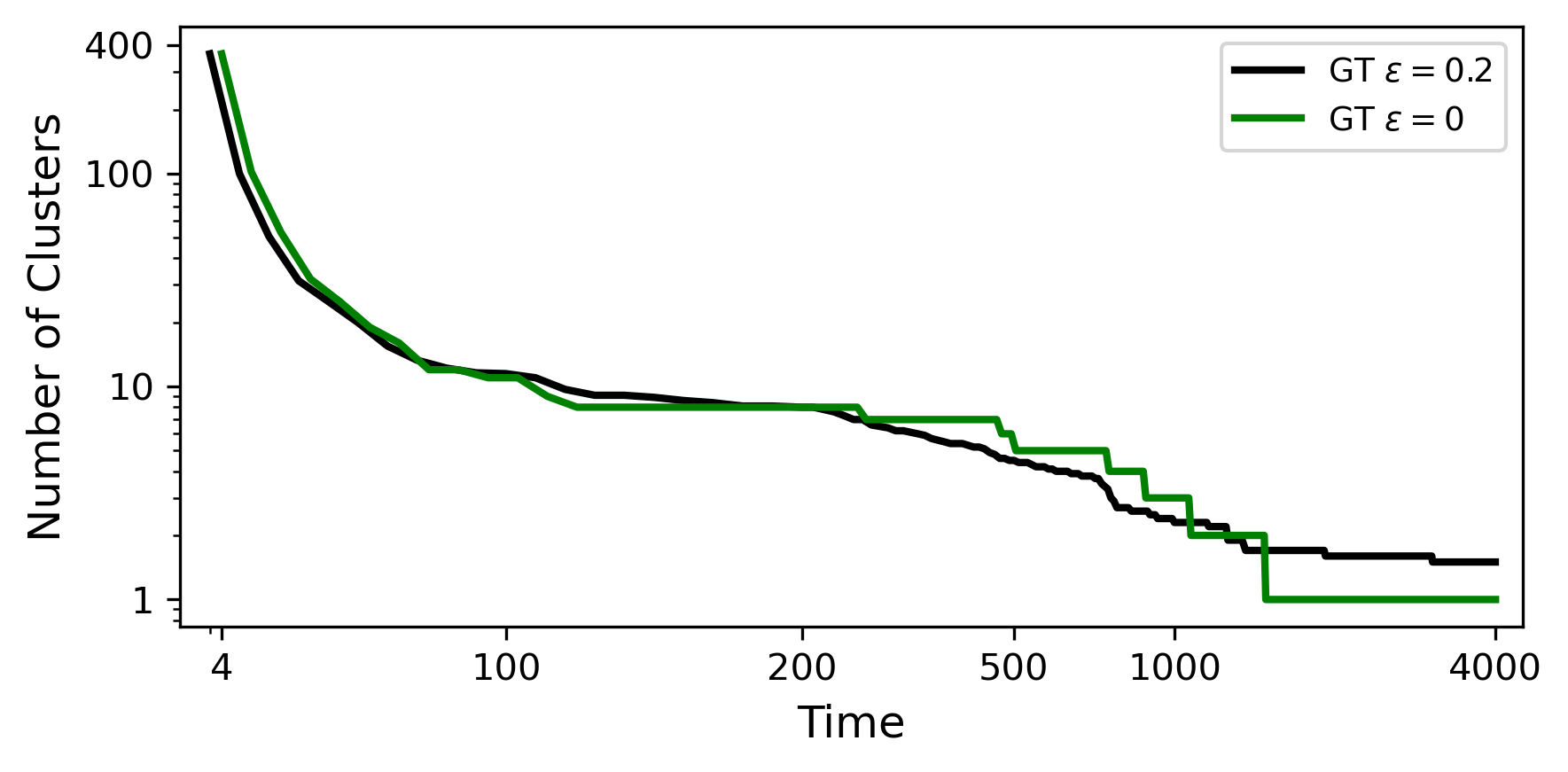}} 
        & \multicolumn{3}{c}{\includegraphics[width=.47\textwidth]{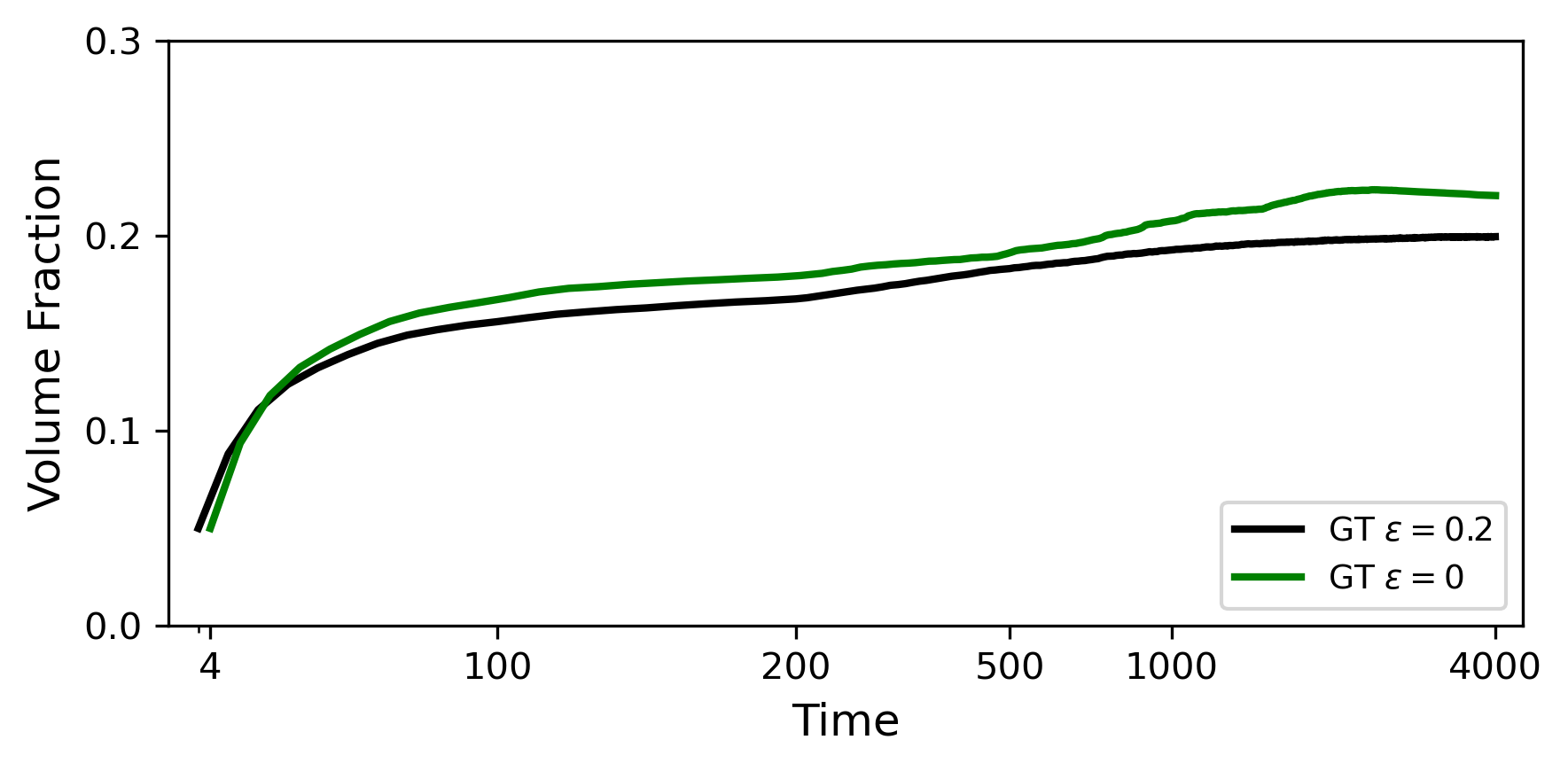}} \\
        & \multicolumn{6}{c}{(a)} \\[0.3em]
        & $t=4$ & $100$ & $200$ & $500$ & $1000$ & $4000$ \\ 
        \rotatebox[origin=l]{90}{$\epsilon=0$} &
        \raisebox{-0.2\height}{\includegraphics[width = 0.15\textwidth]{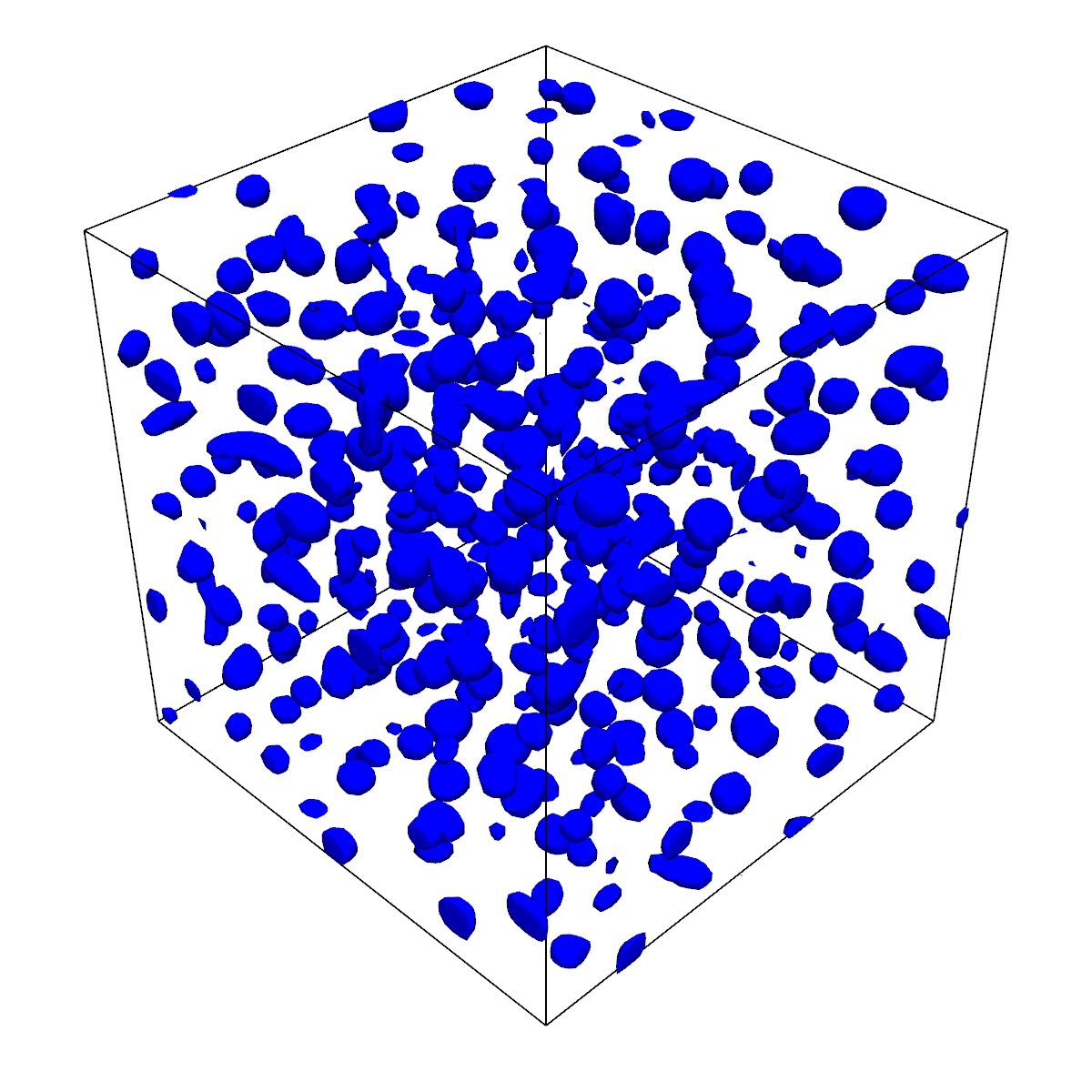}} &
        \raisebox{-0.2\height}{\includegraphics[width = 0.15\textwidth]{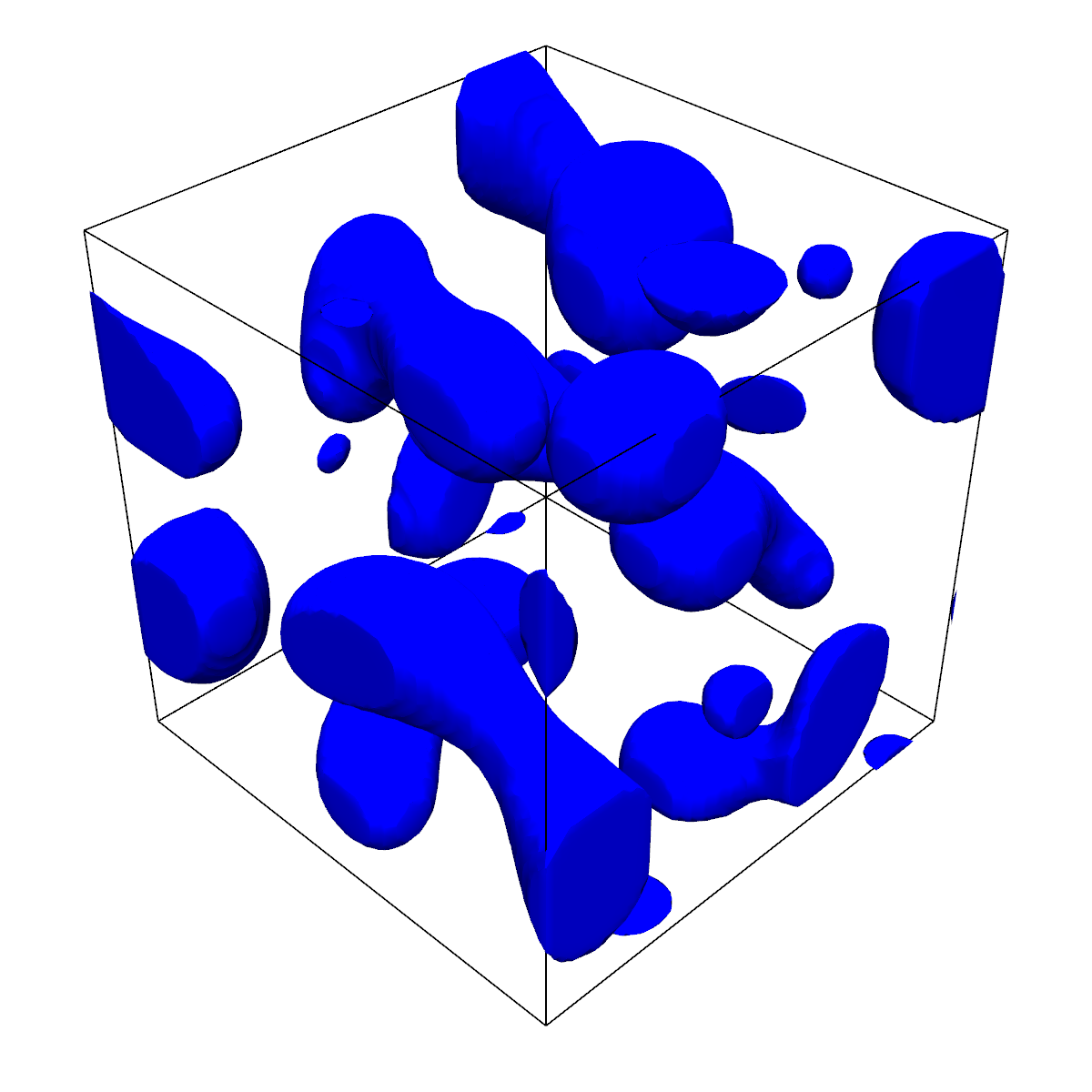}} &
        \raisebox{-0.2\height}{\includegraphics[width = 0.15\textwidth]{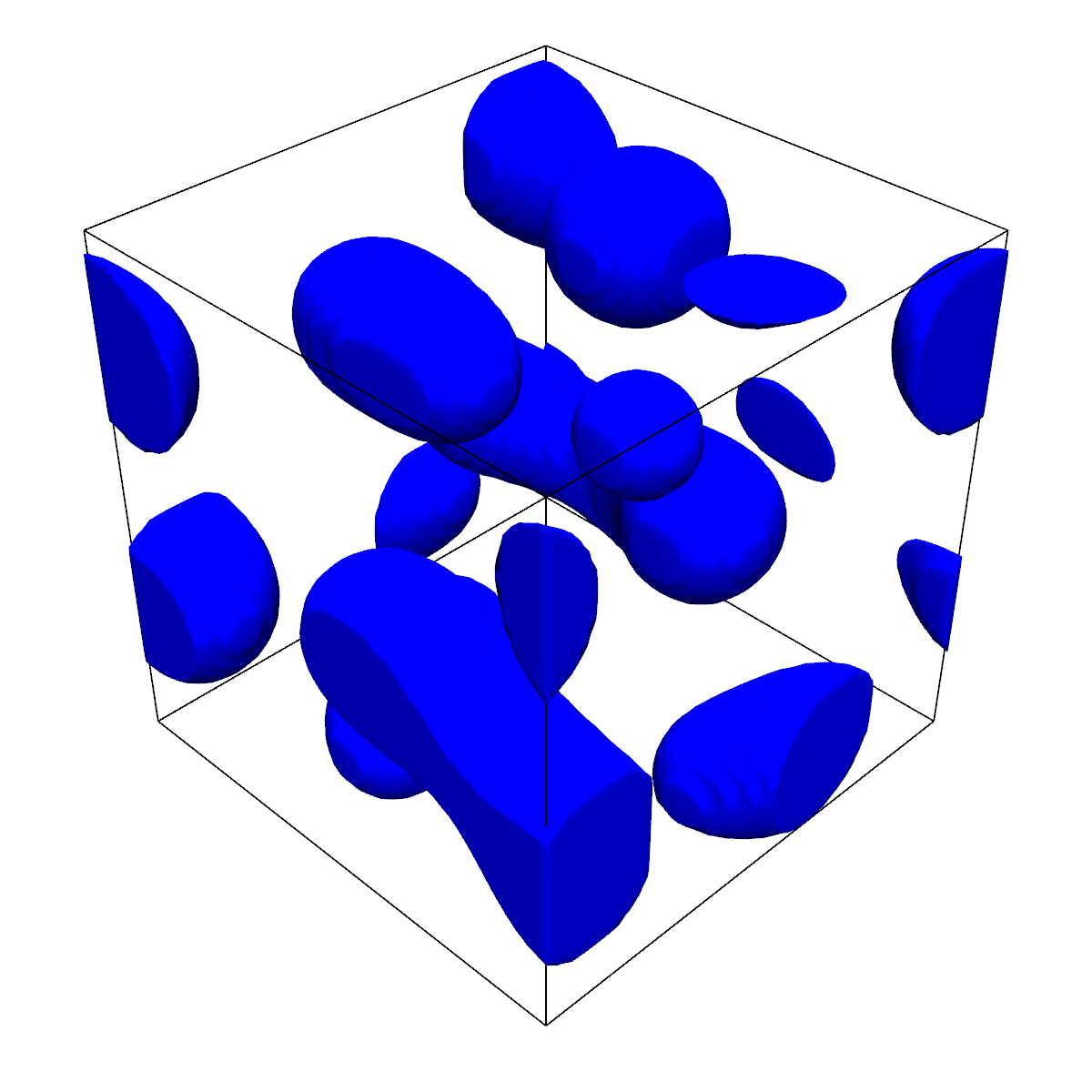}} &
        \raisebox{-0.2\height}{\includegraphics[width = 0.15\textwidth]{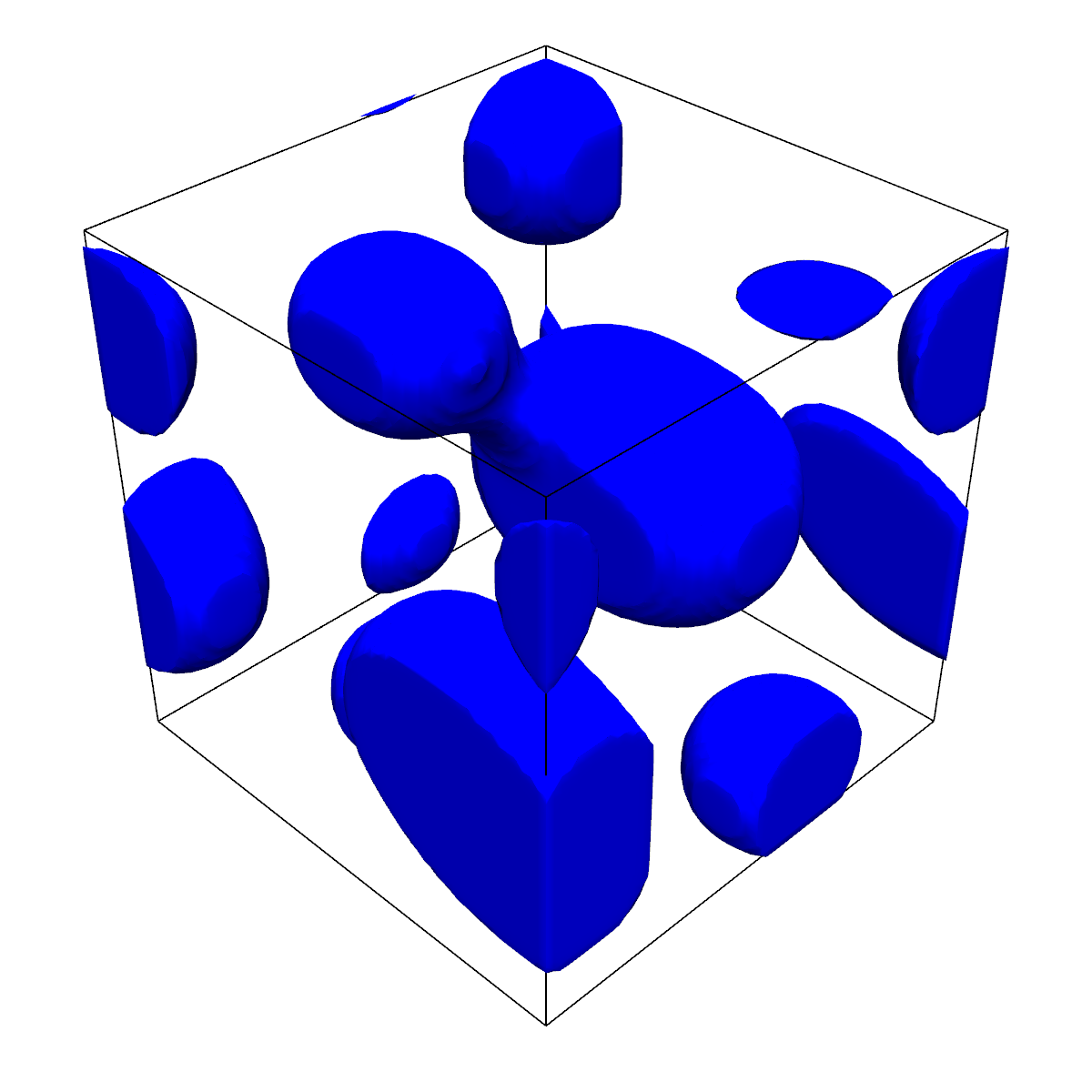}} &
        \raisebox{-0.2\height}{\includegraphics[width = 0.15\textwidth]{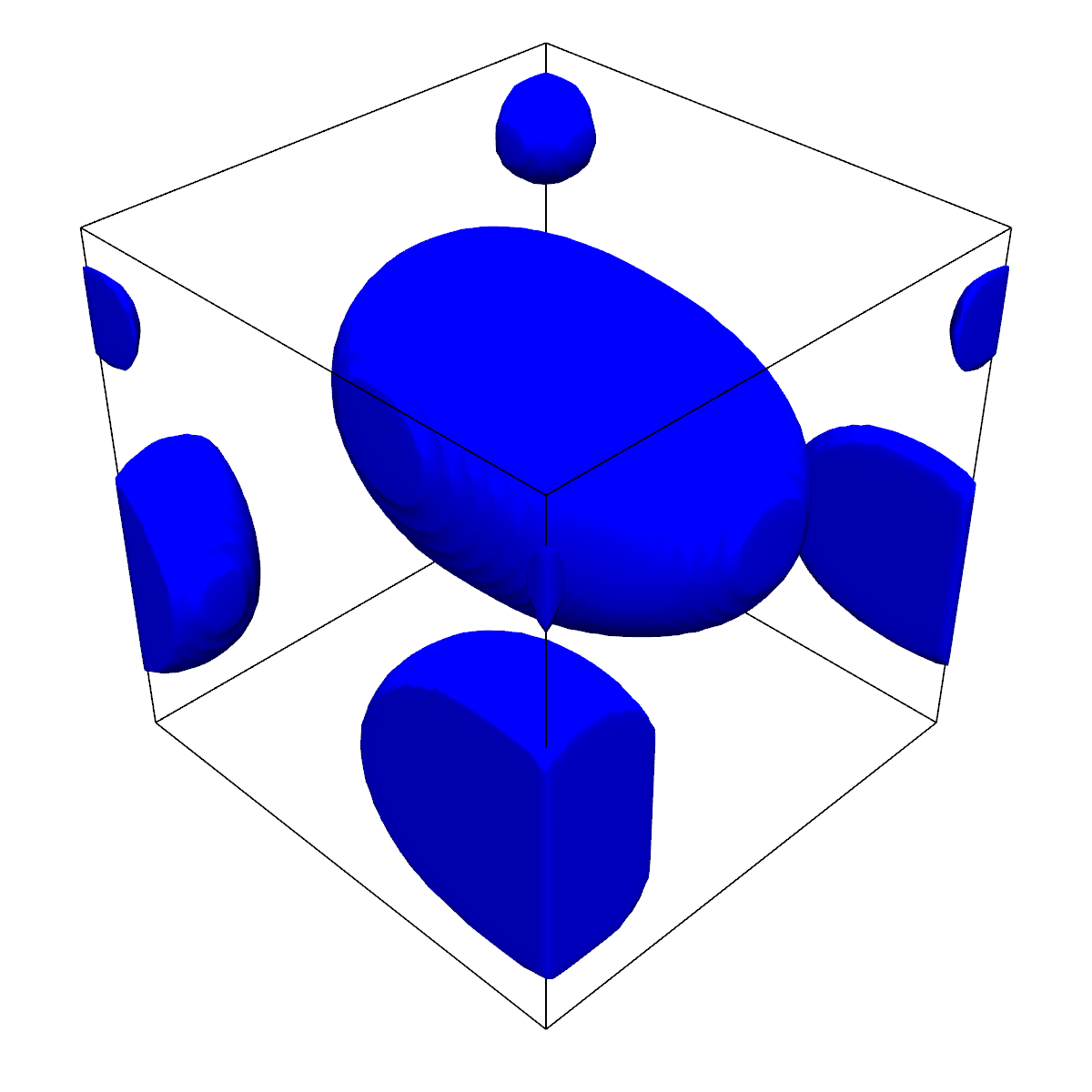}} &
        \raisebox{-0.2\height}{\includegraphics[width = 0.15\textwidth]{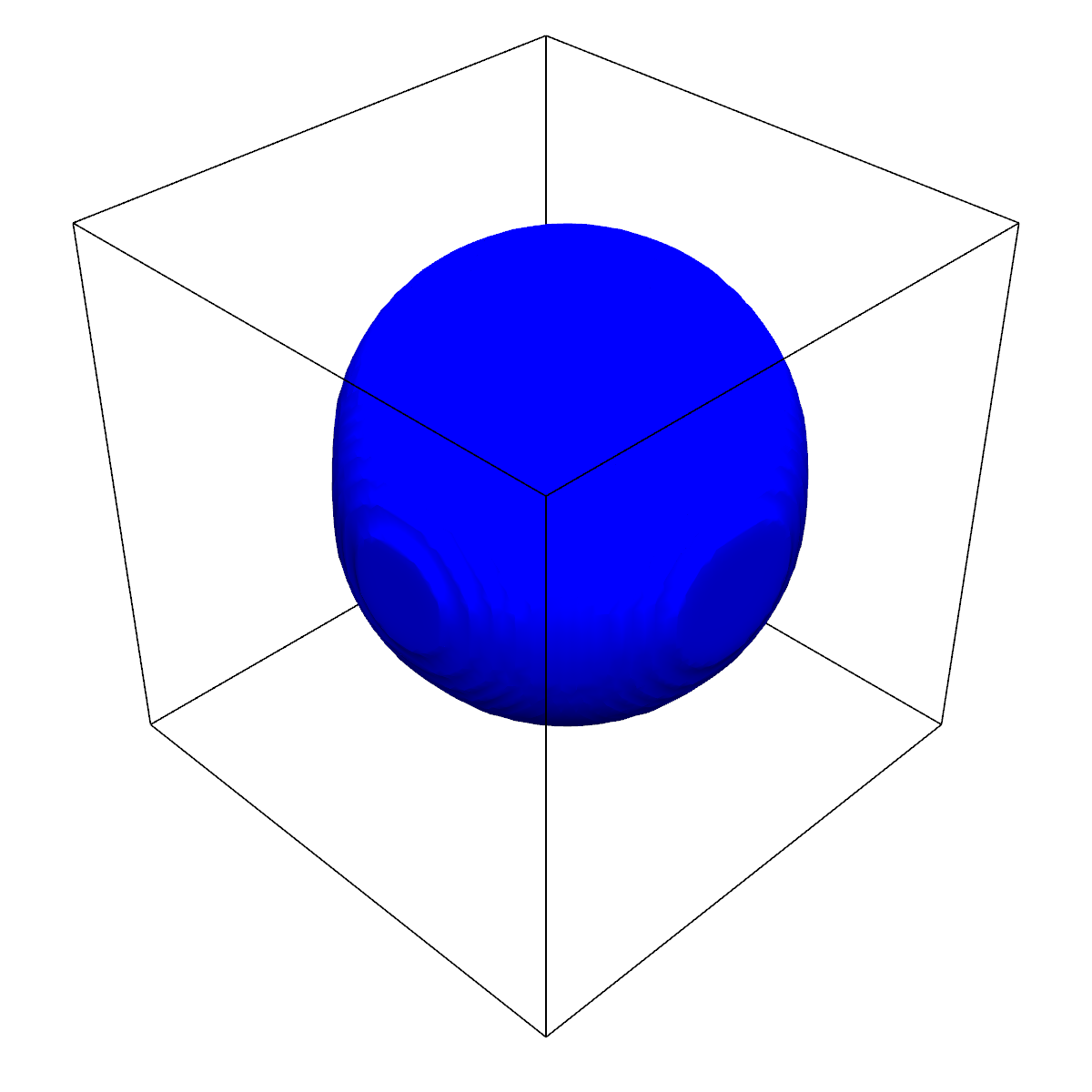}} \\
        & \multicolumn{6}{c}{(b)} \\[0.3em]
        \rotatebox[origin=l]{90}{$\epsilon=0.2$} &
        \raisebox{-0.2\height}{\includegraphics[width = 0.15\textwidth]{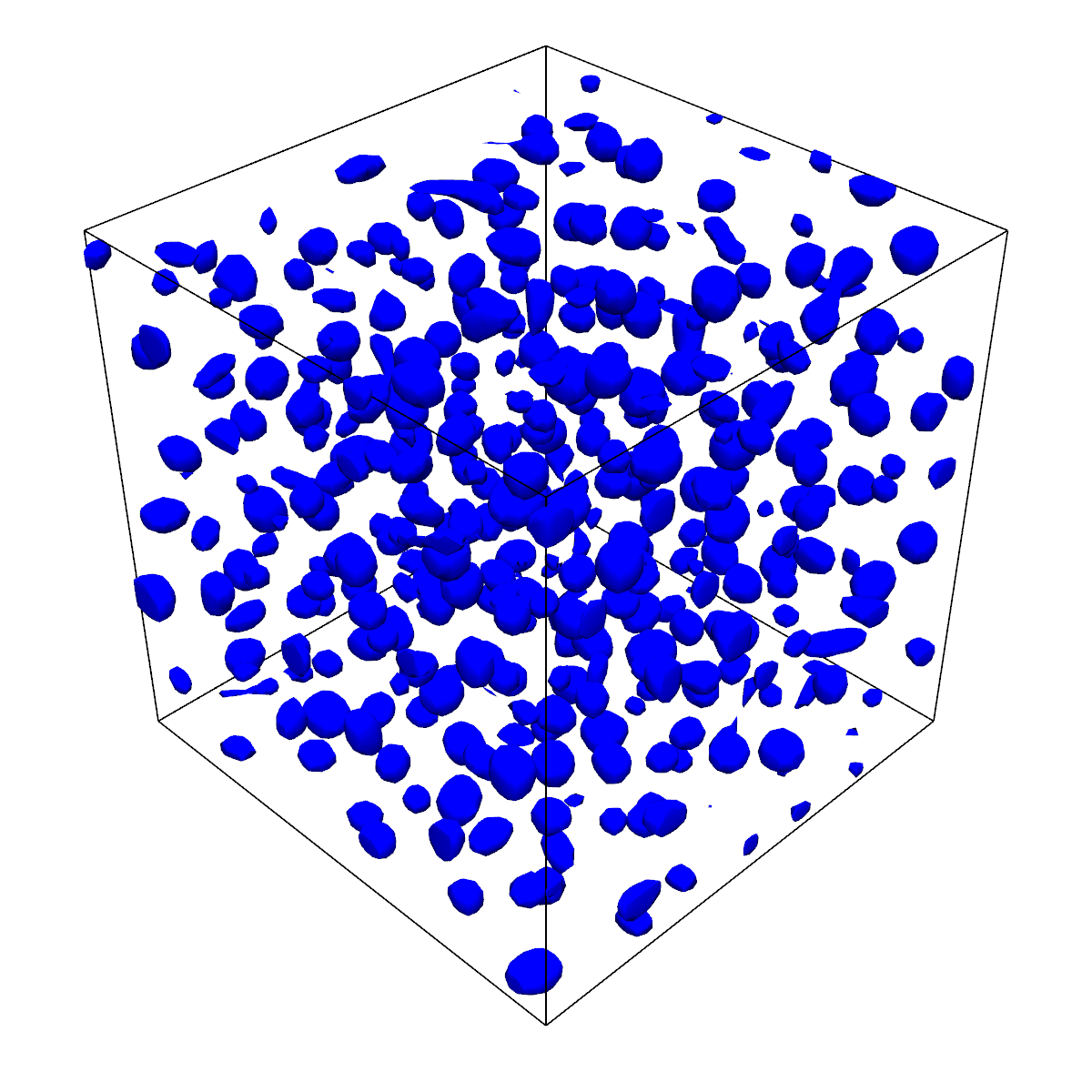}} &
        \raisebox{-0.2\height}{\includegraphics[width = 0.15\textwidth]{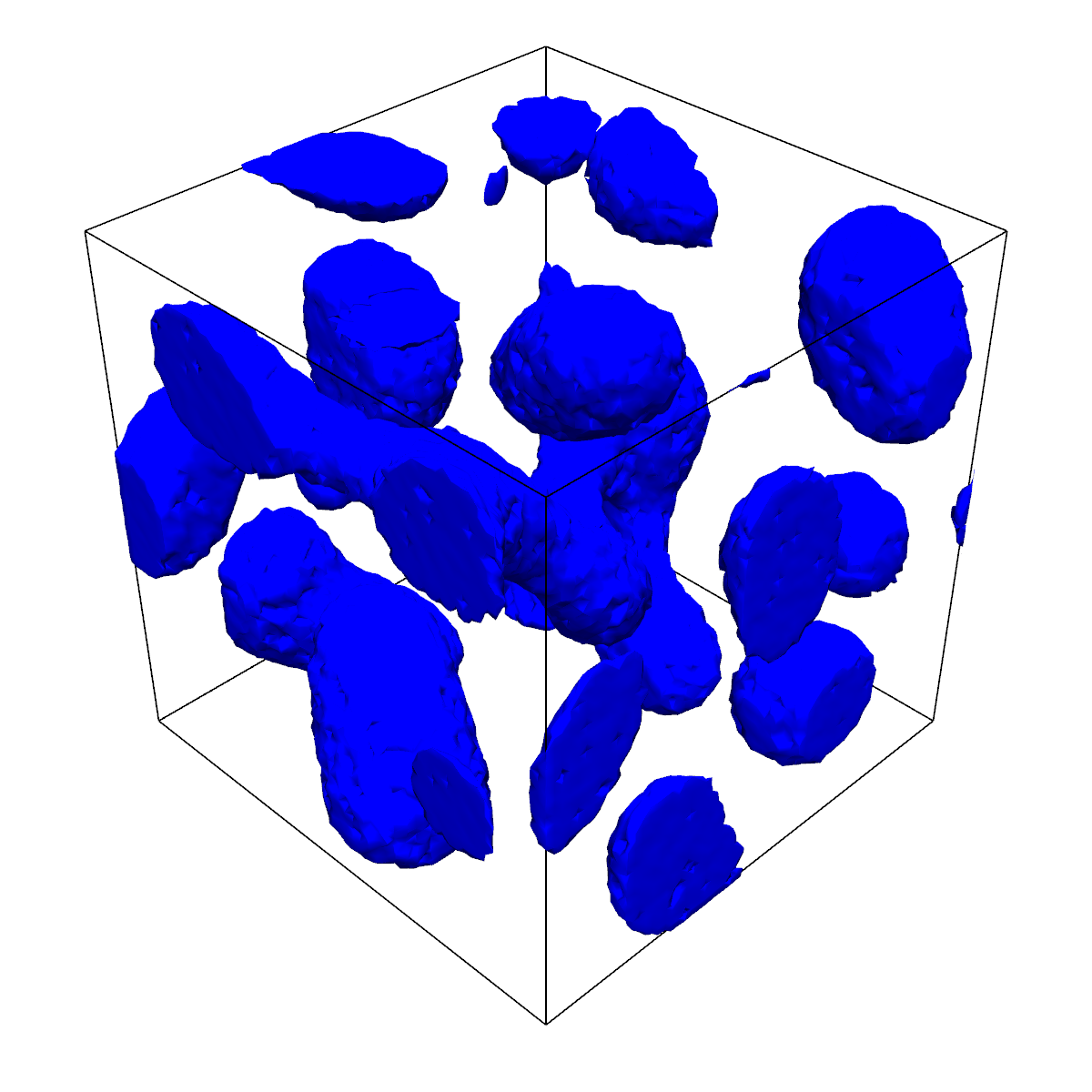}} &
        \raisebox{-0.2\height}{\includegraphics[width = 0.15\textwidth]{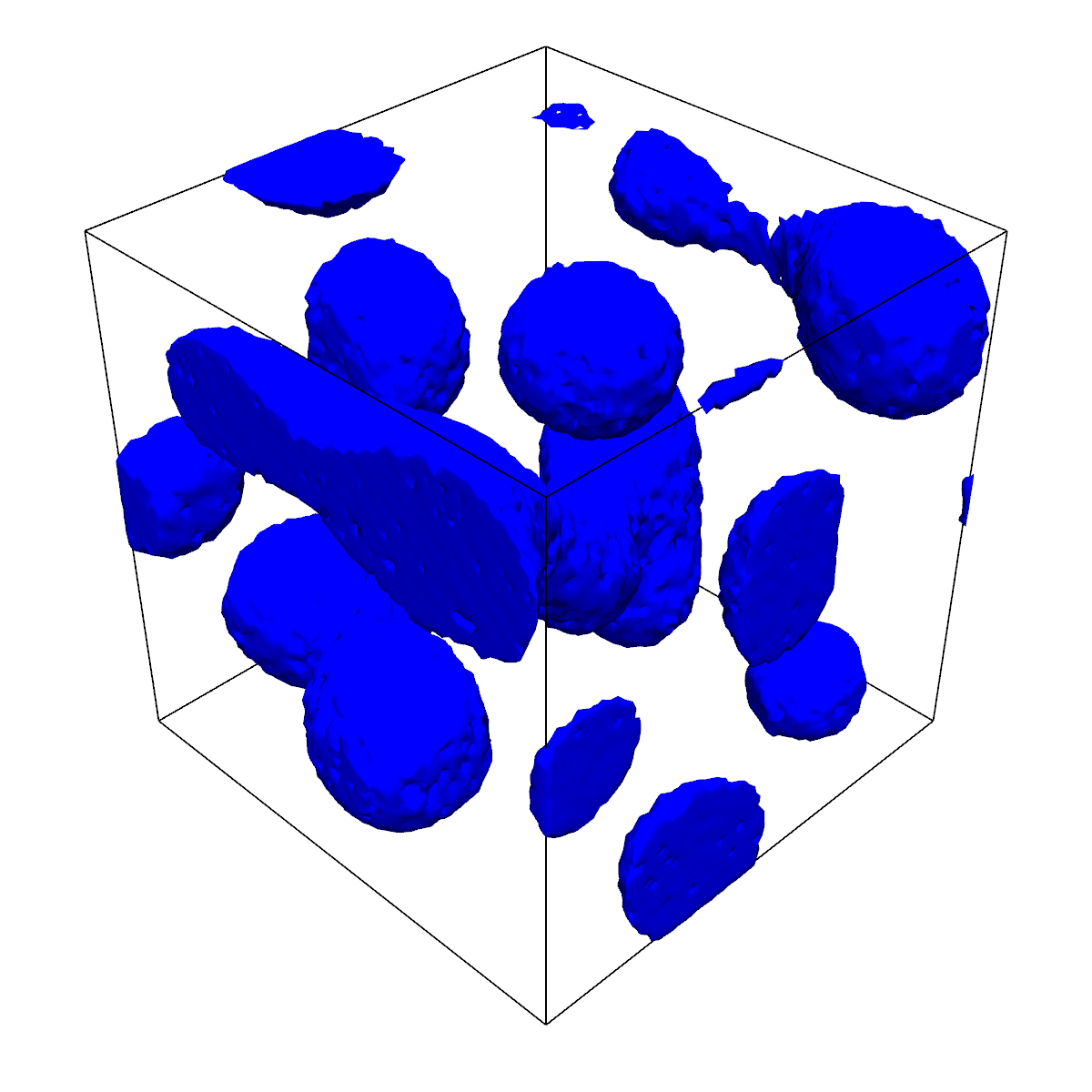}} &
        \raisebox{-0.2\height}{\includegraphics[width = 0.15\textwidth]{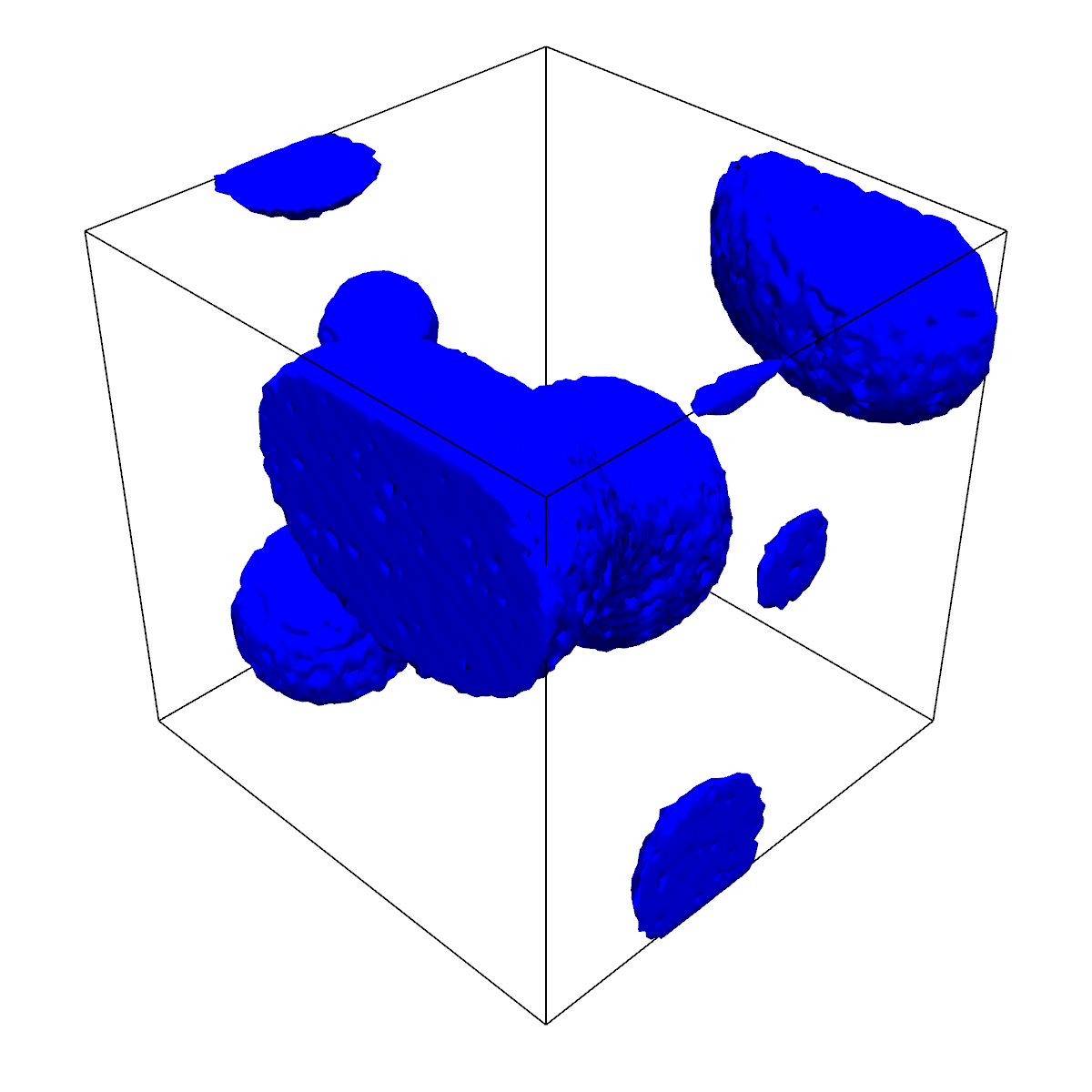}} &
        \raisebox{-0.2\height}{\includegraphics[width = 0.15\textwidth]{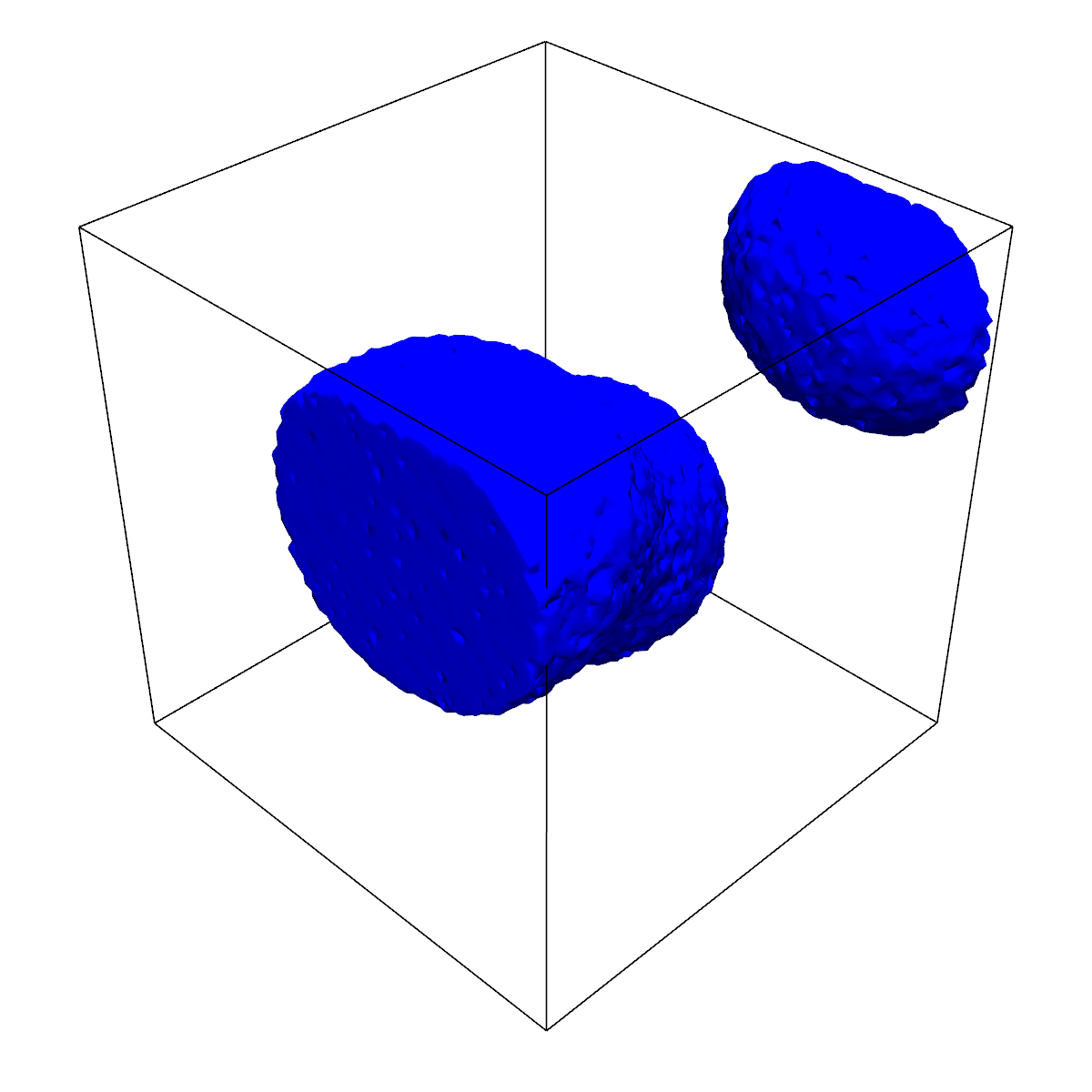}} &
        \raisebox{-0.2\height}{\includegraphics[width = 0.15\textwidth]{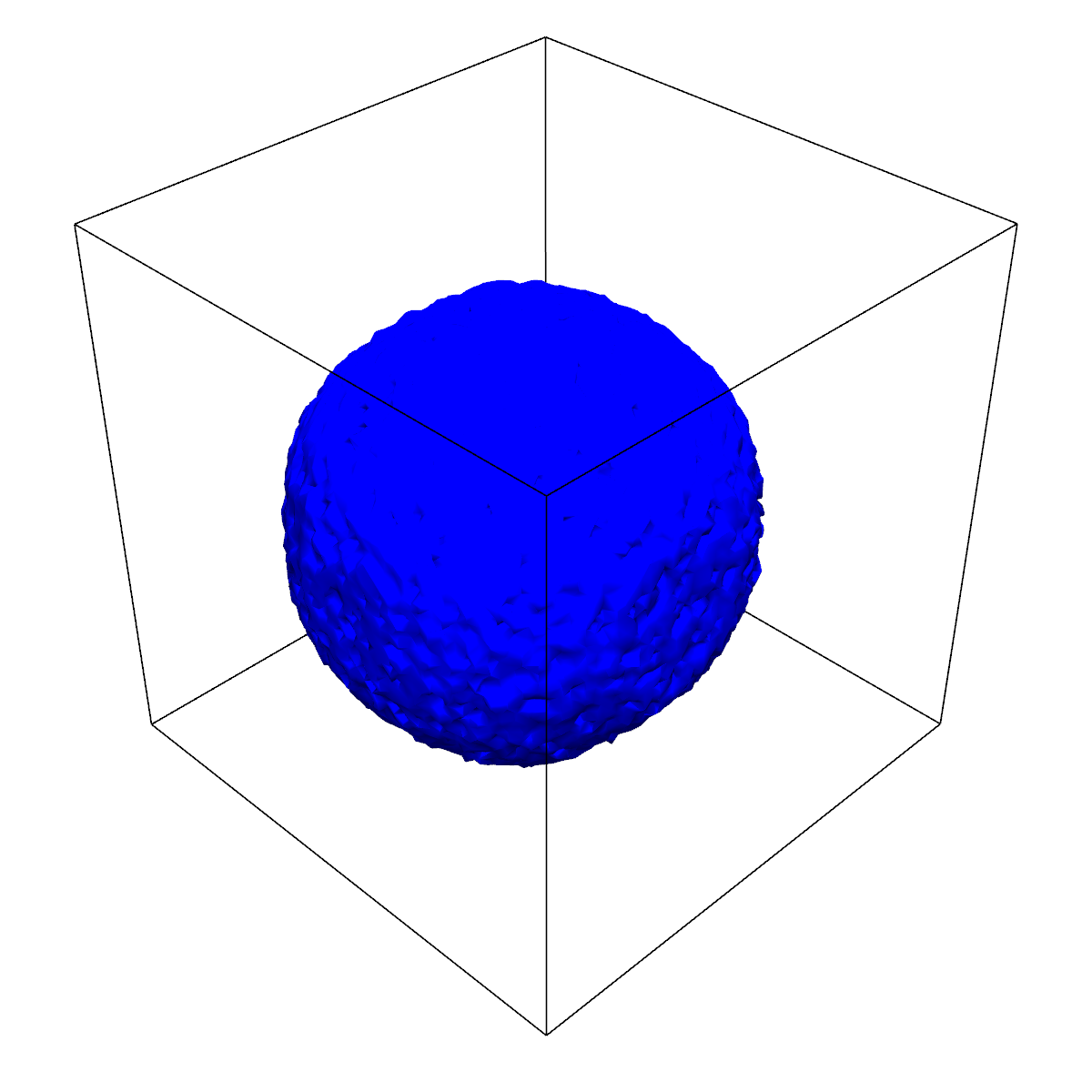}} \\
        & \multicolumn{6}{c}{(c)}
    \end{tabular}
    \caption{
    Example 3D SPDE trajectories of 70:30 mixture ($\langle c \rangle=0.4$) on a $64^3$ grid with $\epsilon=0.2$, starting from the same initial condition with slight perturbation $\mathcal{U}(-0.01,0.01)$. (a) Evolution of the number of precipitate clusters (voxels with $c \leq -0.8$) and the precipitate volume fraction. (b) Precipitate structure evolution for $\epsilon=0$ and (c) $\epsilon=0.2$.}
    \label{fig:GT_glomeration_GT}
\end{figure}

The second case on the more dilute 70:30 mixture ($\langle c \rangle = 0.4$) provides quantitative analysis for the number of clusters and their volume fractions of the precipitated minority species (A).  Here the volume is defined as the count of the phase-separated $c$ minority voxels with $c<-0.8$, and a corresponding minority cluster is a set of such voxels connected via voxel face, edge or corner while accounting for the periodic boundary condition.
For better statistics, the simulations were performed on a larger periodic $64^3$ grid for $4\times10^6$ steps with time interval $1\times 10^{-3}$ and total simulation time $4\times 10^{3}$.  
%And the connectivity is in three dimensional coordinates counts 26 neighbors, including face, edge and corner neighbors. 
%Moroever, volume fraction is the fraction of the domain that belongs to the phase separated regions (threshold $c<-0.8$). 
Fig.~\ref{fig:GT_glomeration_GT}a shows the average over 10 trajectories $\epsilon=0$) and $\epsilon=0.2$. 
The number of clusters decreases faster in the presence of noise up to time $t \approx 10$ with about 10 precipitate clusters remaining, suggesting that the system evolves faster with stochastic fluctuations. We refrain from drawing quantitative conclusions beyond $t \approx 10$ due to insufficient statistics. The volume fraction (sum of $c$) of the minority clusters is higher with noise in the very beginning, then becomes lower than $\epsilon=0$.
%, meaning that the average $c$ of these ``noisy'' clusters is lower than without noise. This is understandable as the simulation without noise is  
Fig.~\ref{fig:GT_glomeration_GT}b ($\epsilon=0$) and c ($\epsilon=0.2$) show example snapshots from $t=4$ to $t=4000$. 
%The snapshot at $t=500$ clearly shows the $\epsilon=0.2$ case has less number of clusters, while in the end ($t=4000$) both cases converge to spherical surfaces given the initial conditions.
%and the volume fraction in noisy setting is slightly smaller for $\epsilon=0.2$, which also indicates the system evolves faster when adding stochastic noise terms. Subplot (b) and (c) 

\setlength{\tabcolsep}{2pt}

\section{Results on surrogate models}
\label{sec:results_ml}

In this section, we will evaluate the performance of the ML surrogate models outlined in Sec.~\ref{sec:surrogate_model}. The blackbox, non-flux model is first evaluated and ruled out in Sec.~\ref{sec:results-blackbox}.
The flux-based models were analyzed for in-distribution performance in Sec.~\ref{sec:same_mesh_grid_results} and for spatio-temporal extrapolation tests in Sec.~\ref{sec:long_agglo}. Moreover, we validate the learned free energy in Sec.~\ref{sec:ml_free_E}. Lastly, we also explore the extrapolation ability for precipitation at  concentrations unseen in training, in Sec.~\ref{sec:extra_con}.

Due to the stochastic nature of the ground truth data source and the goal of long-term predictions, voxel-wise metrics alone are insufficient for model assessment. We therefore focus on statistical and physics-based criteria, including rollout stability, phase-separation kinetics, learned mobility, free-energy interpretability, and precipitation behavior. The models were trained on trajectories with random initial $\langle c \rangle \sim \mathcal{U}(-0.5, 0.5)$, and tested on two representative cases, $\langle c \rangle=0$ and $\langle c \rangle=0.4$. We first show that the black-box non-flux baseline fails in spatial extrapolation and exclude it from further consideration. We then compare the flux-based surrogates, showing that \femv{} best matches the stochastic dynamics, while the \nfem{} and \nfemv{} are less physically consistent as judged by the learned mobility relation.

\subsection{Blackbox non-conservative baseline vs conservative models}
\label{sec:results-blackbox}
% Show that generalization is poor for blackbox model in just one plot, so that we do not need to include blackbox no flux model results in the following.

\begin{figure}[htp]
    \centering
    \begin{tabular}{c c c c c c c}
        & $t=4$ & $100$ & $200$ & $500$ & $1000$ & $4000$ \\
        \rotatebox[origin=l]{90}{GT \, $\epsilon=0.2$} &
        \raisebox{-0.0\height}{\includegraphics[width = 0.15\textwidth]{fig/Final/GT_64cube_T0s_noise0p2.png}} &
        \raisebox{-0.0\height}{\includegraphics[width = 0.15\textwidth]{fig/Final/GT_64cube_T100s_noise0p2.png}} &
        \raisebox{-0.0\height}{\includegraphics[width = 0.15\textwidth]{fig/Final/GT_64cube_T200s_noise0p2.png}} &
        \raisebox{-0.0\height}{\includegraphics[width = 0.15\textwidth]{fig/Final/GT_64cube_T500s_noise0p2.png}} &
        \raisebox{-0.0\height}{\includegraphics[width = 0.15\textwidth]{fig/Final/GT_64cube_T1000s_noise0p2.png}} &
        \raisebox{-0.0\height}{\includegraphics[width = 0.15\textwidth]{fig/Final/GT_64cube_T4000s_noise0p2.png}} \\
        & \multicolumn{6}{c}{(a)} \\[0.3em]
        \rotatebox[origin=l]{90}{\nflux} &
        \raisebox{-0.1\height}{\includegraphics[width = 0.15\textwidth]{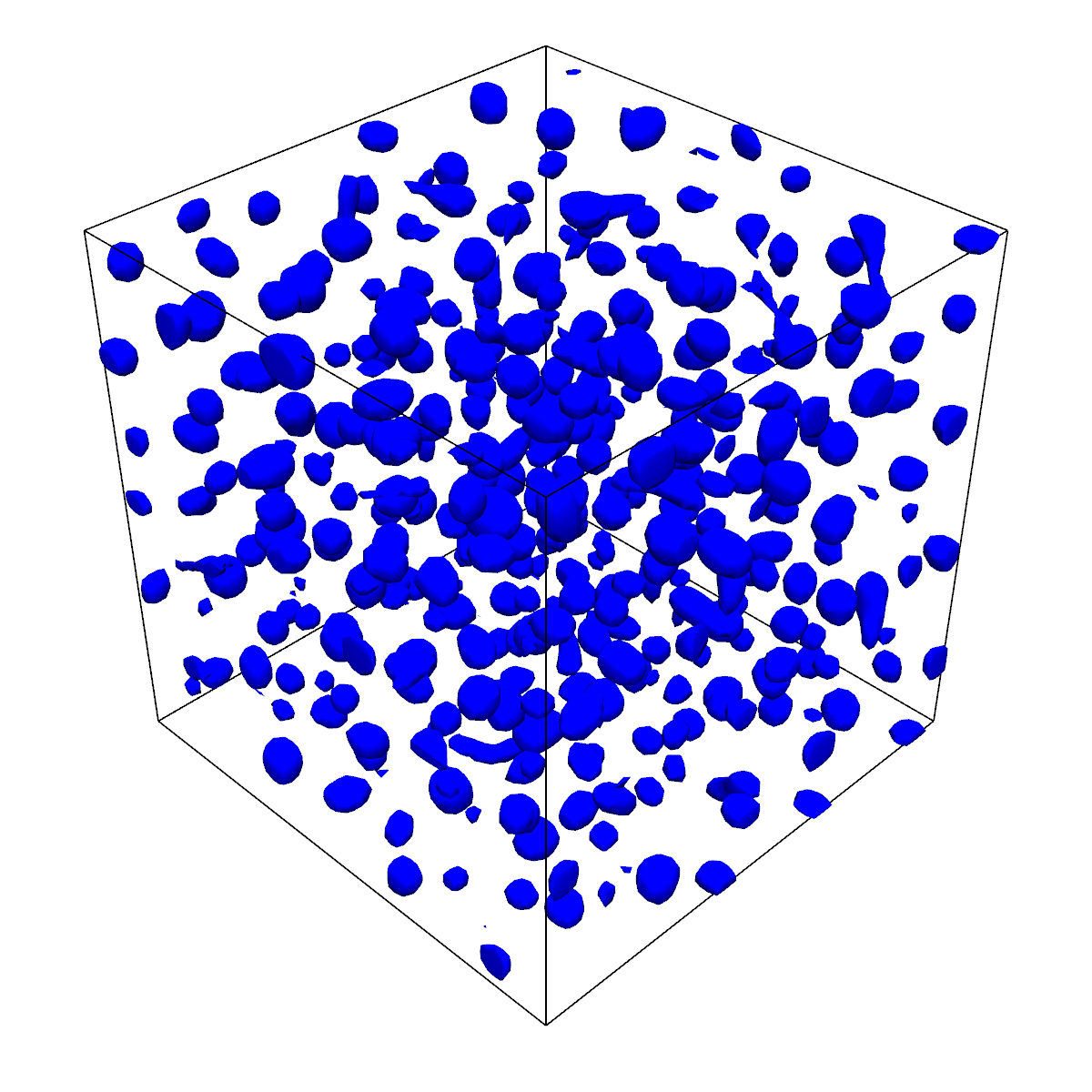}} &
        \raisebox{-0.1\height}{\includegraphics[width = 0.15\textwidth]{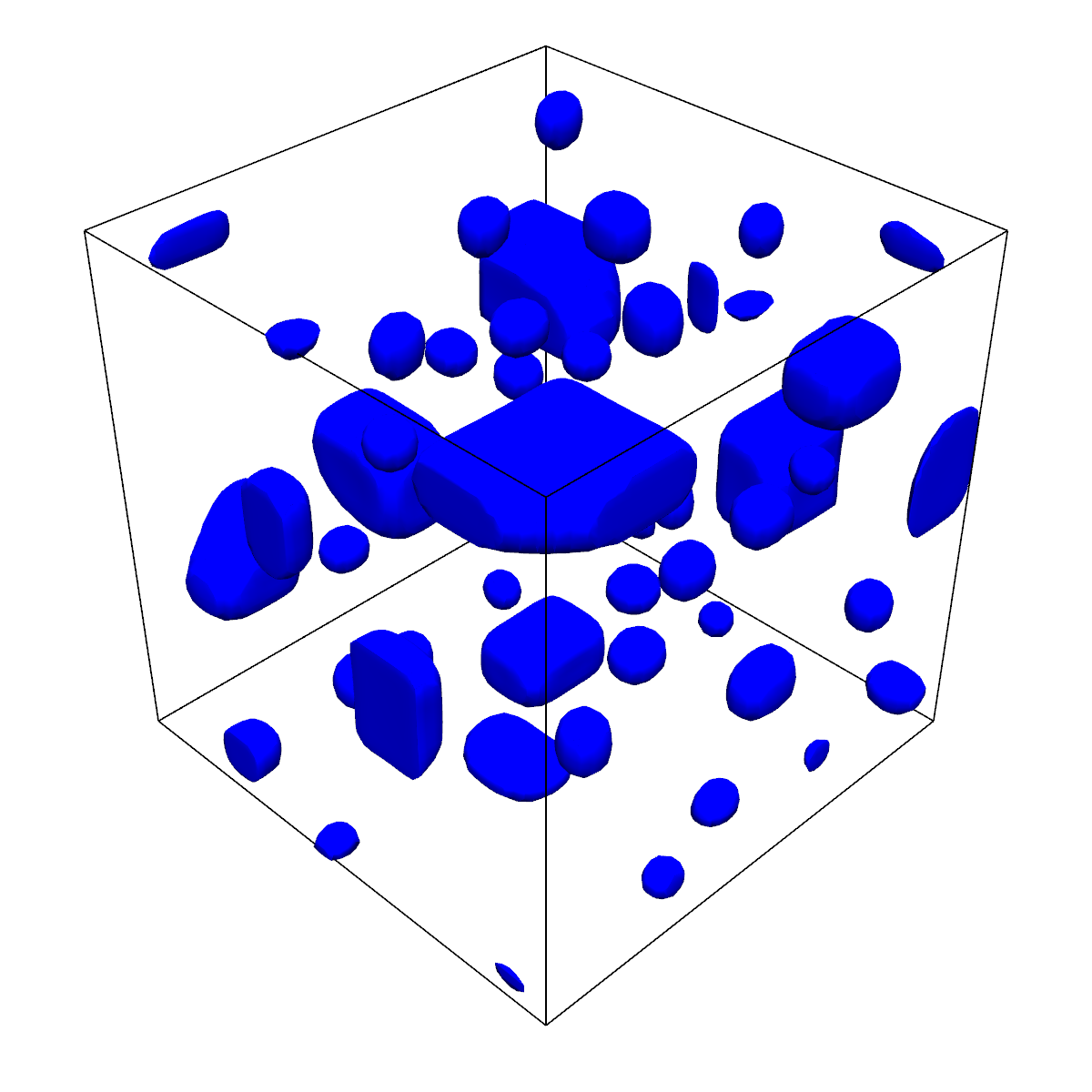}} &
        \raisebox{-0.1\height}{\includegraphics[width = 0.15\textwidth]{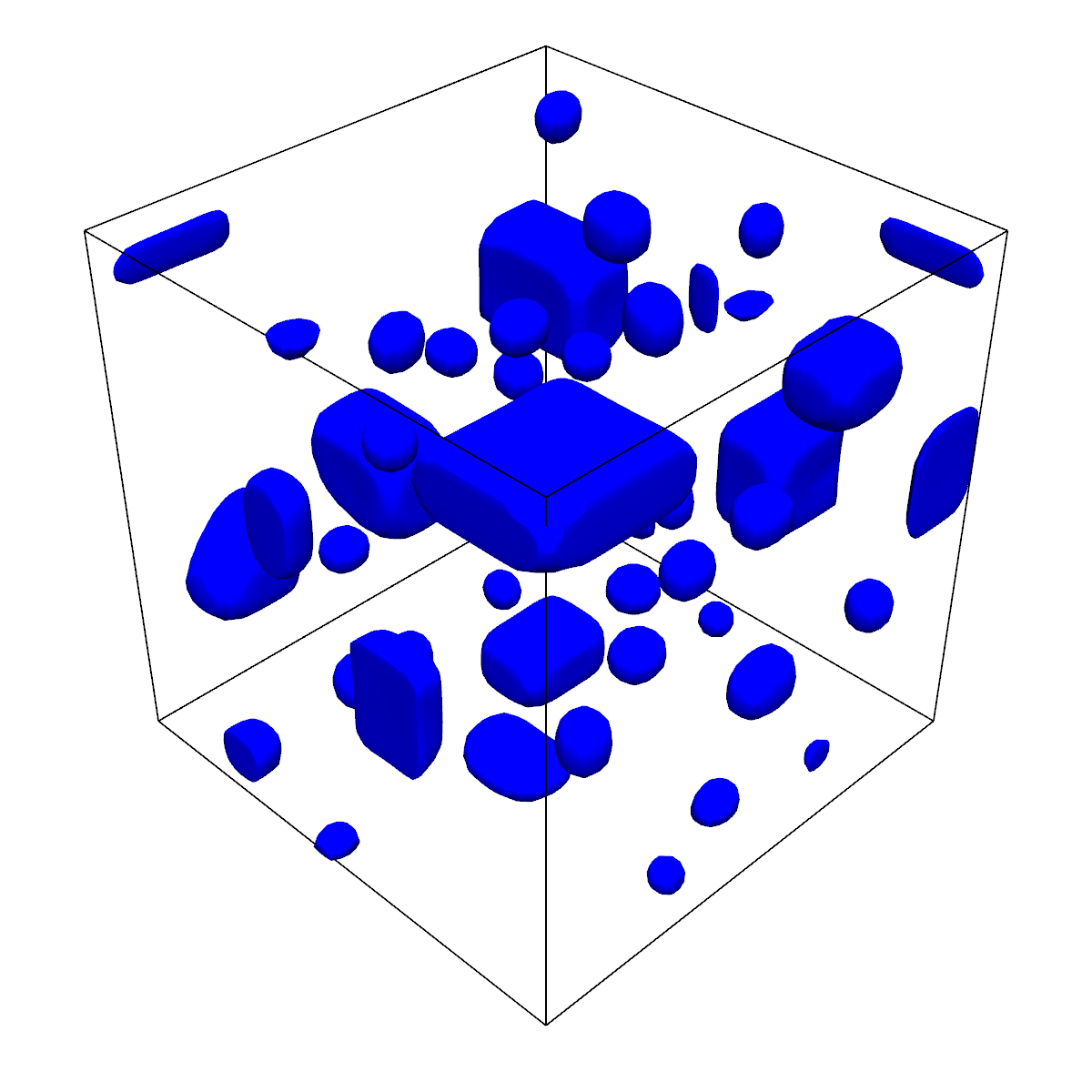}} &
        \raisebox{-0.1\height}{\includegraphics[width = 0.15\textwidth]{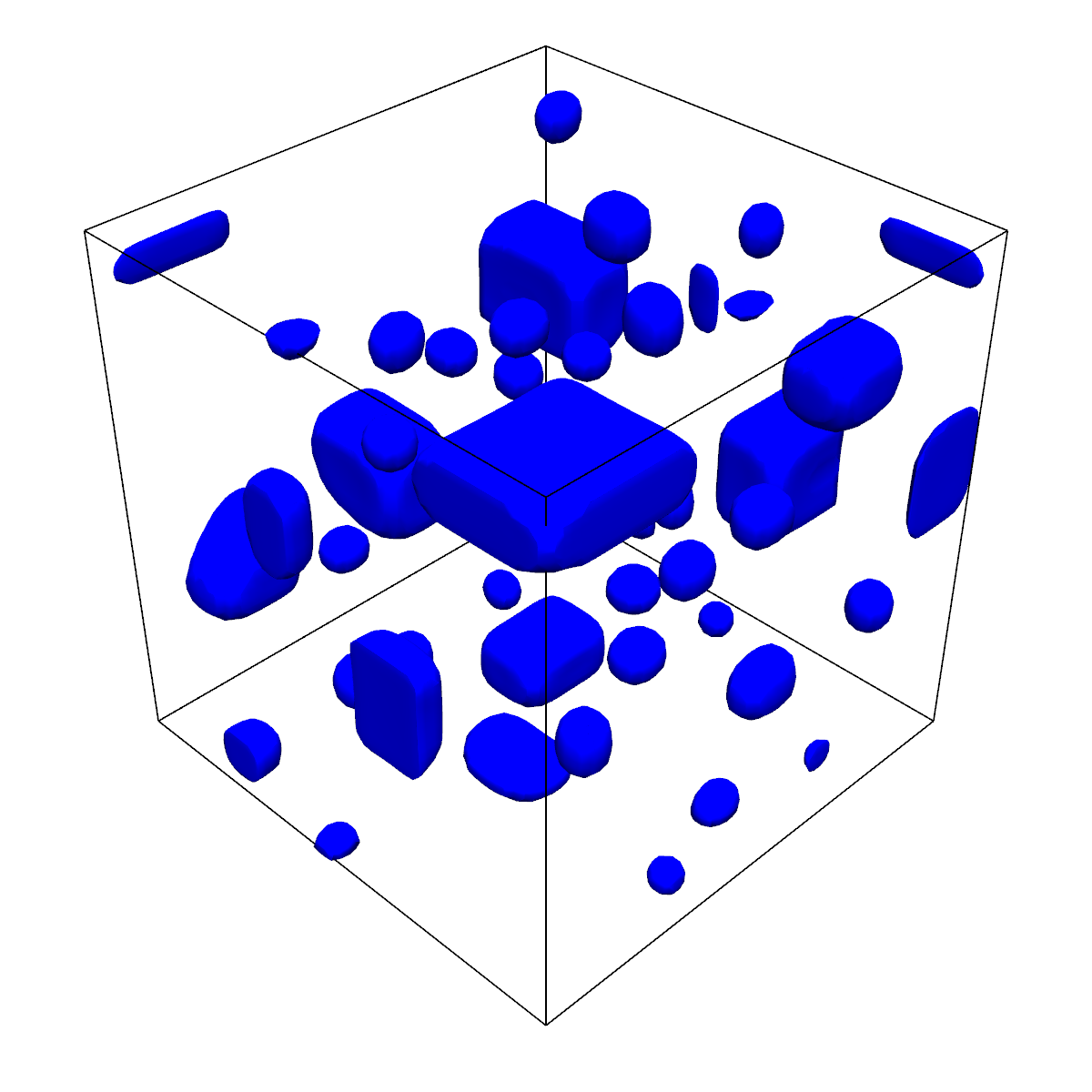}} &
        \raisebox{-0.1\height}{\includegraphics[width = 0.15\textwidth]{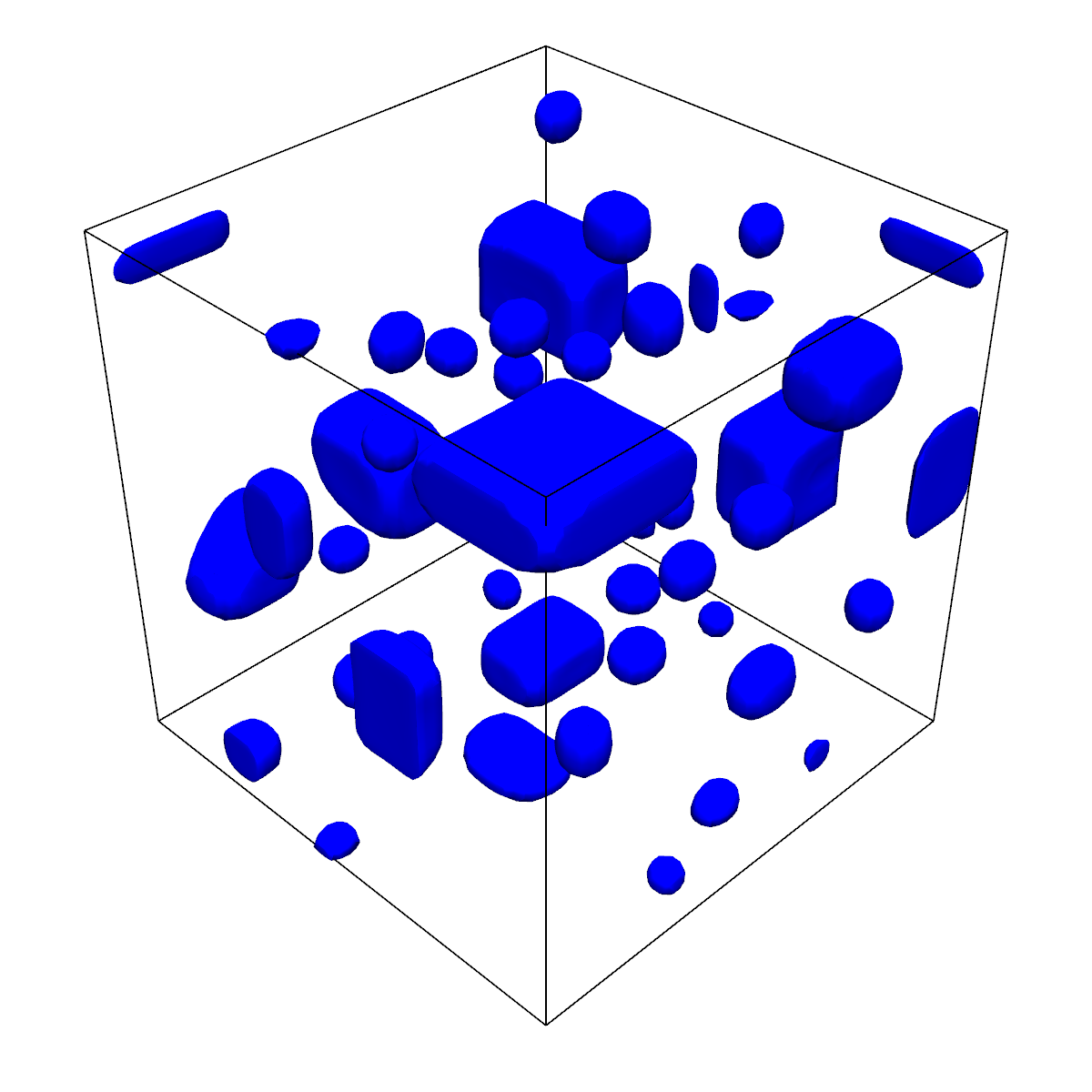}} &
        \raisebox{-0.1\height}{\includegraphics[width = 0.15\textwidth]{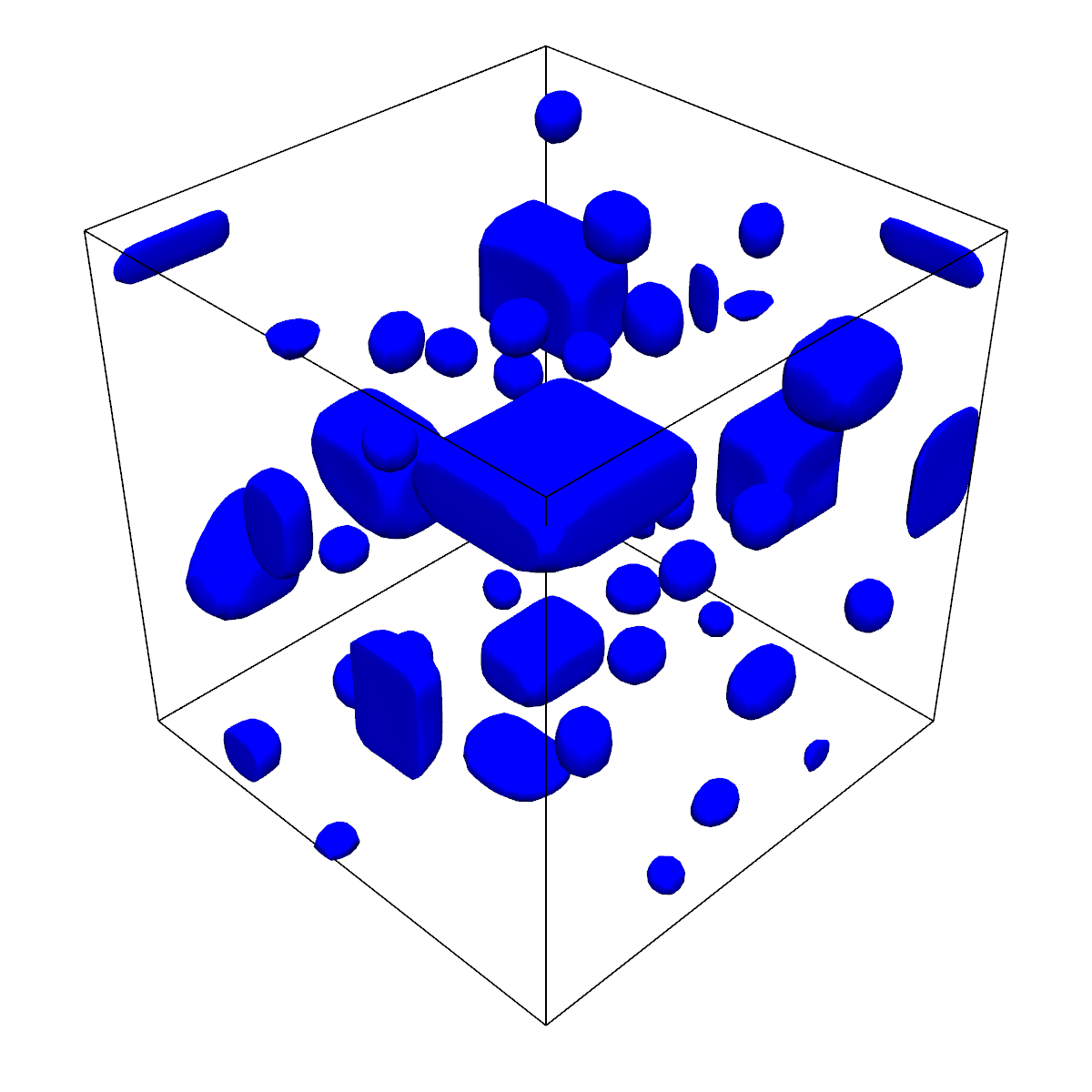}} \\
        & \multicolumn{6}{c}{(b)} \\[0.3em]
        \rotatebox[origin=l]{90}{\femv} &
        \raisebox{-0.1\height}{\includegraphics[width = 0.15\textwidth]{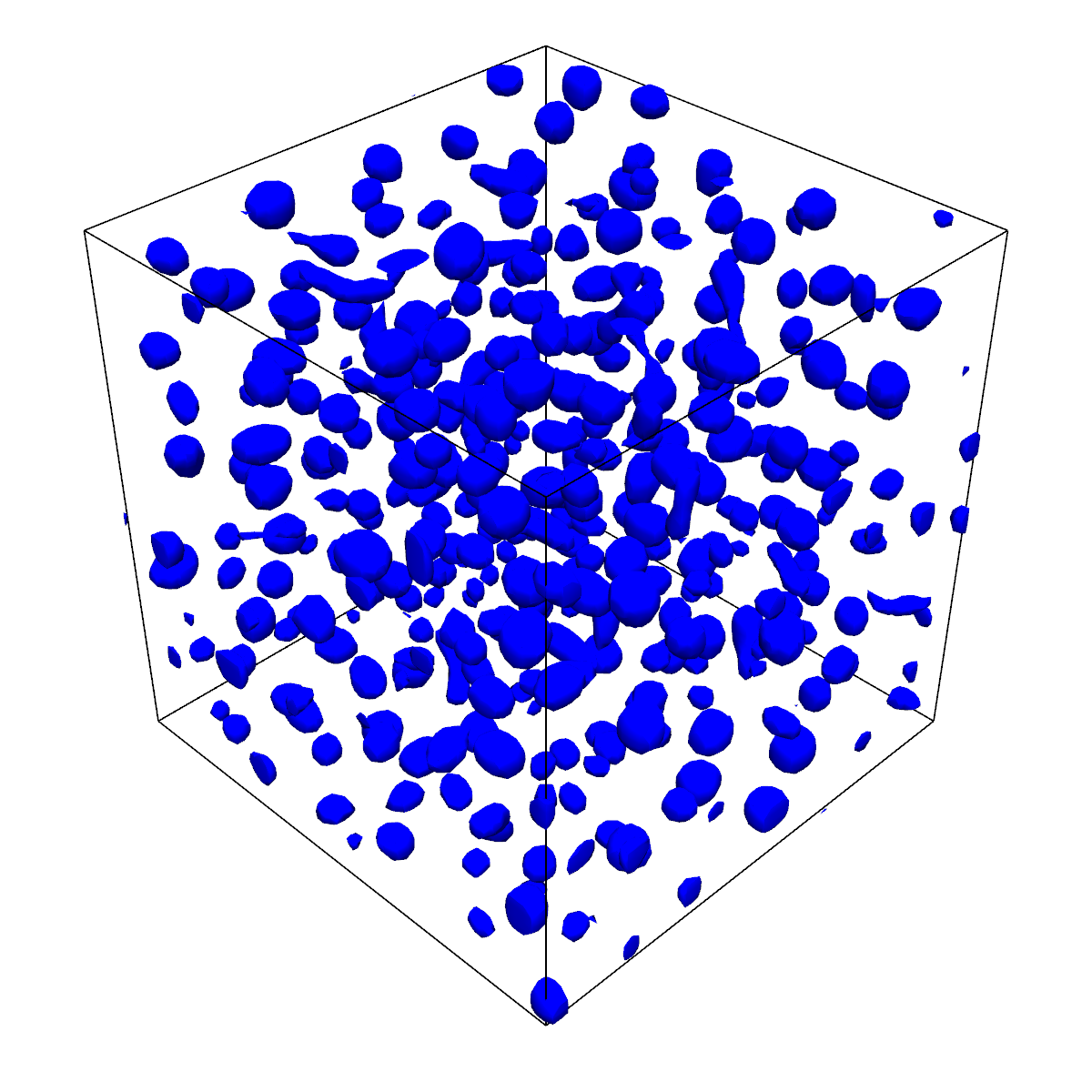}} &
        \raisebox{-0.1\height}{\includegraphics[width = 0.15\textwidth]{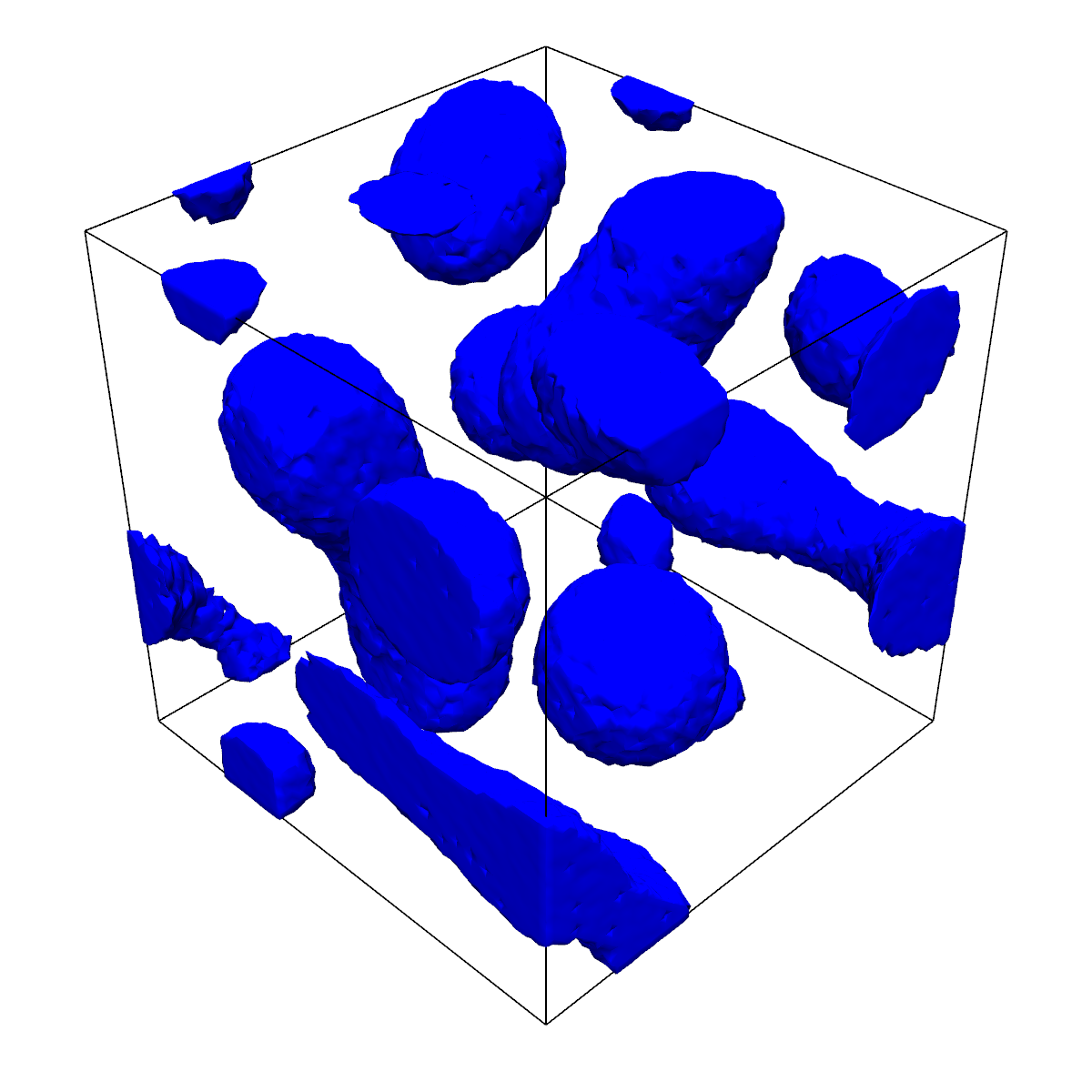}} &
        \raisebox{-0.1\height}{\includegraphics[width = 0.15\textwidth]{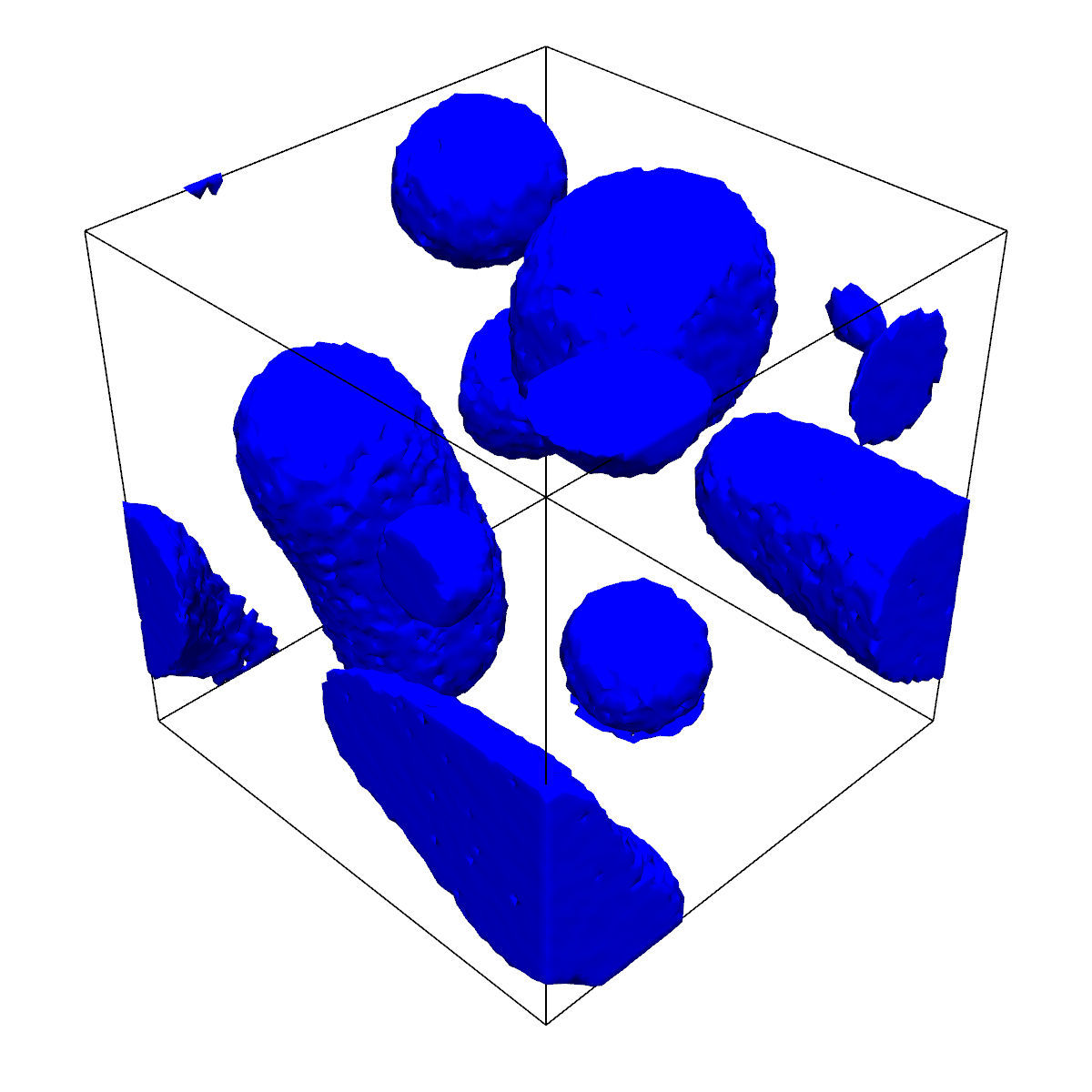}} &
        \raisebox{-0.1\height}{\includegraphics[width = 0.15\textwidth]{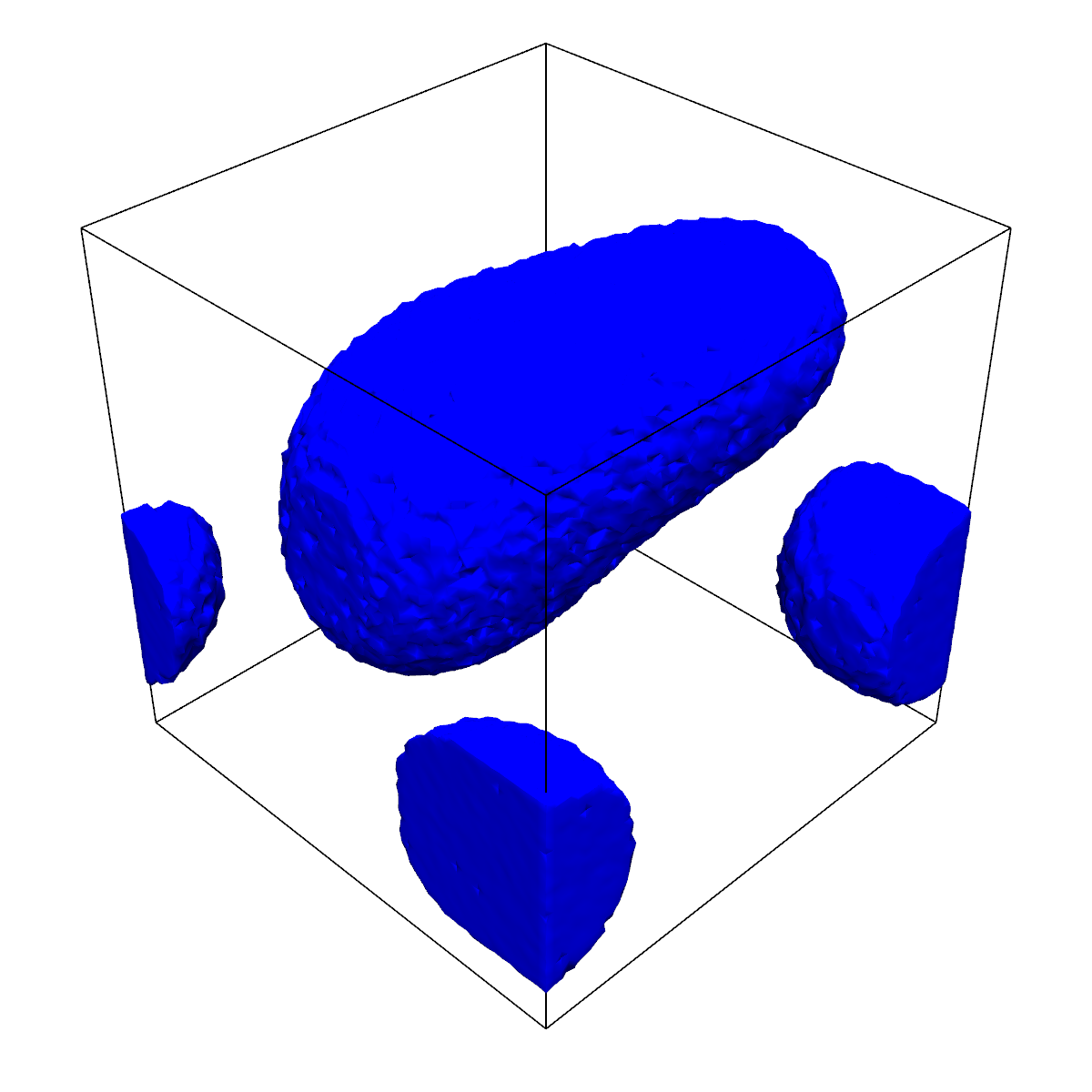}} &
        \raisebox{-0.1\height}{\includegraphics[width = 0.15\textwidth]{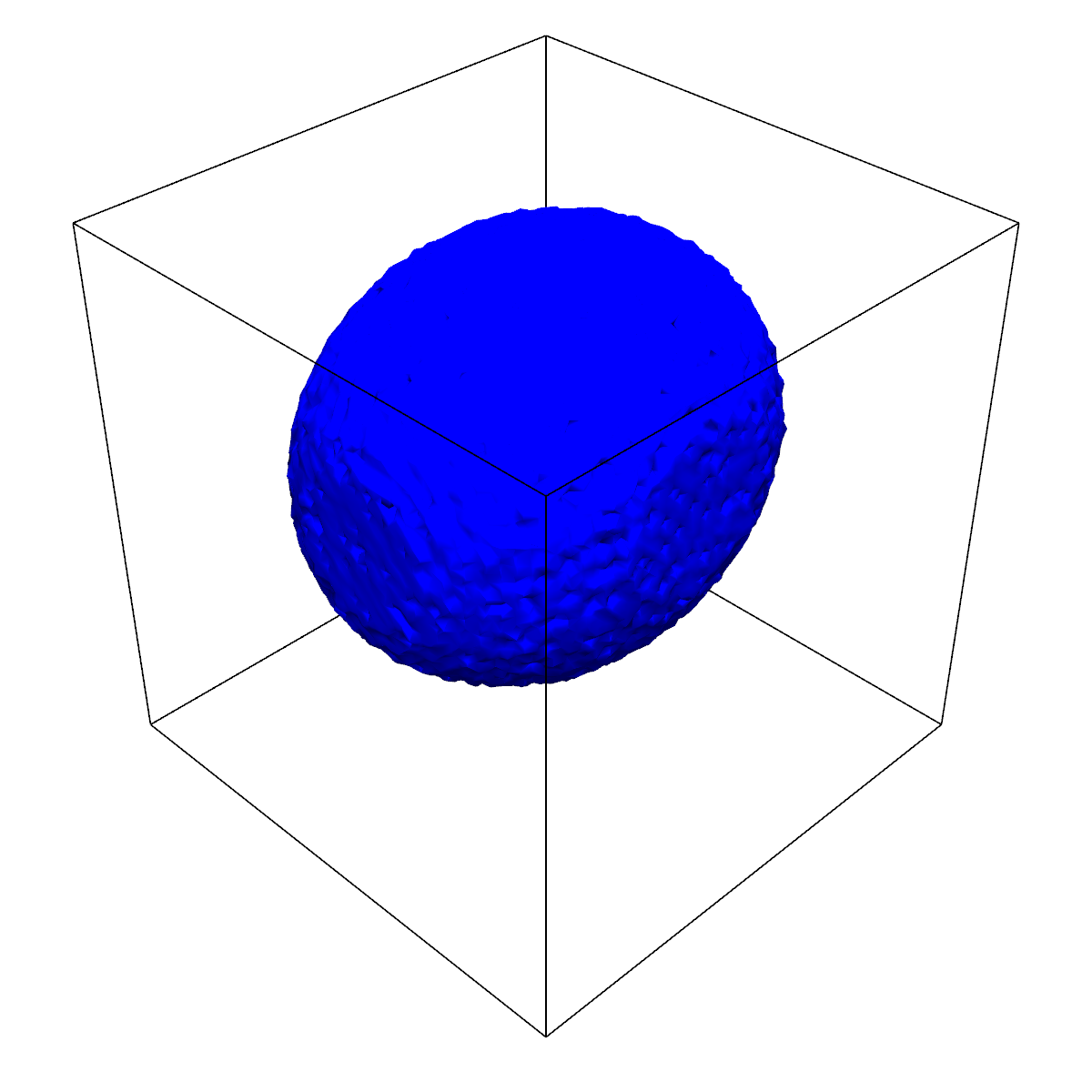}} &
        \raisebox{-0.1\height}{\includegraphics[width = 0.15\textwidth]{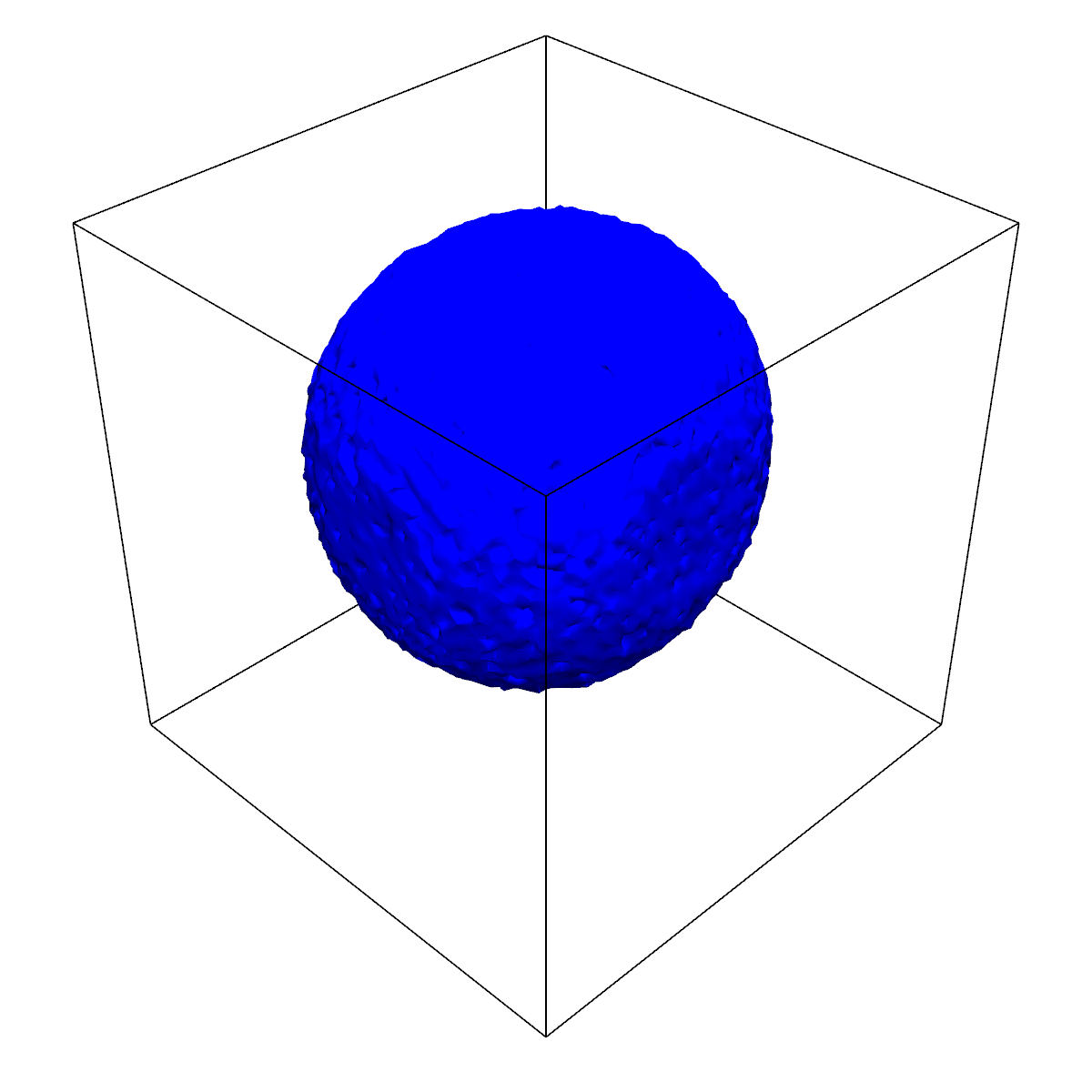}} \\
        & \multicolumn{6}{c}{(c)}\\[0.3em]
        \rotatebox[origin=l]{90}{\fem} &
        \raisebox{-0.1\height}{\includegraphics[width = 0.15\textwidth]{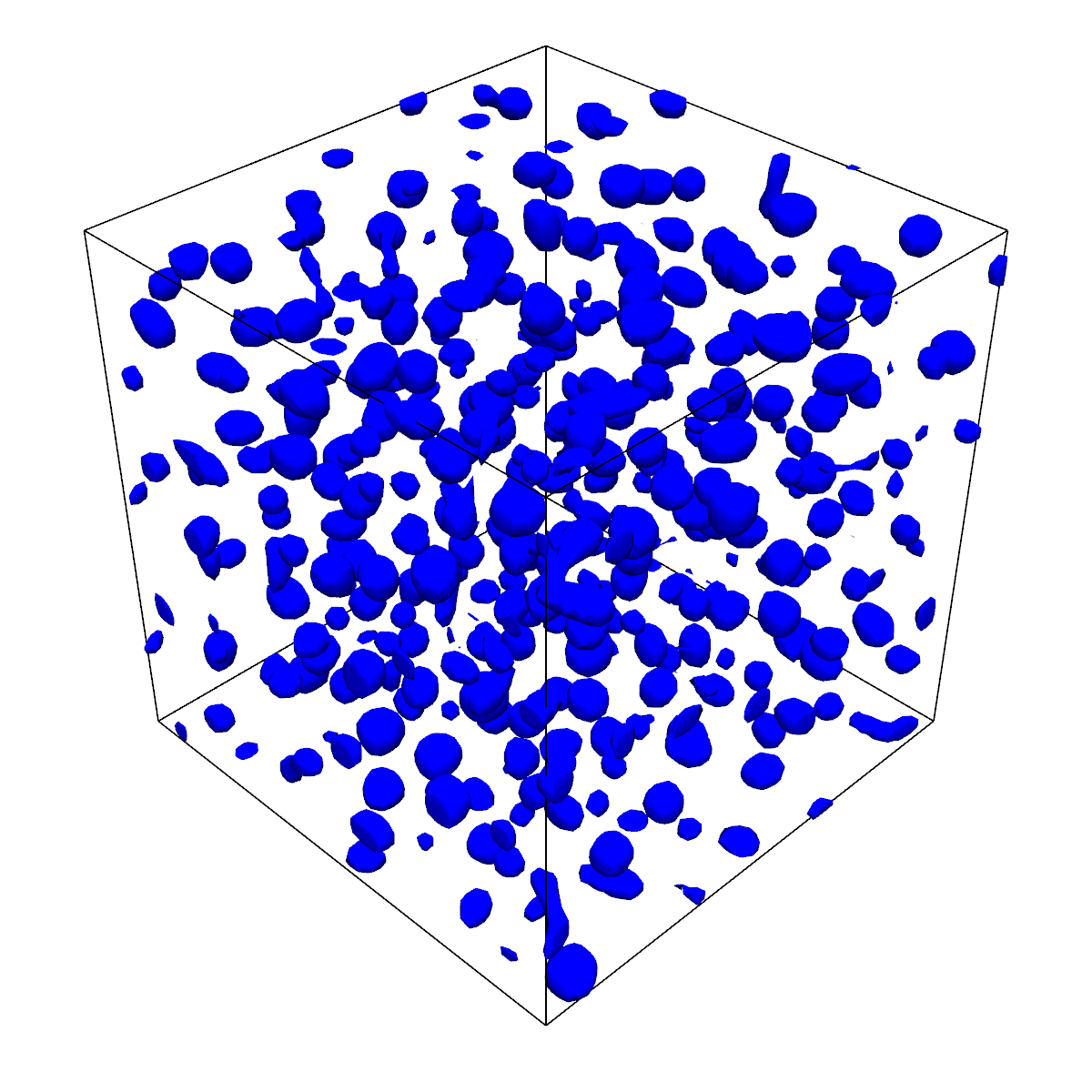}}&
        {\includegraphics[width = 0.15\textwidth]{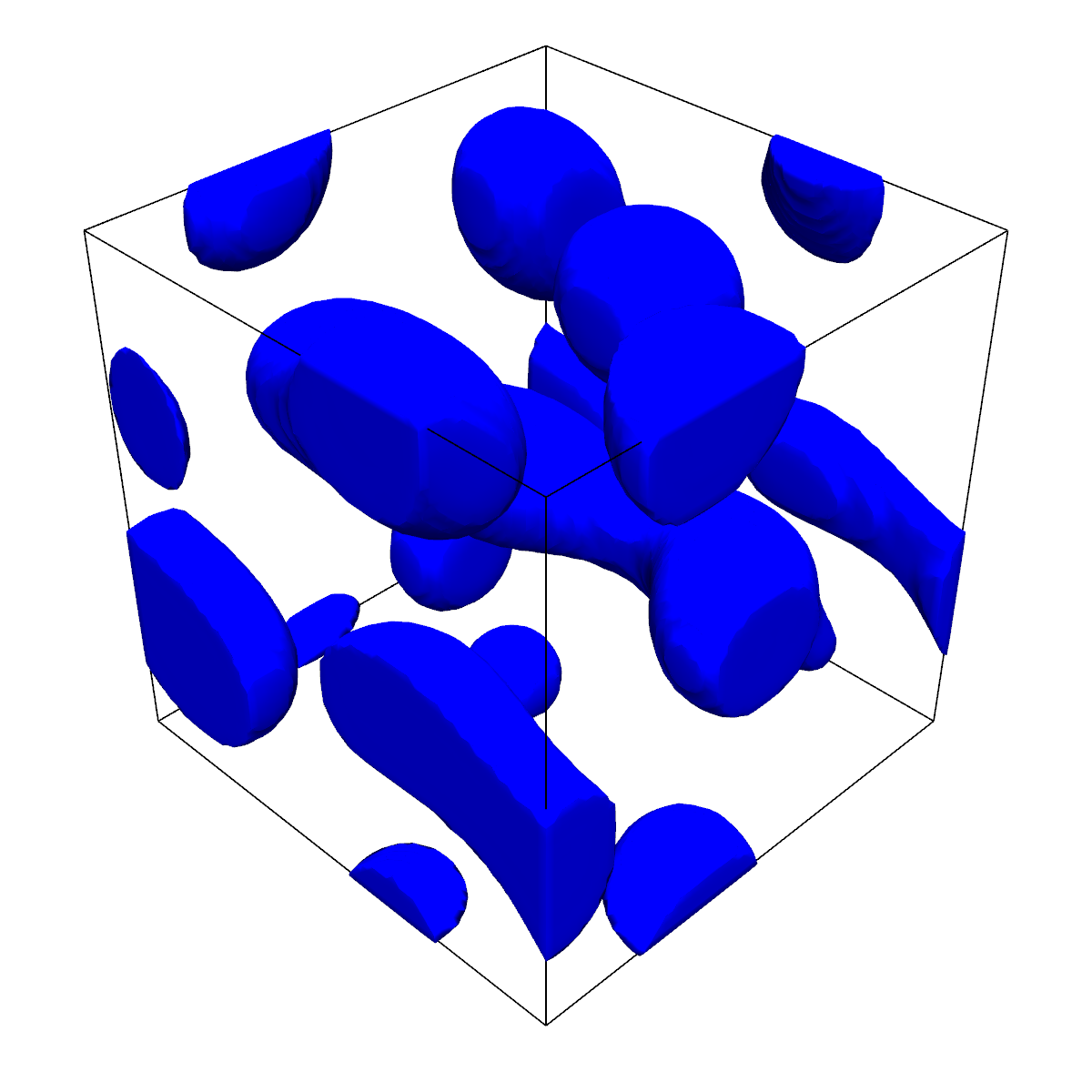}}&
        {\includegraphics[width = 0.15\textwidth]{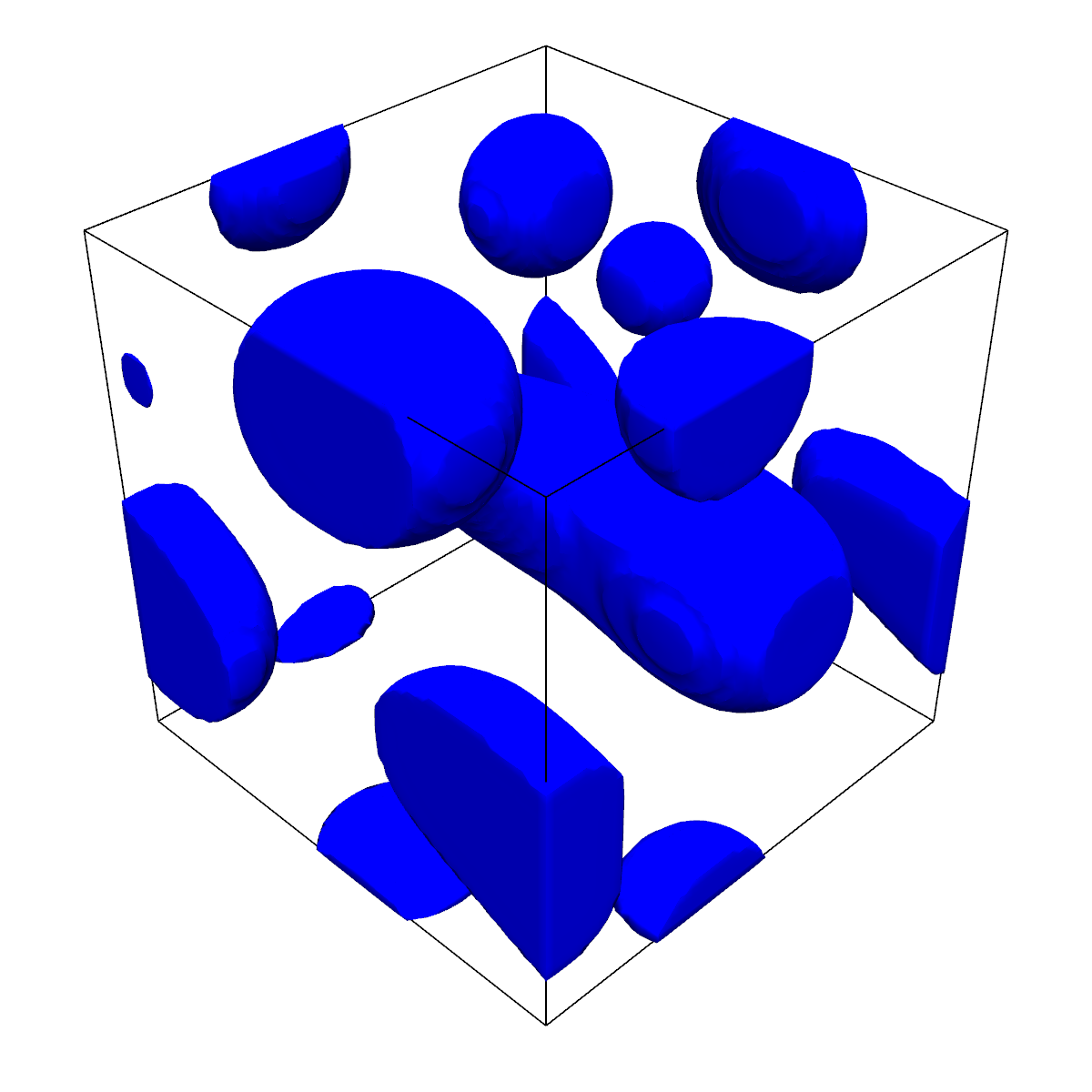}}&
        {\includegraphics[width = 0.15\textwidth]{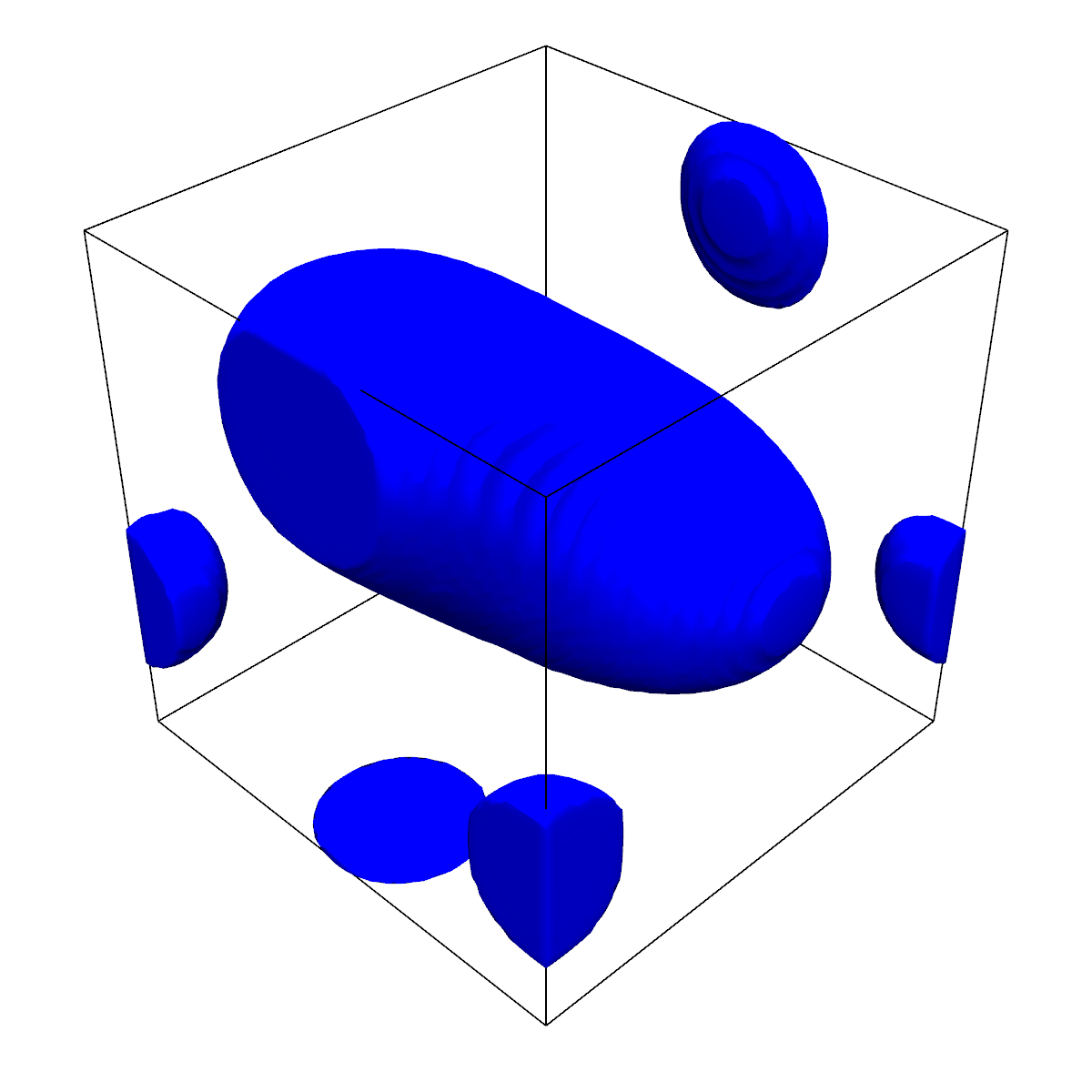}}&
        {\includegraphics[width = 0.15\textwidth]{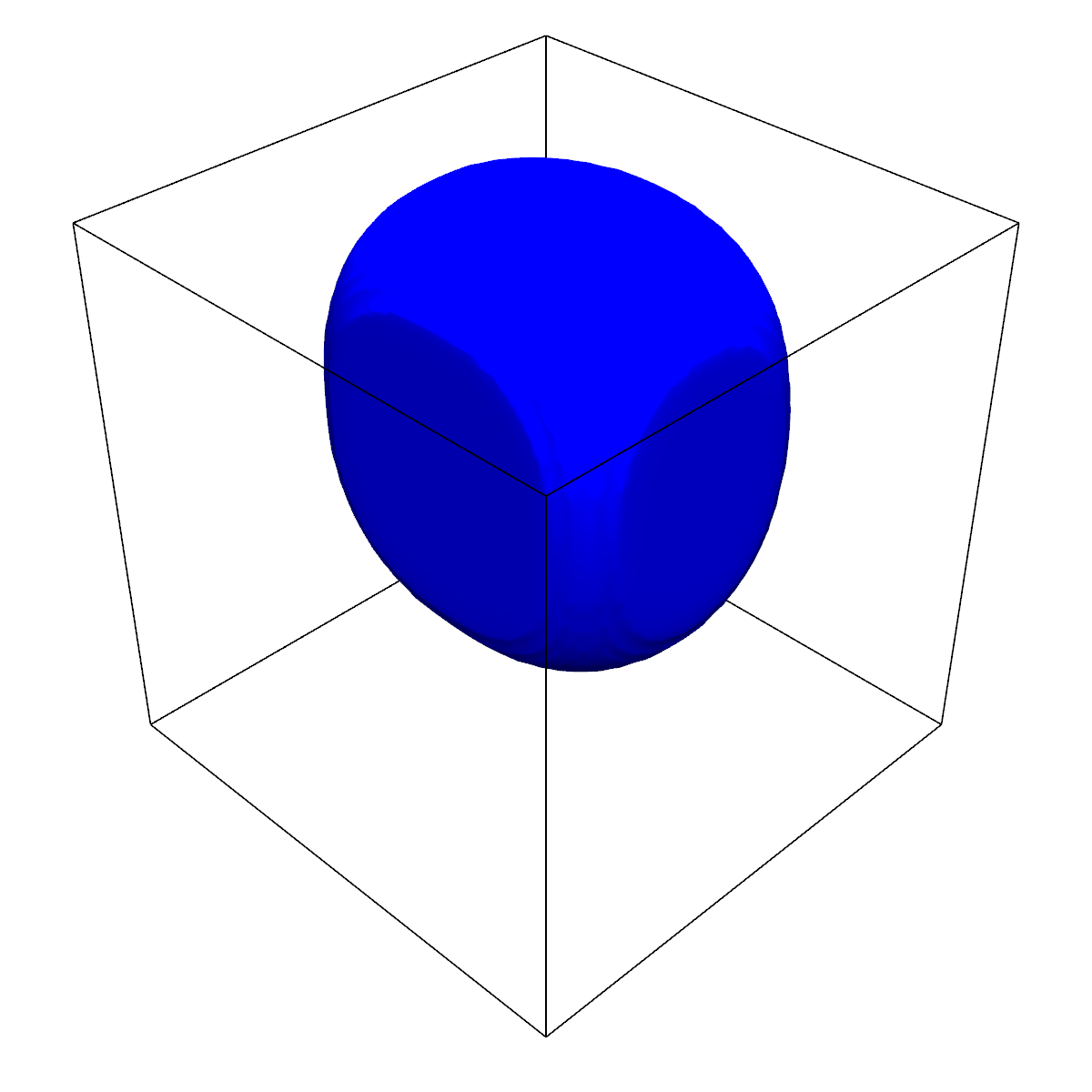}}&
        {\includegraphics[width = 0.15\textwidth]{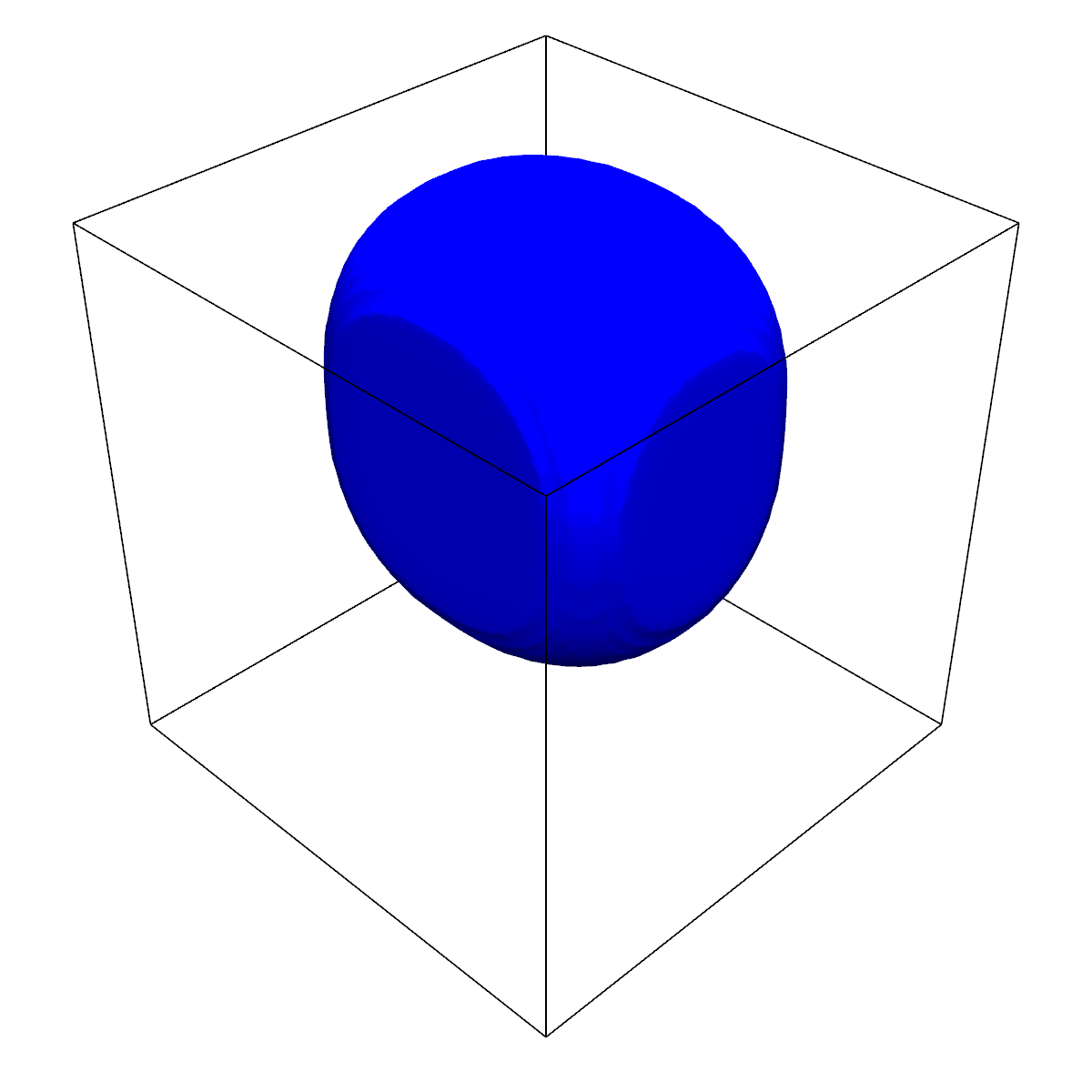}} \\
        & \multicolumn{6}{c}{(d)}\\[0.3em]
        \rotatebox[origin=l]{90}{\nfemv} &
        \raisebox{-0.1\height}{\includegraphics[width = 0.15\textwidth]{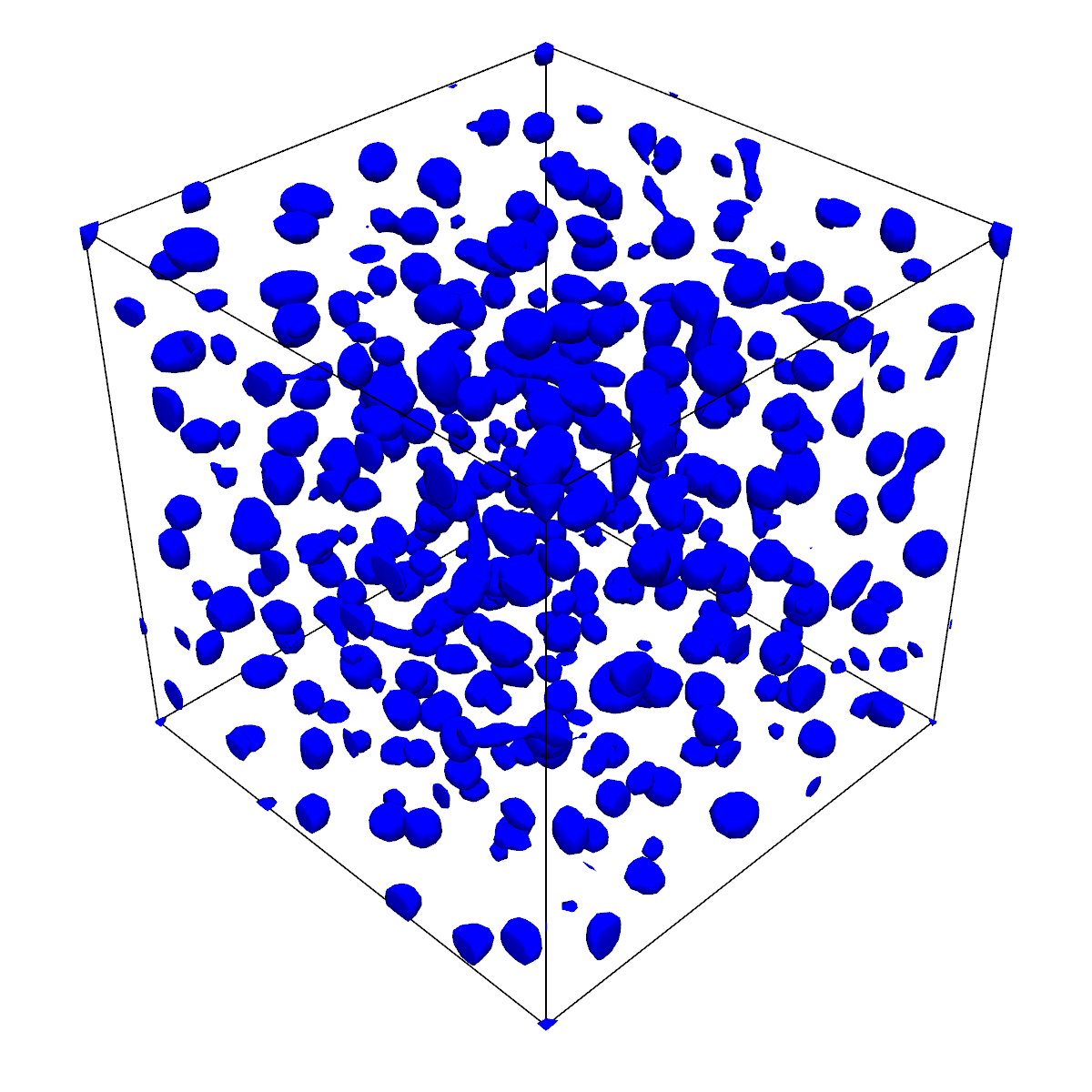}}&
        {\includegraphics[width = 0.15\textwidth]{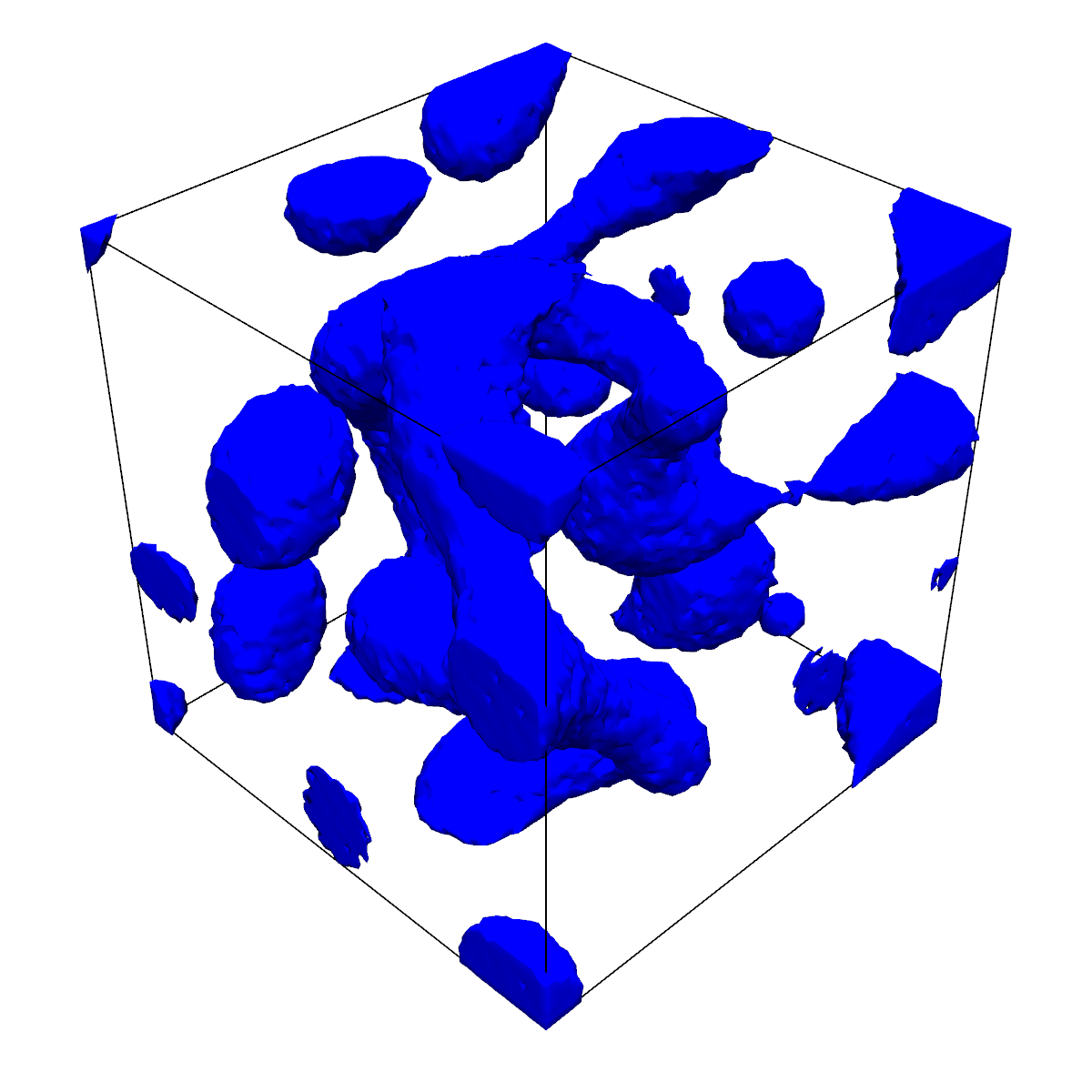}}&
        {\includegraphics[width = 0.15\textwidth]{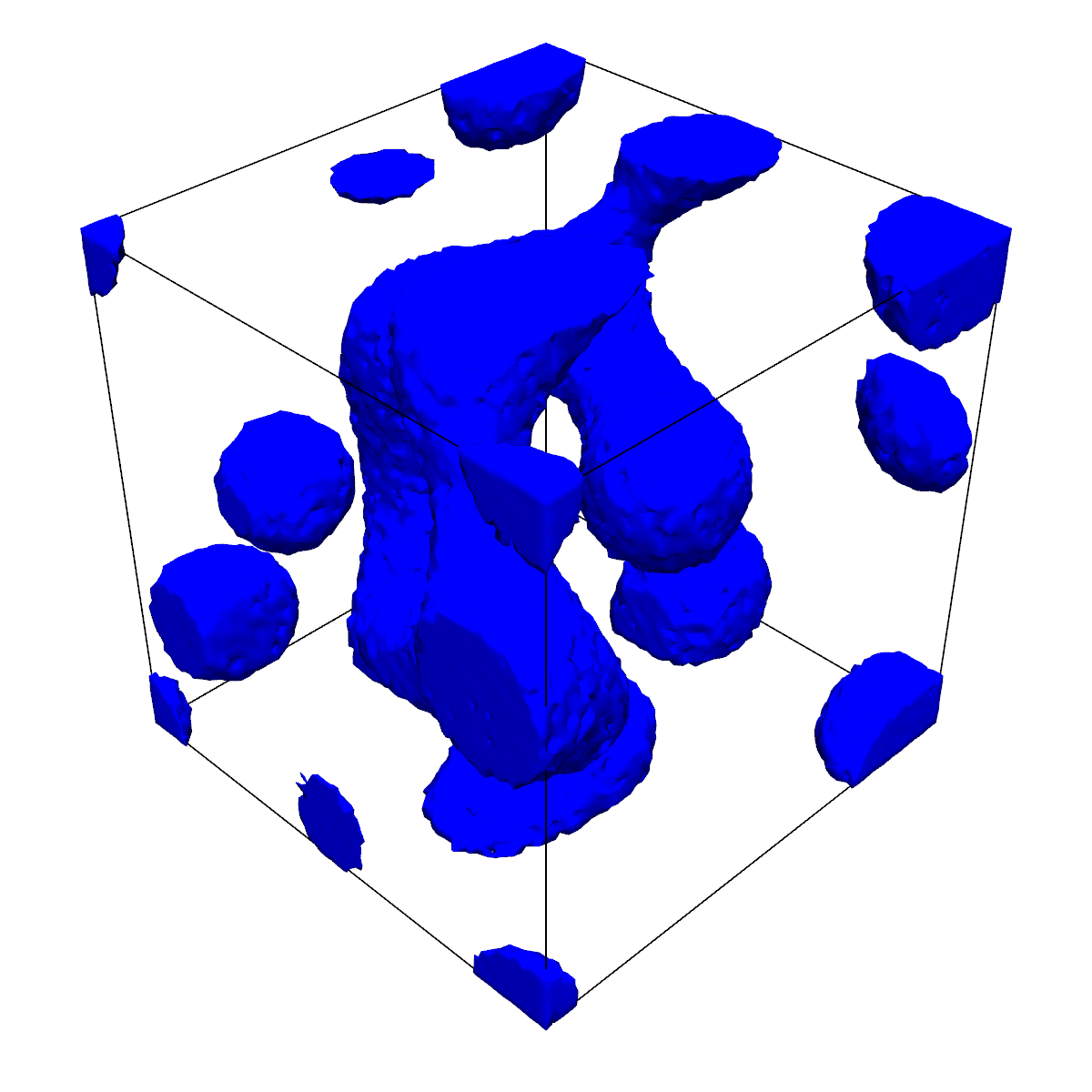}}&
        {\includegraphics[width = 0.15\textwidth]{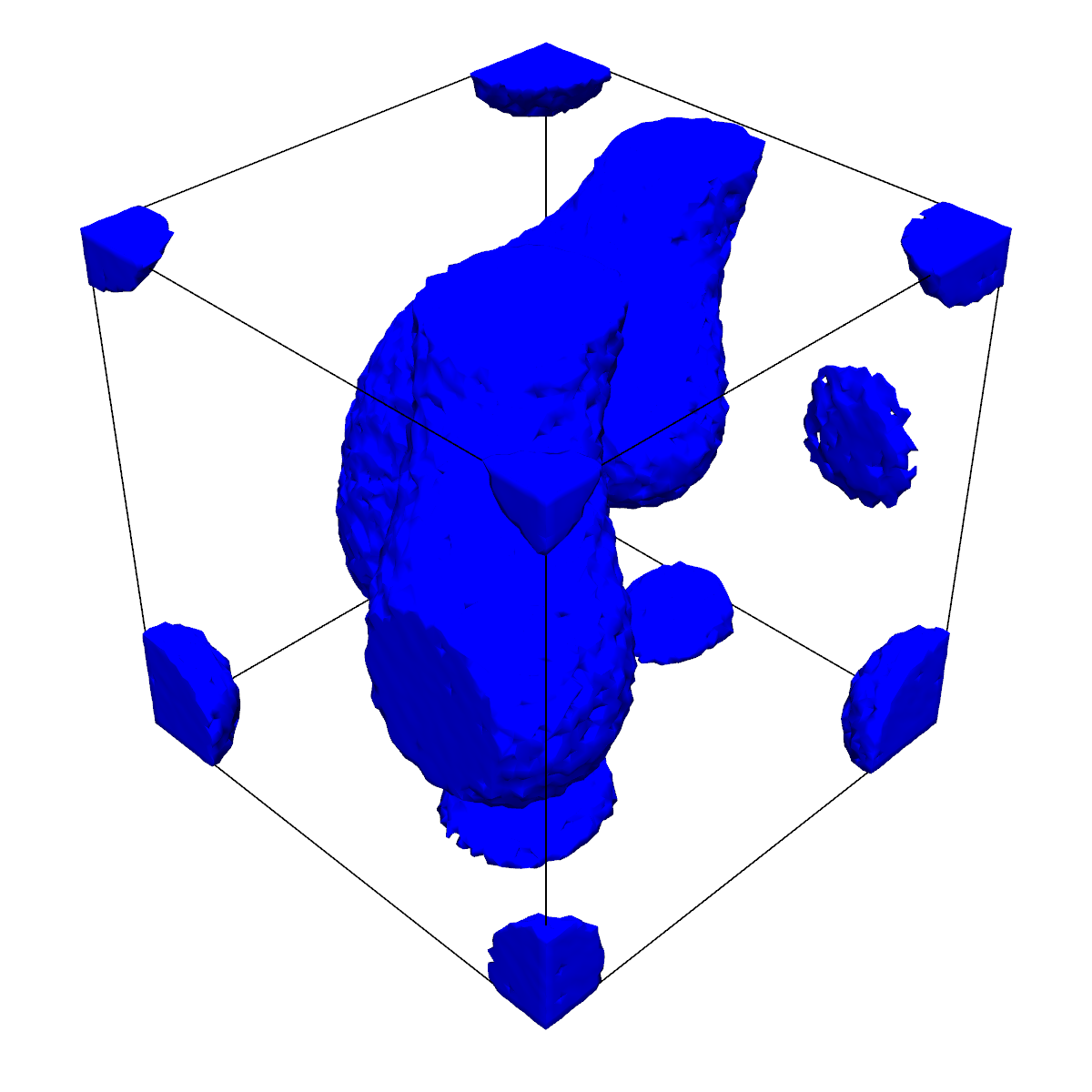}}&
        {\includegraphics[width = 0.15\textwidth]{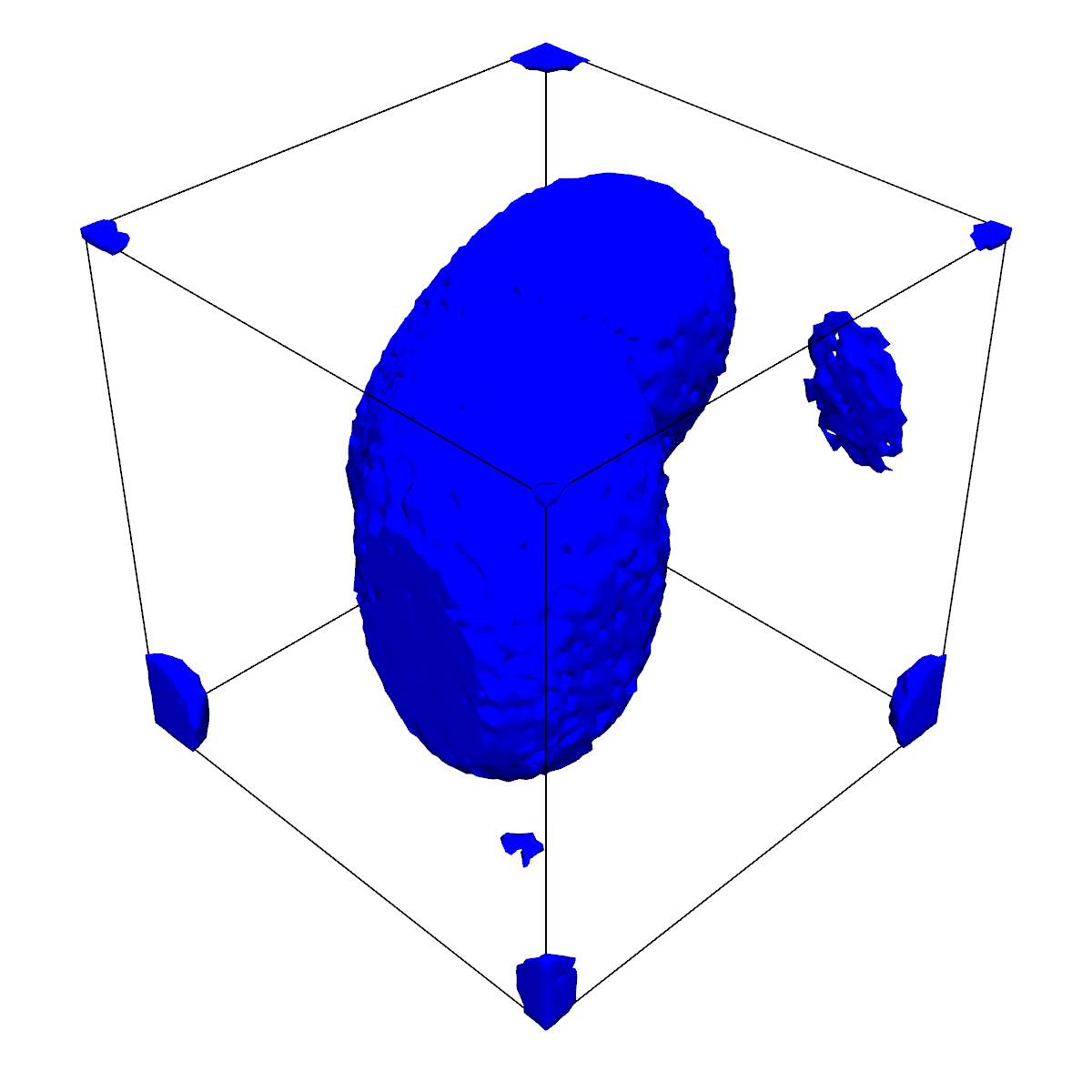}}&
        {\includegraphics[width = 0.15\textwidth]{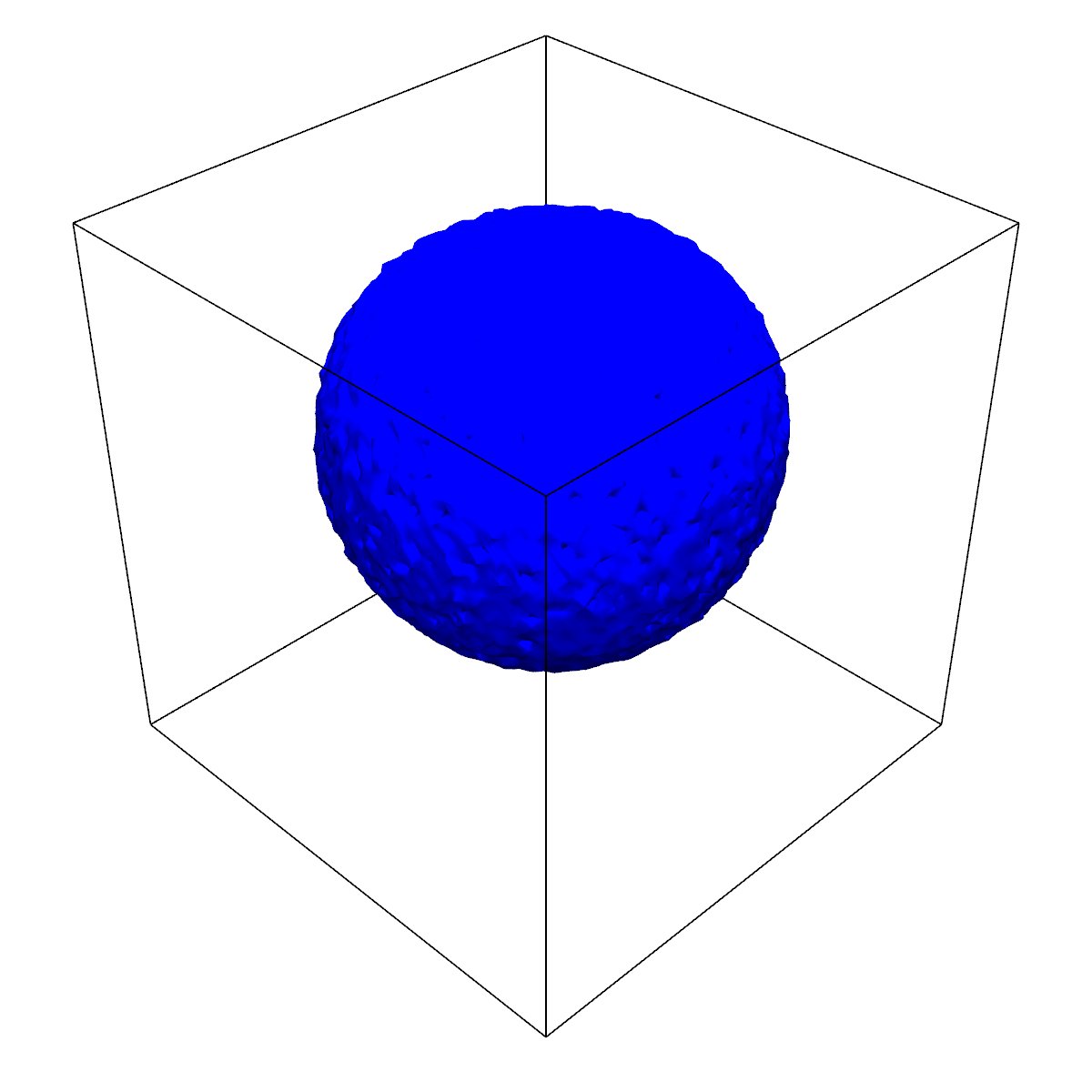}} \\
        & \multicolumn{6}{c}{(e)}\\[0.3em]
        \rotatebox[origin=l]{90}{\nfem} &
        \raisebox{-0.1\height}{\includegraphics[width = 0.15\textwidth]{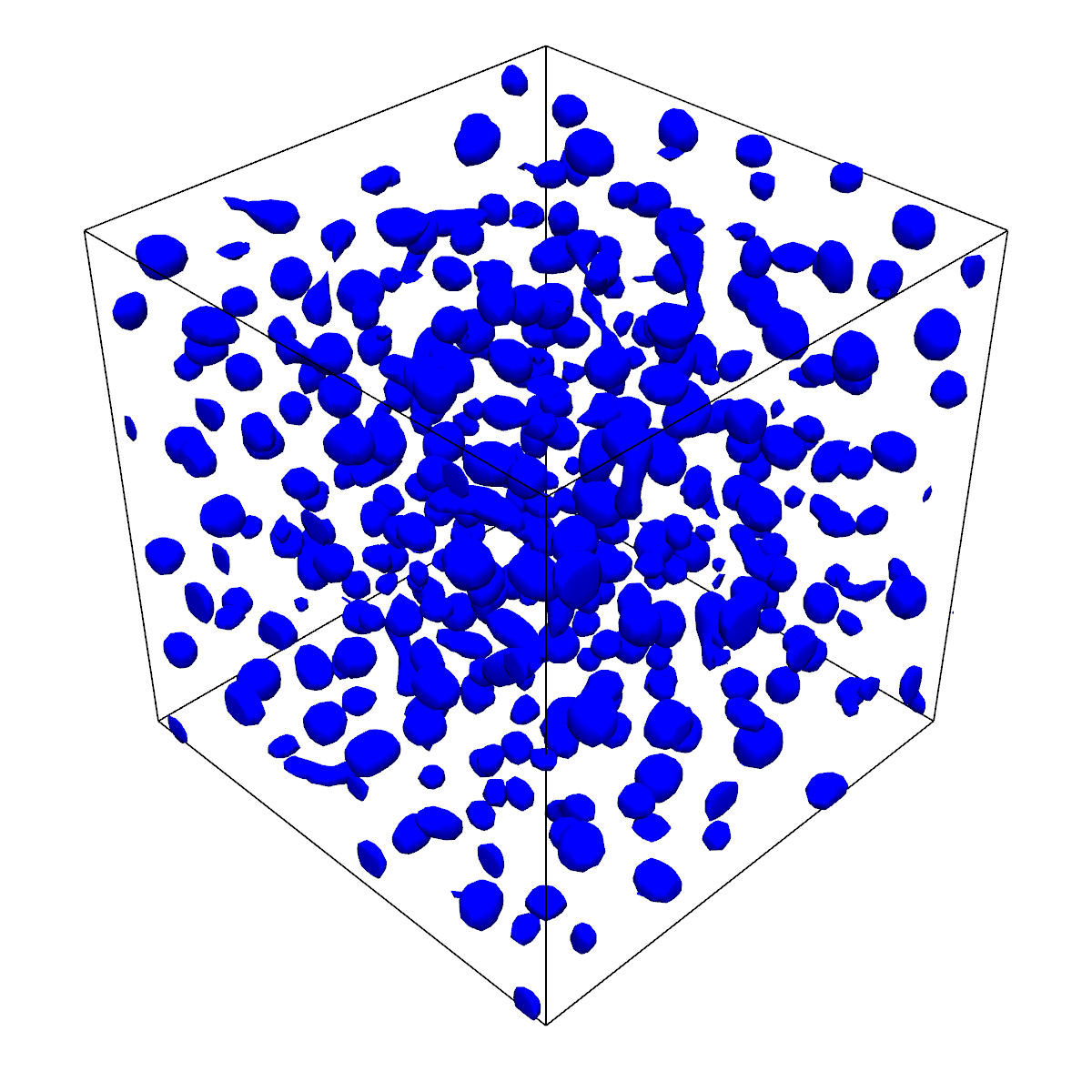}}&
        {\includegraphics[width = 0.15\textwidth]{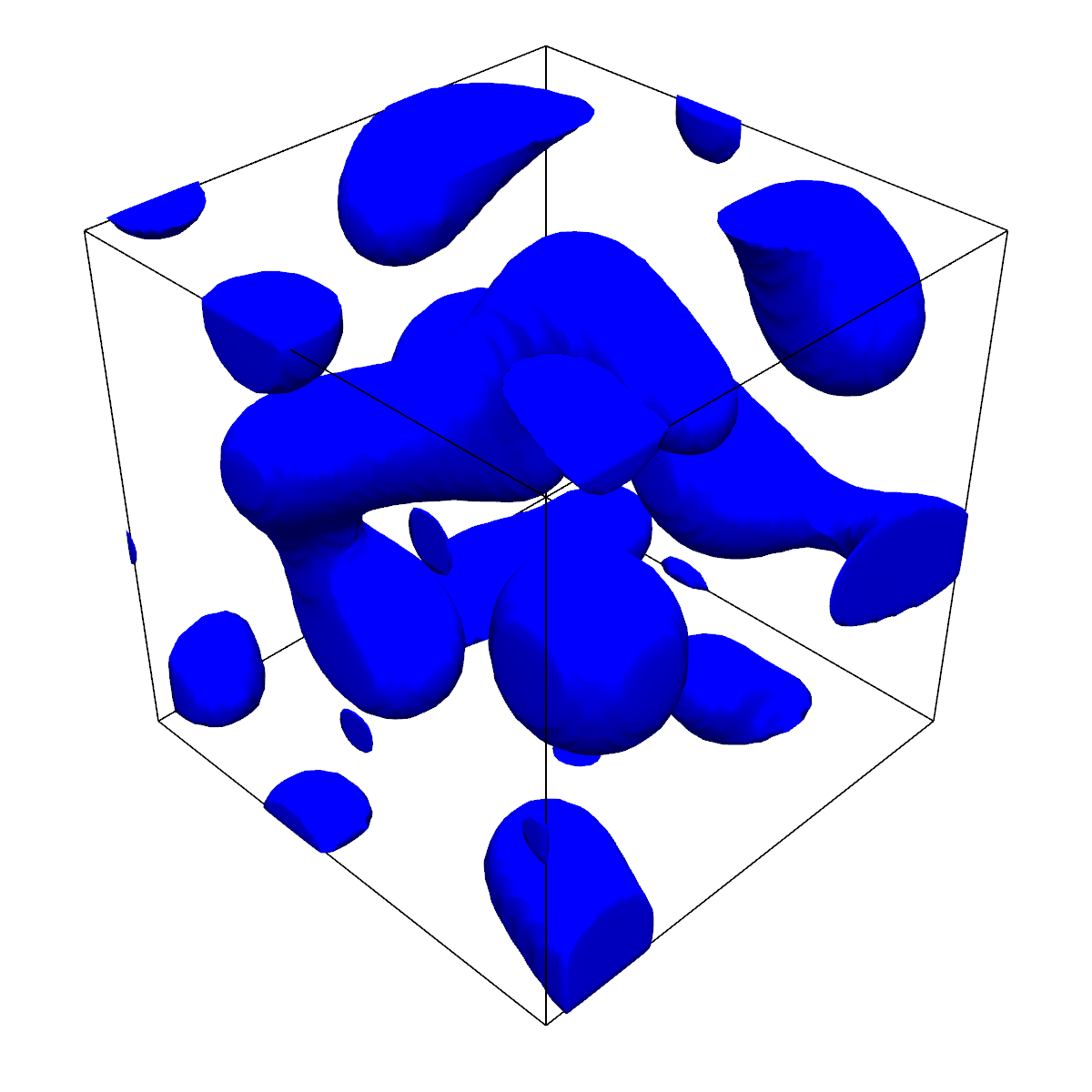}}&
        {\includegraphics[width = 0.15\textwidth]{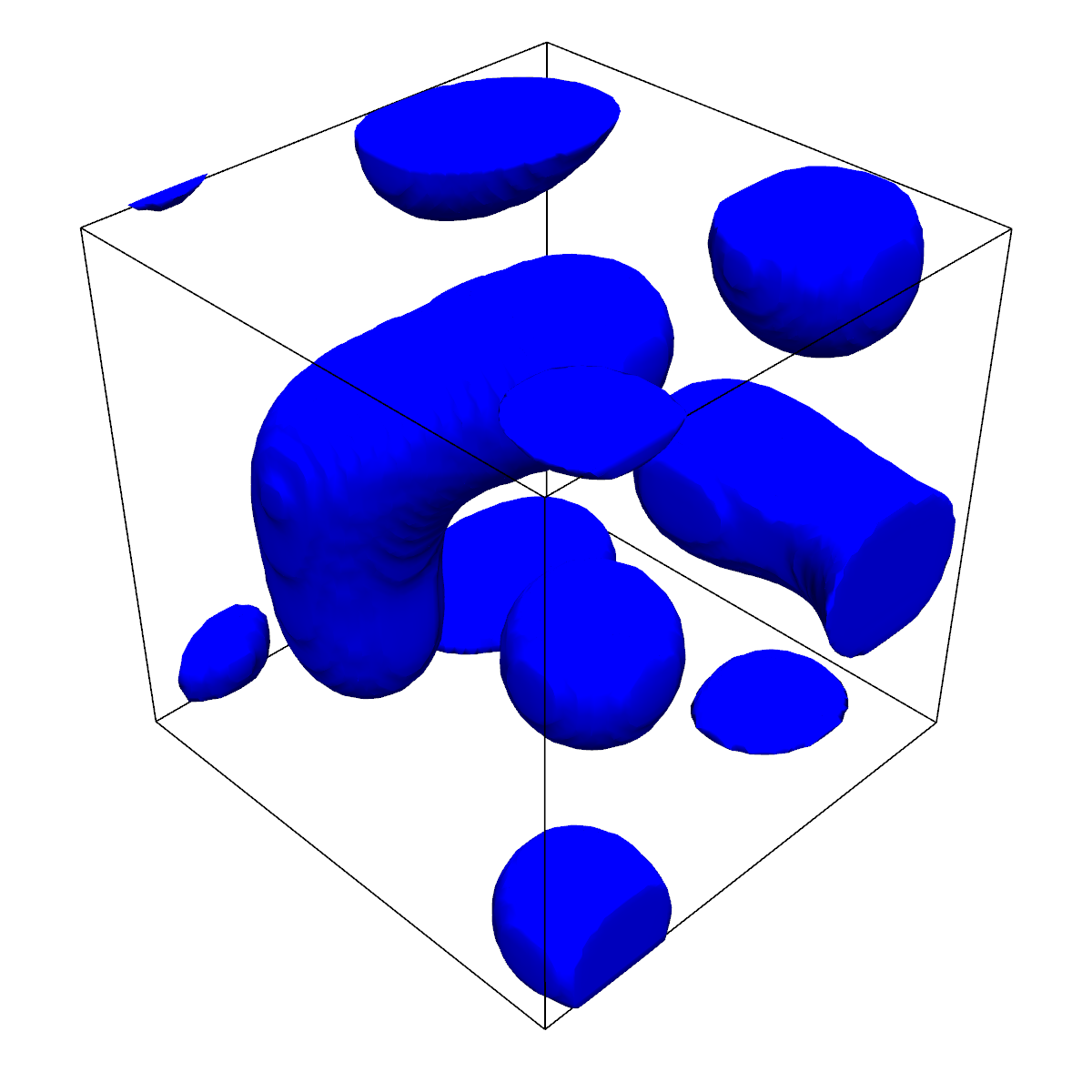}}&
        {\includegraphics[width = 0.15\textwidth]{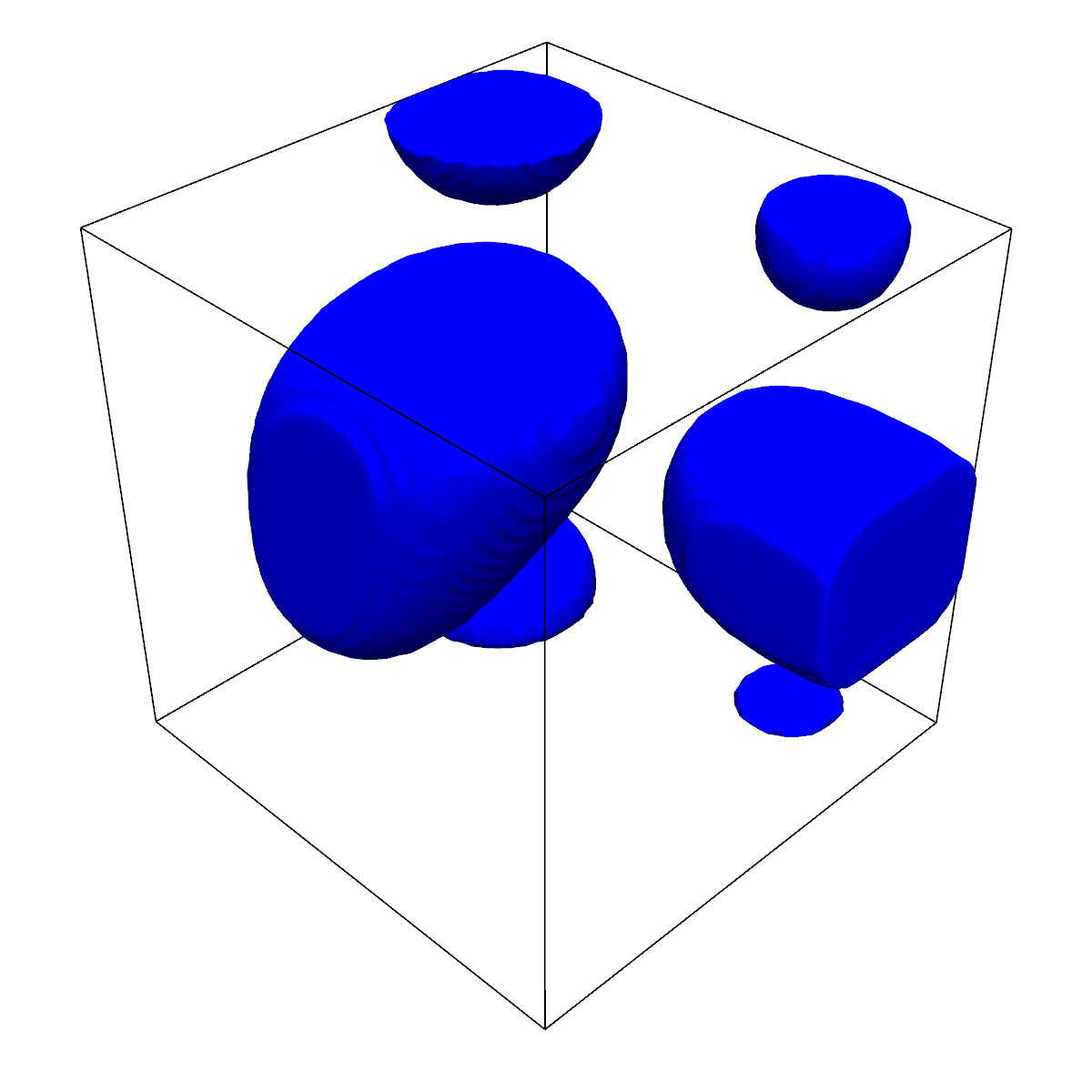}}&
        {\includegraphics[width = 0.15\textwidth]{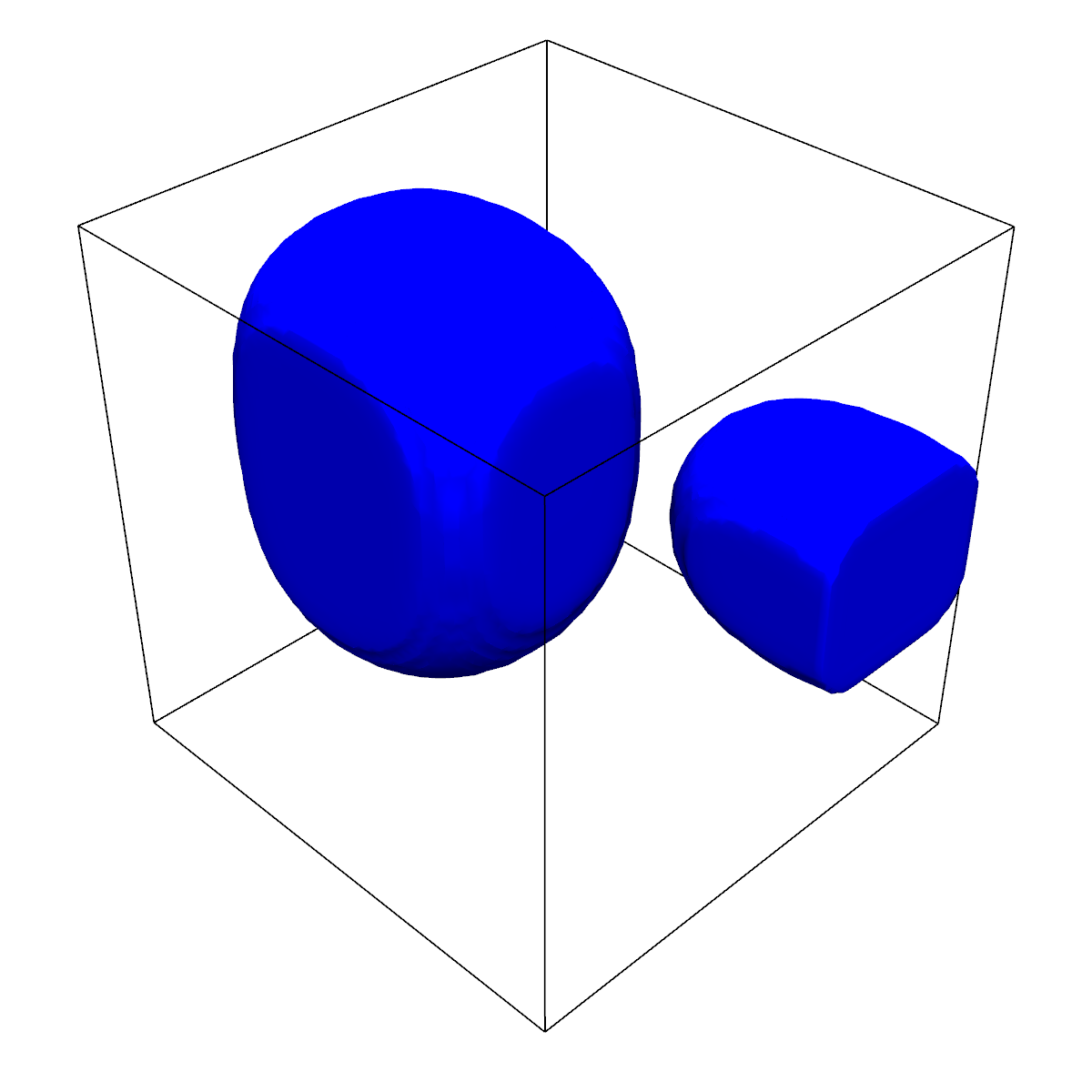}}&
        {\includegraphics[width = 0.15\textwidth]{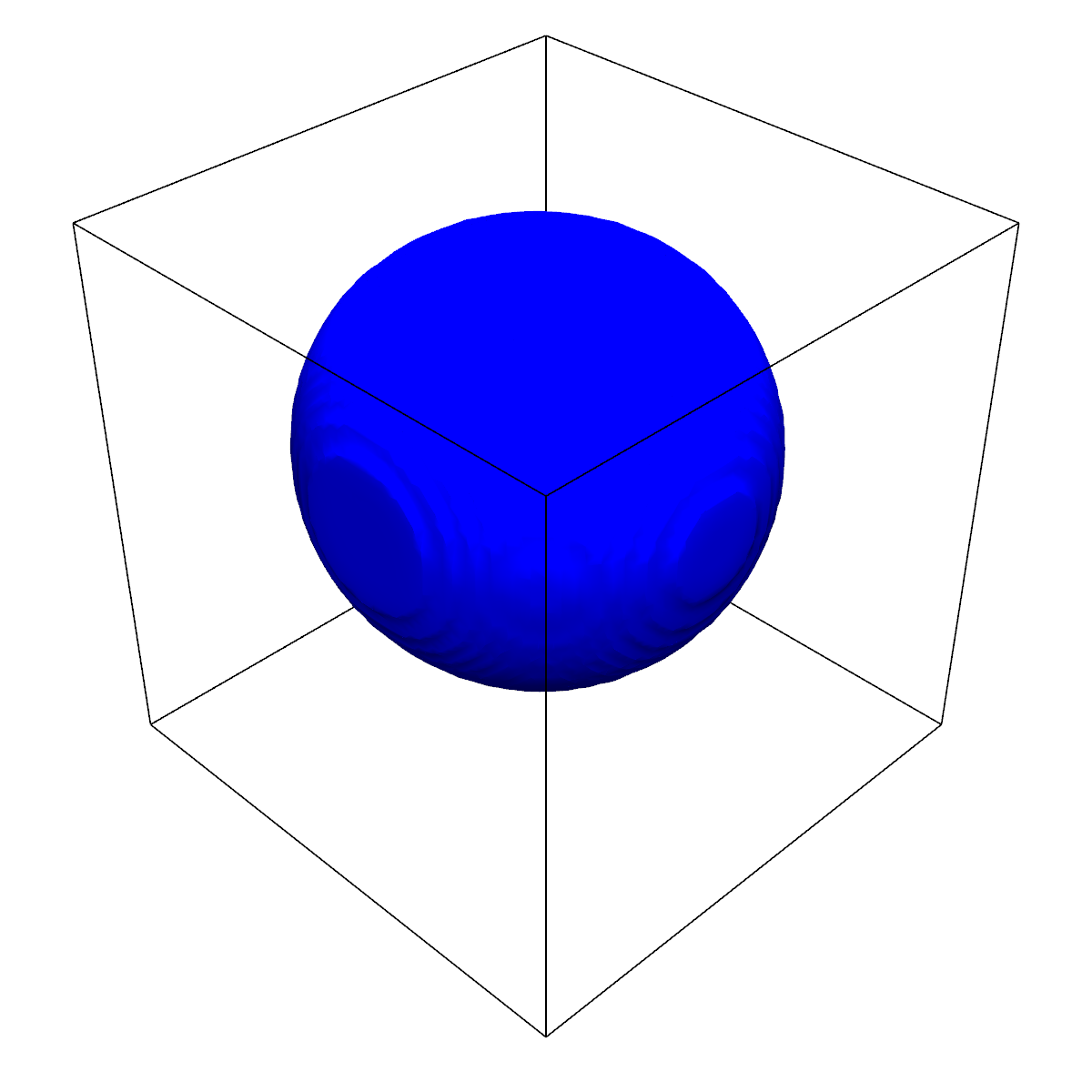}} \\
        & \multicolumn{6}{c}{(f)}\\[0.3em]
           
    \end{tabular}
    \caption{Generalization test of learned surrogate models with $\langle c \rangle = 0.4$ on a $64^3$ grid for a long rollout. (a) Ground truth SPDE trajectory with $\epsilon=0.2$, (b) \nflux{} predictions, (c) \femv{}, (d) \fem{}, (e) \nfemv{}, and (f) \nfem{}.}
    \label{fig:nonflux_glomeration_test}
\end{figure}

All five models defined in Sec.~\ref{sec:surrogate_model} were trained on SPDE trajectories with $\epsilon=0.2$ on a $16^3$ grid. First, the rollout stability was evaluated on test data of unseen trajectories (time duration of 4000) on the same grid that are much longer than the training set (duration of 24). We tested for different initial concentrations $\langle c \rangle =0$ and 0.4 with additional random uniform perturbations $\mathcal{U}(-0.01,0.01)$. As shown in Sec.~\ref{app:indist_nflux}, the black-box non-conservative baseline \nflux{}
surrogate matches the overall trend of temporal evolution and marginal concentration histograms on the training grid. However, when evaluated for spatial extrapolation on a $64^3$ grid, \nflux{} 
fails at late evolution stages and does not preserve spatial structure, as shown in Fig.~\ref{fig:nonflux_glomeration_test}b for a 70:30 mixture ($\langle c \rangle = 0.4$). This shows that without translation equivariance built in through the flux structure, the network cannot generalize to unseen spatial extents or to long time horizon. By contrast, the flux-based conservative surrogates (Fig.~\ref{fig:nonflux_glomeration_test}c-f) all remain stable and generate physically plausible morphologies starting from the same initial configuration. This comparison clearly indicates that a purely black-box autoregressive model is insufficient for robust very-long and large-scale spatiotemporal generalization in this problem. We therefore exclude \nflux{} from the remainder of the discussion and focus on conservative flux-based surrogates.
% \added{Was the blackbox model trained on PDE or SPDE trajectories? 
% \hai{yes, trained with SPDE, replaced GT plots}
% Was the following two figures in this subsection evaluated on 16 or 64 grid points? 
% \hai{the first figure is 16 grid points, the second is 64 grid points}
% Was the story that the blackbox model can fit okay on indistribution trajectories with 16 grid and relatively earth stage, but fails to generalize at all on large cell and late stage?
% \hai{yes, the first figure shows the blackbox model can fit okay on \scube, and the second figure shows it fails to generalize on \lcube.}
% }

% \subsection{Concentration-conserved deterministic and probabilistic models}

% \LS{Metrics of the different surrogate models such as RMSE errors, number of precipitated clusters, volume per cluster vs time, volume histogram at certain $t$, mobility, free energy, and STD plots against GT. Free energy plots in 3D of 3D spheres, cylinders, flat interface.}

\subsection{In-distribution test of conservative models} % on the same grid}
\label{sec:same_mesh_grid_results}
%\LS{Only talk about the FE-M and FE-MV model, this section result too less}

\begin{figure}[htp]
    \centering
    \begin{tabular}{c c}
        \raisebox{-0.14\height}{\includegraphics[width = 0.46\textwidth]{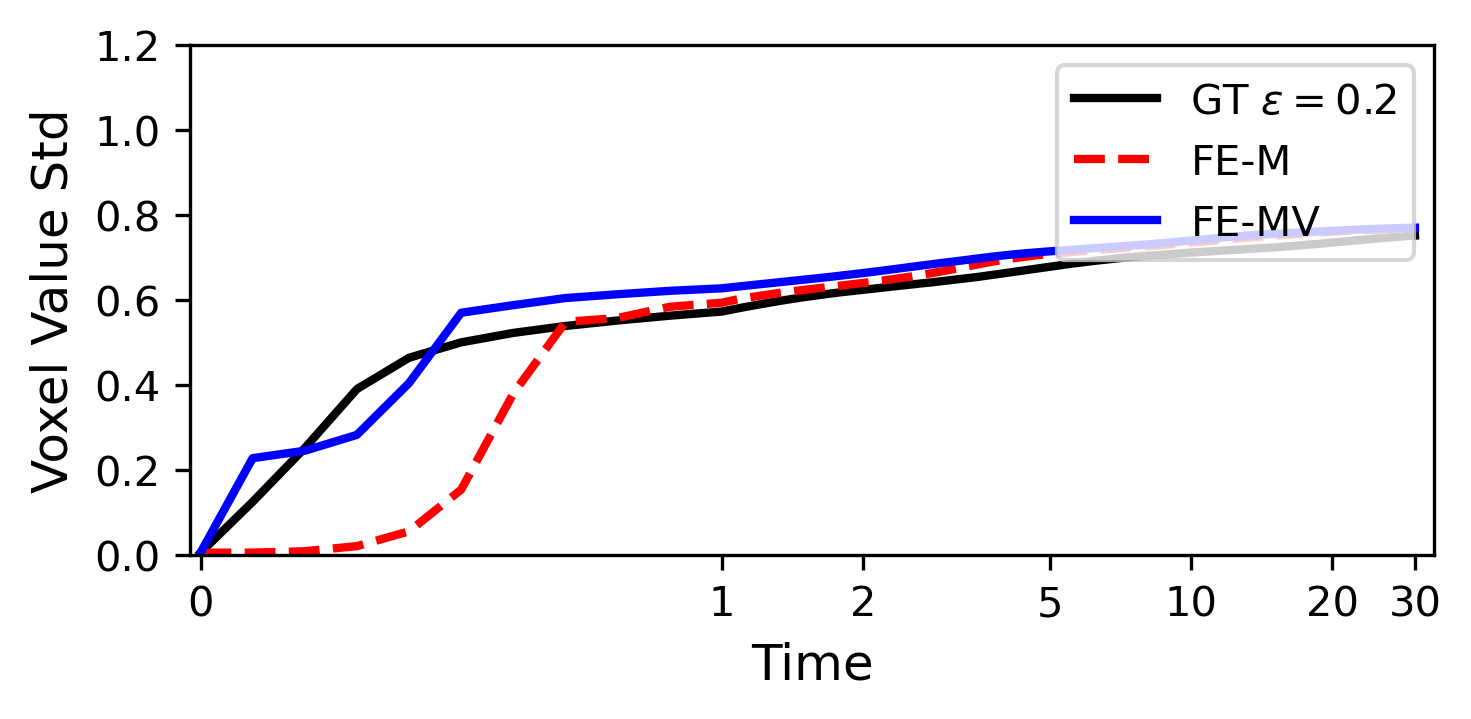}} &
        \raisebox{-0.14\height}{\includegraphics[width = 0.46\textwidth]{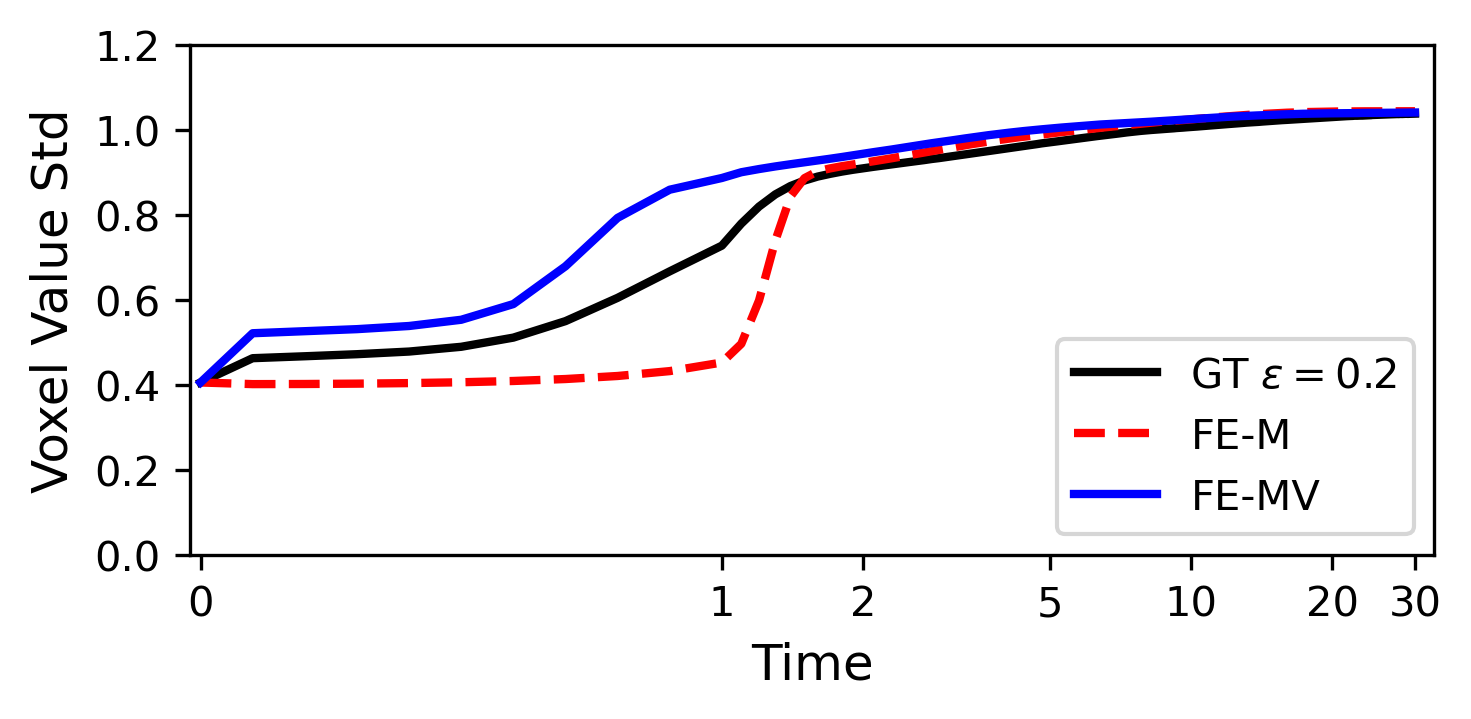}} \\
        (a) & (b)\\
        \raisebox{-0.14\height}{\includegraphics[width = 0.46\textwidth]{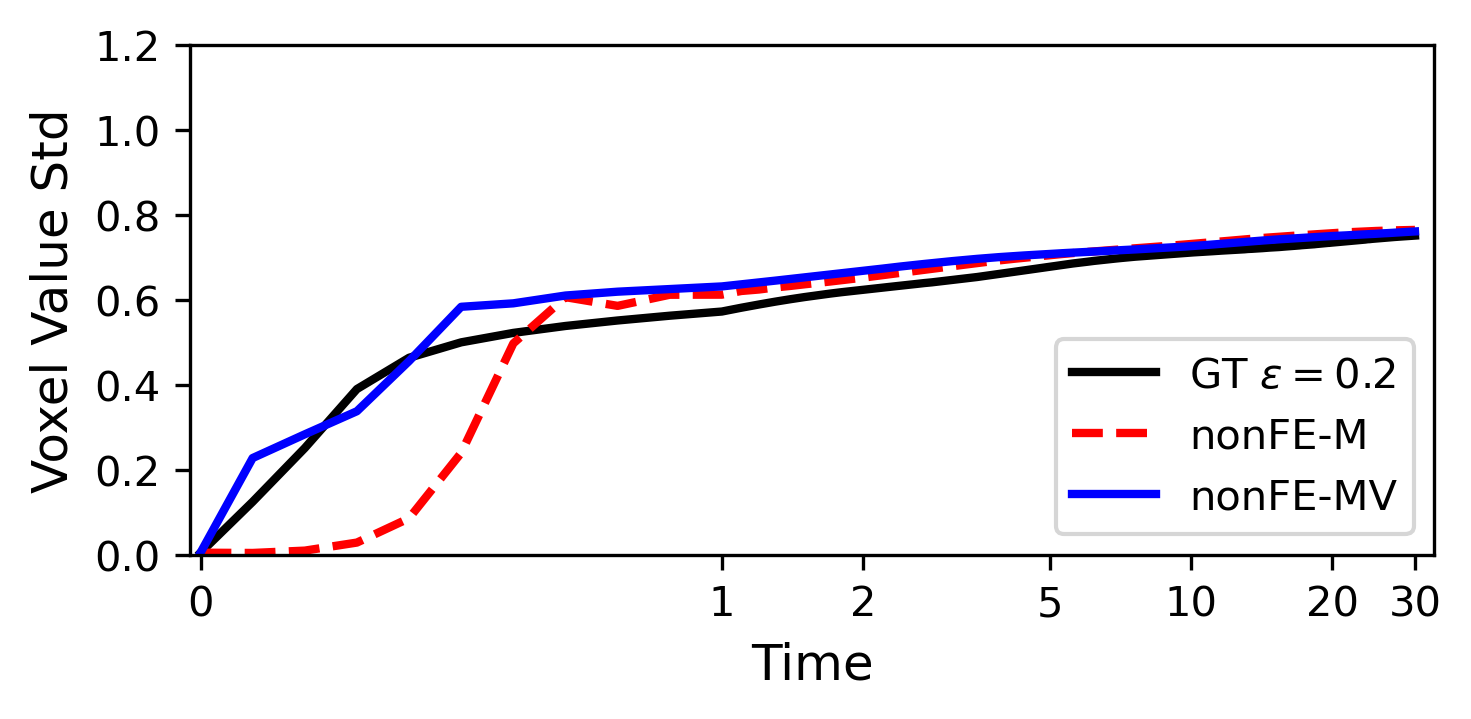}} &
        \raisebox{-0.14\height}{\includegraphics[width = 0.46\textwidth]{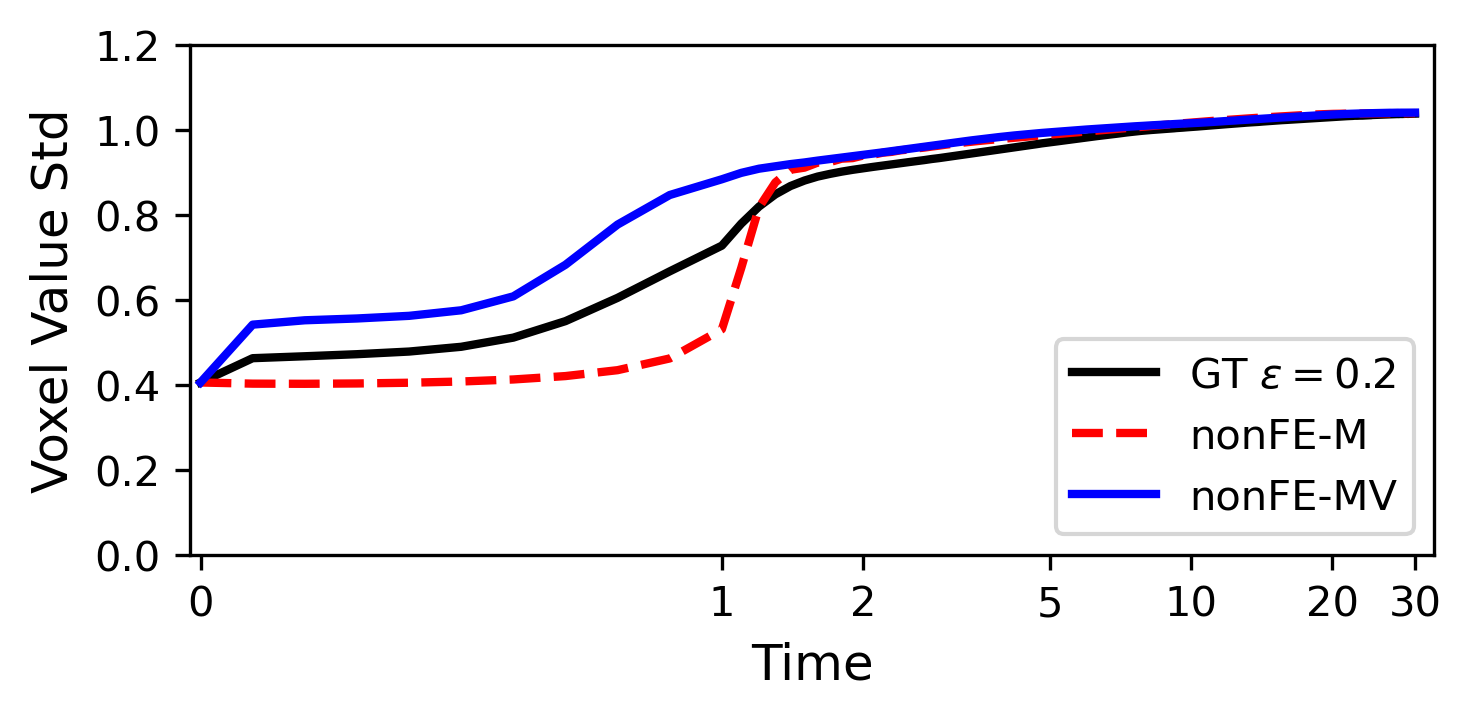}} \\
        (c) & (d)
    \end{tabular}
    \caption{The standard deviation of voxel values of ML models. Free energy based models \femv{}, \fem{} for (a)  $\langle c \rangle = 0$ and (b) $\langle c \rangle = 0.4$, and non free energy \nfemv{}, \nfem{} for (c) $\langle c \rangle = 0$ and (d) $\langle c \rangle = 0.4$. The mean-only models show substantial delay in demixing.}
    %The stochastic or mean-and-variance models show strong alignment with the SPDE ground truth. In contrast, the mean-only models show substantial delay in the early stage of demixing dynamics.}
    %show a delay in the 50:50 mixture configuration ($\langle c \rangle = 0$) and faster evolution in the 70:30 configuration ($\langle c \rangle = 0.4$).}
    \label{fig:in_dist_1}
\end{figure}

Next the conservative surrogate models are tested on the $16^3$ grid, the same as used in training. The standard deviation of voxel values along a trajectory is taken as a statistical measure of demixing, as the mean is constant in these conservative models.
Fig.~\ref{fig:in_dist_1} shows the standard deviation over 10 trajectories for the same two representative compositions 50:50 ($\langle c \rangle = 0$) and 70:30 ($\langle c \rangle = 0.4$). We find that the \femv{} and \nfemv{} models, which predict both a mean and a variance and therefore produce stochastic rollouts, matches the ground truth with $\epsilon=0.2$ better, whereas
the mean-only models \fem{} and \nfem{} show substantial delay in the early stage of demixing dynamics. This is attributed to the physically learned flux fluctuation that accelerates phase separation, in agreement with the observed difference between the SPDE and PDE in Fig.~\ref{fig:SDE_different_equilibrium}. 

% Moreover, the voxel-wise standard deviation is defined in Eq.~\ref{eq:vol_std} and it shows that the \femv{} model is more aligned with the ground truth.%, as shown in Fig.~\ref{fig.in_dist_voxel_hist}. 
% \begin{equation}
% \label{eq:vol_std}
%   \text{Voxel Value Std}(t) = \sqrt{\frac{\sum_{n=1}^{10} \sum_{i,j,k=1}^{16} \left(c_{n,t,i,j,k} - \bar{c}_{t}\right)^2}{10 \times 16^3}}
% \end{equation}
% where $$\bar{c}_{t} = \frac{1}{10 \times 16^3} \sum_{n=1}^{10} \sum_{i,j,k=1}^{16} c_{n,t,i,j,k}$$

\begin{figure}[htp]
    \centering
    {
    \begin{tabular}{c c c c c }
        & \nfem & \fem & \nfemv & \femv
        \\
        \rotatebox[origin=l]{90}{Mobility} &
        \raisebox{-0.2\height}{\includegraphics[width = 0.24\textwidth]{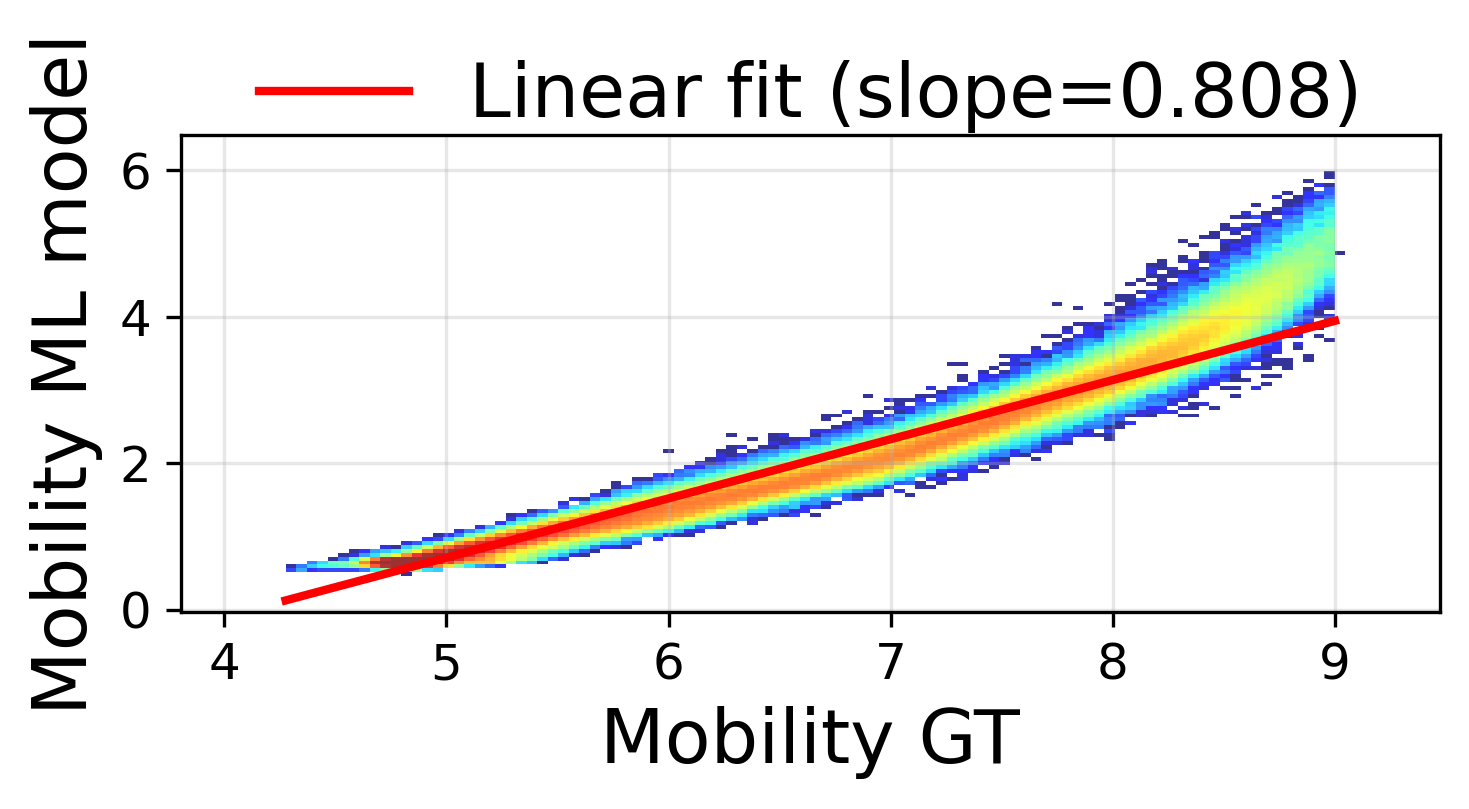}} &
        \raisebox{-0.2\height}{\includegraphics[width = 0.24\textwidth]{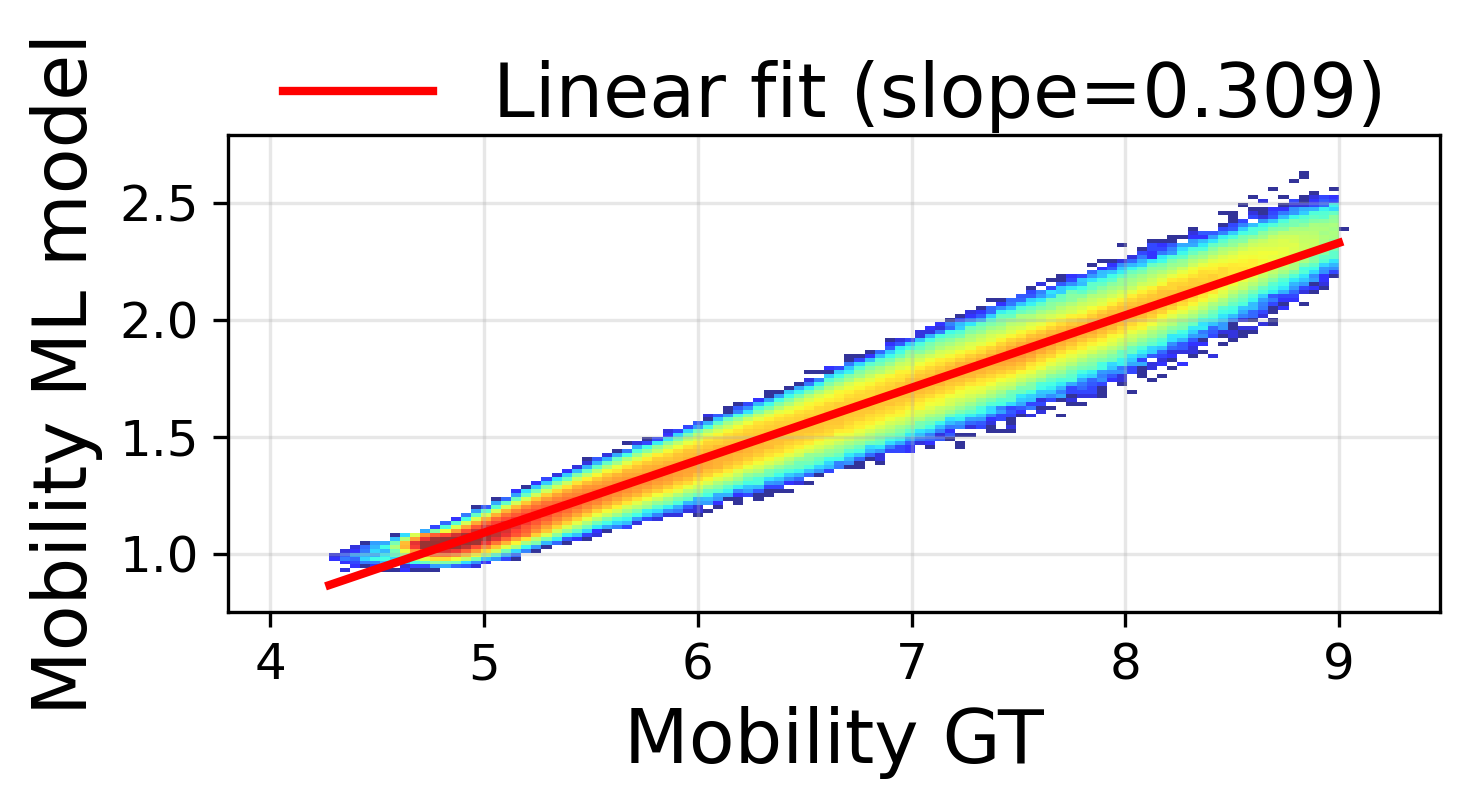}} &
        \raisebox{-0.2\height}{\includegraphics[width = 0.24\textwidth]{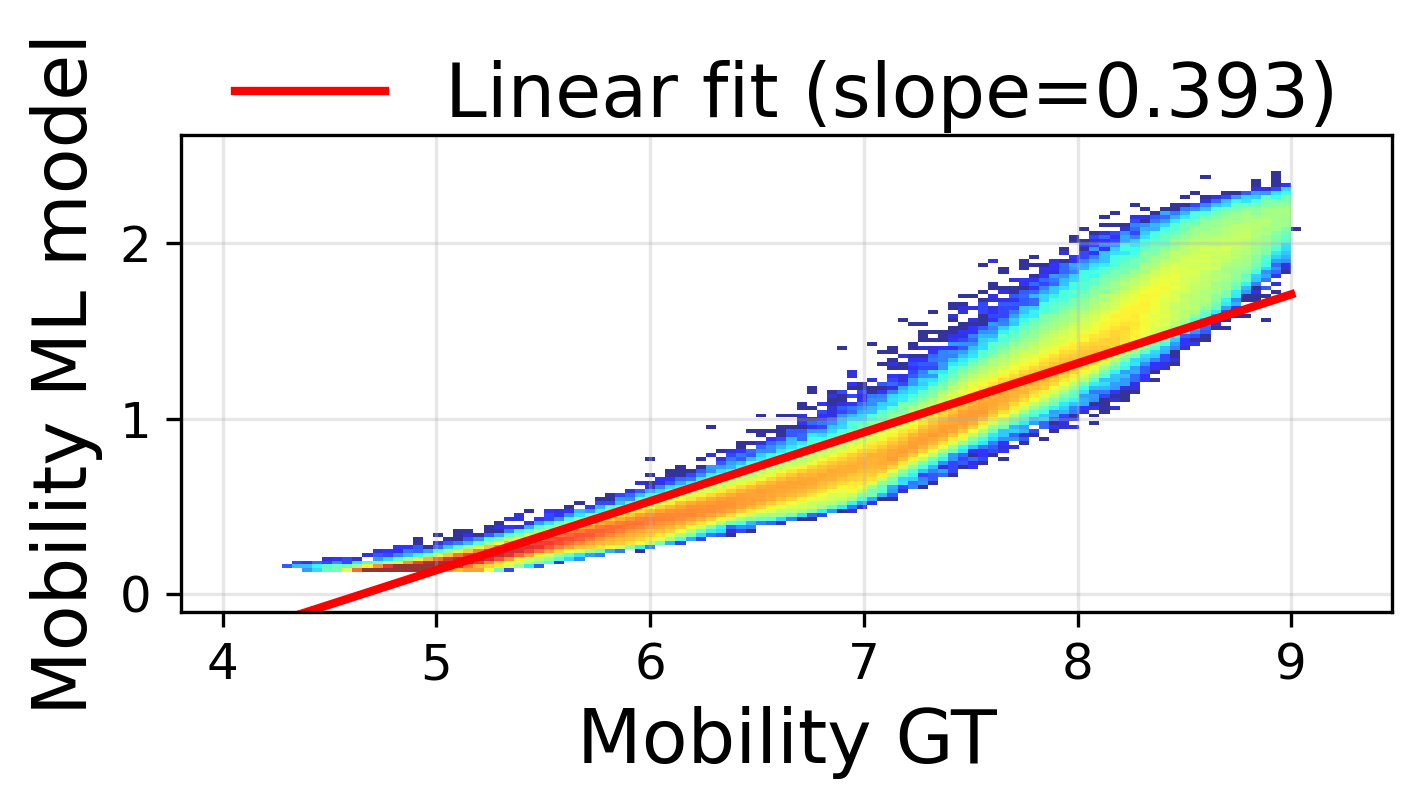}}&
        \raisebox{-0.2\height}{\includegraphics[width = 0.24\textwidth]{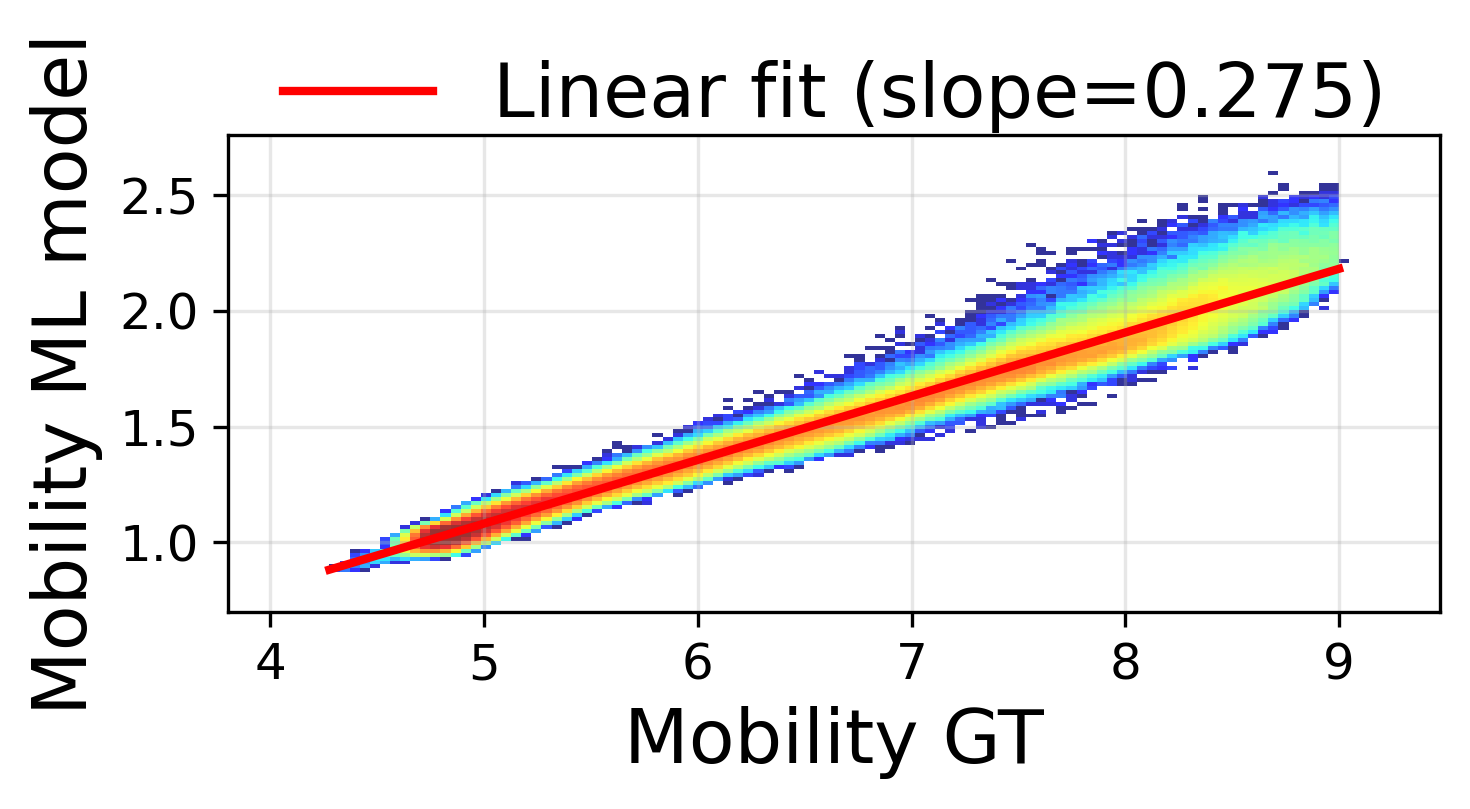}}
        \\
        & \multicolumn{4}{c}{\includegraphics[width=.9\textwidth]{fig/Final/lognorm_colormap.png}} 
    \end{tabular}
    }
    \caption{Machine learned mobility $M$ of different models compared to the ground truth
    %Ground truth (GT) is computed with analytical formulas. 
from 100 snapshots up to $t=10$ with $\langle c \rangle = 0$ on the $16^3$ grid. 
    % (same plots if we use \cofour, thus not shown here).
    }
    % \todo[inline]{Do we still need this mobility comparison since we already have the free energy comparison?}
    \label{fig:indist_ml_mobility}
\end{figure}

% It justifies the need to design a probabilistic or generative model for the stochastic dynamics.
While the free-energy and non-free-energy based variants show similar behavior in the above tests, a difference can be revealed in the learned mobility $M$. Since there is no constraint for the magnitude of the learned $M$, its scale is arbitrary (one may freely scale $M \rightarrow \alpha M$, $F_{ij} \rightarrow \frac{1}{\alpha}\F_{ij}$). Therefore one just expects a linear relationship between a perfectly learned $M$ and the ground truth values.
To verify if the fitted $M$ captures the correct physics, we plot it against the analytical  mobility of Eq.~\ref{eq:flux_spde} in Fig.~\ref{fig:indist_ml_mobility}.  
Fig.~\ref{fig:indist_ml_mobility} indeed shows nearly linear plots for free energy based models \femv{} and \fem{} with conservative driving forces (chemical potential) $\mu = \partial E/\partial c$, as well as substantial non-linear deviation for non-free energy models, which are under non-conservative driving forces.
Similar results were obtained for $\langle c \rangle=0$ and 0.4.
This suggest that while the non-free energy models can reasonably learn the dynamics, they have cancellation of errors in the non-conservative driving force and corresponding kinetic coefficients, resulting in neither quantity being physically meaningful.
%Therefore, a learned relation with the form of $\hat{M}_{ML}=\alpha M_{GT}$ is reasonable.  A linear regression shows that the slope is close to 0.3 for both models. On the contrary, we also plot the the models without free energy components (\nfem{} and \nfemv{}) in Fig.~\ref{fig:indist_ml_mobility}. The comparison shows that without free energy estimation, the purely black box model couldn't predict a physically meaningful mobility. The plot for \cozero and \cofour are identical with respect to mobility plot therefore we only show the result of \cozero  here. 

\setlength{\tabcolsep}{2pt}  % Default is 6pt

%% figure for different time step size

% \begin{figure}[htp]
%     \centering
%     {
%     \begin{tabular}{c c c c c c}
%         &  & \nfem & \fem & \nfemv & \femv 
%         \\
%         \rotatebox[origin=l]{90}{Mobility} &
%         \raisebox{-0.2\height}{} &
%         \raisebox{-0.2\height}{\includegraphics[width = 0.18\textwidth]{fig/Final/fd312_flux_mean_nonFE_t0p5___Mobility_histogram.png}} &
%         \raisebox{-0.2\height}{\includegraphics[width = 0.18\textwidth]{fig/Final/fd312_flux_mean_FE_t0p5___Mobility_histogram.png}} &
%         \raisebox{-0.2\height}{\includegraphics[width = 0.18\textwidth]{fig/Final/fd312_w20_flux_mean_nonFE_t0p5___Mobility_histogram.png}}&
%         \raisebox{-0.2\height}{\includegraphics[width = 0.18\textwidth]{fig/Final/fd312_w20_flux_mean_FE_t0p5___Mobility_histogram.png}}
%         \\
%         & \multicolumn{5}{c}{\includegraphics[width=.85\textwidth]{fig/Final/lognorm_colormap.png}} 
%     \end{tabular}
%     }
%     \caption{{\bf time size 0.05, tskip 50} Mobility (1st row) and Std Heatmaps (2nd row), by different models. GT is computed with analytical formulas based on a pair of the current voxel and its neighboring voxel. ML is predicted by mobility and stochastic modules. Dataset includes 100 concentration frames with \cozero on \scube{} at the time $t=10$ (same plots if we use \cofour, thus not show here). 3rd row and 4th row are the histograms of GT mobility and std verus ML models, respectively. Please comment out the row that you do not need!
%     }
% \end{figure}

\subsection{Generalization tests: long-term agglomeration}
\label{sec:long_agglo}
\begin{figure}[H]
    \centering
    \begin{tabular}{c c c}
        & \fem & \femv \\ 
        \rotatebox[origin=l]{90}{Clusters} &
        \raisebox{-0.1\height}{\includegraphics[width = 0.40\textwidth]{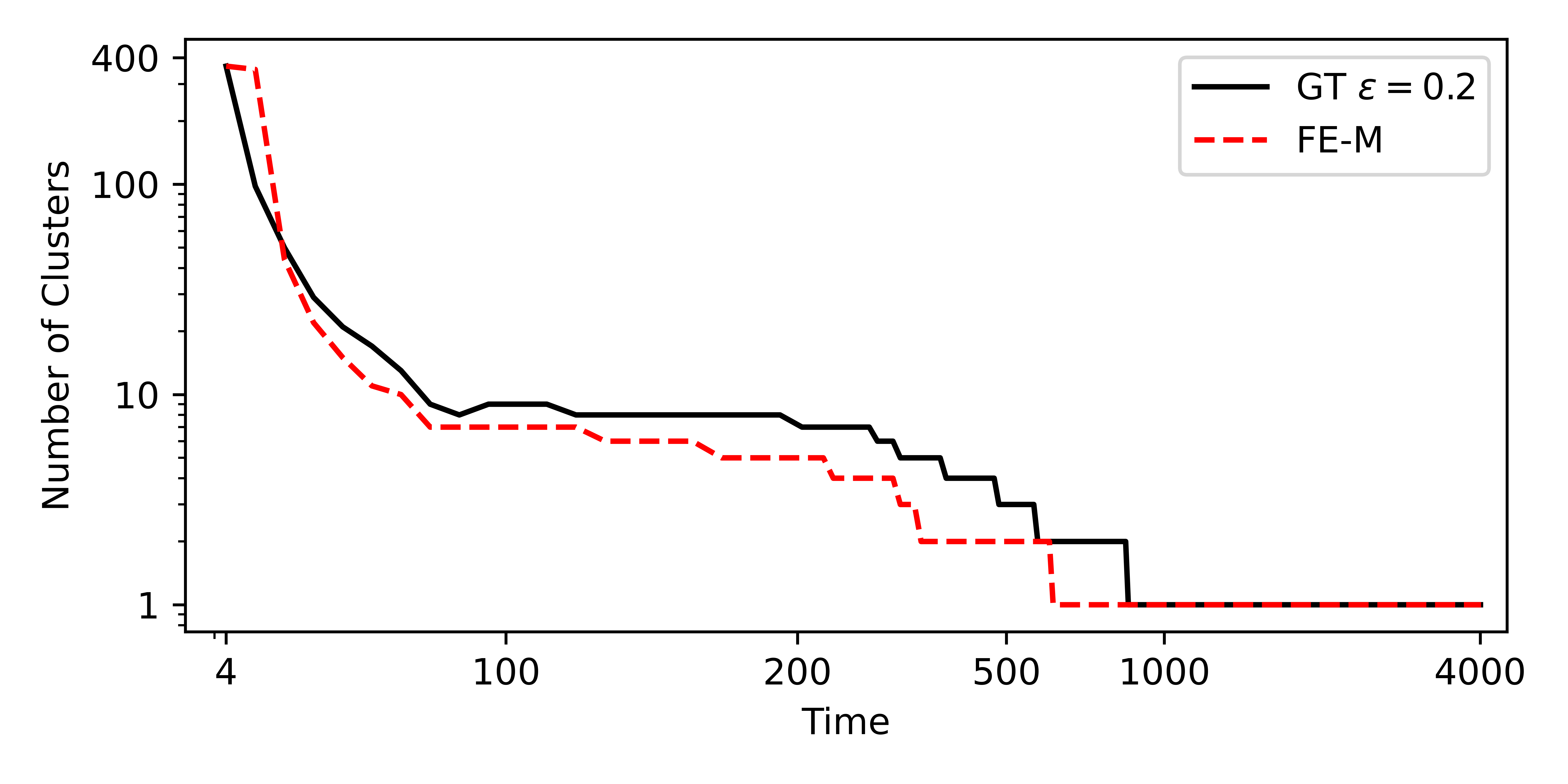}} &
        \raisebox{-0.1\height}{\includegraphics[width = 0.40\textwidth]{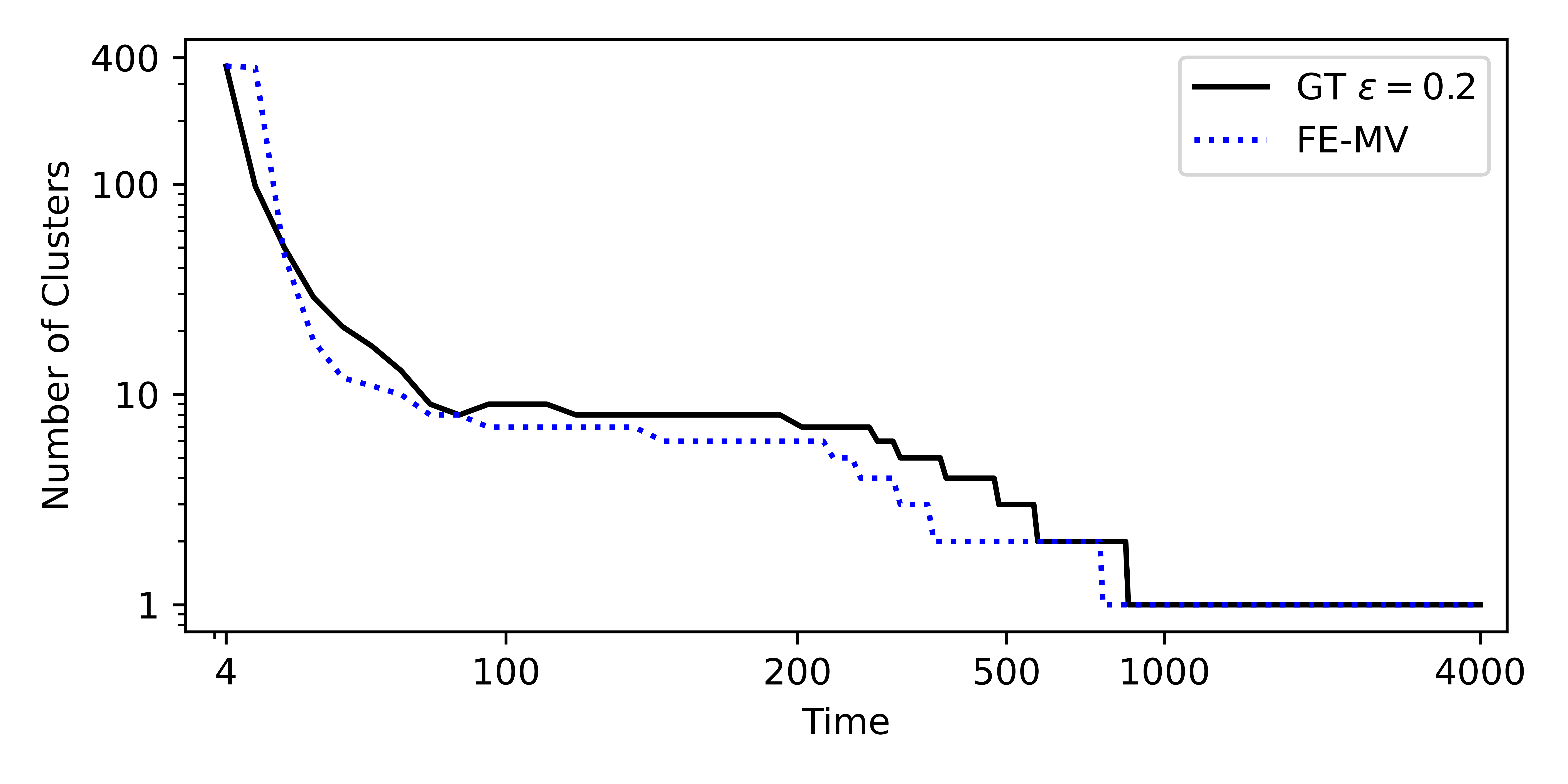}}
        \\
        \rotatebox[origin=l]{90}{Volume} &
        \raisebox{-0.1\height}{\includegraphics[width = 0.40\textwidth]{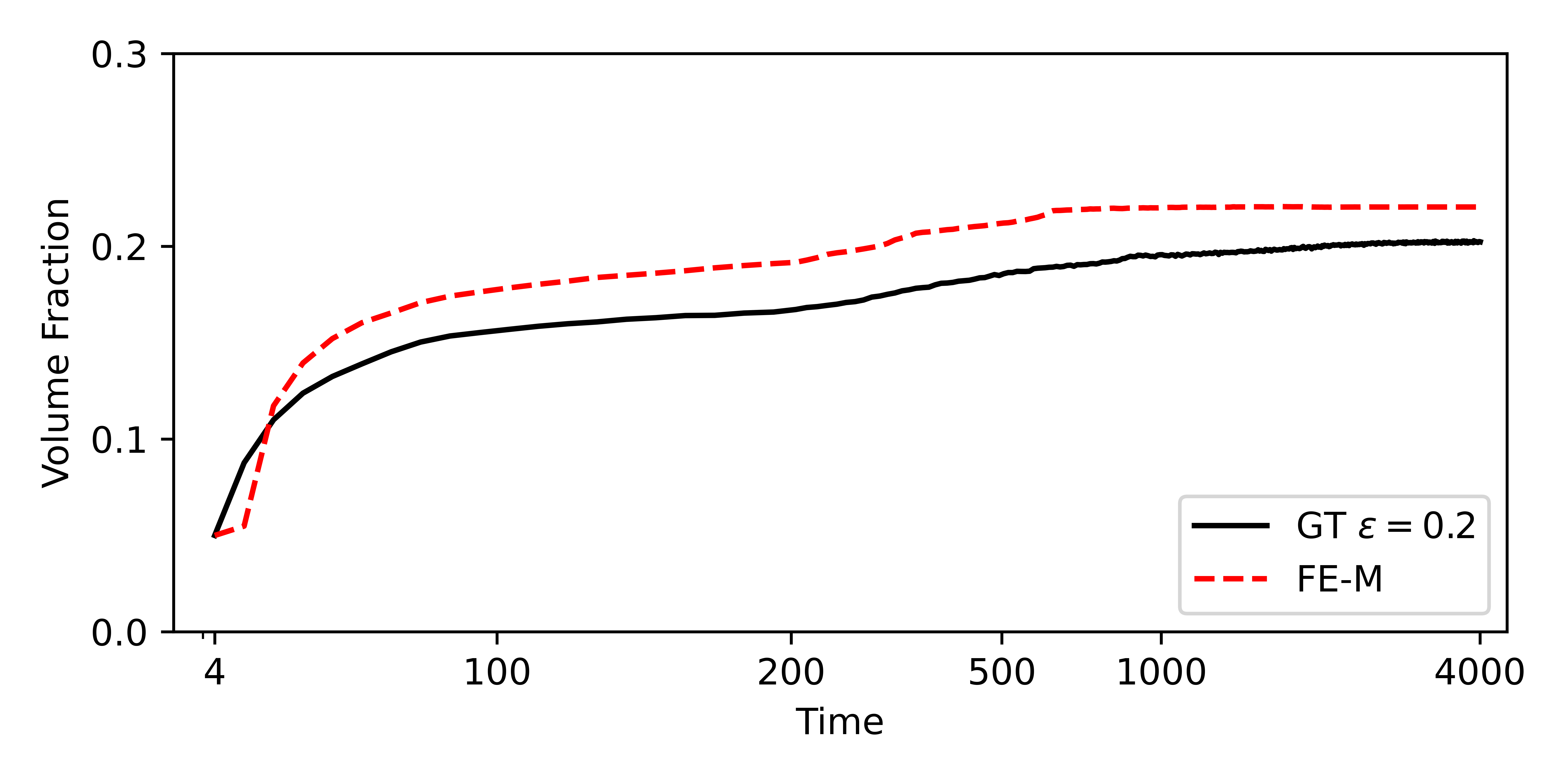}} &
        \raisebox{-0.1\height}{\includegraphics[width = 0.40\textwidth]{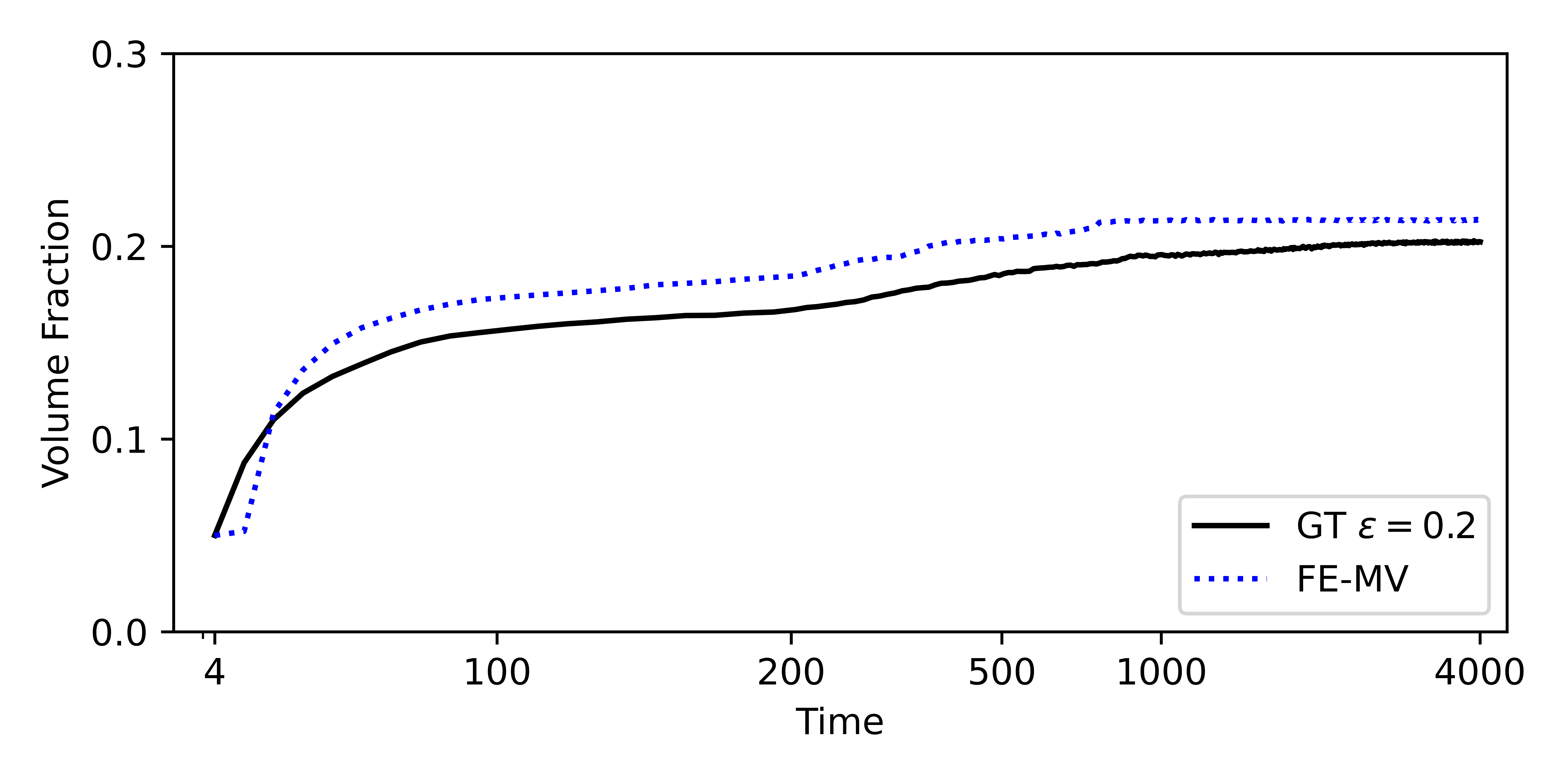}}
    \end{tabular}
    \caption{%{\bf time size 0.1, tskip 100} 
    Extrapolated agglomeration test for clusters and volume fraction evolution of $\langle c \rangle =0.4$  on a  $64^3$ grid}
    \label{fig:agg_clusters}
\end{figure}

\begin{figure}[H]
    \centering
    {
    \begin{tabular}{c} 
        Total energy curves
        \\ 
        % \raisebox{0.0\height}{\includegraphics[width = 0.48\textwidth]{fig/Final/energy_curves.png}} &
        \raisebox{0.0\height}{\includegraphics[width = 0.7\textwidth]{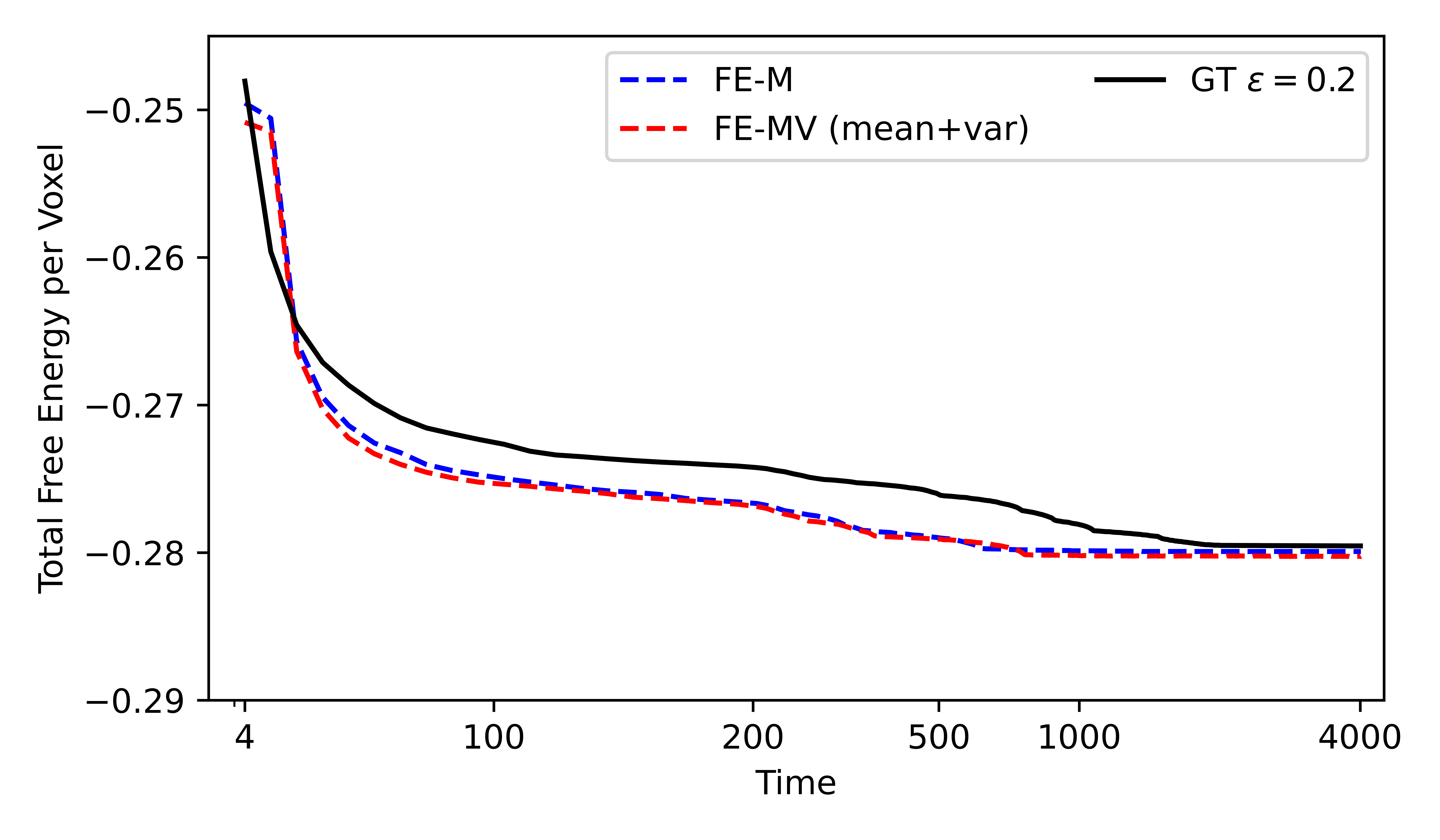}} 
    \end{tabular}
    }
    % \todo[inline]{How is the total energy computed?}
    \caption{%{\bf time size 0.1, tskip 100} 
    Total free energy evolution for the same tests.}
    \label{fig:agg_fe}
\end{figure}

Beyond in-distribution test, the extrapolation test is also important to check if the model learns the physics well. We performed  a long-term agglomeration test for $\langle c \rangle = 0.4$ on a $64^3$ grid. As shown previously in Fig.~\ref{fig:nonflux_glomeration_test},
% and \ref{fig:agg_test}, 
all the conservative surrogate models are stable compared to the ground truth up to $t=4000$ and reach similar stationary state of a single (nearly) spherical precipitate. 
%Given the initial concentration profile, the ground truth would have a spherical interface and we also see that both free energy-based model could predict the spherical interface. However, the evolving dynamics is faster to generate spherical surface for the free energy based learning models when $t=1000$. 
Moreover, we calculated the number of clusters and volume fractions. As shown in Fig.~\ref{fig:agg_clusters},  both free-energy based variants' predictions are generally aligned with the ground truth. We also plot the total energy per voxel versus time and show it in Fig.~\ref{fig:agg_fe}. This curve is computed as $A/\Omega$ where $A$ is the total free energy defined in Eq.~\ref{eq:def_energy} and $\Omega$ is the volume. Due to the arbitrary scaling for multiplicative mobility $M$ and energy $E$, the model predictions are scaled to match the overall scale of the ground truth.
The good alignment between machine learning predictions and the true data shows that our free energy model has learned the real physics well. %Therefore, our model has a good physics interpretability compared to black-box surrogates. 

\setlength{\tabcolsep}{2pt}

\setlength{\tabcolsep}{2pt}

\subsection{Machine learned free energy verification}
\label{sec:ml_free_E}
\begin{figure}[H]
    \centering
    \begin{tabular}{c}
        \raisebox{-0.05\height}{\includegraphics[width = 0.8\textwidth]{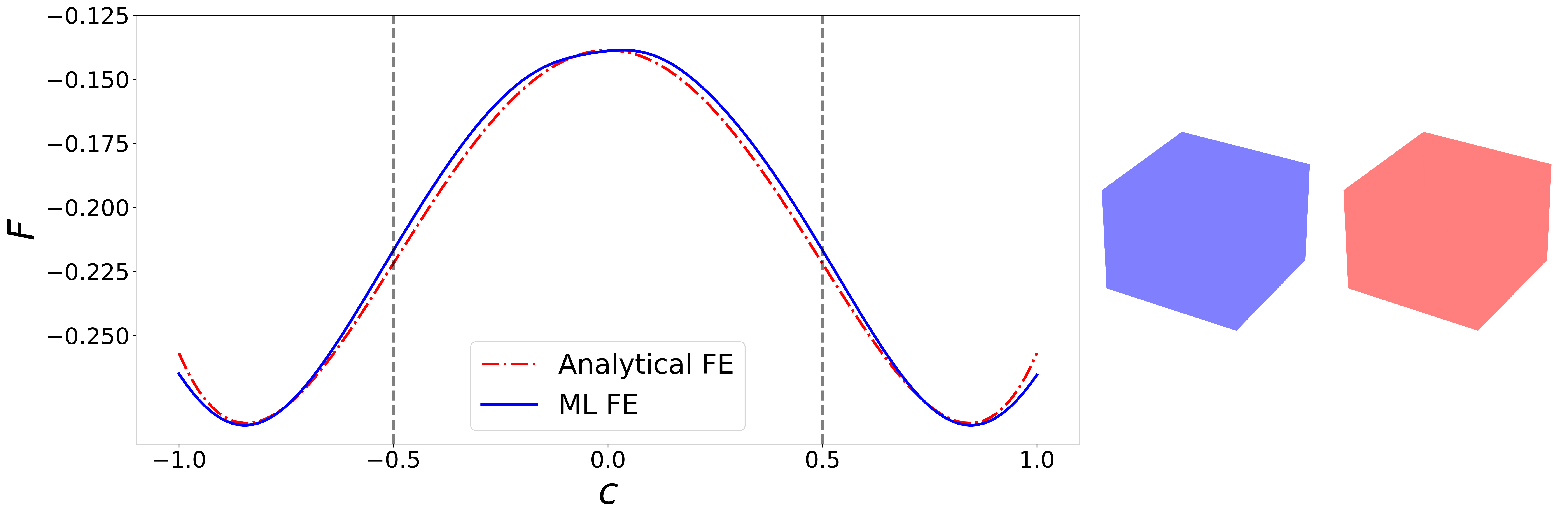}} \\
        (a) \\[0.3em]
        \raisebox{-0.05\height}{\includegraphics[width = 0.8\textwidth]{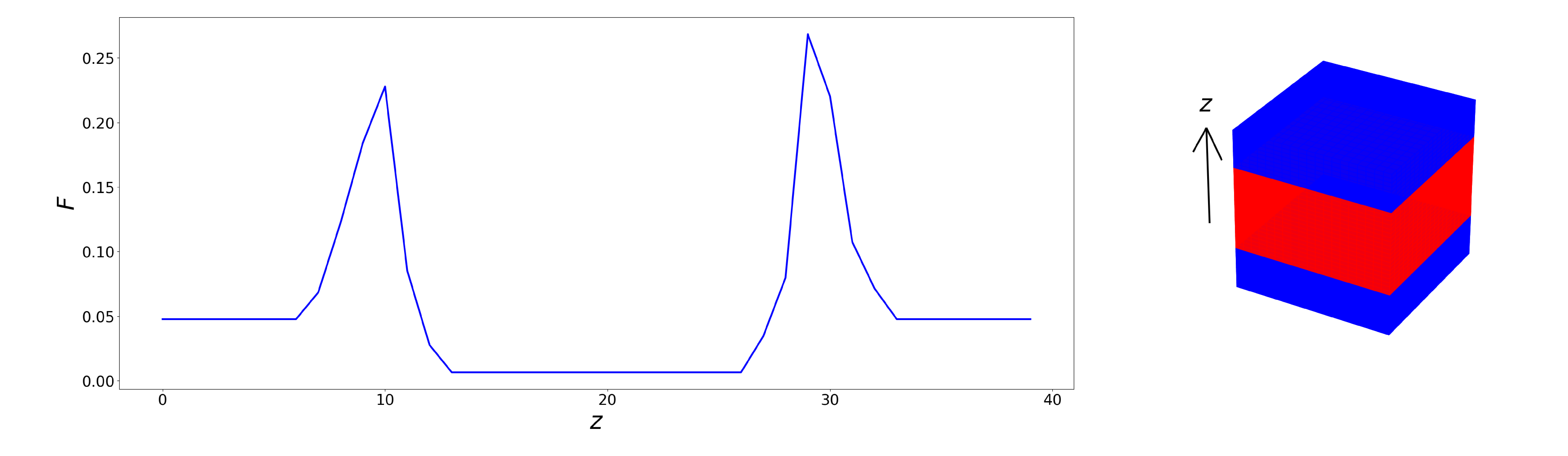}} \\
        (b) \\[0.3em]
        \raisebox{-0.05\height}{\includegraphics[width = 0.8\textwidth]{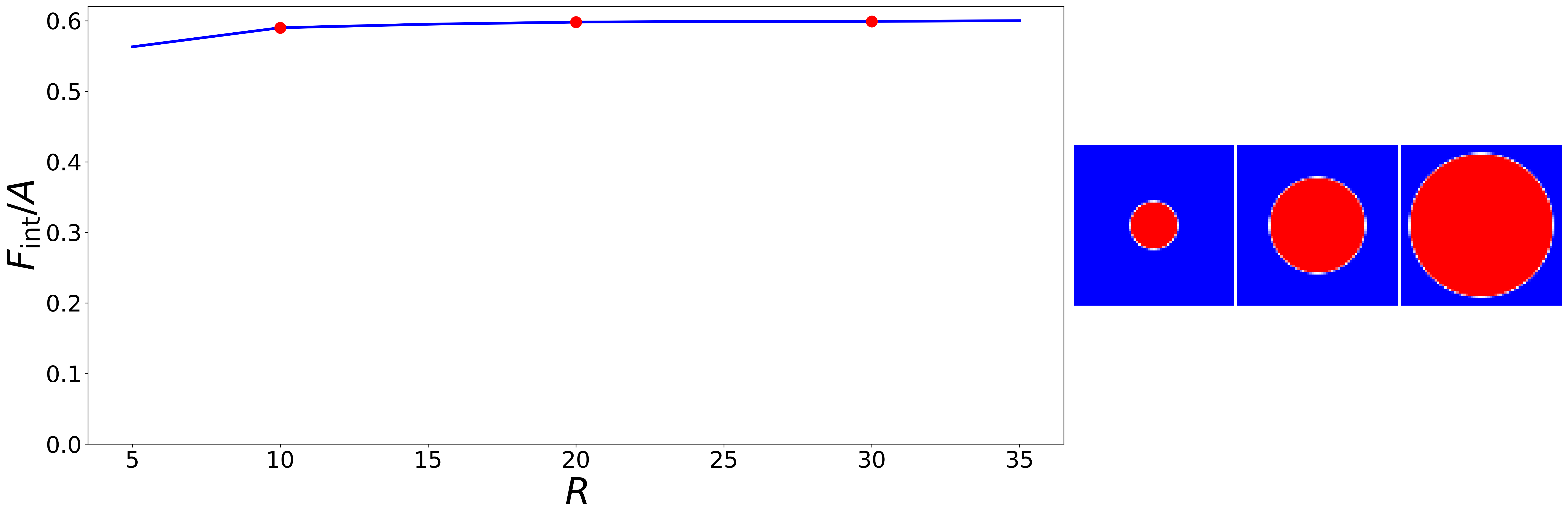}} \\
        (c)
    \end{tabular}
    \caption{\texttt{FE-MV} learned free energy curves. (a) Bulk free energy landscape. (b) Free energy for flat interface along the $z$ axis. (c) Interfacial free energy per area vs radius (spherical interface).}
    \label{fig:flat_interface}
\end{figure}

Since the learned energy model $\hat{E}(c)$ is a complicated, high-dimensional neural network rather than simple expressions and hard to digest directly, we query its behavior to see if it actually makes physical sense.
We plot the learned $\hat{E}(c)$ of \femv{} in Fig.~\ref{fig:flat_interface}. Subplot (a) shows the bulk free energy landscape of uniform field $E(\mathbf{c} \equiv c)$   versus $c$. With an affine calibration to match the scales (same as in Fig.~\ref{fig:agg_fe}), the learned bulk free energy closely follows the double well features of the ground truth in Fig.~\ref{fig:bulk_free_energy}.
%We rescale the learned bulk free energy curve by  so it can be directly compared with the analytical reference. 
% Let $c_L,c_R$ be the left/right minima, and $c_B$ the barrier location of the learned curve. 
% We first apply a linear ``tilt'' correction $\tilde a_b(c)=a_b^{\mathrm{ML}}(c)+k c$ with $k=\big(a_b^{\mathrm{ML}}(c_R)-a_b^{\mathrm{ML}}(c_L)\big)/(c_L-c_R)$ to enforce equal well depths. 
% We then scale and shift $\tilde a_b$ to match the reference barrier height (relative to the mean of the two minima) and align the mean well depth with the reference curve, yielding the calibrated landscape $a_b^{\mathrm{aligned}}(c)$ used for visualization and subsequent energy diagnostics. The input concentration is a uniform constant field for this subplot and we also visualizes the field when $c=0.25$ and $c=0.75$ and plot them in the right part of the subplot.

Fig.~\ref{fig:flat_interface}b shows the free energy profile for a manually created flat interface unit-cell of two-phase coexistence. The total height of the cell is 40 and the two sharp peaks at around $z=10$ and $z=30$ indicate the interfacial regions where $c$ transitions between A and B phases. The plotted energy was averaged on the $x-y$ plane. The energy peaks represent the excess free energy cost of creating the interface. These results indicate that our machine-learned free energy model correctly grasped the key physics involved in immiscible phases.
%And we attached the vertical flat interface in the right part of this subplot.

Fig.~\ref{fig:flat_interface}c shows the interfacial free energy change versus radius. Each data point was evaluated with a sphere of minority of various radius embedded in the majority phase. The model predicts an interfacial tension between $0.56-0
.60$, which is approximately independent of curvature, in agreement with the ground truth isotropic model.
%The absolute value may differ from the ground truth by a constant scaling factor, but 
%The model correctly captures the lack of curvature dependence. 
%We also plot cross-section curves when the radius $R=10, 20$ and $30$ respectively. 
In short, all the three plots confirm the proposed model's ability to learn the correct thermodynamic behavior for phase separation dynamics. Remarkably, the model learned the thermodynamics purely from observing the trajectories of order parameters with no knowledge of energy or chemical potential.
% \begin{figure}[H]
% \centering
% \includegraphics[width = 0.9\textwidth]{fig/FreeEnergy/analysis_LJ_v3_struct_3d_22.png}

\begin{figure}[htp!]
    \centering
    \begin{tabular}{c c c}
         & $t=50$ & $t=2400$ \\
        \rotatebox[origin=l]{90}{GT $\epsilon=0.2$} &
        \raisebox{-0.2\height}{\includegraphics[width = 0.3\textwidth]{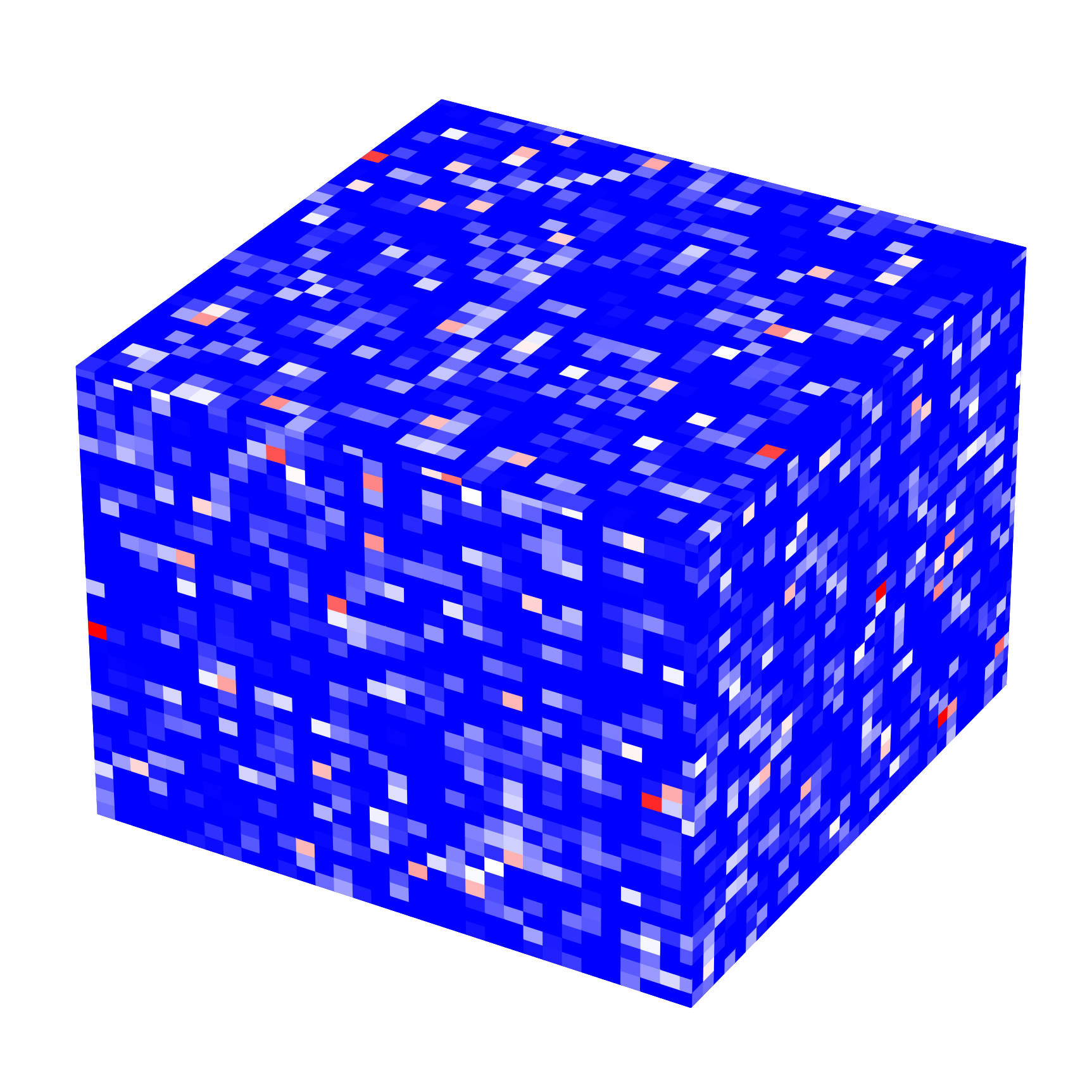}} &
        \raisebox{-0.2\height}{\includegraphics[width = 0.3\textwidth]{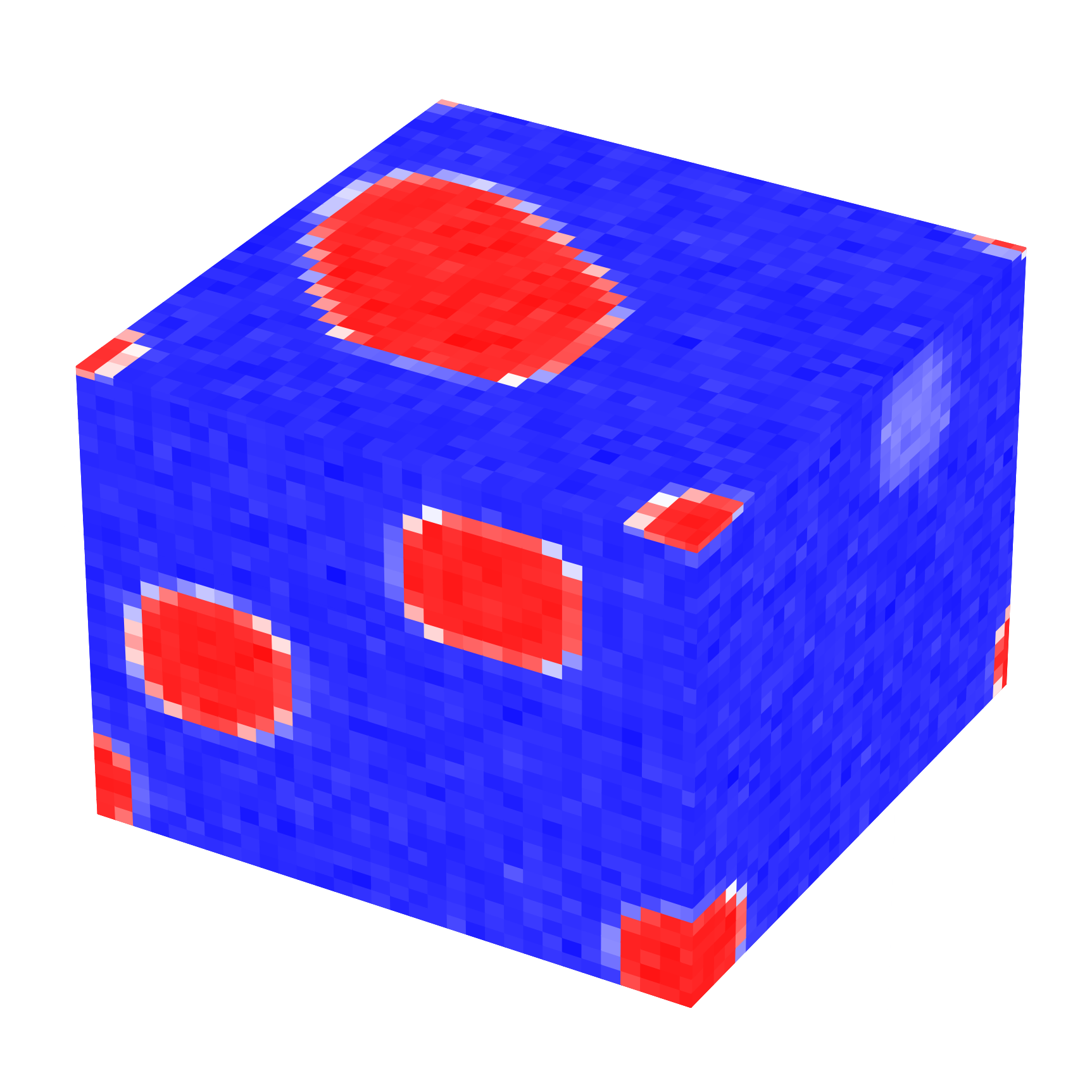}} \\
        & \multicolumn{2}{c}{(a)} \\[0.3em]
        \rotatebox[origin=l]{90}{\femv{}} &
        \raisebox{-0.2\height}{\includegraphics[width = 0.3\textwidth]{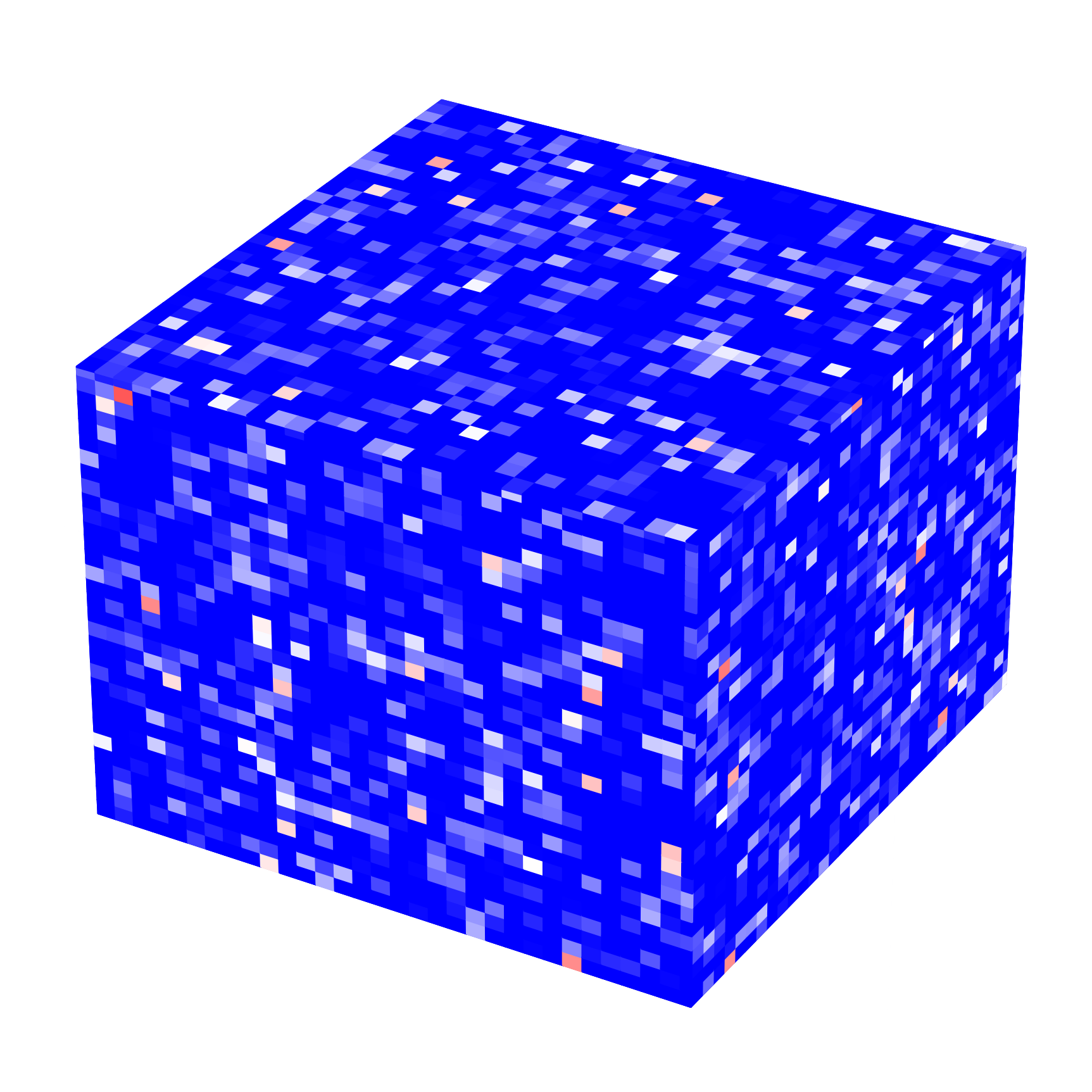}} &
        \raisebox{-0.2\height}{\includegraphics[width = 0.3\textwidth]{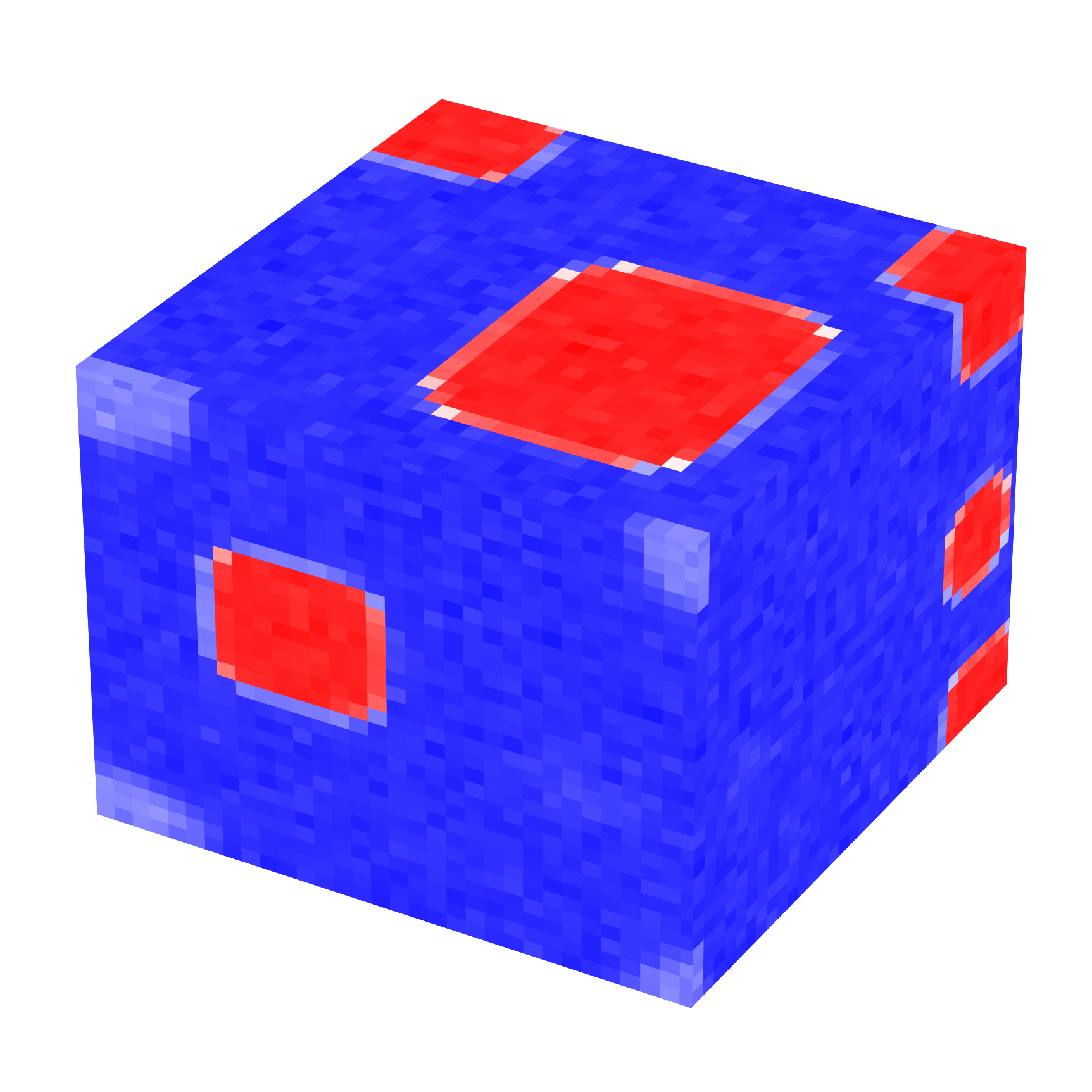}} \\
        & \multicolumn{2}{c}{(b)} \\[0.3em]
        & \multicolumn{2}{c}{\includegraphics[width=.7\textwidth]{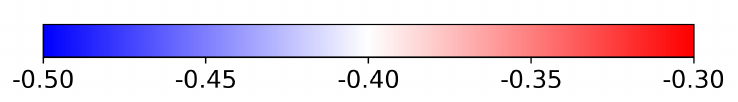}}
    \end{tabular}
    \caption{Comparison for extrapolation trajectory $\langle c \rangle =-0.51$. (a) Ground truth value $\epsilon=0.2$. (b) \femv{} model.}
    \label{fig:extra_c051}
\end{figure}

\begin{figure}[htp!]
\centering
\includegraphics[width=0.5\textwidth]{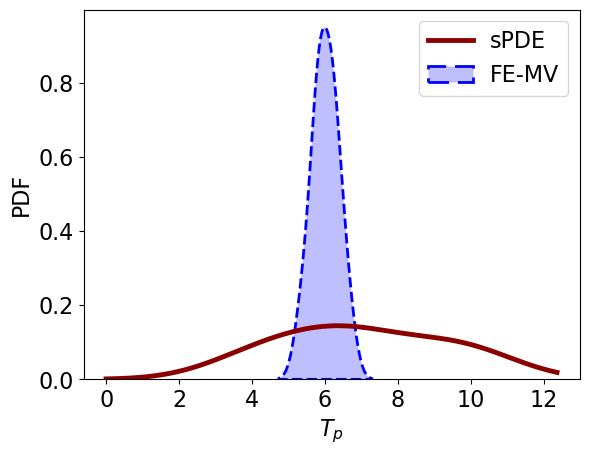}
\caption{Distribution of precipitation time $t_p$ over 40 independent runs, \femv{} VS stochastic PDE simulation.}
\label{fig:extra_c051_dist}
\end{figure}

\subsection{Generalization to nucleation dynamics: the case for stochastic surrogates}
\label{sec:extra_con}

The spinodal decomposition regime used for training provides an ideal setting for learning and validating the surrogate architecture, but it does not expose the most fundamental advantage of flux-level stochasticity: the ability to simulate thermally activated, barrier-crossing events that are entirely inaccessible to deterministic models. We now demonstrate this directly by evaluating both \femv{} and \fem{} on initial concentrations in the metastable region, outside the spinodal training region $|\langle c \rangle| < 0.5$.

At $\langle c \rangle =-0.51$, just outside the spinodal boundary, the deterministic \fem{} surrogate produces no nucleation events at all without seeds. This is not a failure of training or architecture tuning; it is a fundamental limitation of any deterministic model, which cannot spontaneously generate the rare thermal fluctuation required to cross the nucleation barrier. By contrast, the stochastic \femv{} successfully nucleates without seeds, in qualitative agreement with the ground truth SPDE (Fig.~\ref{fig:extra_c051}).
To assess this quantitatively, we define the precipitation time $t_p$ as the first time at which one or more voxels appear with $|\langle c \rangle |>0.8$, and evaluate an ensemble of 40 independent \femv{} rollouts. The surrogate yields a mean precipitation time $\langle \hat{t}_p \rangle=6.2$, close to the ground truth value of 7.0. The predicted distribution is narrower than the ground truth (Fig.~\ref{fig:extra_c051_dist}), which is expected: the surrogate was trained exclusively on spinodal trajectories, so it has no direct exposure to the statistics of barrier-crossing fluctuations in the metastable region. Extending deeper into the metastable regime, where barrier-crossing nucleation events are rarer in the training data, requires a dedicated data strategy, such as active learning or importance sampling of rare trajectories, which is deferred to future work.

These results establish a clear distinction between the deterministic and stochastic variants: in the spinodal regime both perform comparably on long-time statistics, though the former exhibits delayed early-time dynamics. But only the stochastic surrogate is a viable model class for the metastable regime. This distinction is not a quantitative difference in accuracy, but a qualitative one. A surrogate without flux-level noise cannot model nucleation at all, regardless of how well it is trained. This section therefore serves as the clearest justification for the architectural choices made in this work.

% \section{Ablation Study}
% \label{sec:discussion}

% \LS{maybe show metrics? plots?}

\section{Conclusion}
\label{sec:conclusion}
In this work, we developed and evaluated a family of physics-aware surrogate models for the stochastic Cahn–Hilliard equation, a prototypical model of noise-driven conserved phase separation dynamics. The central architectural innovation is to parameterize the surrogate at the level of inter-cell fluxes rather than predicted states, which guarantees exact mass conservation by construction and allows stochasticity to be introduced at its physically correct scale — the fluctuating current between neighboring cells.

Through systematic ablation across five model variants, we find that all three design choices are necessary for a physically consistent surrogate: flux-level conservation, a learnable free energy function, and explicit stochastic flux amplitude. Removing conservation (non-flux baseline) produces a model that fails catastrophically under spatiotemporal extrapolation. Removing the free energy constraint (nonFE-MV, nonFE-M) yields models that reproduce coarsening statistics reasonably well but learn a mobility field that is physically inconsistent, exhibiting error cancellation between a non-conservative driving force and miscalibrated kinetic coefficients. Removing stochasticity (FE-M) produces a deterministic surrogate that matches long-time ensemble statistics in the spinodal regime but introduces systematic delay in early-time demixing kinetics. More critically, it is entirely unable to model nucleation or barrier crossing in the metastable regime. Only FE-MV, the full stochastic free-energy variant, satisfies all three physical requirements simultaneously and generalizes across regimes.

As an early attempt at quantitative surrogate modeling of stochastic dynamical systems, this work made several simplifying assumptions.
The covariance matrix for fluxes was assumed to be diagonal. More complicated generative models with nonvanishing covariance between different fluxes or even non-Gaussian noises will be interesting subjects for the future. We also fixed the temperature to a single value. The nucleation and growth mechanism for metastable concentrations beyond the present unstable spinodal decomposition reactions will also be the subject of future studies. 
The most important generalization of this paper is towards applications in molecular dynamics data. Work is underway to develop surrogate models trained on trajectories of molecular dynamics simulations and will be published separately.

\section*{Acknowledgements}
This work was performed under the auspices of the
U.S. Department of Energy by Lawrence Livermore
National Laboratory under Contract DE-AC52-07NA27344. This work was funded by the
Laboratory Directed Research and Development
(LDRD) Program at LLNL under project tracking
code 25-ERD-002 and 23-ERD-029. Computing support for this work came from the LLNL Institutional Computing Grand Challenge program.

% Supplemental information to this work is available free of charge at https://pubs.acs.org

\section*{Author Contributions}
VHN and LS performed dataset set generation, model refinement and training, and data analysis. FZ initiated the project, wrote the code base, and supervised the research. SL and JK secured the funding. LS, VHN and FZ wrote the paper with inputs from all authors.

\section*{Data Availability}
The data required to reproduce this work can be requested from FZ.

\section*{Code Availability}
The source code and a demo for implementing the SD rollout are available at \url{http://www.github.com/llnl/NPS}.

\section*{Competing interests}
The authors state that there is no conflict of interest.

\printbibliography

@article{convnext,
archivePrefix = {arXiv},
arxivId = {2201.03545},
author = {Liu, Zhuang and Mao, Hanzi and Wu, Chao-Yuan and Feichtenhofer, Christoph and Darrell, Trevor and Xie, Saining},
eprint = {2201.03545},
journal = {arxiv: 2201.03545},
title = {{A ConvNet for the 2020s}},
url = {http://arxiv.org/abs/2201.03545},
year = {2022}
}

@article{adamw,
archivePrefix = {arXiv},
arxivId = {1711.05101},
author = {Loshchilov, Ilya and Hutter, Frank},
eprint = {1711.05101},
journal = {arxiv: 1711.05101},
title = {{Decoupled weight decay regularization}},
year = {2017}
}

@article{Bronchart2008PRL-Finel,
author = {Bronchart, Q and {Le Bouar}, Y and Finel, A},
doi = {10.1103/PhysRevLett.100.015702},
issn = {0031-9007},
journal = {Physical Review Letters},
month = {jan},
number = {1},
pages = {015702},
publisher = {APS},
title = {{New Coarse-Grained Derivation of a Phase Field Model for Precipitation}},
url = {https://link.aps.org/doi/10.1103/PhysRevLett.100.015702},
volume = {100},
year = {2008}
}

@article{raissi2019physics,
  title={Physics-informed neural networks: A deep learning framework for solving forward and inverse problems involving nonlinear partial differential equations},
  author={Raissi, Maziar and Perdikaris, Paris and Karniadakis, George E},
  journal={Journal of Computational physics},
  volume={378},
  pages={686--707},
  year={2019},
  publisher={Elsevier}
}

@article{sun2020surrogate,
  title={Surrogate modeling for fluid flows based on physics-constrained deep learning without simulation data},
  author={Sun, Luning and Gao, Han and Pan, Shaowu and Wang, Jian-Xun},
  journal={Computer Methods in Applied Mechanics and Engineering},
  volume={361},
  pages={112732},
  year={2020},
  publisher={Elsevier}
}

@article{wang2021understanding,
  title={Understanding and mitigating gradient flow pathologies in physics-informed neural networks},
  author={Wang, Sifan and Teng, Yujun and Perdikaris, Paris},
  journal={SIAM Journal on Scientific Computing},
  volume={43},
  number={5},
  pages={A3055--A3081},
  year={2021},
  publisher={SIAM}
}

@article{lu2019deeponet,
  title={Deeponet: Learning nonlinear operators for identifying differential equations based on the universal approximation theorem of operators},
  author={Lu, Lu and Jin, Pengzhan and Karniadakis, George Em},
  journal={arXiv preprint arXiv:1910.03193},
  year={2019}
}

@article{li2020fourier,
  title={Fourier neural operator for parametric partial differential equations},
  author={Li, Zongyi and Kovachki, Nikola and Azizzadenesheli, Kamyar and Liu, Burigede and Bhattacharya, Kaushik and Stuart, Andrew and Anandkumar, Anima},
  journal={arXiv preprint arXiv:2010.08895},
  year={2020}
}

@article{vaswani2017attention,
  title={Attention is all you need},
  author={Vaswani, Ashish and Shazeer, Noam and Parmar, Niki and Uszkoreit, Jakob and Jones, Llion and Gomez, Aidan N and Kaiser, {\L}ukasz and Polosukhin, Illia},
  journal={Advances in neural information processing systems},
  volume={30},
  year={2017}
}

@article{han2022predicting,
  title={Predicting physics in mesh-reduced space with temporal attention},
  author={Han, Xu and Gao, Han and Pfaff, Tobias and Wang, Jian-Xun and Liu, Li-Ping},
  journal={arXiv preprint arXiv:2201.09113},
  year={2022}
}

@article{sun2023unifying,
  title={Unifying predictions of deterministic and stochastic physics in mesh-reduced space with sequential flow generative model},
  author={Sun, Luning and Han, Xu and Gao, Han and Wang, Jian-Xun and Liu, Liping},
  journal={Advances in Neural Information Processing Systems},
  volume={36},
  pages={60636--60660},
  year={2023}
}

@article{gao2024bayesian,
  title={Bayesian conditional diffusion models for versatile spatiotemporal turbulence generation},
  author={Gao, Han and Han, Xu and Fan, Xiantao and Sun, Luning and Liu, Li-Ping and Duan, Lian and Wang, Jian-Xun},
  journal={Computer Methods in Applied Mechanics and Engineering},
  volume={427},
  pages={117023},
  year={2024},
  publisher={Elsevier}
}

@article{du2024conditional,
  title={Conditional neural field latent diffusion model for generating spatiotemporal turbulence},
  author={Du, Pan and Parikh, Meet Hemant and Fan, Xiantao and Liu, Xin-Yang and Wang, Jian-Xun},
  journal={Nature Communications},
  volume={15},
  number={1},
  pages={10416},
  year={2024},
  publisher={Nature Publishing Group UK London}
}

@incollection{voter2007introduction,
  title={Introduction to the kinetic Monte Carlo method},
  author={Voter, Arthur F},
  booktitle={Radiation effects in solids},
  pages={1--23},
  year={2007},
  publisher={Springer}
}

@article{binder1985monte,
  title={The Monte Carlo method for the study of phase transitions: A review of some recent progress},
  author={Binder, Kurt},
  journal={Journal of Computational Physics},
  volume={59},
  number={1},
  pages={1--55},
  year={1985},
  publisher={Elsevier}
}

@article{cahn1958free,
  title={Free energy of a nonuniform system. I. Interfacial free energy},
  author={Cahn, John W and Hilliard, John E},
  journal={The Journal of chemical physics},
  volume={28},
  number={2},
  pages={258--267},
  year={1958},
  publisher={American Institute of Physics}
}

@article{chen2002phase,
  title={Phase-field models for microstructure evolution},
  author={Chen, Long-Qing},
  journal={Annual review of materials research},
  volume={32},
  number={1},
  pages={113--140},
  year={2002},
  publisher={Annual Reviews 4139 El Camino Way, PO Box 10139, Palo Alto, CA 94303-0139, USA}
}

@article{cook1970brownian,
  title={Brownian motion in spinodal decomposition},
  author={Cook, HE},
  journal={Acta metallurgica},
  volume={18},
  number={3},
  pages={297--306},
  year={1970},
  publisher={Elsevier}
}

@article{Frishman2020PRX-SFI,
  author  = {Frishman, Anna and Ronceray, Pierre},
  title   = {Learning Force Fields from Stochastic Trajectories},
  journal = {Phys.\ Rev.\ X},
  volume  = {10},
  pages   = {021009},
  year    = {2020},
  doi     = {10.1103/PhysRevX.10.021009}
}

@article{Haas2013JPCB,
  author  = {Haas, Kevin R. and Yang, Haw and Chu, Jhih-Wei},
  title   = {Expectation-Maximization of the Potential of Mean Force and
             Diffusion Coefficient in {Langevin} Dynamics from Single-Molecule
             {FRET} Data Photon by Photon},
  journal = {J.\ Phys.\ Chem.\ B},
  volume  = {117},
  number  = {49},
  pages   = {15591--15605},
  year    = {2013},
  doi     = {10.1021/jp405983d}
}

@article{Yang2021P,
author = {Yang, Kaiqi and Cao, Yifan and Zhang, Youtian and Fan, Shaoxun and Tang, Ming and Aberg, Daniel and Sadigh, Babak and Zhou, Fei},
doi = {10.1016/j.patter.2021.100243},
journal = {Patterns},
month = {apr},
pages = {100243},
publisher = {Elsevier Inc.},
title = {{Self-supervised learning and prediction of microstructure evolution with convolutional recurrent neural networks}},
volume = {2},
year = {2021}
}

@article{Fan2024MLST,
archivePrefix = {arXiv},
arxivId = {2310.15153},
author = {Fan, Shaoxun and Hitt, Andrew L. and Tang, Ming and Sadigh, Babak and Zhou, Fei},
doi = {10.1088/2632-2153/ad3e4b},
eprint = {2310.15153},
issn = {2632-2153},
journal = {Machine Learning: Science and Technology},
month = {jun},
number = {2},
pages = {025027},
title = {{Accelerate microstructure evolution simulation using graph neural networks with adaptive spatiotemporal resolution}},
volume = {5},
year = {2024}
}

@article{MontesdeOcaZapiain2021nCM,
author = {{Montes de Oca Zapiain}, David and Stewart, James A. and Dingreville, R{\'{e}}mi},
journal = {npj Computational Materials},
number = {1},
pages = {1--11},
title = {{Accelerating phase-field-based microstructure evolution predictions via surrogate models trained by machine learning methods}},
volume = {7},
year = {2021}
}

@article{Bertin2024nCM-Learning,
author = {Bertin, Nicolas and Bulatov, Vasily V. and Zhou, Fei},
doi = {10.1038/s41524-024-01378-4},
eprint = {2309.14450},
issn = {2057-3960},
journal = {npj Computational Materials},
month = {aug},
number = {1},
pages = {192},
title = {{Learning dislocation dynamics mobility laws from large-scale MD simulations}},
volume = {10},
year = {2024}
}

@misc{Tian2025-Scaling,
  title = {Scaling {{Kinetic Monte-Carlo Simulations}} of {{Grain Growth}} with {{Combined Convolutional}} and {{Graph Neural Networks}}},
  author = {Tian, Zhihui and Suwandi, Ethan and Oppelstrup, Tomas and Bulatov, Vasily V. and Harley, Joel B. and Zhou, Fei},
  year = 2025,
  number = {2511.17848},
  eprint = {2511.17848},
  publisher = {arXiv},
  doi = {10.48550/ARXIV.2511.17848},
  urldate = {2026-03-05},
  archiveprefix = {arXiv}
}

@article{Ji2025-Scalable,
archivePrefix = {arXiv},
arxivId = {2511.03884},
author = {Ji, Kaihua and Sun, Luning and Liu, Shusen and Zhou, Fei and Heo, Tae Wook},
eprint = {2511.03884},
pages = {1--29},
title = {{Scalable Autoregressive Deep Surrogates for Dendritic Microstructure Dynamics}}
}

@article{Becker1935AP-CNT,
  title = {Kinetische Behandlung der Keimbildung in {\"u}bers{\"a}ttigten D{\"a}mpfen},
  author = {Becker, R. and D{\"o}ring, W.},
  journal = {Annalen der Physik},
  volume = {416},
  number = {8},
  pages = {719--752},
  year = {1935},
  doi = {10.1002/andp.19354160806}
}

% \section{Appendix}
\newpage
\setcounter{figure}{0}
\renewcommand{\figurename}{Supplementary Figure}
\renewcommand{\thefigure}{S\arabic{figure}}

\setcounter{table}{0}
\renewcommand{\tablename}{Supplementary Table}
\renewcommand{\theHtable}{Supplement.\thetable}

\appendix

\section{Phase diagram}
Phase diagram details are shown in Supplementary Fig.~\ref{fig:bulk_free_energy_app}.
\begin{figure}[htp]
%Plot[ c^4/4 -c^2/2 + 0.2(((1+c)/2)*log((1+c)/2) + ((1-c)/2)*log((1-c)/2)), {c, -1,1}]
    \centering
    % \includegraphics[width=0.7\textwidth]{fig/F_mu_comparison}
%%%%%%%%%%%%%%%%%%%%%%%%%%% move below to SI
    % \includegraphics[width=0.7\textwidth]{fig/toy2d-free-energy.png}
    \begin{tabular}{c| c c c c c}
        % \multicolumn{3}{c}{Free energy curves at different temperature T} & 
        % \multicolumn{3}{c}{Free energy landscape and solubility limit}
        % \\
        \multicolumn{3}{c}{\includegraphics[width=.47\textwidth]{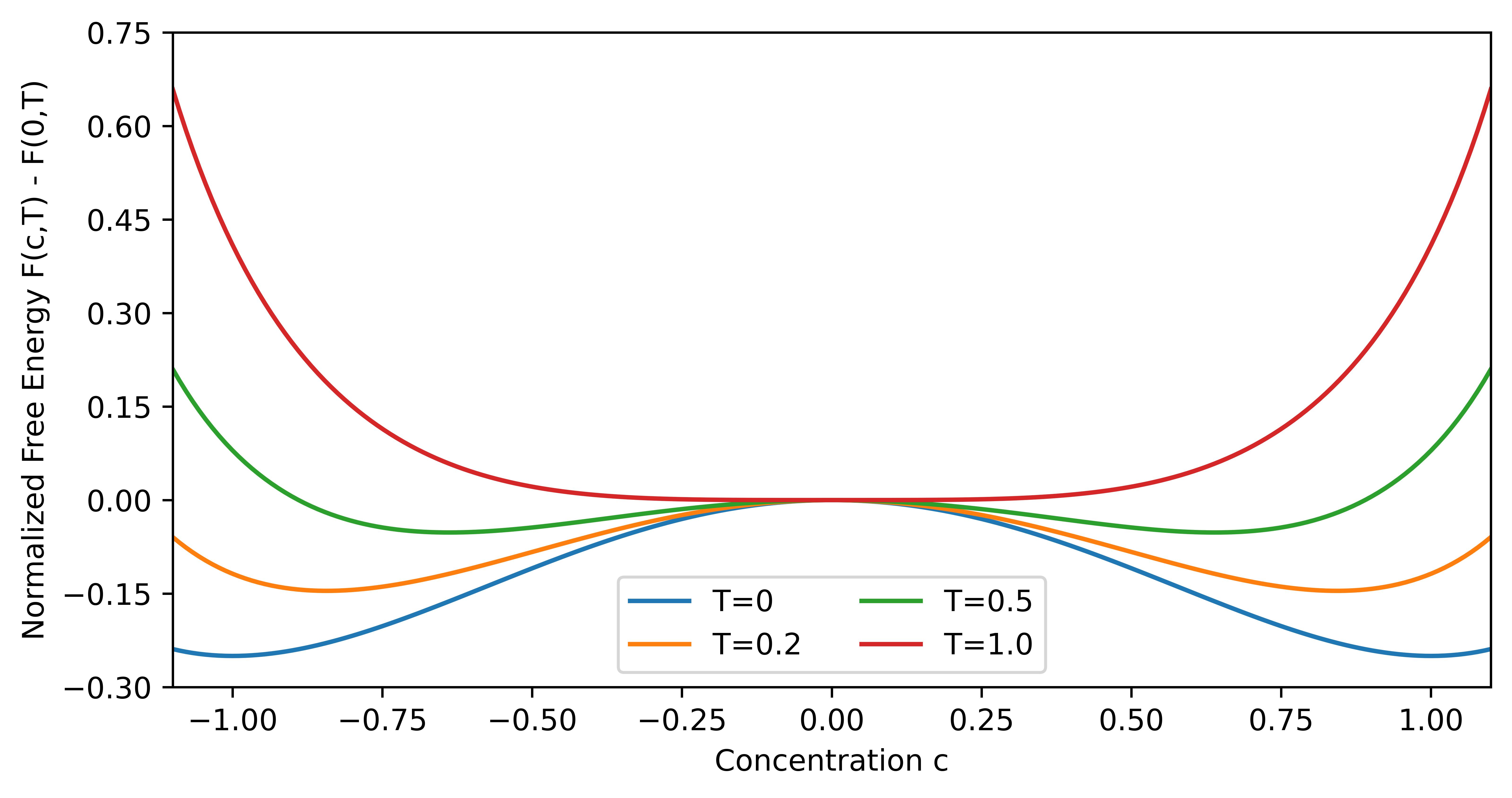}} 
        & 
        \multicolumn{3}{c}{\includegraphics[width=.47\textwidth]{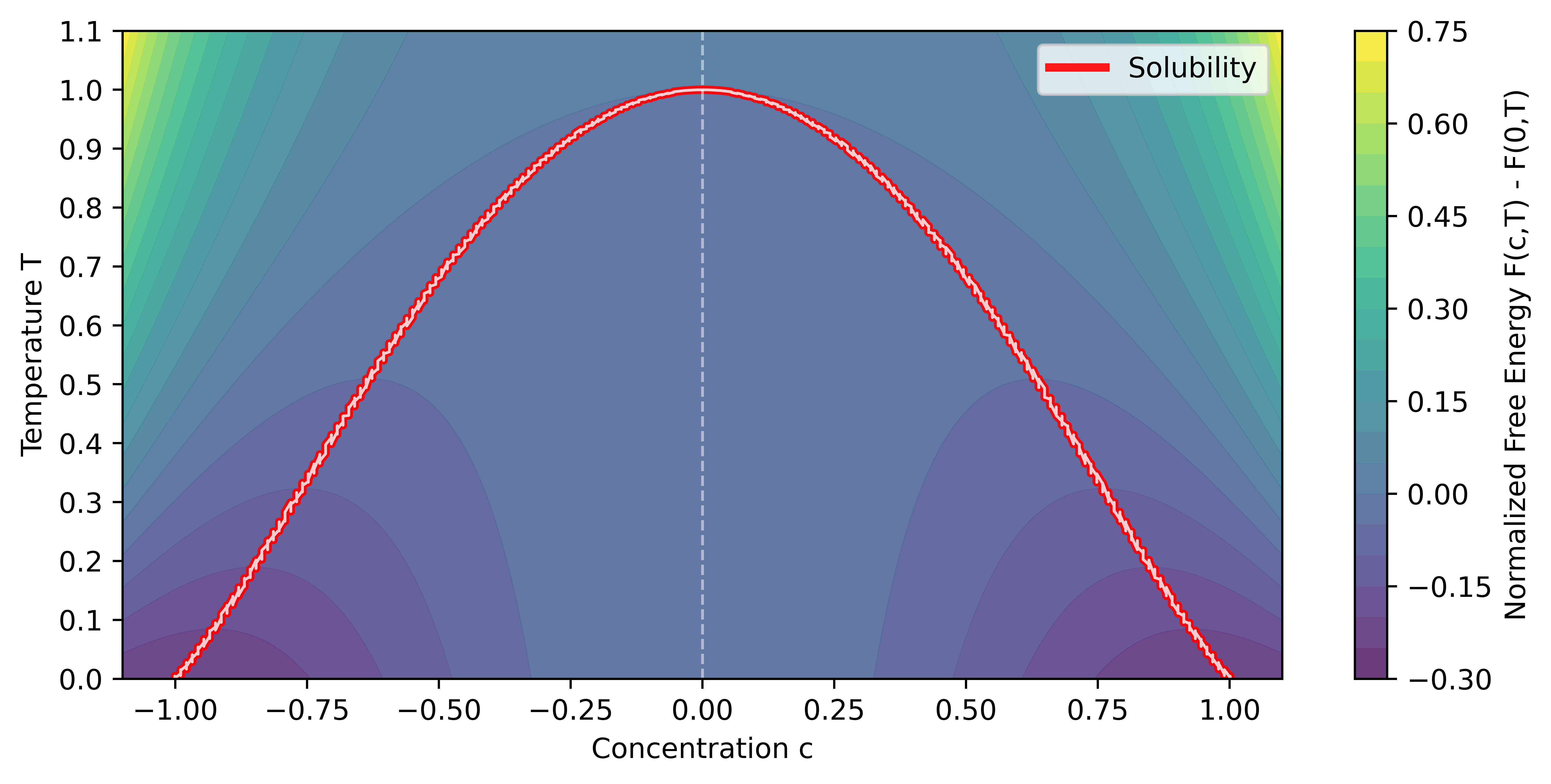}} 
    \end{tabular}
%%%%%%%%%%%%%%%%%%%%%%%%%%% END
    \caption{     Bulk free energy and solubility limit at different temperature. 
    %The concentrations of solubility ($\pm 0.85$) and spinodal ($\pm 0.5$) were highlighted.
 % \added{[Suggestion: move original figures (bottom rows) to appendix ]}
    %, and replace with: subfig (a), FE curve with T=0.2 only (these whole T range is not really considered and only distraction.) (b) a schematic highlighting that we are using GT of SPDE, and surrogate are spatiotemporally downsampled generative NN predicting mean and variance.]}
    }
    \label{fig:bulk_free_energy_app}
\end{figure}

\section{Additional details on 1D and 3D SPDE solutions}

\subsection{1D SPDE analysis for \cozero}
\label{sec:1D_SPDE_32}
Additional 1D statistics on the 32-grid case are shown in Supplementary Fig.~\ref{fig:1d_frames_fourier_copo_32_grid}.
\begin{figure}[htp]
    \centering
    \begin{tabular}{c c}
        no noise $\epsilon=0$ & no noise $\epsilon=0.2$ 
        \\
        \raisebox{-0.5\height}{\includegraphics[width = 0.5\textwidth]{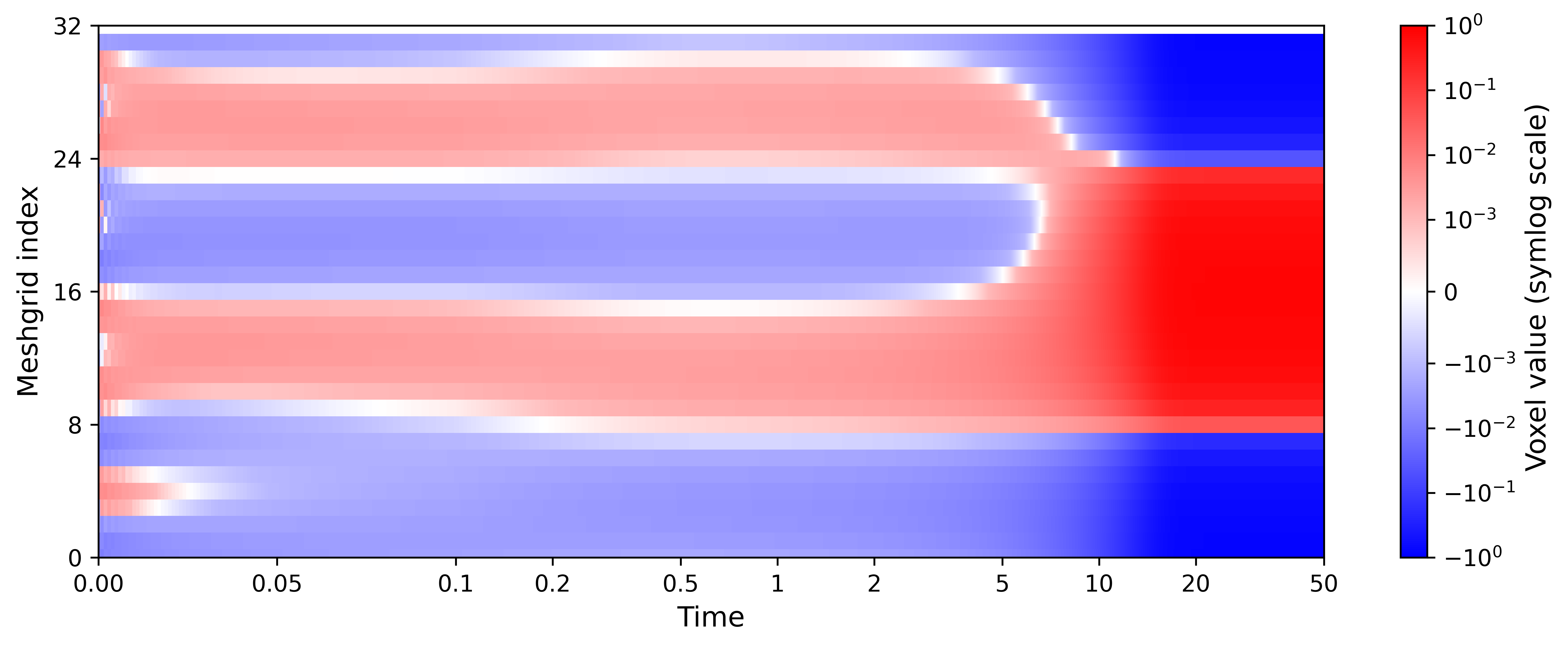}} &
        \raisebox{-0.5\height}{\includegraphics[width = 0.5\textwidth]{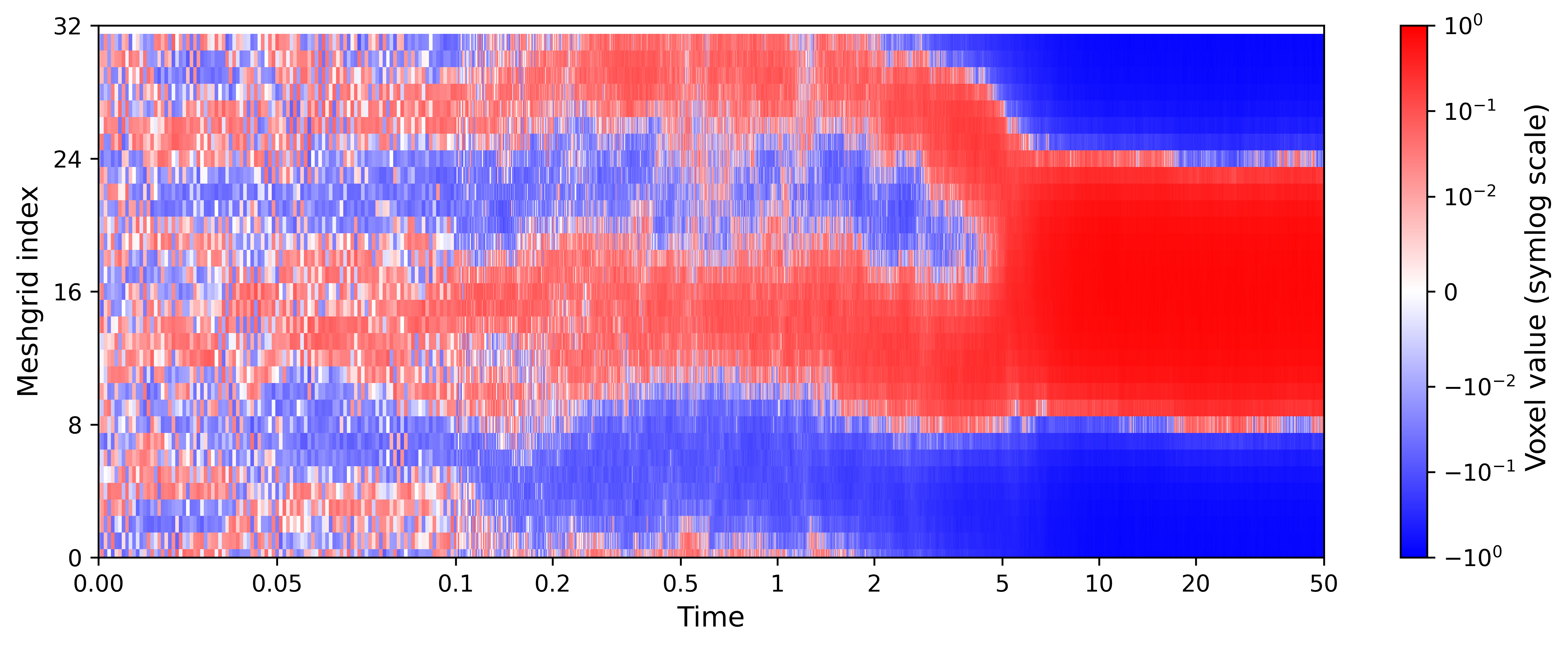}}
        \\
        \multicolumn{2}{c}{\includegraphics[width=1\textwidth]{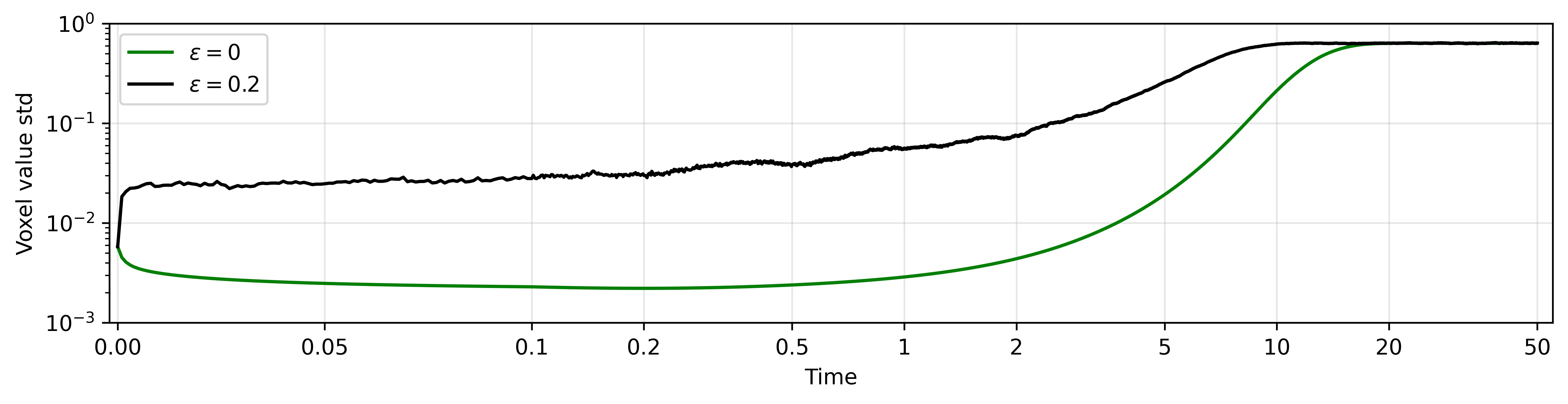}}
        \\
        no noise $\epsilon=0$ & no noise $\epsilon=0.2$ 
        \\
        \raisebox{-0.5\height}{\includegraphics[width = 0.5\textwidth]{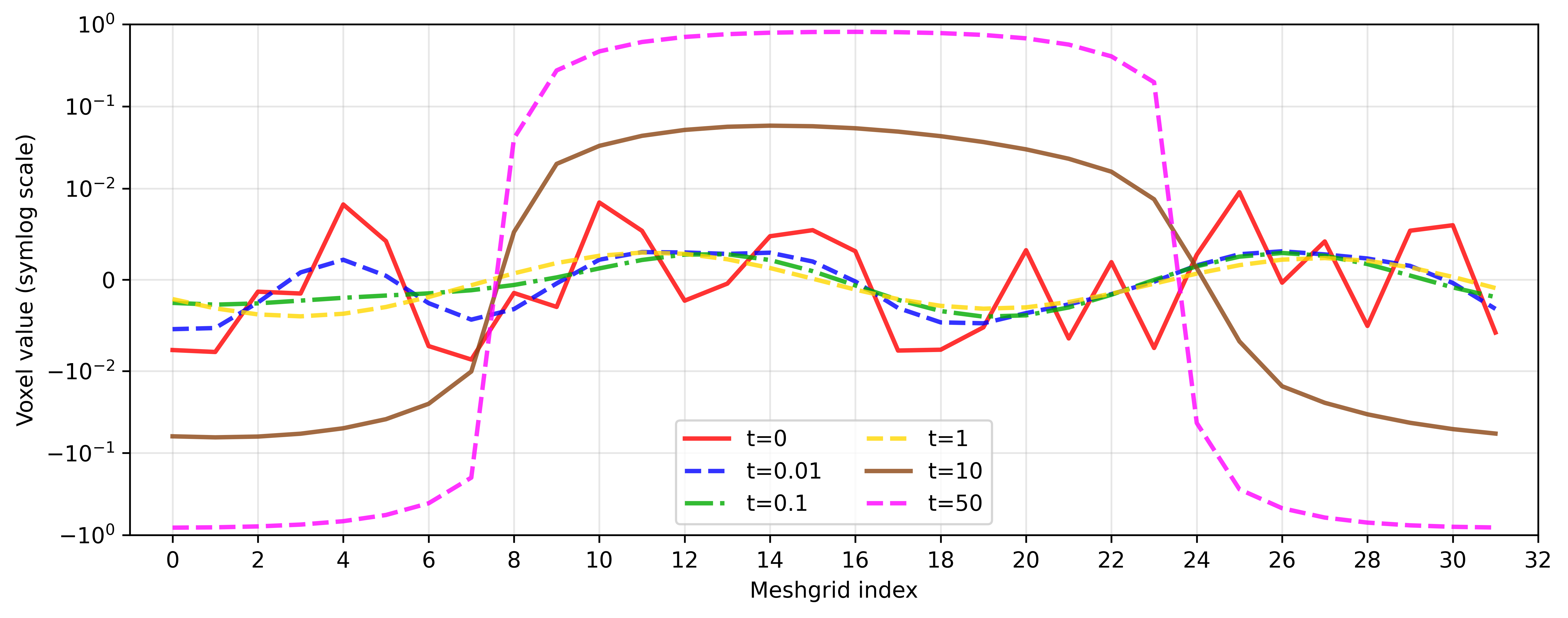}} &
        \raisebox{-0.5\height}{\includegraphics[width = 0.5\textwidth]{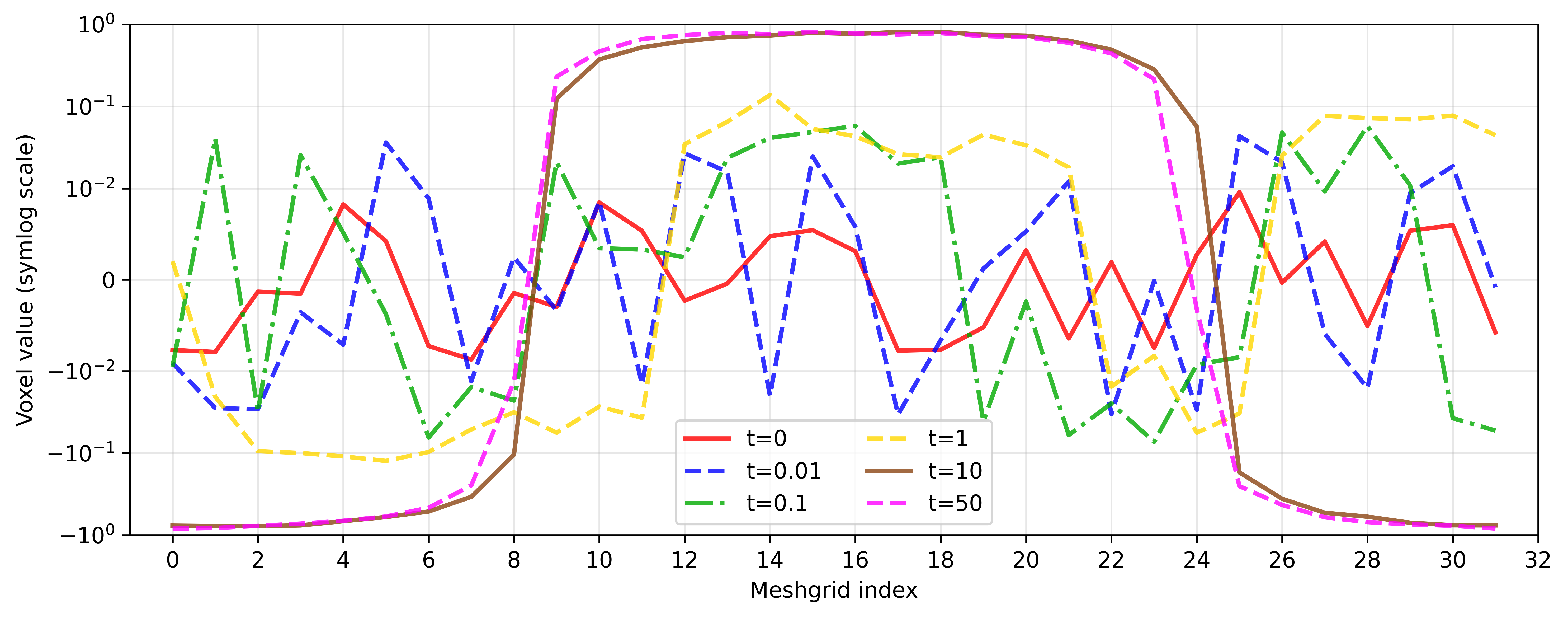}}
        \\
        % \raisebox{-0.5\height}{\includegraphics[width = 0.5\textwidth]{fig/Final/SDE_1d_fourier_noise_0p0.png}} &
        % \raisebox{-0.5\height}{\includegraphics[width = 0.5\textwidth]{fig/Final/SDE_1d_fourier_noise_0p2.png}}
        % \\
        \raisebox{-0.5\height}{\includegraphics[width = 0.5\textwidth]{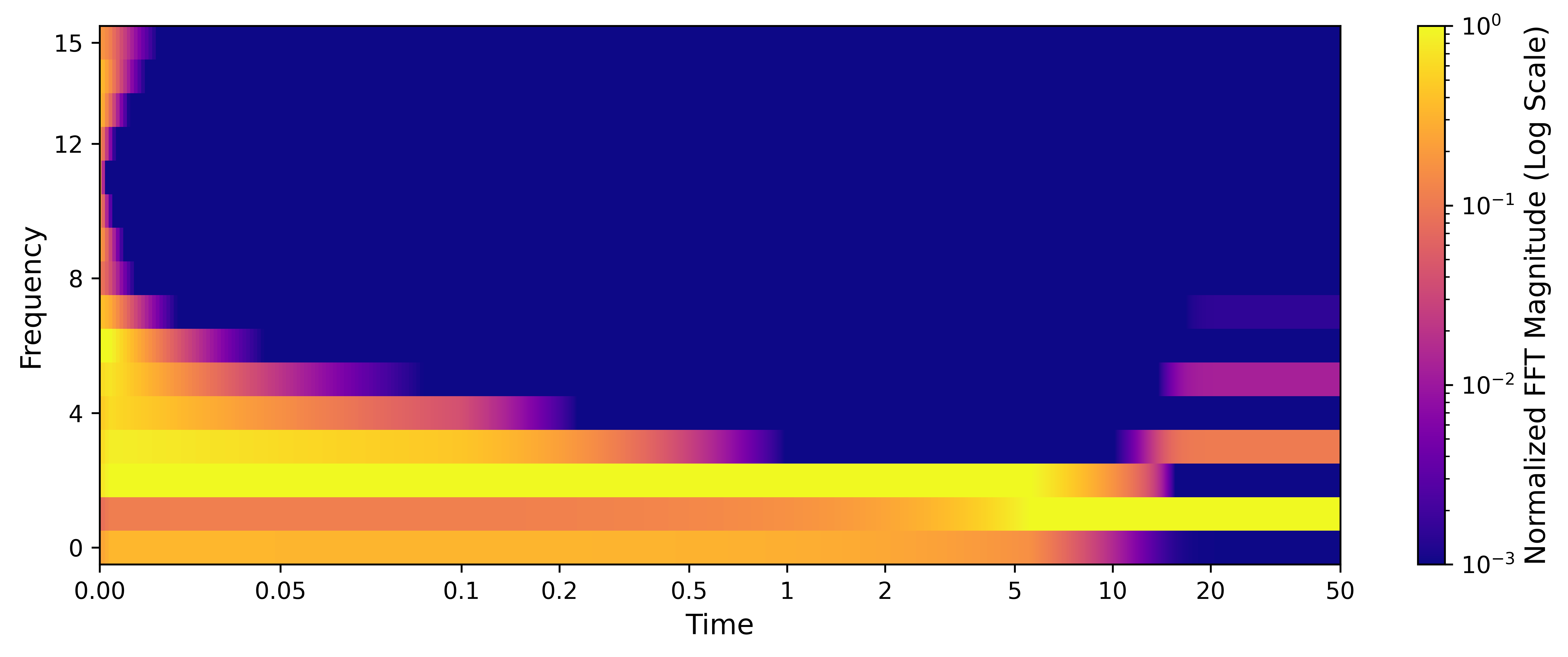}} &
        \raisebox{-0.5\height}{\includegraphics[width = 0.5\textwidth]{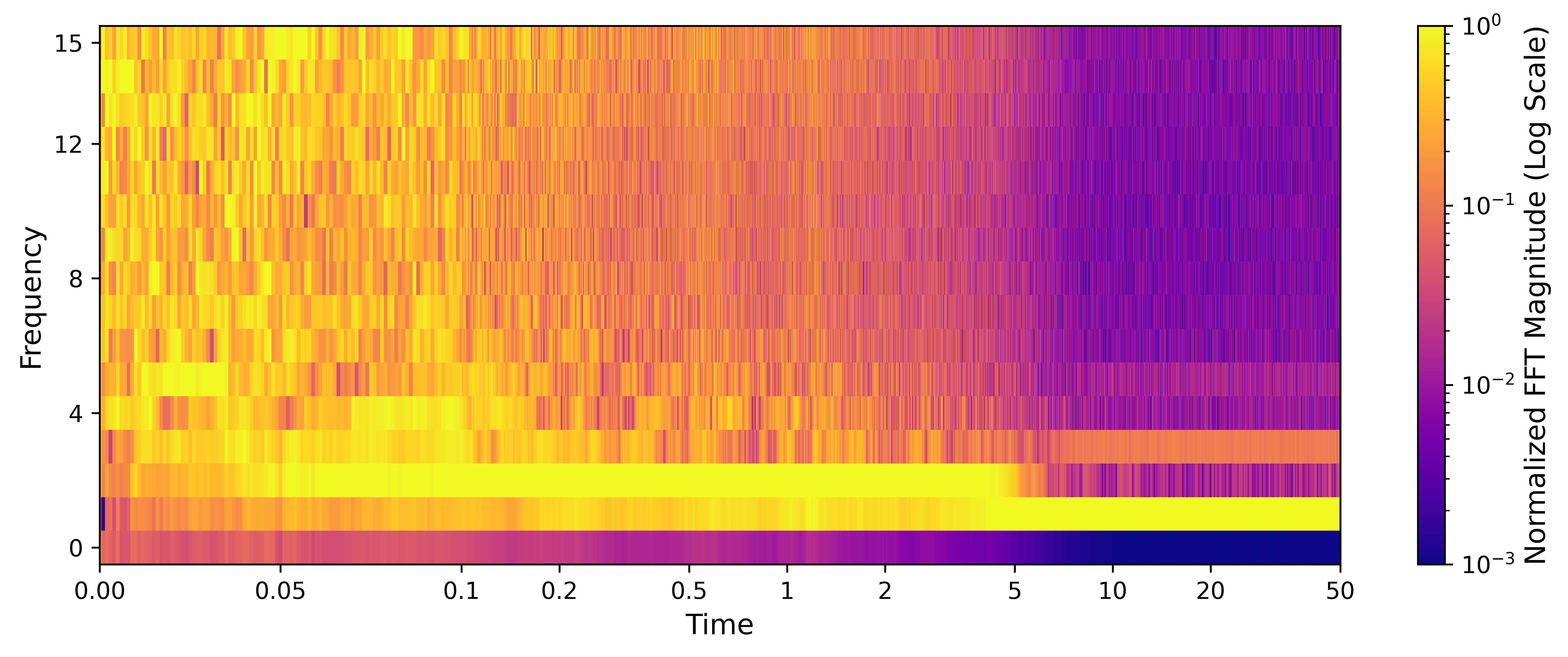}}
    \end{tabular}
    \caption{
    Results for a $32^2$ grid with $\langle c \rangle = 0$ (TOP) The standard deviation voxel (1st MIDDLE) value frames (2nd MIDDLE), Fourier spectrum (BOTTOM) normalized Fourier spectrum.}
    \label{fig:1d_frames_fourier_copo_32_grid}
\end{figure}

\subsection{1D SPDE analysis for \cofour{}}
\label{sec:1D_SPDE_256}
\begin{figure}[htp]
    \centering
    \begin{tabular}{c c}
        no noise $\epsilon=0$ & no noise $\epsilon=0.2$
        \\
        \raisebox{-0.5\height}{\includegraphics[width = 0.5\textwidth]{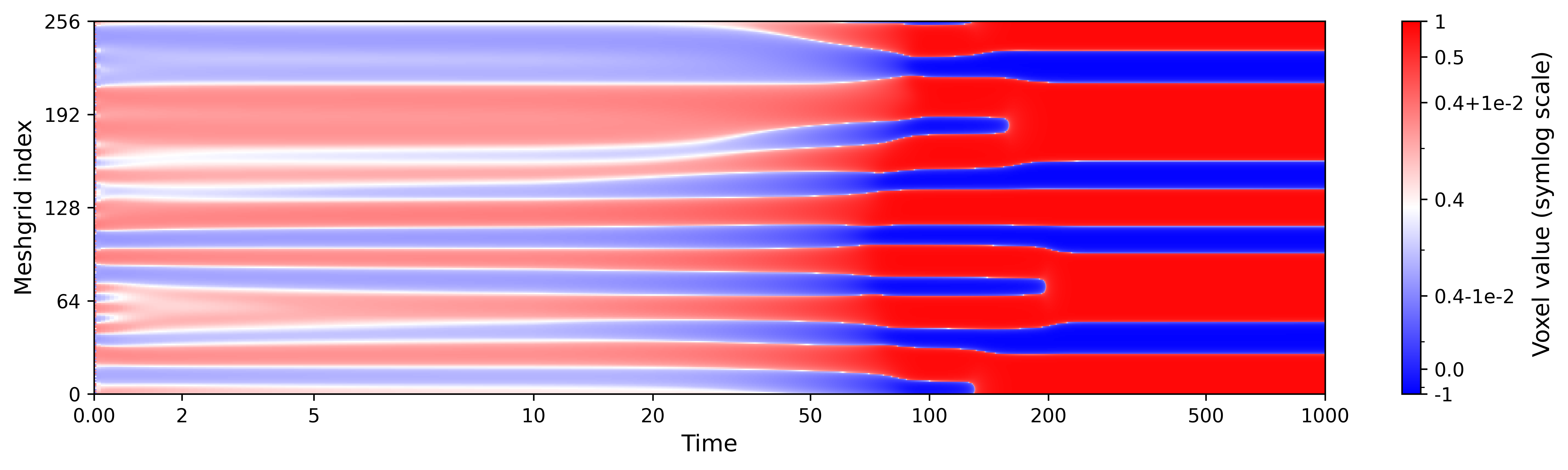}} &
        \raisebox{-0.5\height}{\includegraphics[width = 0.5\textwidth]{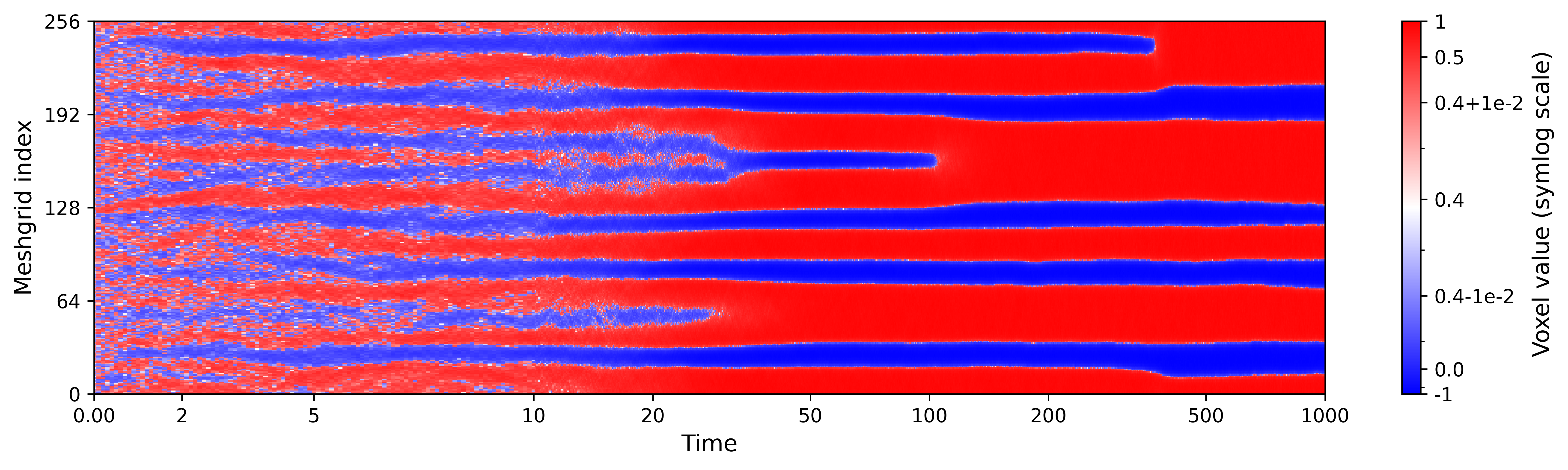}}
        \\
        \multicolumn{2}{c}{\includegraphics[width=1\textwidth]{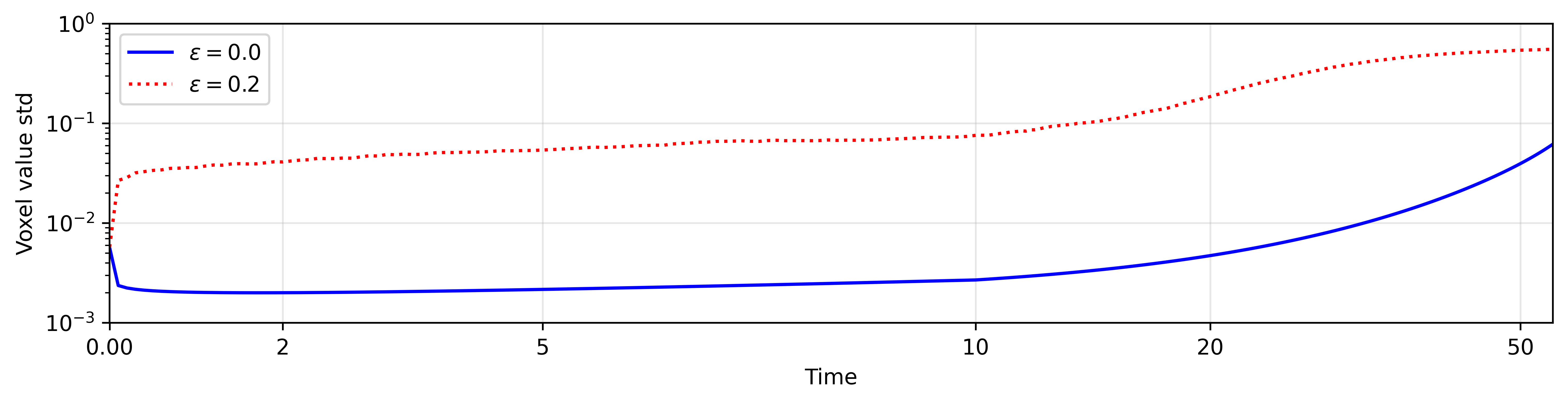}}
        \\
        no noise $\epsilon=0$ & no noise $\epsilon=0.2$ 
        \\
        \raisebox{-0.5\height}{\includegraphics[width = 0.5\textwidth]{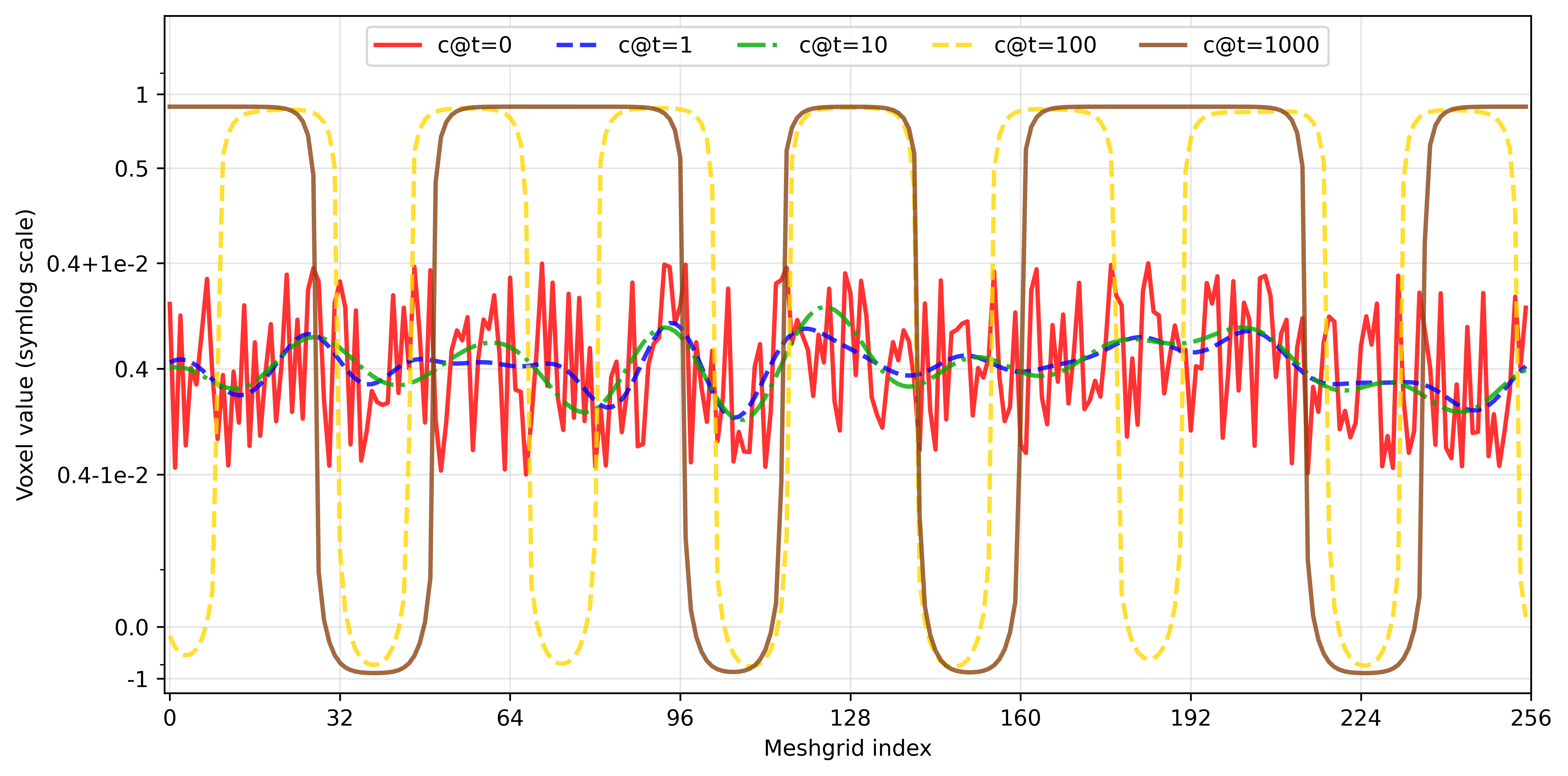}} &
        \raisebox{-0.5\height}{\includegraphics[width = 0.5\textwidth]{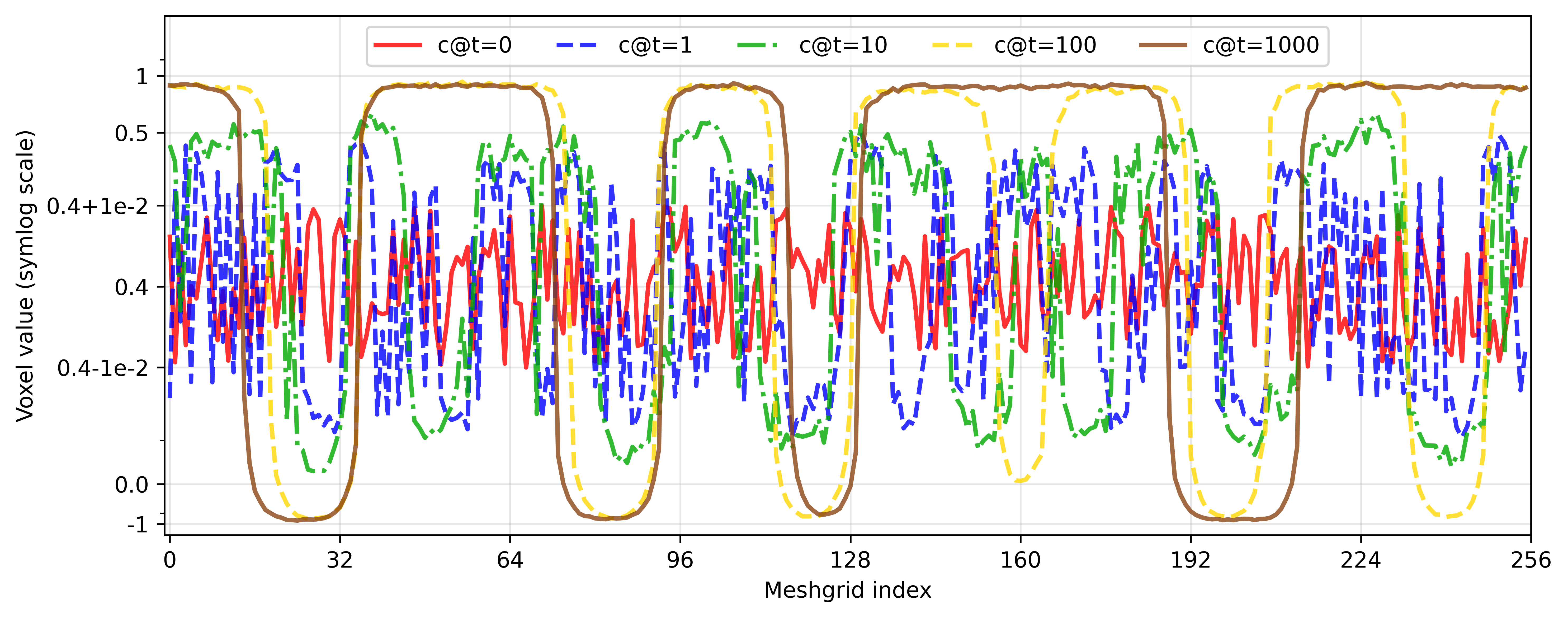}}
        % \\
        % \raisebox{-0.5\height}{\includegraphics[width = 0.5\textwidth]{fig/Final/SDE_1d_fourier_noise_0p0256_grid_c0p4.png}} &
        % \raisebox{-0.5\height}{\includegraphics[width = 0.5\textwidth]{fig/Final/SDE_1d_fourier_noise_0p2256_grid_c0p4.png}}
        \\
        \raisebox{-0.5\height}{\includegraphics[width = 0.5\textwidth]{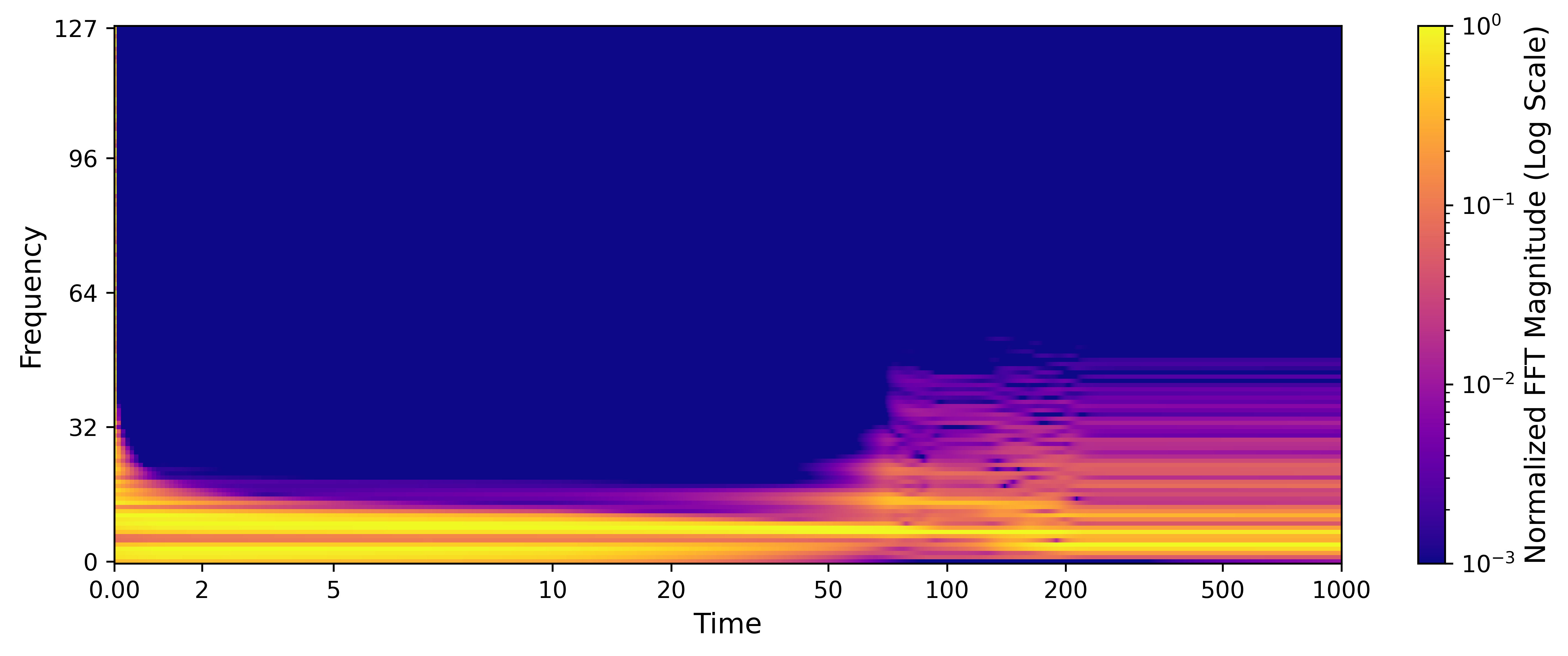}} &
        \raisebox{-0.5\height}{\includegraphics[width = 0.5\textwidth]{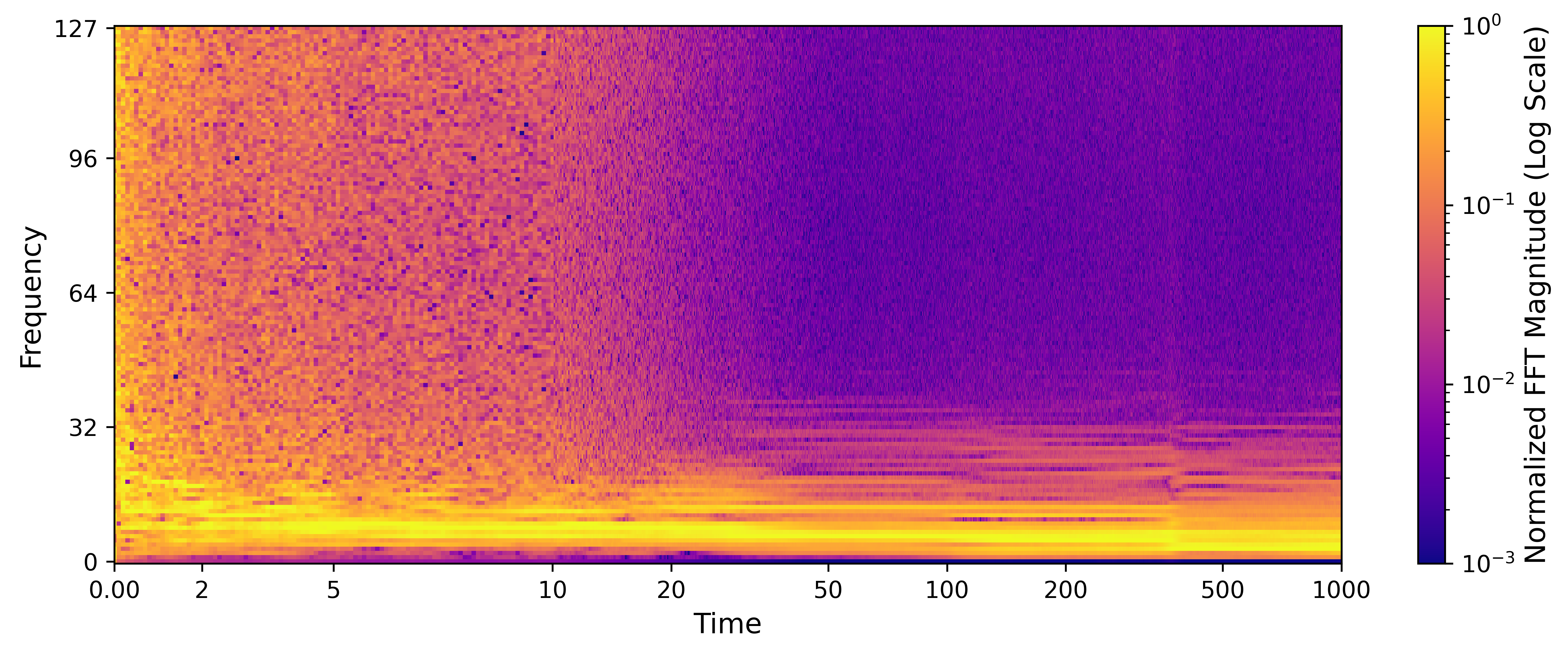}}
    \end{tabular}
    \caption{{\bf 256 meshgrid with \cofour{}} (TOP) The standard deviation voxel (1st MIDDLE) value frames (2nd MIDDLE), Fourier spectrum (BOTTOM) normalized Fourier spectrum.}
    \label{fig:1d_frames_fourier}
\end{figure}

Results for the shifted-initial-condition \cofour case with grid size $32$ and $256$ are shown in Supplementary Fig.~\ref{fig:1d_frames_fourier_cop4_32_grid} to Fig.~\ref{fig:1d_frames_fourier}, respectively.

\begin{figure}[htp]
    \centering
    \begin{tabular}{c c}
        no noise $\epsilon=0$ & no noise $\epsilon=0.2$ 
        \\
        \raisebox{-0.5\height}{\includegraphics[width = 0.5\textwidth]{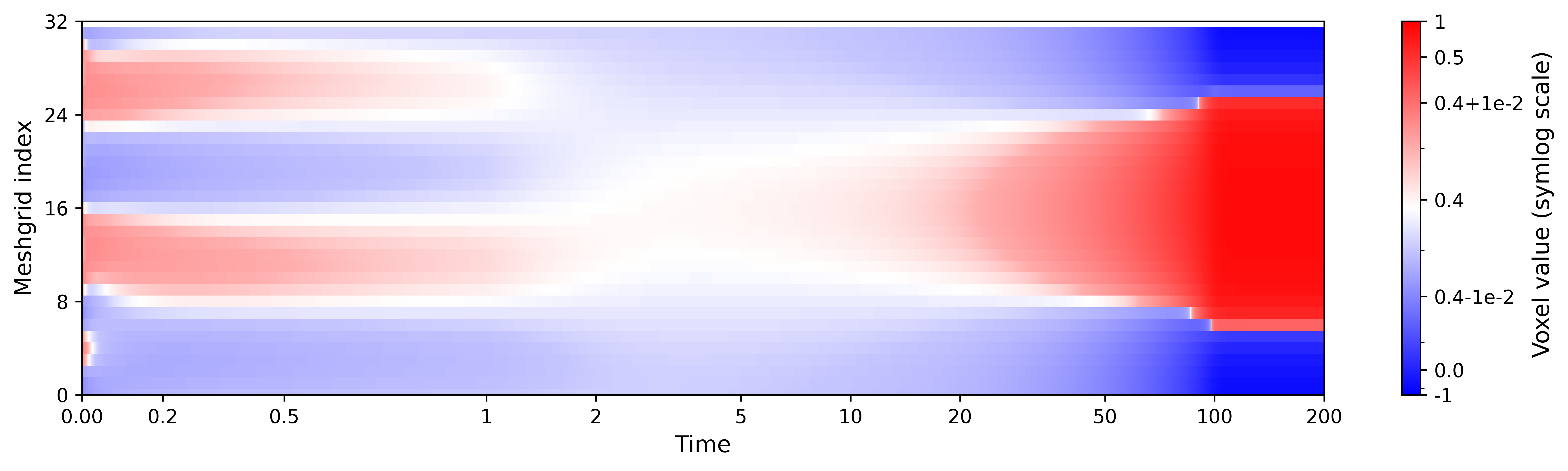}} &
        \raisebox{-0.5\height}{\includegraphics[width = 0.5\textwidth]{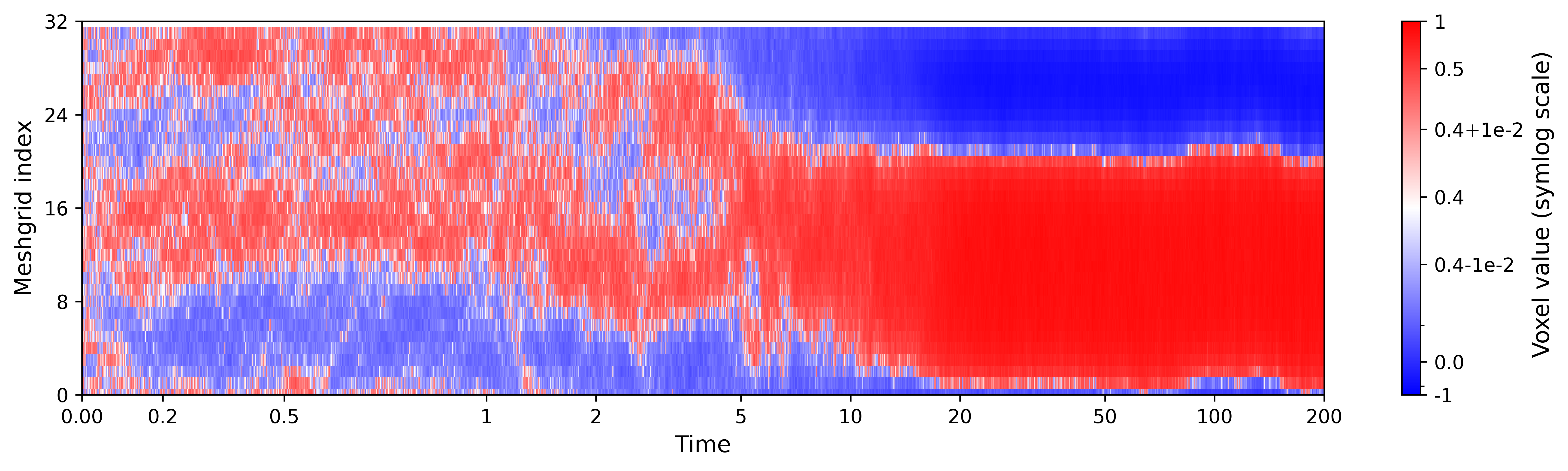}}
        \\
        \multicolumn{2}{c}{\includegraphics[width=1\textwidth]{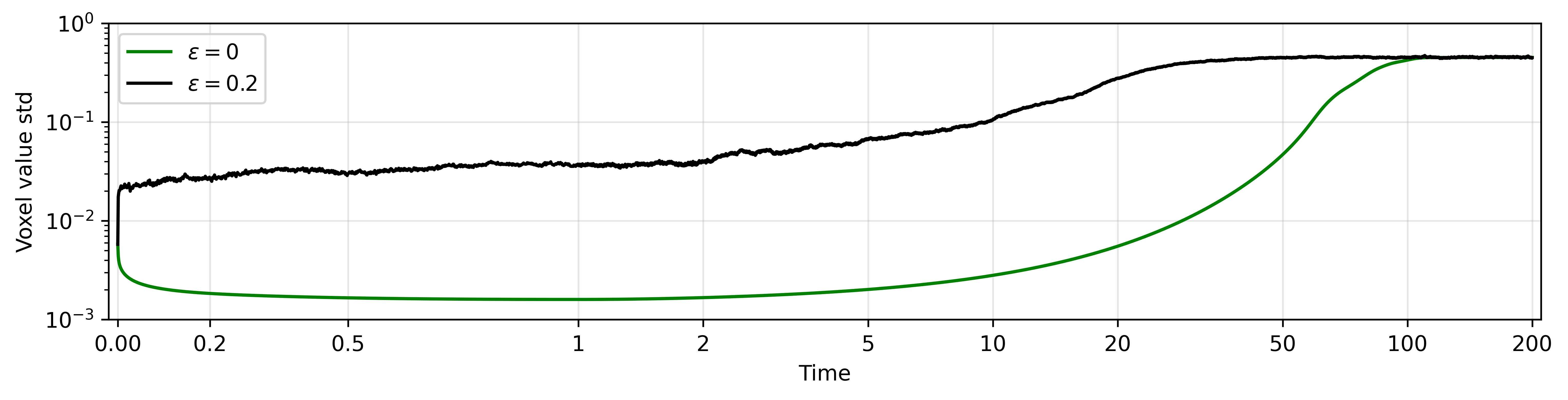}}
        \\
        no noise $\epsilon=0$ & no noise $\epsilon=0.2$ 
        \\
        \raisebox{-0.5\height}{\includegraphics[width = 0.5\textwidth]{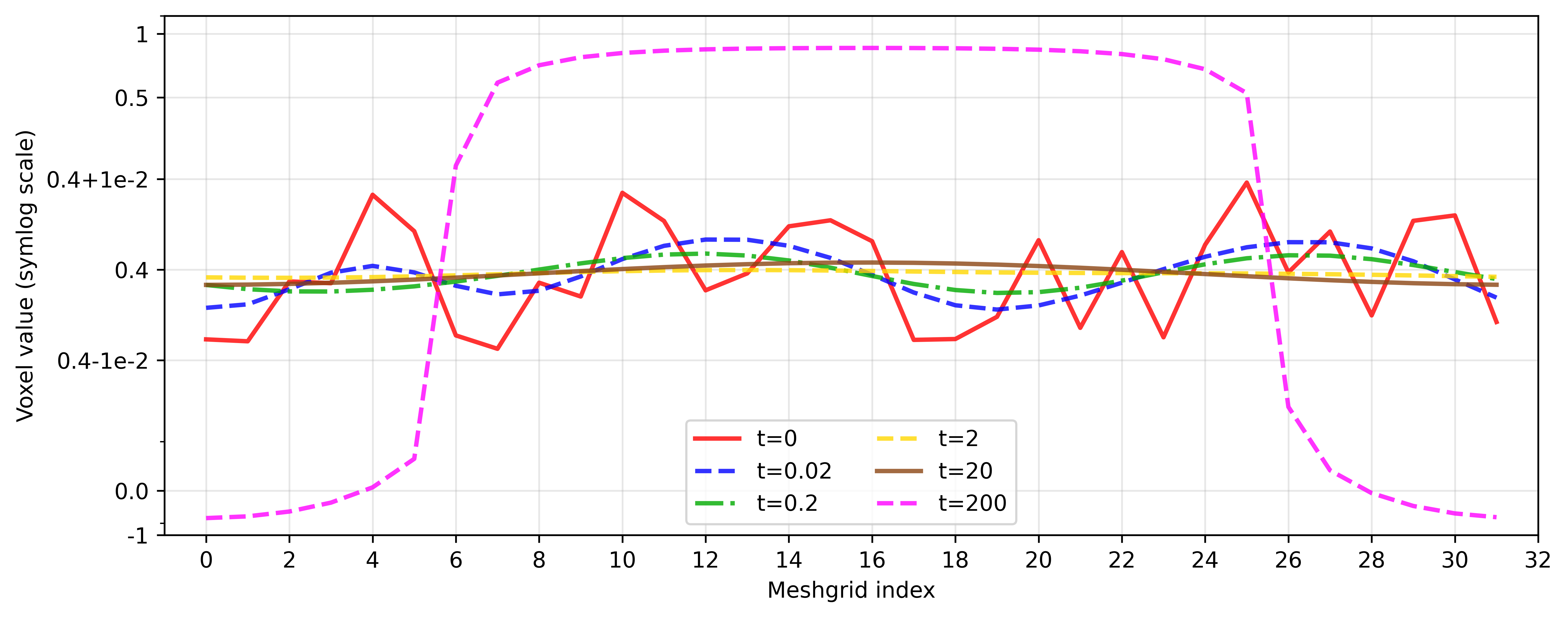}} &
        \raisebox{-0.5\height}{\includegraphics[width = 0.5\textwidth]{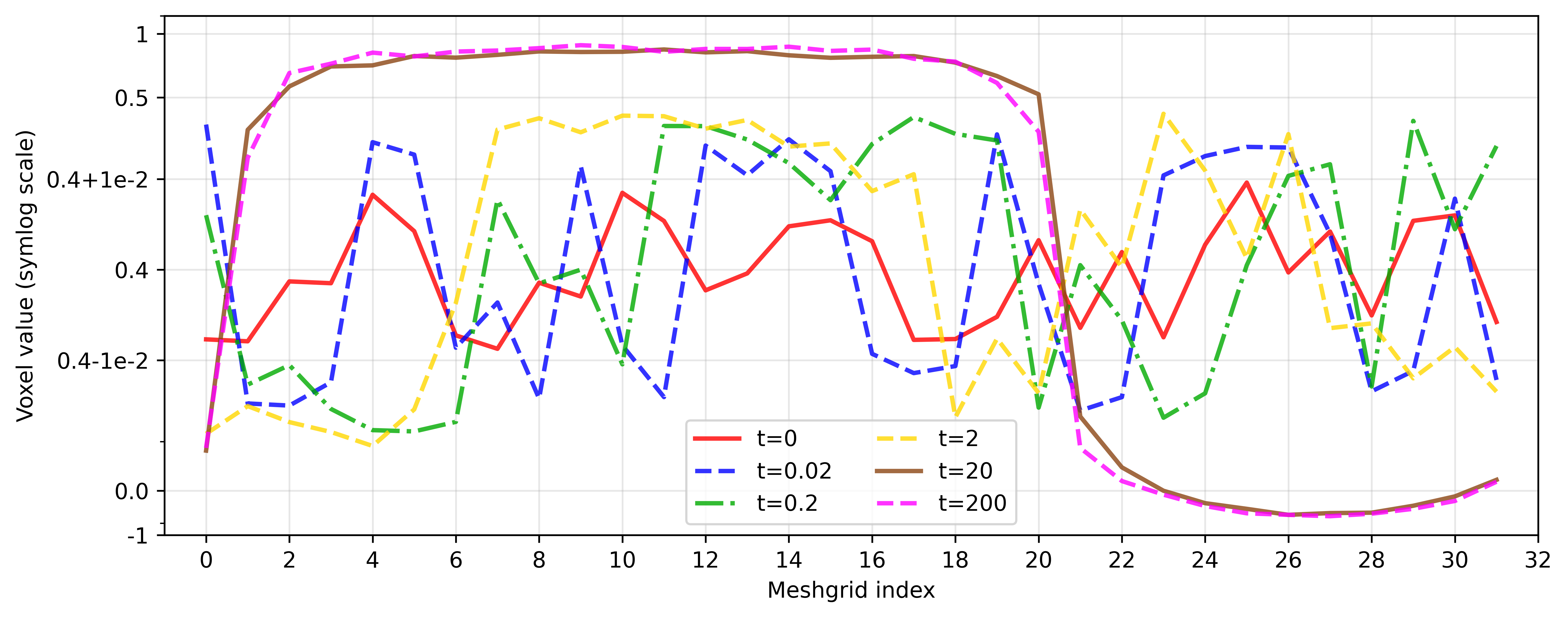}}
        \\
        % \raisebox{-0.5\height}{\includegraphics[width = 0.5\textwidth]{fig/Final/SDE_1d_fourier_noise_0p0_c0p4.png}} &
        % \raisebox{-0.5\height}{\includegraphics[width = 0.5\textwidth]{fig/Final/SDE_1d_fourier_noise_0p2_c0p4.png}}
        % \\
        \raisebox{-0.5\height}{\includegraphics[width = 0.5\textwidth]{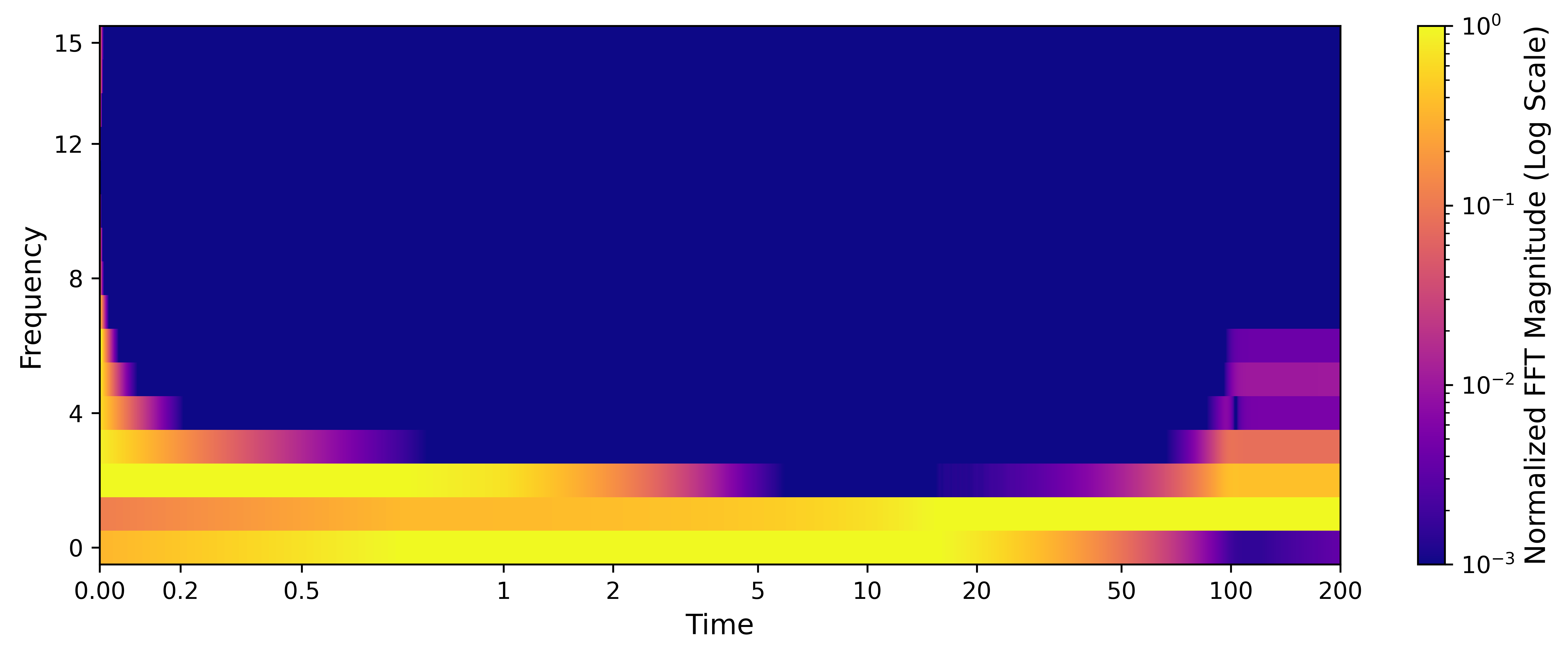}} &
        \raisebox{-0.5\height}{\includegraphics[width = 0.5\textwidth]{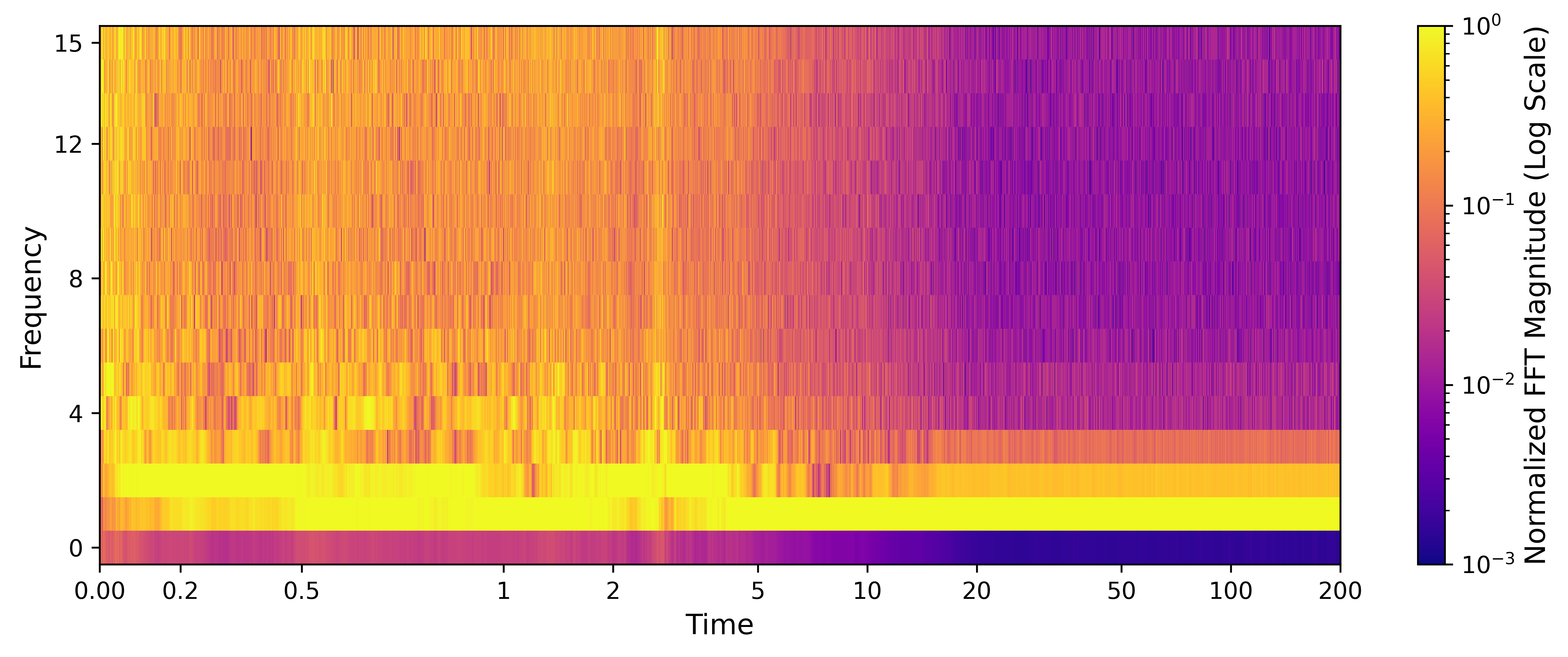}}
    \end{tabular}
    \caption{{\bf 32 meshgrid with \cofour{}} (TOP) The standard deviation voxel (1st MIDDLE) value frames (2nd MIDDLE), Fourier spectrum (BOTTOM) normalized Fourier spectrum.}
    \label{fig:1d_frames_fourier_cop4_32_grid}
\end{figure}

\subsection{Convergence of 3D SPDE with different time step size}
\label{sec:convergence_dt}

The SPDE system exhibits almost similar convergence behavior as either using the smaller time step size $10^{-4}$ or a larger one $10^{-3}$as shown in Supplementary Fig.~\ref{fig:SDE_with_smaller_dt}. In our experiments, we opt to generate SPDE data using the $10^{-3}$ time step size, as it is  more computationally efficient while still yielding similarly satisfactory results. Note that, the SPDE solution is unstable as solving with the time step size of $10^{-2}$.

% \begin{figure}[htp]
%     \centering
%     \begin{tabular}{c c}
%         Time size $10^{-3}$ & Time size $10^{-4}$ \\
%         \raisebox{-0.5\height}{\includegraphics[width = 0.48\textwidth]{fig/Final/SDE_stdofVoxel_c0p0.png}} &
%         \raisebox{-0.5\height}{\includegraphics[width = 0.48\textwidth]{fig/Final/SDE_stdofVoxel_c0p0_dt_1eminus4.png}}
%     \end{tabular}
%     \caption{The standard deviation of voxel values from 10 SPDE simulations solved with different time step size. {\bf Left:} the time size of $10^{-3}$, {\bf Right:} the time size of $10^{-4}$.}
%     \label{fig:SDE_with_smaller_dt}
% \end{figure}

\begin{figure}[htp]
    \centering
    \begin{tabular}{c c c c}
        \multicolumn{2}{c}{Time size $10^{-3}$} &
        \multicolumn{2}{c}{ Time size $10^{-4}$} 
        \\
        \multicolumn{2}{c}{\raisebox{-0.35\height}{\includegraphics[width = 0.45\textwidth]{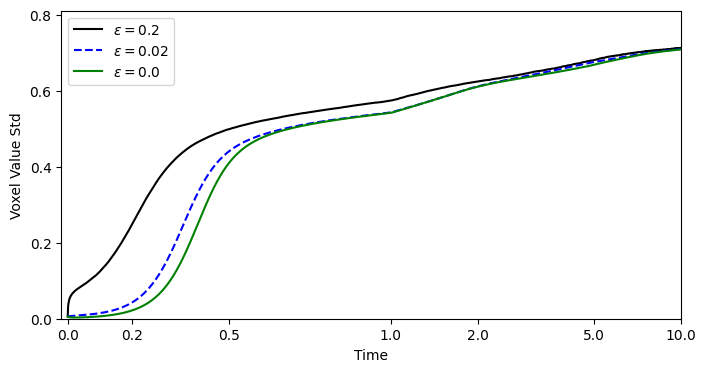}}} &
        \multicolumn{2}{c}{\raisebox{-0.35\height}{\includegraphics[width = 0.45\textwidth]{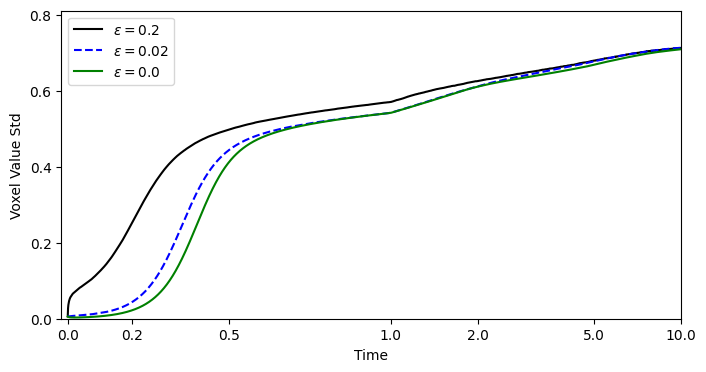}}}
        \\
        \multicolumn{4}{c}{\raisebox{-0.35\height}{\includegraphics[width = 0.9\textwidth]{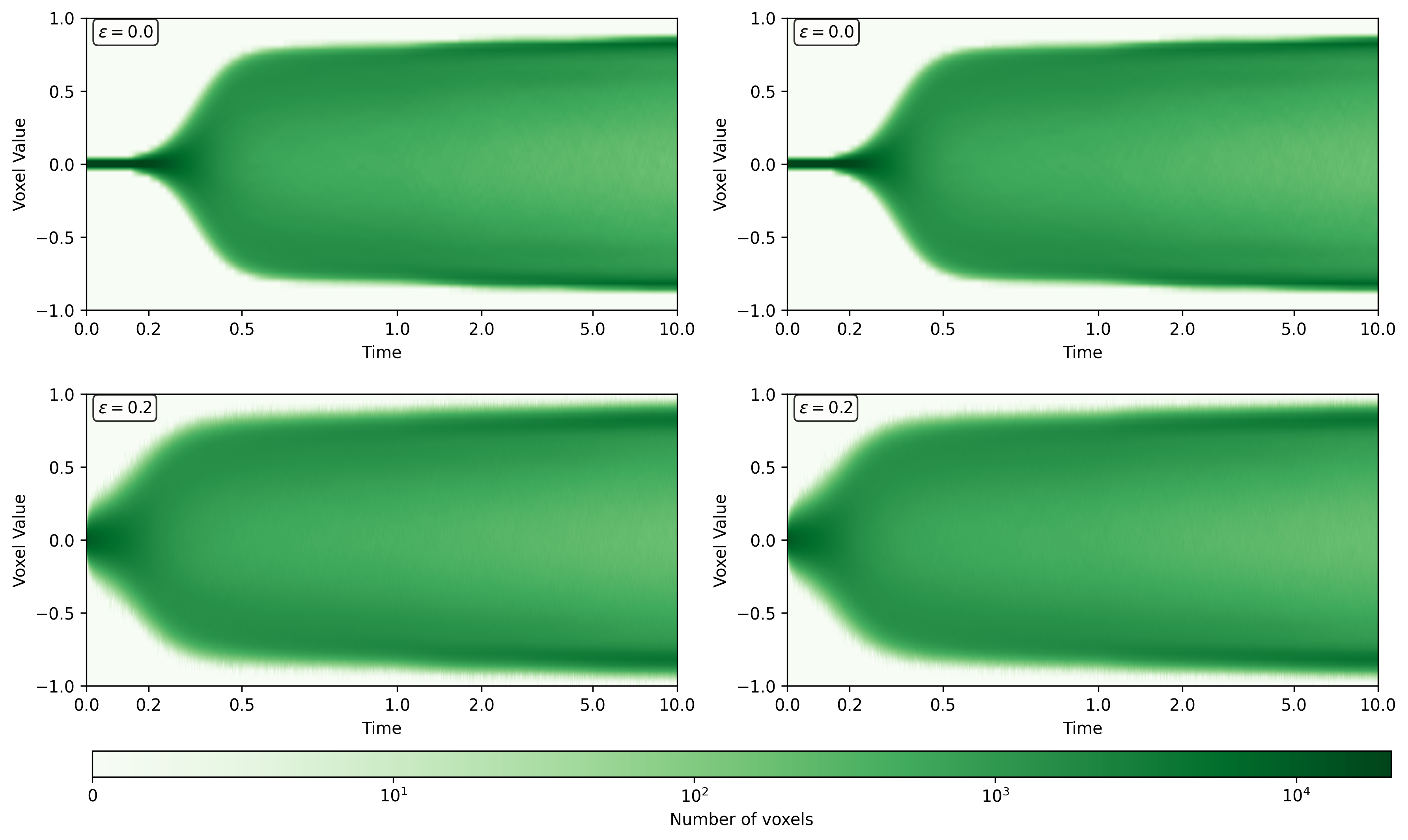}}}
        % \\
        % \multicolumn{2}{c}{\raisebox{-0.35\height}{\includegraphics[width = 0.45\textwidth]{fig/Final/SDE_c0p0_noise0p0_histogram.png}}} &
        % \multicolumn{2}{c}{\raisebox{-0.35\height}{\includegraphics[width = 0.45\textwidth]{fig/Final/SDE_c0p0_noise0p0_histogram_dt_1eminus4.png}}}
        % \\
        % \multicolumn{2}{c}{\raisebox{-0.35\height}{\includegraphics[width = 0.45\textwidth]{fig/Final/SDE_c0p0_noise0p2_histogram.png}}} &
        % \multicolumn{2}{c}{\raisebox{-0.35\height}{\includegraphics[width = 0.45\textwidth]{fig/Final/SDE_c0p0_noise0p2_histogram_dt_1eminus4.png}}}
    \end{tabular}
    \caption{
        {\bf Top} The standard deviation of voxel values from 10 independent 3D simulations at   $\langle c \rangle = 0$ with different noise levels $\epsilon=$0, 0.02, 0.2 and integrated with time step size of $10^{-3}$ and $10^{-4}$.
        {\bf Middle} The histogram of voxel values from 10 PDE simulations ($\epsilon=0$) solved with different time step size.
        {\bf Bottom} Same for for SPDE simulations ($\epsilon=0.2$).
        {\bf Left} column: time step of $10^{-3}$, {\bf right:} time step of $10^{-4}$.}
    \label{fig:SDE_with_smaller_dt}
\end{figure}

\subsection{3D SPDE analysis}
\label{sec:3D_SPED_app}
\begin{figure}[htp!]
    \centering
        \begin{tabular}{c c c c}
        \multicolumn{2}{c}{\cozero} &
        \multicolumn{2}{c}{ \cofour} 
        \\
        \multicolumn{2}{c}{\raisebox{-0.35\height}{\includegraphics[width = 0.45\textwidth]{fig/Final/SDE_stdofVoxel_c0p0.png}}} &
        \multicolumn{2}{c}{\raisebox{-0.35\height}{\includegraphics[width = 0.45\textwidth]{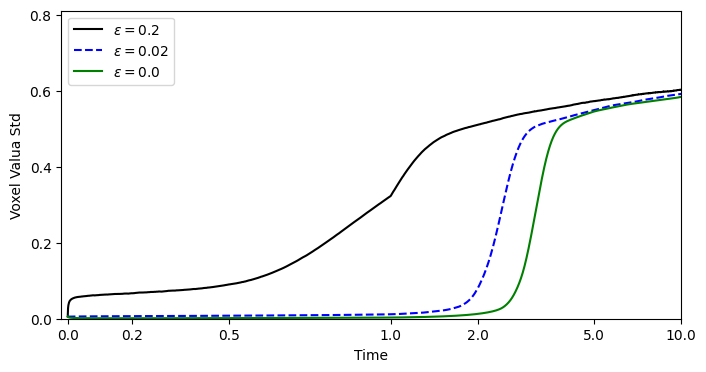}}}
        \\
        \multicolumn{4}{c}{\raisebox{-0.35\height}{\includegraphics[width = 0.9\textwidth]{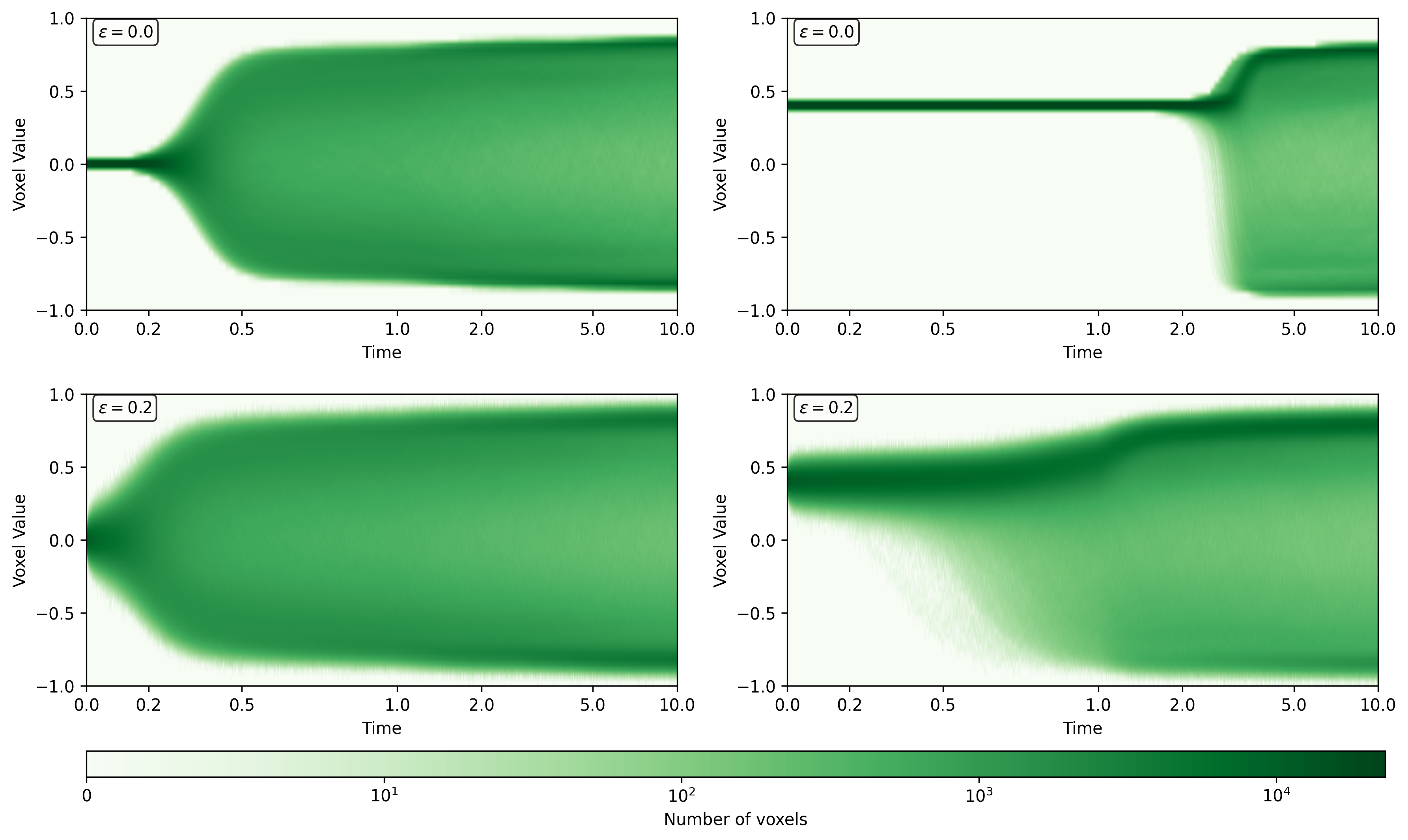}}}
        % \\
        % \multicolumn{2}{c|}{\raisebox{-0.35\height}{\includegraphics[width = 0.45\textwidth]{fig/Final/SDE_c0p0_noise0p0_histogram.png}}} & \quad &
        % \multicolumn{2}{c}{\raisebox{-0.35\height}{\includegraphics[width = 0.45\textwidth]{fig/Final/SDE_c0p4_noise0p0_histogram.png}}}
        % \\
        % \multicolumn{2}{c|}{\raisebox{-0.35\height}{\includegraphics[width = 0.45\textwidth]{fig/Final/SDE_c0p0_noise0p2_histogram.png}}} & \quad &
        % \multicolumn{2}{c}{\raisebox{-0.35\height}{\includegraphics[width = 0.45\textwidth]{fig/Final/SDE_c0p4_noise0p2_histogram.png}}}
        \\
        $\epsilon=0, t=0.4$ & $\epsilon=0.2, t=0.4$ &
        $\epsilon=0, t=2$ & $\epsilon=0.2, t=2$ 
        \\
        \raisebox{-0.24\height}{\includegraphics[width = 0.2\textwidth]{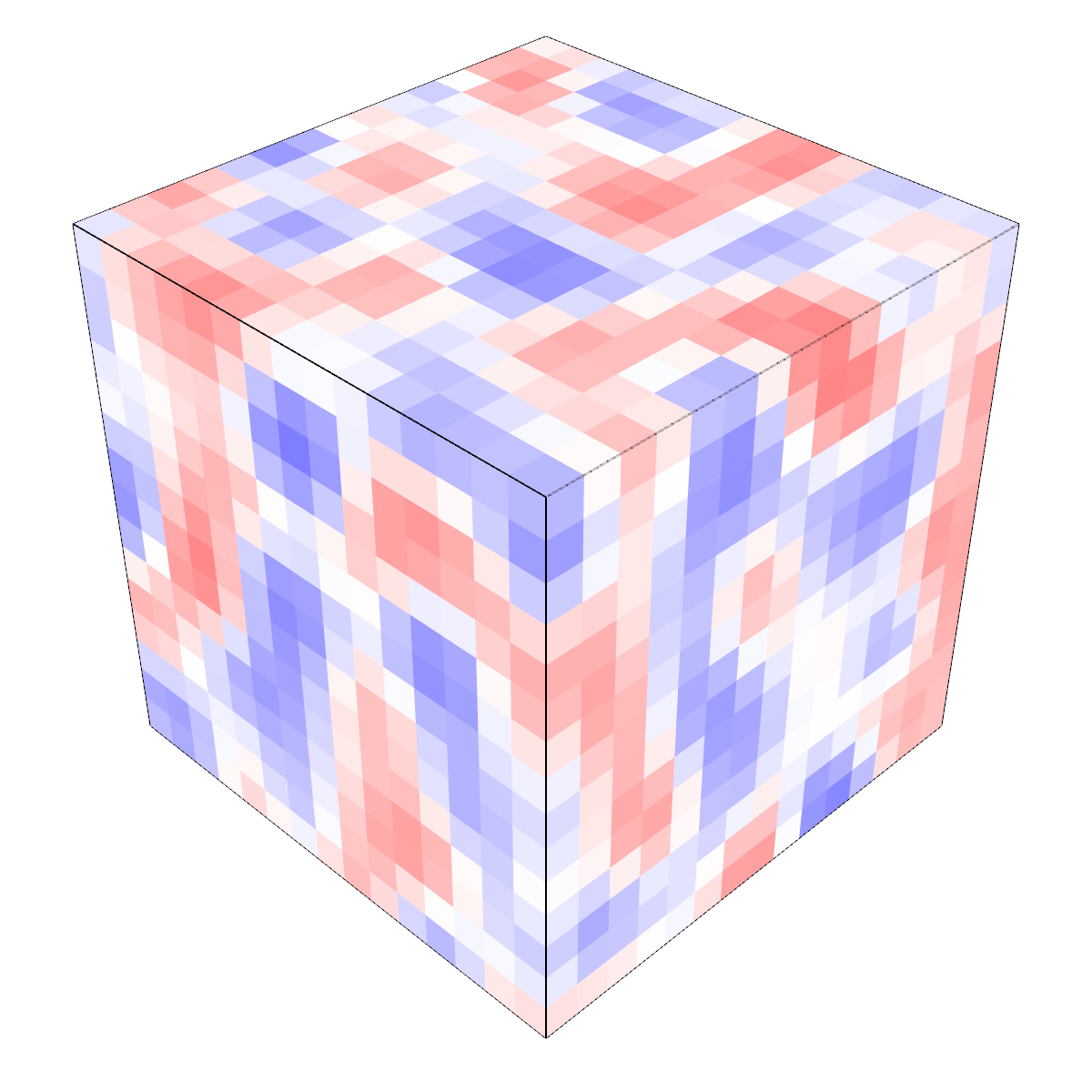}} &
        \raisebox{-0.24\height}{\includegraphics[width = 0.2\textwidth]{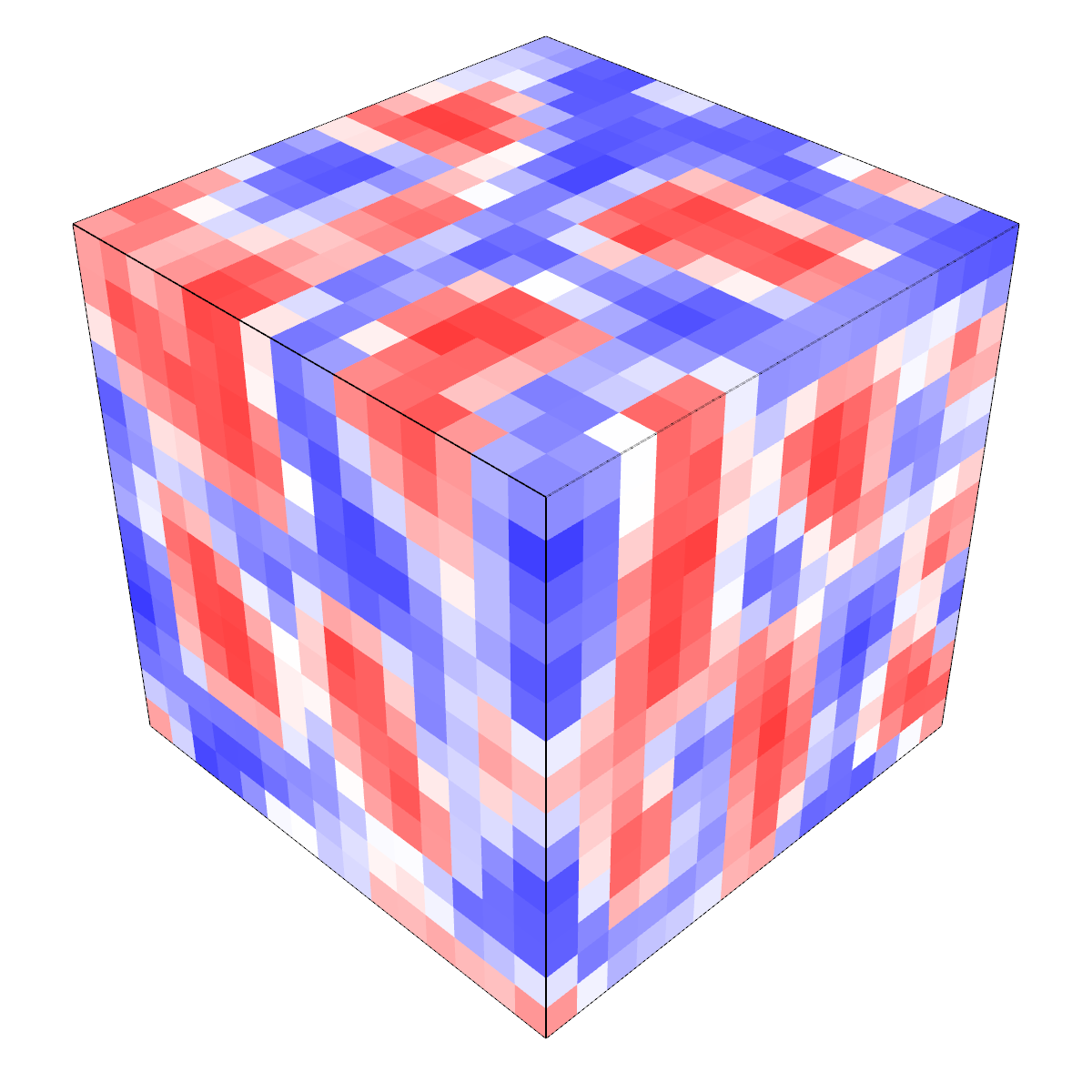}} & 
        \raisebox{-0.24\height}{\includegraphics[width = 0.2\textwidth]{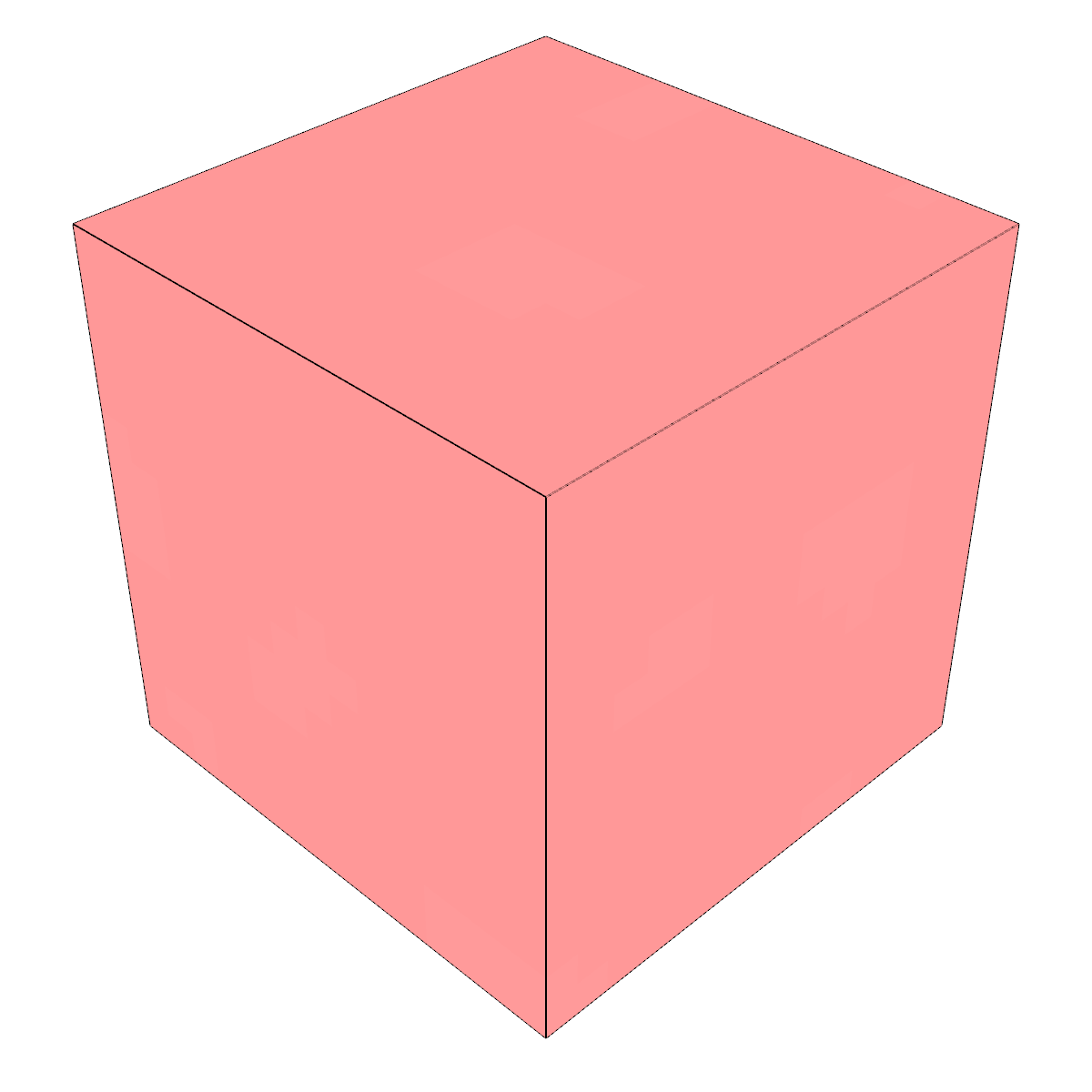}} &
        \raisebox{-0.24\height}{\includegraphics[width = 0.2\textwidth]{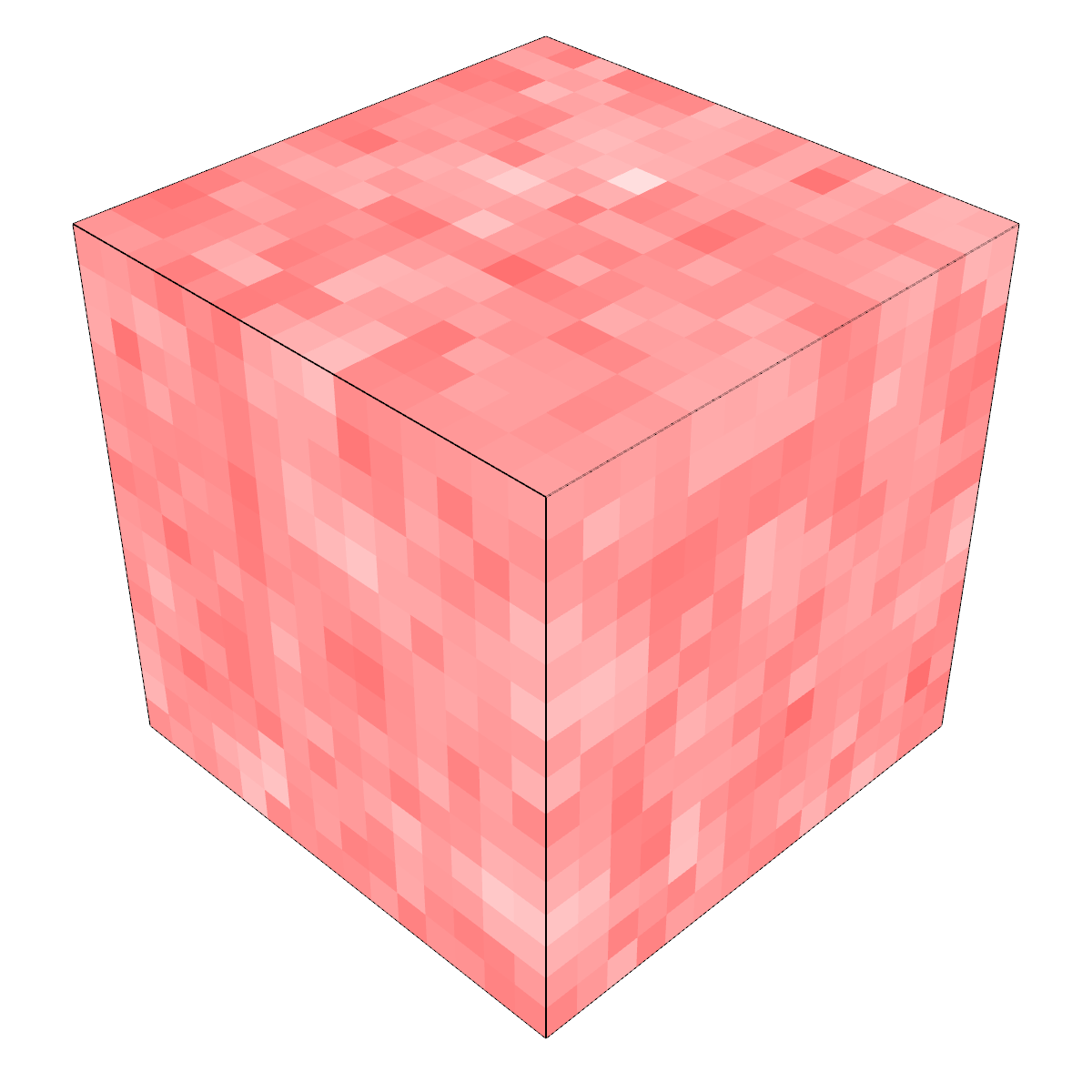}}
        \\
        \multicolumn{4}{c}{\includegraphics[width=.8\textwidth]{fig/Final/SDE_different_samples_from_same_init_colormap_withlabels.png}} 
    \end{tabular}
    \caption{
    % \added{[Can you remove the left move text (rotated), and put the labels $\epsilon=0$ etc into the figure of the 2nd and 3rd rows? \hai{you meant put it inside like legend? [fz: yes, like legend or a text label inside the frame]}
    % Is it possible to adjust the color bar to increase the contrast of the light gray, dark gray, black, etc? \hai{need to check [fz: can you just keep one color bar and make it larger?]}]}
    % \added{Q1. What exactly is depicted here? Mean and plus/minus std? Is it more meaningful or informative to draw a contour plot of the histogram/KDE of the distributions, say at a certain contour value? You will see a single mode to bi-modal transition.  Q2 I feel like figure 2 and 3 should be somehow combined. They are very connected. How about a 2x5 table, each column for c0 in [0, 0.4]; first row: c0 snapshot; 2nd: eps=0; 3rd row: eps=0.2; 4: histogram; 5th: histogram map from figure 2 as a summary}
    % \added{[FZ: can you change the legend scale so that the middle values at c around 0 look lighter? Like more contrast so that it becomes obviously bimodal?]}
    {\bf Top:} The evolution of mean $\pm$ standard deviation of voxel values of 10 random SPDE samples with respect to different stochastic noise levels, $\epsilon=0$ and $\epsilon=0.2$. {\bf Middle:} The corresponding histogram of voxel values. {\bf Bottom:} The concentration frames from SPDE  $\epsilon=0$ and $\epsilon=0.2$  
    {\bf Left:} Initial concentration field \cozero. {\bf Right:} Initial concentration field \cofour.
    }
    \label{fig:SDE_different_noise_levels_variance_voxel_new_app}
\end{figure}

Supplementary Fig.~\ref{fig:SDE_different_noise_levels_variance_voxel_new_app} shows the growth speed of voxel value mean $\pm$ standard deviation evolution, averaging over 10 SPDE trajectories, for both $\epsilon=0$ and $\epsilon=0.2$. The corresponding histograms of voxel value evolution are shown in the second and third rows, respectively. It can be seen that stochastic noise promotes faster phase separationn for both initial concentration fields, $\langle c \rangle =0$ and 0.4.  Concentration field samples at the time point of 0.4 and 2 for $\langle c \rangle =0$ and 0.4, respectively, are presented in the bottom row of Fig.~\ref{fig:SDE_different_noise_levels_variance_voxel_new_app}. The concentration state simulated with noise reaches the coarsening phase much earlier than the one without noise. 

\setlength{\tabcolsep}{2pt}

\subsection{In-distribution test for \nflux{} model}

\label{app:indist_nflux}
Additional in-distribution test results for \nflux model are shown in Supplementary Fig.~\ref{app_in_dist_no_flux}.

\setlength{\tabcolsep}{2pt}  % Default is 6pt
\begin{figure}[htp]
    \centering
    \begin{tabular}{c c| c c}
        \multicolumn{2}{c}{\cozero}  &  \multicolumn{2}{c}{\cofour}  \\ 
        \raisebox{-0.14\height}{\includegraphics[width = 0.265\textwidth]{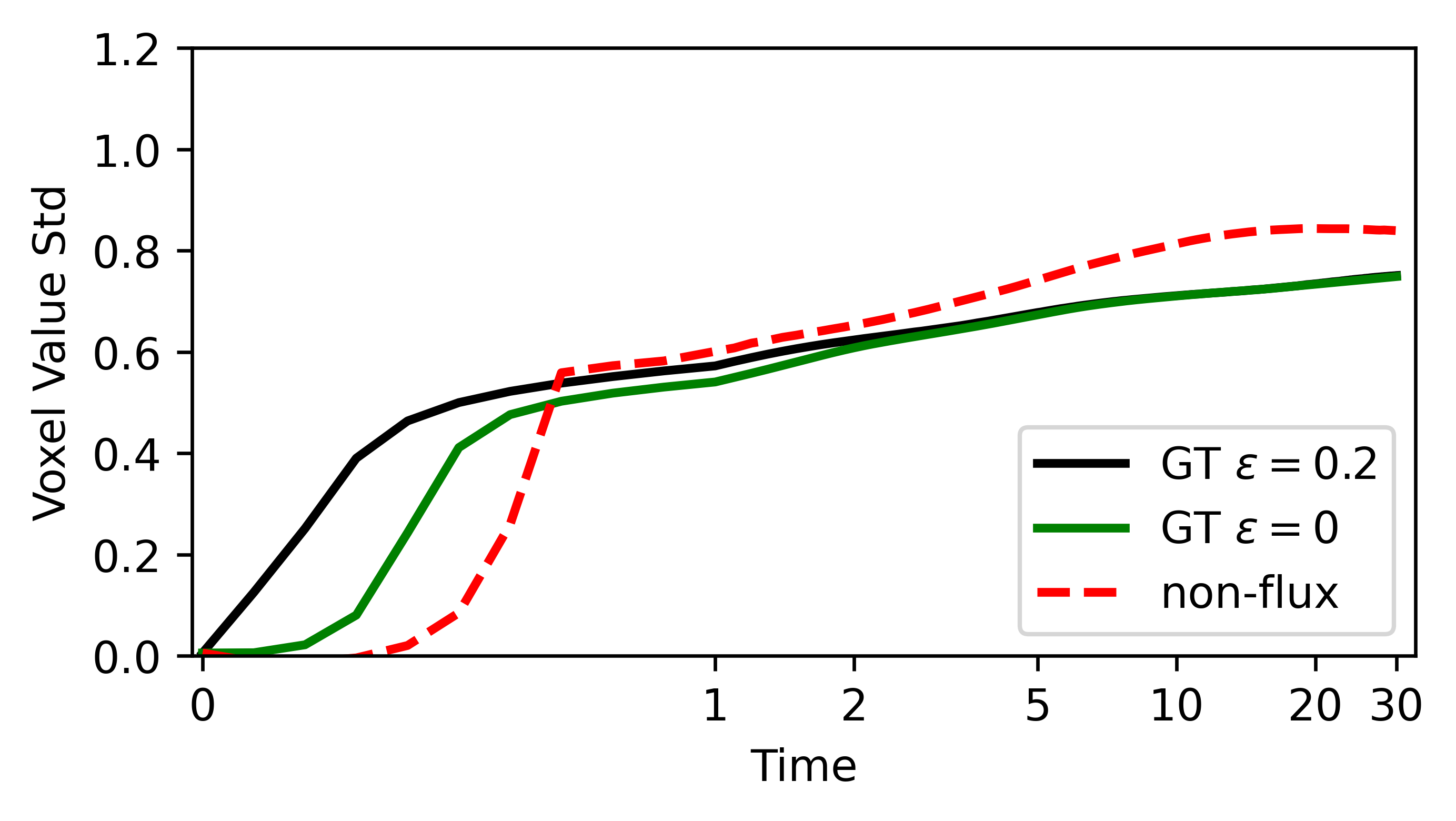}} &
        \raisebox{-0.14\height}{\includegraphics[width = 0.215\textwidth]{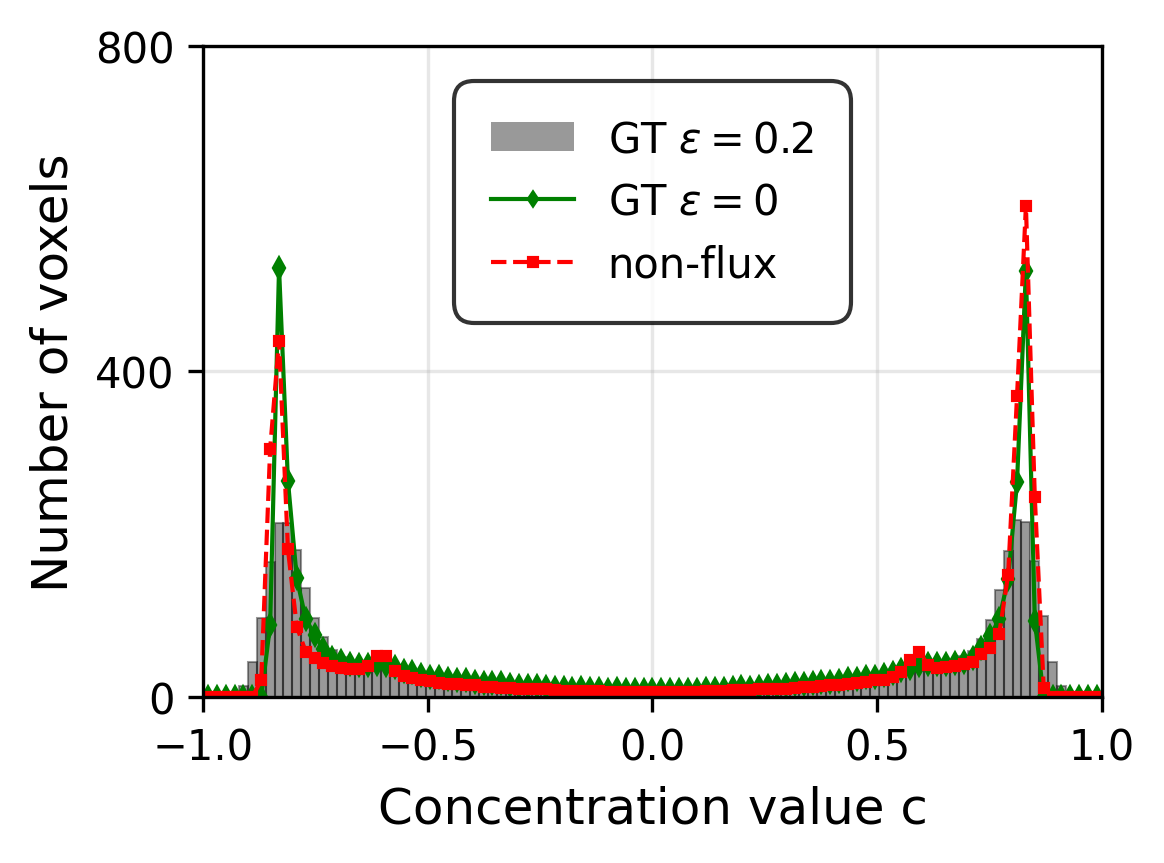}} &
        \raisebox{-0.14\height}{\includegraphics[width = 0.265\textwidth]{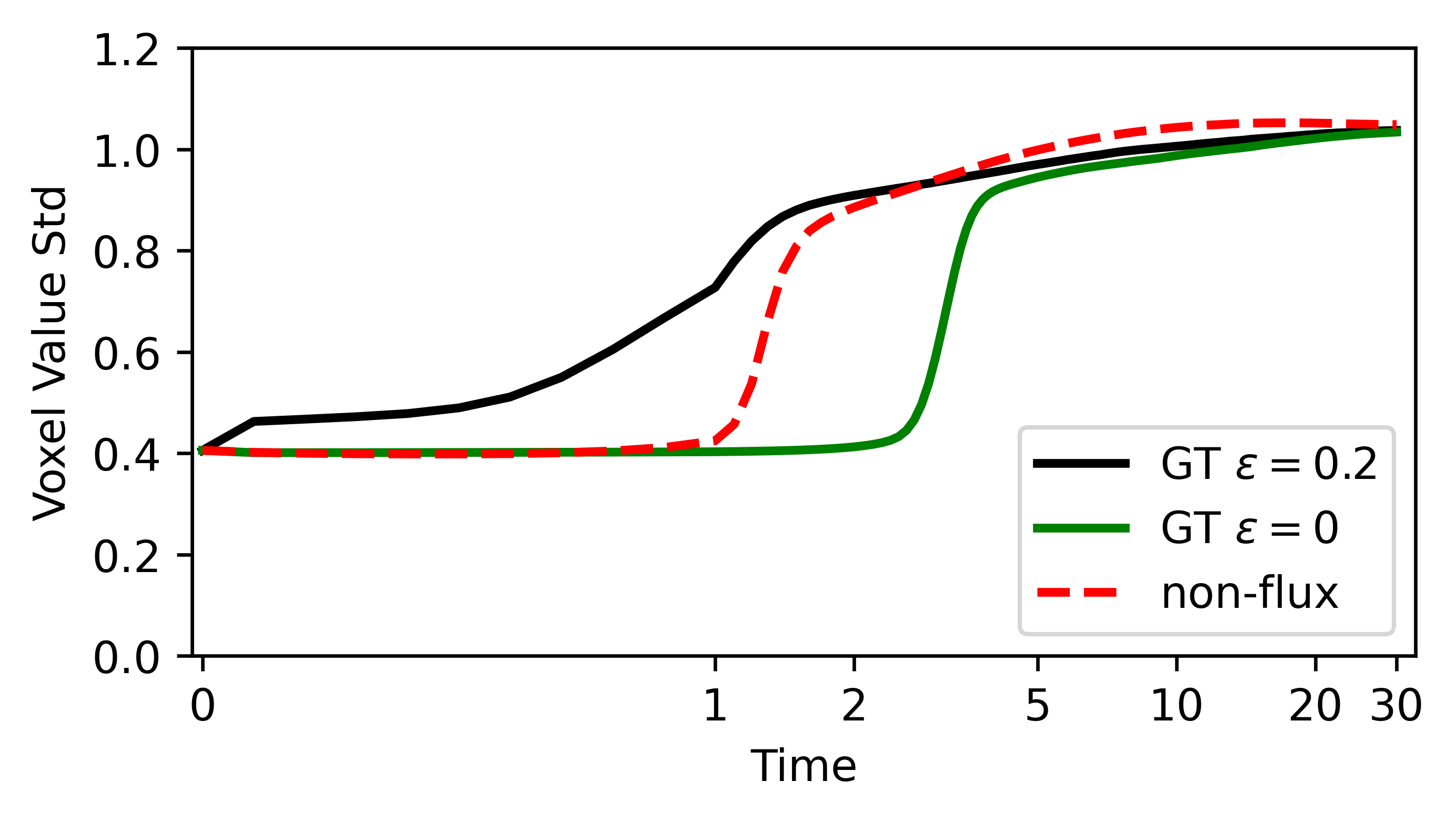}} &
        \raisebox{-0.14\height}{\includegraphics[width = 0.215\textwidth]{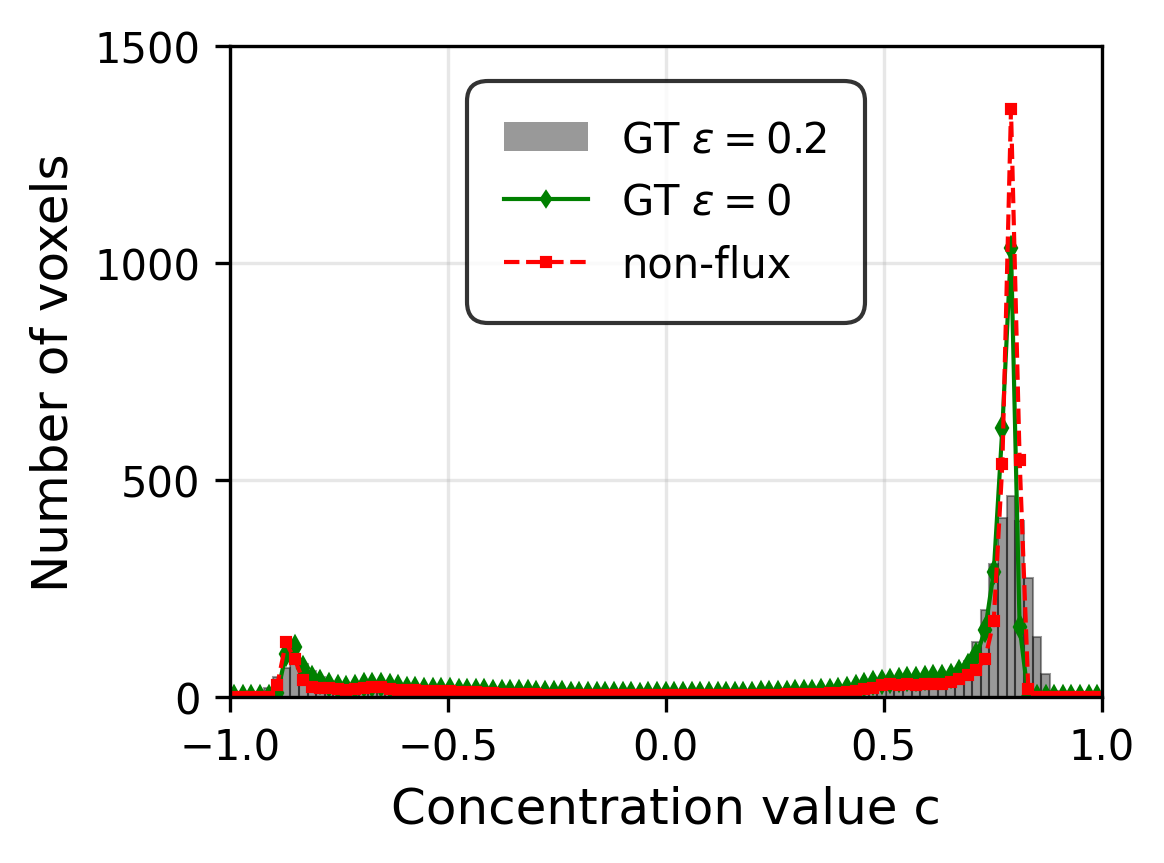}} 
    \end{tabular}
    \caption{The standard deviation of voxel value of the \nflux{} model averaged over 100 {validation} trajectories on a $16^3$ grid and the corresponding $c$ value histogram at $t=10$. {\bf Left:} \cozero{} {\bf Right:} \cofour{}.}
    \label{app_in_dist_no_flux}
\end{figure}

\subsection{Additional in-distribution result}
\label{app:indist_nvoxel_hist}
Additional results for flux-based NN models are shown in Supplementary Fig.~\ref{fig.in_dist_voxel_hist}.
\setlength{\tabcolsep}{2pt}  % Default is 6pt
\begin{figure}[htp]
    \centering
    \begin{tabular}{c c c c c}
        % & 
        \nfem & \fem & \nfemv & \femv \\ 
        % \rotatebox[origin=l]{90}{c = -0.4} &
        % \raisebox{-0.14\height}{\includegraphics[width = 0.24\textwidth]{fig/report9/f_non_flux_nonFE__16cube_sample0_T10s_hist_noise0p2_based.png}} &
        % \raisebox{-0.14\height}{\includegraphics[width = 0.24\textwidth]{fig/report9/fd312_flux_mean_nonFE__16cube_sample0_T10s_hist_noise0p2_based.png}} &
        % \raisebox{-0.14\height}{\includegraphics[width = 0.24\textwidth]{fig/report9/fd312_flux_mean_FE__16cube_sample0_T10s_hist_noise0p2_based.png}} &
        % \raisebox{-0.14\height}{\includegraphics[width = 0.24\textwidth]{fig/report9/fd312_w20_flux_mean_var_nonFE__16cube_sample0_T10s_hist_noise0p2_based.png}}&
        % \raisebox{-0.14\height}{\includegraphics[width = 0.24\textwidth]{fig/report9/fd312_w20_flux_mean_var_FE__16cube_sample0_T10s_hist_noise0p2_based.png}}
        % \\
        % \rotatebox[origin=l]{90}{\cozero} &
        % \raisebox{-0.14\height}{\includegraphics[width = 0.24\textwidth]{fig/report9/f_non_flux_nonFE__16cube_sample1_T10s_hist_noise0p2_based.png}} &
        \raisebox{-0.14\height}{\includegraphics[width = 0.24\textwidth]{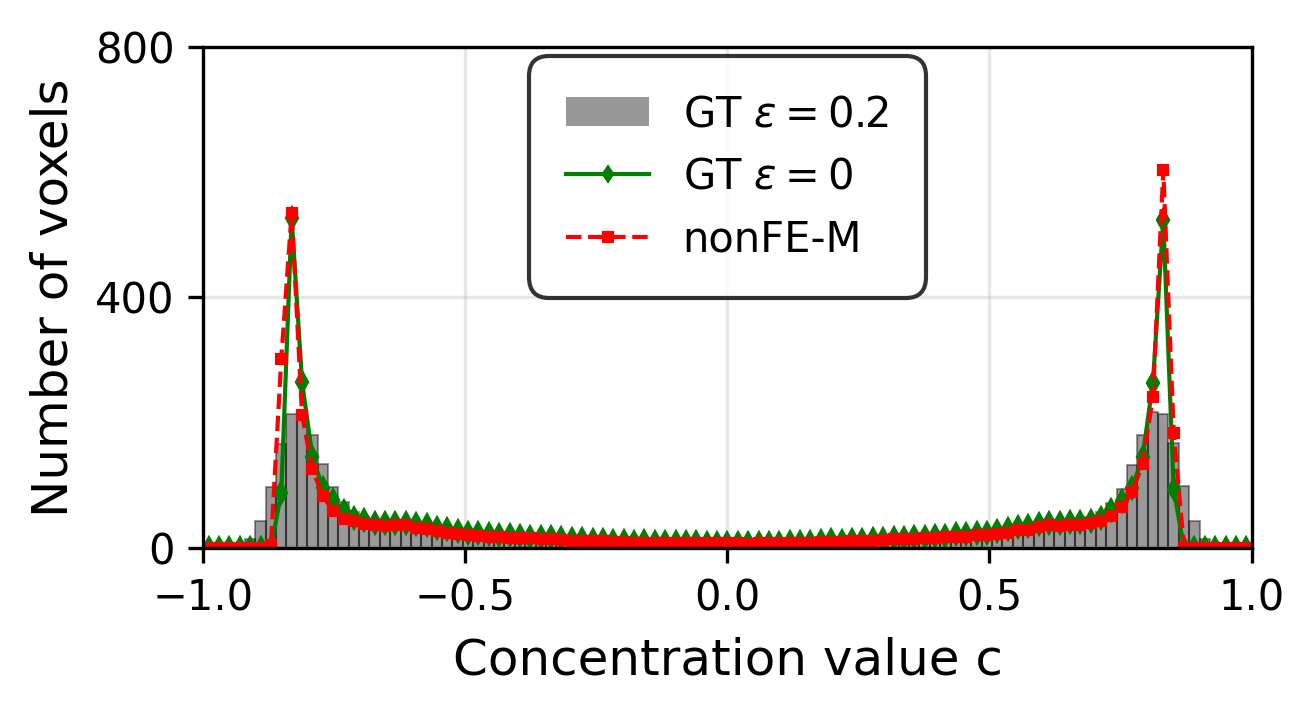}} &
        \raisebox{-0.14\height}{\includegraphics[width = 0.24\textwidth]{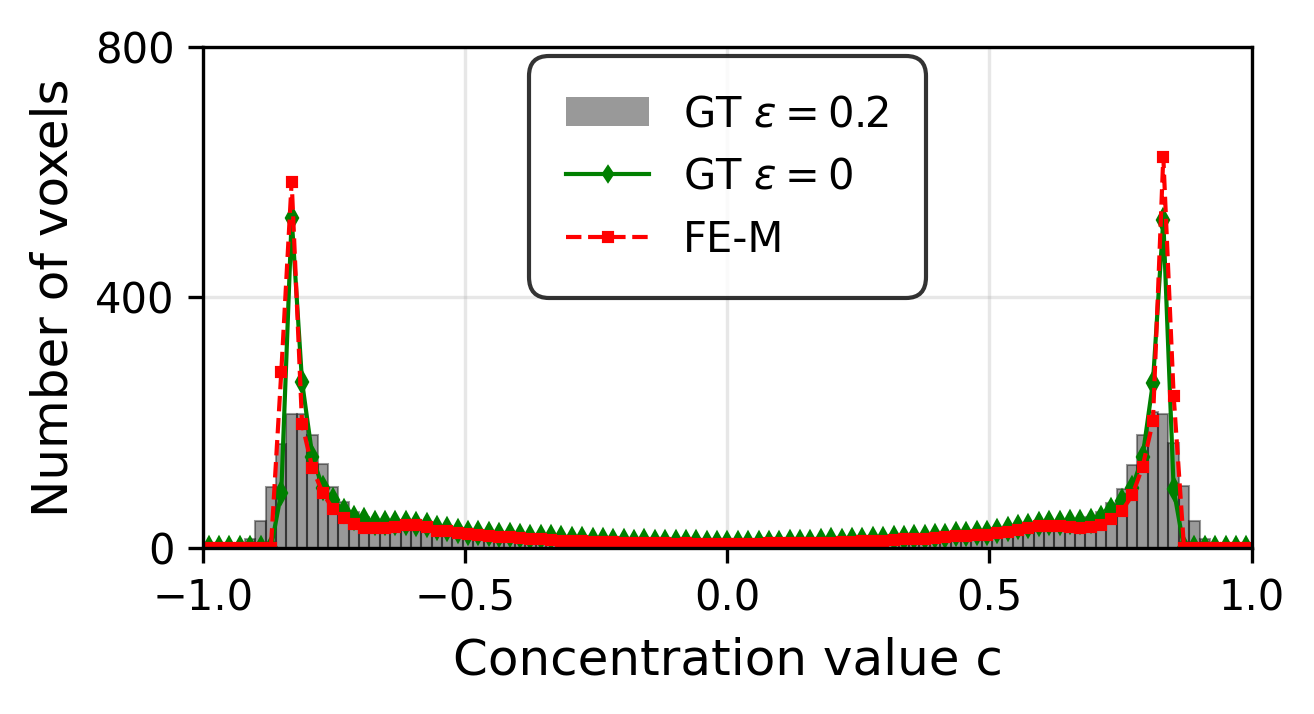}} &
        \raisebox{-0.14\height}{\includegraphics[width = 0.24\textwidth]{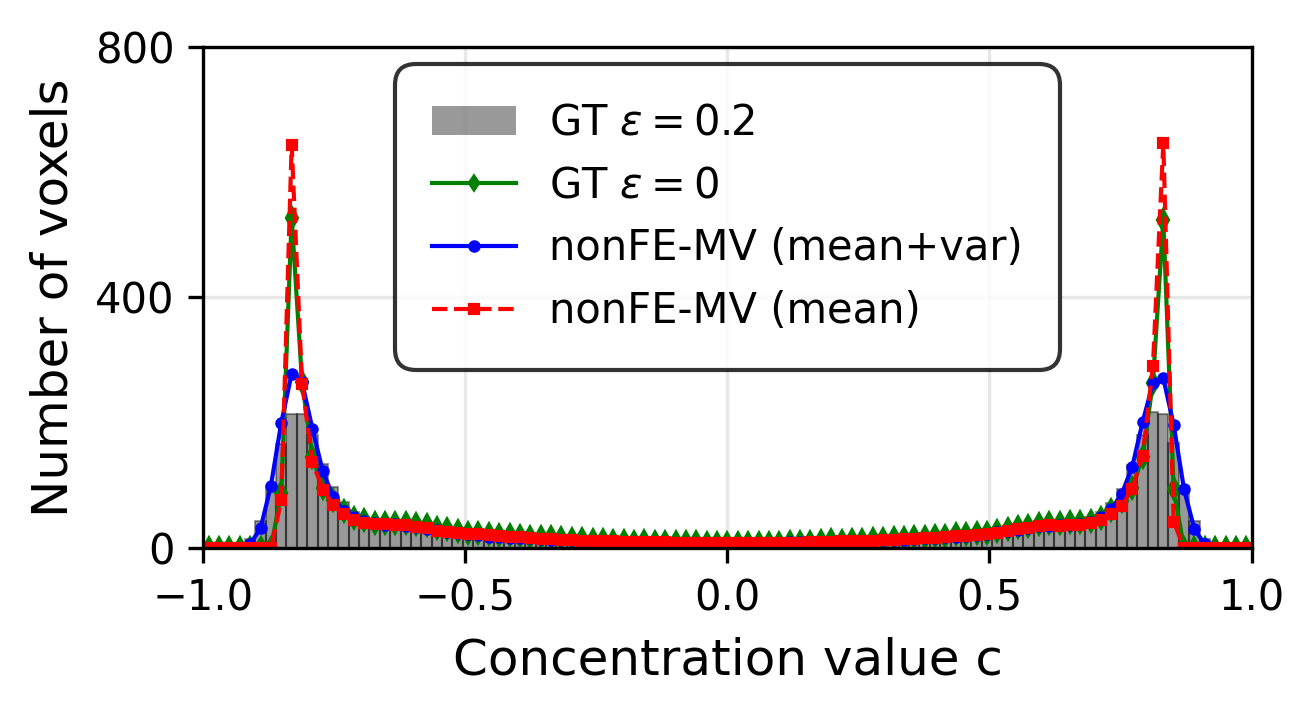}}&
        \raisebox{-0.14\height}{\includegraphics[width = 0.24\textwidth]{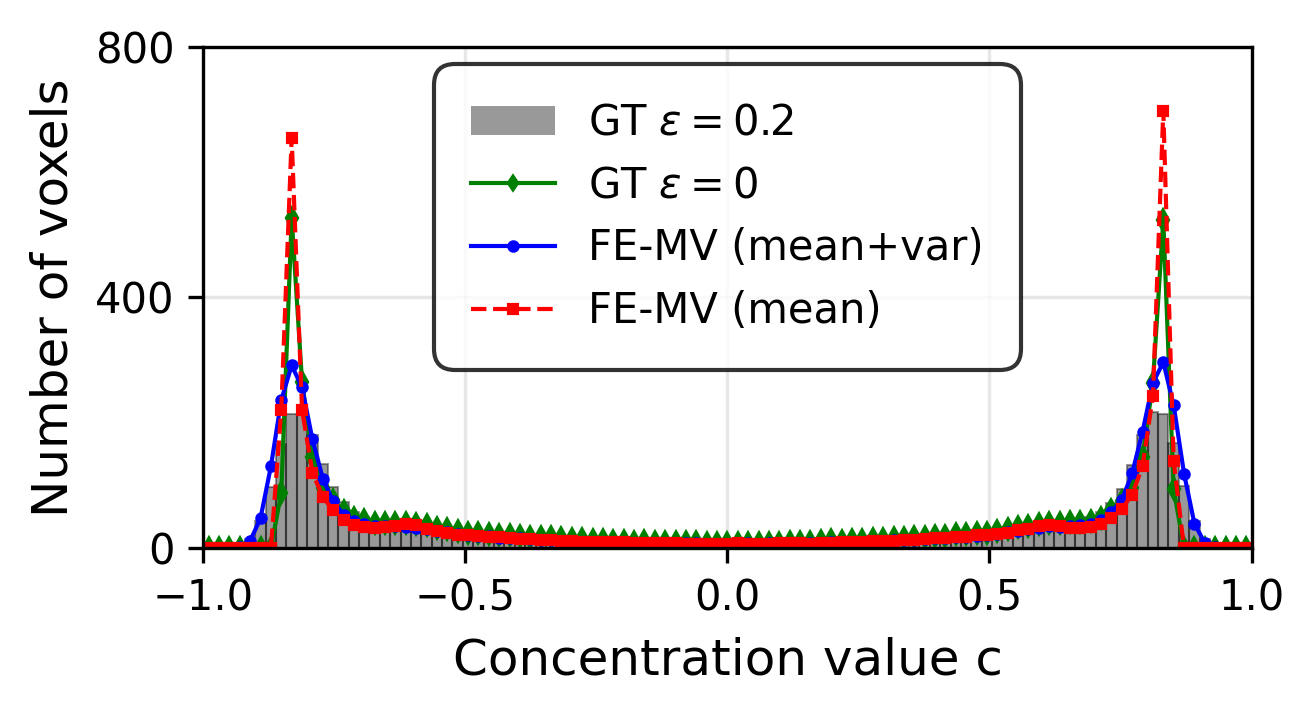}}
        \\
        % \rotatebox[origin=l]{90}{\cofour} &
        % \raisebox{-0.14\height}{\includegraphics[width = 0.24\textwidth]{fig/report9/f_non_flux_nonFE__16cube_sample2_T10s_hist_noise0p2_based.png}} &
        \raisebox{-0.14\height}{\includegraphics[width = 0.24\textwidth]{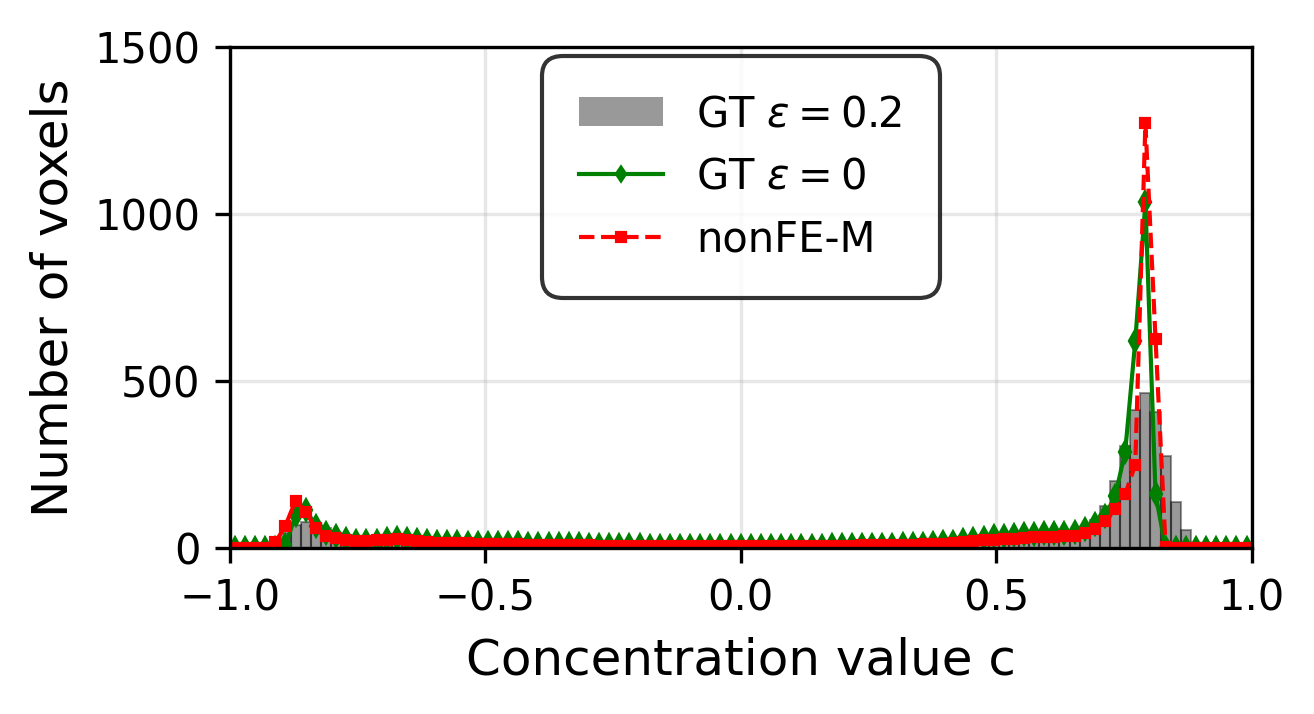}} &
        \raisebox{-0.14\height}{\includegraphics[width = 0.24\textwidth]{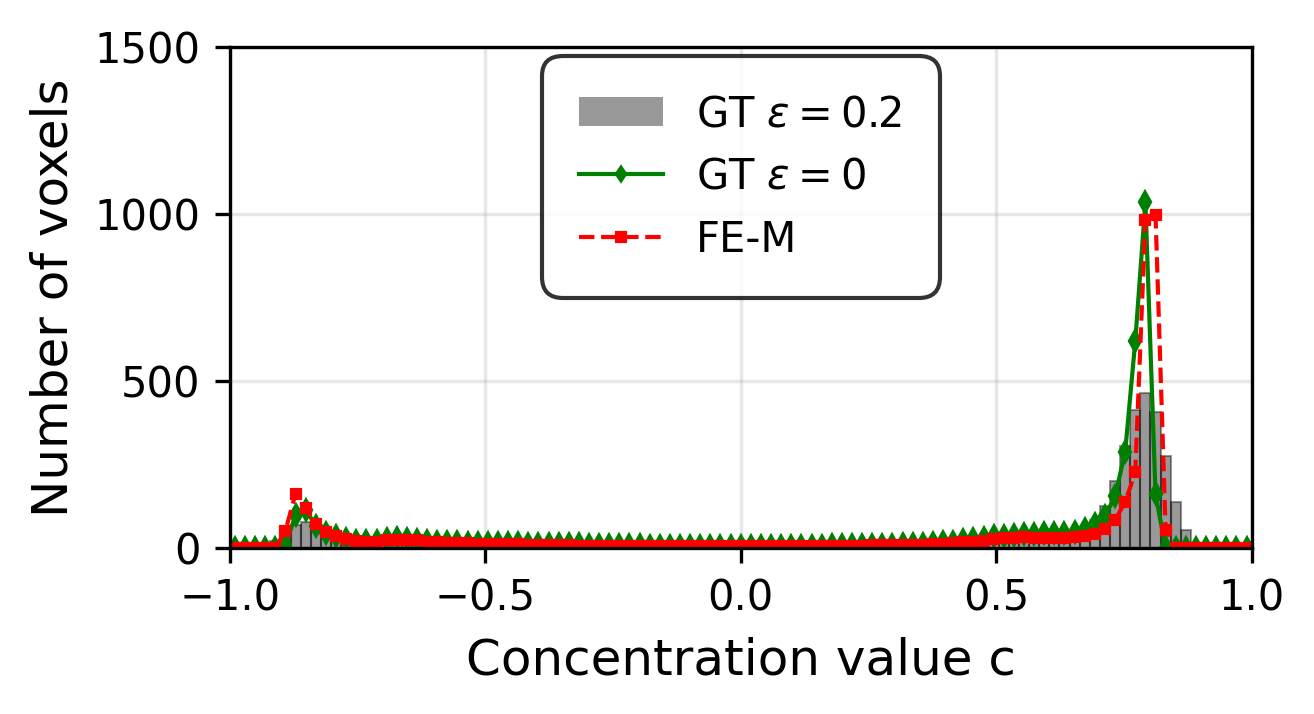}} &
        \raisebox{-0.14\height}{\includegraphics[width = 0.24\textwidth]{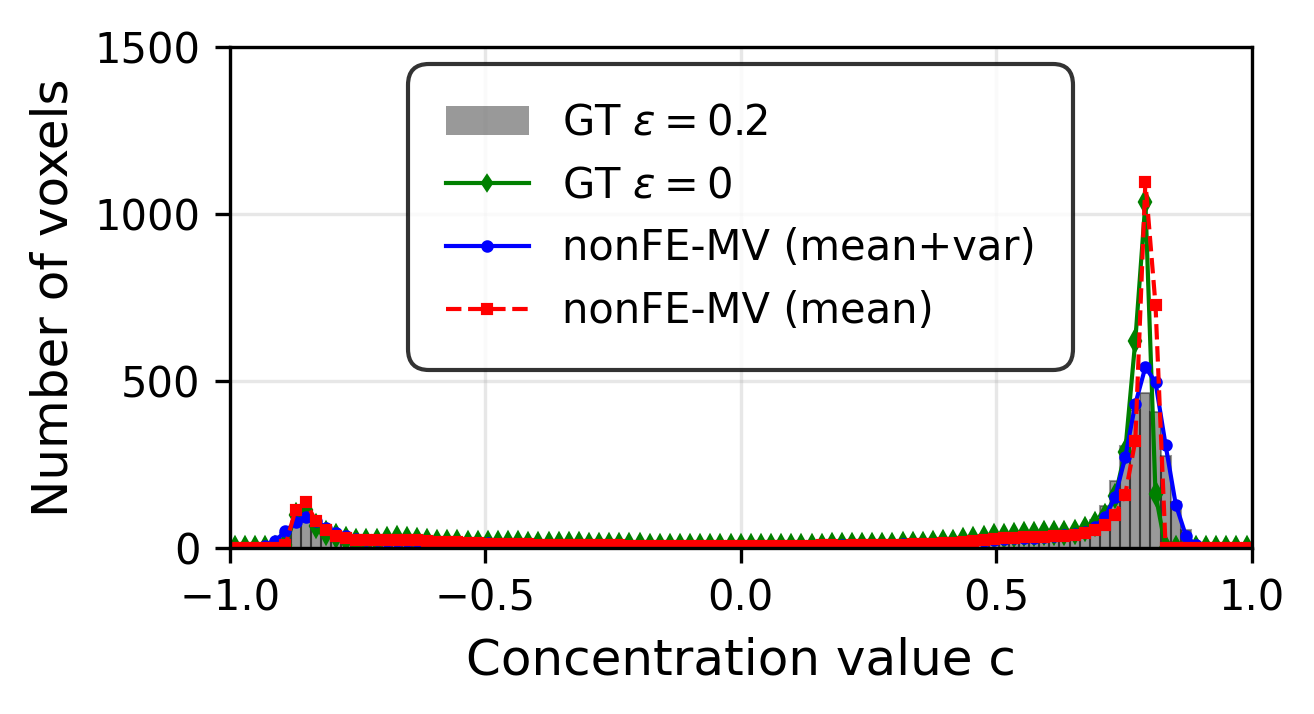}}&
        \raisebox{-0.14\height}{\includegraphics[width = 0.24\textwidth]{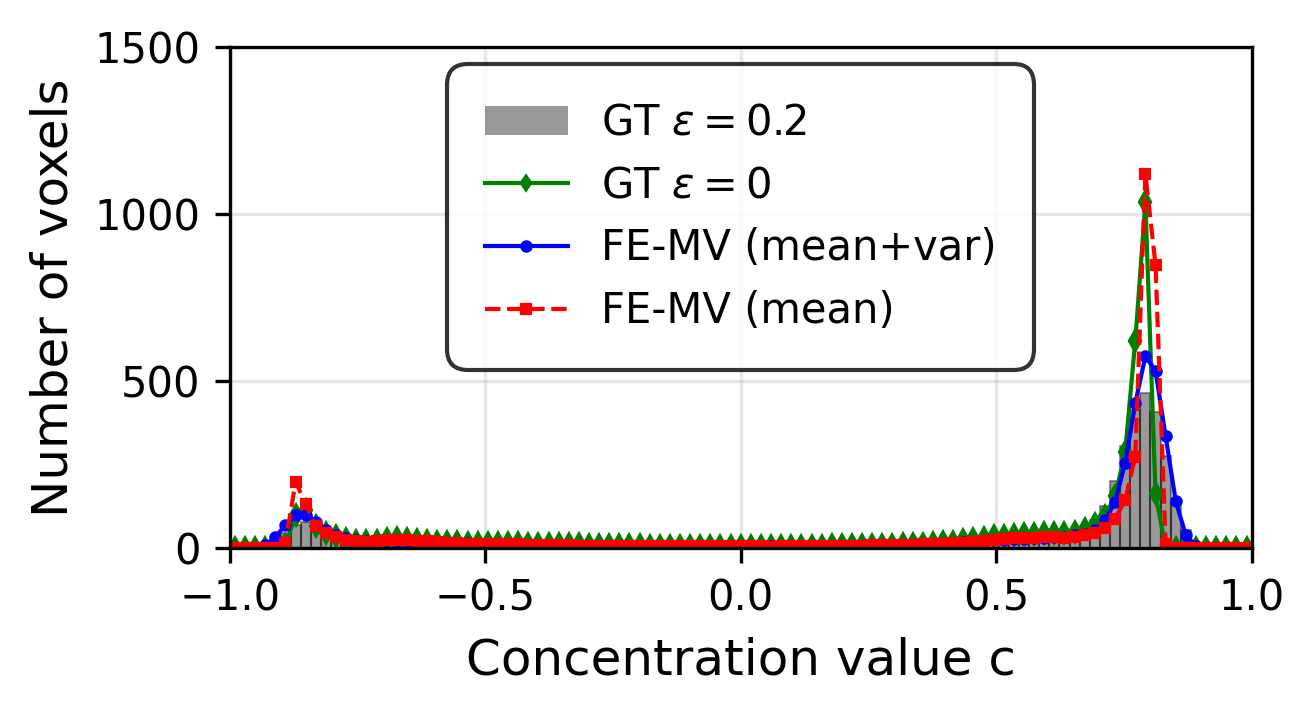}}
    \end{tabular}
    \caption{Prediction and GT (no noise and noise $\epsilon=0.2$) histograms at $t=10s$ on mesh grid $16 \times 16 \times 16$. {\bf Top row:} \cozero{} {\bf Bottom row:} \cofour}
    \label{fig.in_dist_voxel_hist}
\end{figure}

\end{document}